\documentclass[rmp,aps,reprint,notitlepage,superscriptaddress,noeprint,nolongbibliography]{revtex4-2}

\usepackage[utf8]{inputenc}
\usepackage[english]{babel}
\usepackage[dvipsnames]{xcolor}
\usepackage{tabularx, svg, graphicx, multirow}
\usepackage[normalem]{ulem}

\usepackage{amsmath,amssymb,bbm,mathrsfs,mathtools,graphicx,comment,textcomp,amsfonts,dsfont,lettrine,units,txfonts}
\definecolor{darkblue}{rgb}{0.,0.,0.5}
\usepackage[colorlinks,linkcolor=darkblue,citecolor=darkblue,urlcolor=darkblue]{hyperref}

\newcommand{\id}{\mathbbm{1}}

\newcommand{\Z}{\mathbb{Z}}
\newcommand{\C}{\mathbb{C}}
\newcommand{\R}{\mathbb{R}}

\newcommand{\e}{e}
\newcommand{\diff}{d}
\newcommand{\Diff}{D}
\newcommand{\imag}{i}
\newcommand{\ii}{i}
\newcommand{\ket}[1]{\lvert #1 \rangle}
\newcommand{\bra}[1]{\langle #1 \rvert}
\newcommand{\braket}[1]{\langle #1 \rangle}
\newcommand{\bvec}[1]{\mathbf{#1}}
\newcommand{\unitvec}[1]{\hat{\mathbf{#1}}}
\newcommand{\hatcal}[1]{\hat{\mathcal{#1}}}

\makeatletter
\newcommand*{\transpose}{%
  {\mathpalette\@transpose{}}%
}
\newcommand*{\@transpose}[2]{%
  \raisebox{\depth}{$\m@th#1\intercal$}%
}
\makeatother

\newcommand{\abs}[1]{\left\lvert #1 \right\rvert}

\newcommand{\tr}{\mathop{\mathrm{tr}}}

\renewcommand{\Im}{\mathop{\mathrm{Im}}}

\begin{document}

\title{Universality in driven open quantum matter}

\author{Lukas M. Sieberer}
\affiliation{Institute for Theoretical Physics, University of Innsbruck, 6020 Innsbruck, Austria}

\author{Michael Buchhold}
\affiliation{Institute for Theoretical Physics, University of Cologne, Z\"{u}lpicher Stra\ss{}e 77, 50937 Cologne, Germany}

\author{Jamir Marino}
\affiliation{Institute for Physics, Johannes Gutenberg University of Mainz, 55099 Mainz, Germany}

\author{Sebastian Diehl}
\affiliation{Institute for Theoretical Physics, University of Cologne, Z\"{u}lpicher Stra\ss{}e 77, 50937 Cologne, Germany}

\date{\today{}}

\begin{abstract}
  Universality is a powerful concept, which enables making qualitative and
  quantitative predictions in systems with extensively many degrees of
  freedom. It finds realizations in almost all branches of physics, including in
  the realm of nonequilibrium systems. Our focus here is on its manifestations
  within a specific class of nonequilibrium stationary states: driven open
  quantum matter. Progress in this field is fueled by a number of uprising
  platforms ranging from light-driven quantum materials over synthetic quantum
  systems like cold atomic gases to the functional devices of the noisy
  intermediate scale quantum era. These systems share in common that, on the
  microscopic scale, they obey the laws of quantum mechanics, while detailed
  balance underlying thermodynamic equilibrium is broken due to the simultaneous
  presence of Hamiltonian unitary dynamics and nonunitary drive and
  dissipation. The challenge is then to connect this microscopic physics to
  macroscopic observables, and to identify universal collective phenomena that
  uniquely witness the breaking of equilibrium conditions, thus having no
  equilibrium counterparts. In the framework of a Lindblad-Keldysh field theory,
  we discuss on the one hand the principles delimiting thermodynamic equilibrium
  from driven open stationary states, and on the other hand show how unifying
  concepts such as symmetries, the purity of states, and scaling arguments are
  implemented. We then present instances of universal behavior structured into
  three classes: new realizations of paradigmatic nonequilibrium phenomena,
  including a survey of first experimental realizations; novel instances of
  nonequilibrium universality found in these systems made of quantum
  ingredients; and genuinely quantum phenomena out of equilibrium, including in
  fermionic systems. We also discuss perspectives for future research on driven
  open quantum matter.
\end{abstract}

\maketitle

\tableofcontents

\section{Introduction}
\label{sec:introduction}

Universality underlies the possibility to perform the transition from micro- to
macrophysics quantitatively by systematically discarding irrelevant
information. Historically, the concept of universality has guided and enabled
progress in many-particle physics in thermodynamic equilibrium. But it is
operative more generally in Nature, in systems with an extensive number of
degrees of freedom. This work surveys expressions of universality in driven open
quantum matter. Systems that belong to this class are defined by the appearance
of coherent and driven-dissipative dynamics on an equal footing, placing them
far from thermodynamic equilibrium, even if they reach a stationary state.

\subsection{Universality} 

The most dramatic expression of universality occurs close to the critical point
of a second order phase transition: There, the loss of memory about the
microscopic physics is so strong, that the long-wavelength physics is fully
determined by the dimensionality, symmetries and conservation laws, as well as
the range of interactions. A system with extensively many degrees of freedom is
then characterized by few universal critical exponents defining the universality
class. Only a handful of such universality classes are found in Nature, despite
the plethora of materials and tailor-made platforms that make up the world
around us \cite{Goldenfeld1992, Zinn-Justin,
  Hohenberg1977}.

Universality, however, manifests in a variety of forms. In its weakest form,
universality may be viewed as the fact that Nature is organized
hierarchically. Focusing on universality more quantitatively, a suitable
starting point is a model of a system which applies on length and time scales
that can be considered as microscopic. The goal is then to bridge to the
macroscopic physics emerging on scales that are a few (but not many) orders of
magnitude larger. In that setting, universality often results from degeneracies,
giving rise to gapless phases of matter that can exist without fine-tuning to a
critical point across an extended parameter range. Such situations are
characterized by large-scale fluctuations dominating the macroscopic
behavior. Mechanisms enabling degeneracies are manifold. A case in point is the
spontaneous breakdown of continuous global symmetries, leading to gapless
Goldstone modes. Examples include phonons in solids or phase modes in atomic
Bose-Einstein condensates. Additionally, soft modes can occur in situations
where symmetry breaking is prohibited due to low dimensionality. Universal
effects are even enhanced due to the restricted available phase space, as seen
in the Kosterlitz-Thouless critical phase in two dimensions. Another class of
stable gapless modes appears in the form of hydrodynamics, associated with
global conservation laws. Examples include particle and heat diffusion in number
and energy-conserving systems, as well as the response of topological
insulators.

All the examples mentioned above are consistent with thermodynamic equilibrium
conditions, but their fundamental characteristics are not limited to that
context. Both criticality and weaker realizations of universal behavior have
been established in nonequilibrium statistical mechanics. A paradigmatic example
of a nonequilibrium critical point occurs in directed percolation, a variant of
the percolation problem that violates detailed
balance \cite{SchloeglDP,GRASSBERGER1978,GrassbergerTDP,Cardy_1980,Janssen1976a,Hinrichsen2000,OdorRev}. Universal
nonequilibrium phases facilitated by robust gapless modes are exemplified by the
 Kardar-Parisi-Zhang equation \cite{Kardar1986, Krug1997,
  Halpin-Healy1995, Takeuchi2018}. It was introduced to describe the roughening of
driven interfaces observed in phenomena like the spreading of fire
fronts \cite{Maunuksela1997} or the growth of bacterial
colonies \cite{Allen2019}. The phenomenon of self-organized criticality stands
in between these cornerstones: A system self-tunes to criticality, and due to
this mechanism exhibits stable universal behavior in extended parameter regimes
\cite{Bak1988,Bak1987,Sneppen2,Olami_Feder_Chris1,Drossel1992,Grinstein1990,Watkins2016,Aschwanden2016,pruessner2012self}

\subsection{Driven open quantum matter}

This article focuses on universality in driven open quantum matter, representing
a novel class of nonequilibrium many-body systems. The distinguishing feature of
systems in this class is the simultaneous presence of coherent quantum dynamics,
external drive, and dissipation. This circumstance is typically realized when
matter is strongly coupled to external light fields, like lasers. The ability to
drive thermodynamically large systems with such fields without compromising
their intricate many-body behavior due to noise and heating has emerged
relatively recently. Yet it already includes a broad spectrum of platforms for
atoms, light, and solids. Among these are exciton-polariton systems in
semiconductor heterostructures \cite{Carusotto2013}, atoms \cite{Mivehvar_2021}
or solids \cite{Huebener2021} immersed into cavities, photonic platforms and
microcavity arrays \cite{Hartmann_2008,Noh_2016,Noh_2017,Ozawa_2019}, circuit
quantum electrodynamics \cite{Vool_2017,Blais_2021}, ultracold atoms and ions
\cite{muller2012aamop,Harrington_2022}, Rydberg gases
\cite{Browaeys_2020,SaffmanRev,MorschDisRydRev}, and light-induced
superconductors \cite{Mitrano2016,Cavalleri2018}. Driven open quantum matter
also encompasses functionalized matter, such as noisy intermediate-scale quantum
\cite{Preskill2018} devices of superconducting circuits \cite{Satzinger_2021}
and Rydberg tweezer arrays \cite{Semeghini_2021}. The need to conceptualize
these systems as instances of driven open quantum matter is clear; this need
also starts to get recognized more widely for systems such as cold atoms in
optical lattices, specifically when long time scales are being considered
\cite{Pichler2010}.

All of the systems in the class of driven open quantum matter have two common
characteristics:

\paragraph*{Nonequilibrium conditions.}

The combination of external driving and the coupling to dissipative reservoirs,
inducing an open system character, pushes driven open quantum matter out of
equilibrium. Without driving, open systems may still reach thermodynamic
equilibrium with their surroundings, and thus obey the principle of detailed
balance. The study of such equilibrium open systems has been pioneered by
Caldeira and Leggett
\cite{CaldeiraLeggett1983a,CaldeiraLeggett1983b,WeissDissipative}, building on
the influence functional techniques introduced by Feynman and Vernon
\cite{FeynmanVernon1963}. However, adding a drive generically induces stationary
fluxes of energy, particle number, entropy etc., and thus leads to a violation
of detailed balance between system and environment. The presence of
nonequilibrium conditions even in the stationary state distinguishes driven open
many-body systems from closed and undriven ones. Even though the latter can
exhibit nonequilibrium behavior in their time evolution, they generically reach
a stationary state of thermodynamic equilibrium.

\paragraph*{Quantum dynamics.}

On the microscopic scale, instances of driven
open quantum matter need to be described as quantum systems---quantum mechanical
effects such as phase coherence and entanglement cannot be discarded. However,
given the fragility of quantum mechanical correlations, this does not
imply that quantum mechanical effects persist up to macroscopic
scales. In many instances, an effective (semi-)classical description of the
macroscopic behavior is possible---in parallel to finite temperature quantum
systems. Still, drive and dissipation need not act destructively on quantum
many-body correlations, and can even be harnessed to induce
them \cite{Diehl2008,Verstraete2009}. Identifying and describing universal
effects of this type is both a unique challenge and an opportunity offered by
driven open quantum systems. This holds the potential to spark a new field of
nonequilibrium \textit{quantum} statistical mechanics.

\subsection{Synopsis}

\begin{table*}
\begin{ruledtabular}
\begin{tabular}{ccc}
  Realizations of paradigmatic
  & \multirow{2}{*}{Novel nonequilibrium universality}
  & \multirow{2}{*}{Quantum nonequilibrium phenomena}
  \\  nonequilibrium universality & & 
  \\[.1cm] \hline \\[-.2cm]
  \hyperref[Sec:DP]{V.~Absorbing state phase transitions}
  & \multirow{2}{*}{\hyperref[sec:driv-open-crit]{VIII.~Driven open criticality}}
  & \multirow{2}{*}{\hyperref[sec:quantum-criticality]{XI.~Quantum criticality in driven open systems}} \\
  \hyperref[Sec:DP]{and directed percolation} & & \\[.1cm]
  \hyperref[sec:SOC]{VI.~Self-organized criticality and}
  & \multirow{2}{*}{\hyperref[sec:slowly-rapidly-driven-open-systems]{IX.~Slowly and rapidly driven open systems}}
  & \multirow{2}{*}{\hyperref[sec:imp]{XII.~Universality in dissipative quantum impurities}} \\
  \hyperref[sec:SOC]{Rydberg experiments} & & \\[.1cm]
  & \multirow{2}{*}{\hyperref[sec:firstorder]{X.~Nonequilibrium first-order phase transitions}}
  & \multirow{2}{*}{\hyperref[sec:fermions]{XIII.~Universality in fermion systems}} \\ & & \\[.1cm]
  \multicolumn{2}{c}{\hyperref[sec:driven-open-BEC-2D-1D]{VII.~Driven open condensates in low spatial dimensions}}
\end{tabular}
\end{ruledtabular}
\caption{Manifestations of universality in driven open quantum
  matter are grouped into three main directions in this review.}
  \label{fig:overview}
\end{table*}

This review both provides the conceptual framework to characterize universality
in driven open quantum matter and highlights key instances of universality in
various physical platforms. It is organized as follows.

\paragraph*{Principles of universality in driven open quantum matter
  (Secs.~\ref{sec:Lindblad-to-Keldysh}--\ref{sec:organ-princ-beyond}).}

Microscopically, the systems described above can be modelled in terms of a
Markovian quantum master equation in Lindblad form. Yet, this is not the ideal
language to distill universality. We thus first introduce an overarching
framework that enables discarding irrelevant information while keeping the
relevant one
(Secs.~\ref{sec:Lindblad-to-Keldysh}--\ref{sec:organ-princ-beyond}). This is
achieved by reformulating the quantum master equation in terms of an equivalent
but more flexible Lindblad-Keldysh field theory
(Sec.~\ref{sec:Lindblad-to-Keldysh}). In particular, this allows us to identify
three principles to preserve the relevant information, and to guide the analysis
of the driven open quantum many-body problem:
\begin{itemize}
\item[(1)] Equilibrium vs.~nonequilibrium stationary states
  (Sec.~\ref{sec:equilibrium-vs-nonequi-steady-states}): Detailed balance
  characteristic of a system in thermodynamic equilibrium is indicated by the
  presence of a discrete symmetry of the Keldysh action. This equips us with a
  criterion to distinguish equilibrium from nonequilibrium conditions in
  practice. By comparing the magnitudes of coupling constants which are
  incompatible with that symmetry to those that are allowed, one can assess
  quantitatively how far a system is away from thermodynamic equilibrium.
\item[(2)] Symmetries and conservation laws (Sec.~\ref{sec:organ-princ-beyond}):
  Symmetries of the Keldysh action come with a fine structure of ``classical''
  and ``quantum'' (or ``weak'' and ``strong'') symmetries, while in equilibrium
  field theories these types of symmetries coincide. Both forms of symmetries,
  if continuous, are tied to the existence of gapless modes, and are thus key to
  the identification of universal phenomena: The spontaneous breakdown of weak
  continuous symmetries is accompanied by the formation of gapless Goldstone
  modes; strong symmetries imply conservation laws, and are related to the
  existence of slow hydrodynamic modes.
\item[(3)] Mixed and pure states, classical and quantum scaling
  (Sec.~\ref{sec:organ-princ-beyond}): Intuitively, one might expect that
  Markovian noise acts similarly to a finite temperature. When the stationary
  state is mixed, this is generally true, and allows us to simplify the Keldysh
  action by taking the semiclassical limit. Critical problems are then described
  by scaling forms of the Keldysh action analogous to finite-temperature
  classical phase transitions. However, under specific conditions, also pure
  stationary states can emerge, similar to the fine-tuning of temperature to
  zero in systems in thermodynamic equilibrium. Critical problems of this class
  exhibit scaling solutions which parallel quantum phase transitions.
\end{itemize}
We then focus on universal phenomena, grouped into three main directions as
illustrated in Tab.~\ref{fig:overview}:

\paragraph*{Realizations of paradigmatic nonequilibrium universality
  (Secs.~\ref{Sec:DP}--\ref{sec:driven-open-BEC-2D-1D}).}

Driven open quantum systems enable the realization of paradigmatic
nonequilibrium scenarios, which hitherto were difficult to implement.  Here we
discuss the above mentioned directed percolation (Sec.~\ref{Sec:DP}) and
self-organized criticality (Sec.~\ref{sec:SOC}), including their observation in
driven Rydberg gases. We also describe the emergence of Kardar-Parisi-Zhang
(KPZ) universality and its recent realization in polariton systems
(Sec.~\ref{sec:driven-open-BEC-2D-1D}). The challenge to theory is to extract
these instances of universality from concrete microscopic platforms via
systematic coarse graining all the way to the macroscale. The recent
experimental breakthroughs highlight driven open quantum platforms as
controllable laboratories for nonequilibrium statistical mechanics.

\paragraph*{Novel nonequilibrium universality
  (Secs.~\ref{sec:driven-open-BEC-2D-1D}--\ref{sec:firstorder}).}

We next discuss universal phenomena which are unique to driven open quantum
systems and have not previously surfaced in nonequilibrium statistical
mechanics. Among these, by analyzing the impact of nonequilibrium conditions on
topological vortex defects, we assess the fate of one of the cornerstones of
low-dimensional statistical mechanics, the Kosterlitz-Thouless transition,
including in the case of strong spatial anisotropy. We furthermore show that
novel kinds of topological phase transitions exist out of equilibrium, such as
vortex turbulence---those transitions occur upon increasing the strength of
nonequilibrium conditions (Sec.~\ref{sec:driven-open-BEC-2D-1D}). Then, we
discuss novel universality in the phase transitions in driven open quantum
matter (Sec.~\ref{sec:driv-open-crit}). A first instance is the identification
of a new independent critical exponent for the Bose condensation transition,
which distinguishes equilibrium from driven open criticality. A novel
nonequilibrium fixed point in nonreciprocally coupled Ising models is also
discussed.  We then turn to the critical behavior in slowly and rapidly driven
open quantum systems (Sec.~\ref{sec:slowly-rapidly-driven-open-systems}). In the
slowly driven limit, the paradigmatic Kibble-Zurek scenario is revealed as the
tip of the iceberg of a more general phenomenology, accessible in a
renormalization group approach, which gives experimental access to the full
spectrum of critical exponents by suitable driving protocols. The opposite limit
of fast, yet not infinitely fast, driving is realized in open Floquet systems,
and is shown to exhibit a similar phenomenology as its slowly driven
counterpart, although with a different physical origin.  Next, we discuss
universal aspects of first-order transitions far from equilibrium
(Sec.~\ref{sec:firstorder}). Here nonthermal noise modifies the formation of
droplets---the universal mechanism behind first-order phase transitions. The
most drastic modification occurs at first-order dark state phase transitions:
The system realizes a bistability between a mixed phase, displaying statistical
fluctuations similar to systems at nonzero temperature, and a dark state phase,
represented by a single, pure quantum state.  All these phenomena are predicted
for concrete platforms such as polariton systems, driven optical lattices for
ultracold atoms, or Rydberg tweezer arrays, but yet await observation. We hope
that our exposition may foster an agenda for future experiments.

\paragraph*{Nonequilibrium quantum phenomena
  (Secs.~\ref{sec:quantum-criticality}--\ref{sec:fermions}).}

This final part reviews progress on one of the unique hallmarks of driven open
quantum systems: the potential for showing collective quantum behavior at the
largest distances. One example of the strong universality encountered near phase
transitions concerns an analog of quantum criticality at equilibrium---Markovian
quantum criticality (Sec.~\ref{sec:quantum-criticality}). Another one focuses on
a class of problems where nonequilibrium perturbations occur via impurities, in
systems otherwise kept in their quantum ground states. The interplay of gapless
many-body modes with the nonequilibrium impurity gives rise to universal
phenomena like the fluctuation induced quantum Zeno effect
(Sec.~\ref{sec:imp}). We also present results on driven open systems composed of
fermions, which do not possess a simple classical---or deterministic
(Sec.~\ref{sec:deterministic-limit})---limit due to Pauli's principle
(Sec.~\ref{sec:fermions}). Here we highlight how fermion systems can be cooled
into pure topological states, and discuss their topological phase
transitions. We also demonstrate that topology provides a strong principle of
universality: the topological response of fermion systems is identical,
irrespective to their dynamics proceeding in- or out of equilibrium. We also
point out a dynamical fine structure in the symmetry classification of
interacting fermion matter, which distinguishes equilibrium from nonequilibrium
conditions. Results on quantum nonequilibrium phenomena are just fledging here;
further research will shape the field of nonequilibrium quantum statistical
mechanics. \\

\begin{figure*}
  \includegraphics[width=\textwidth]{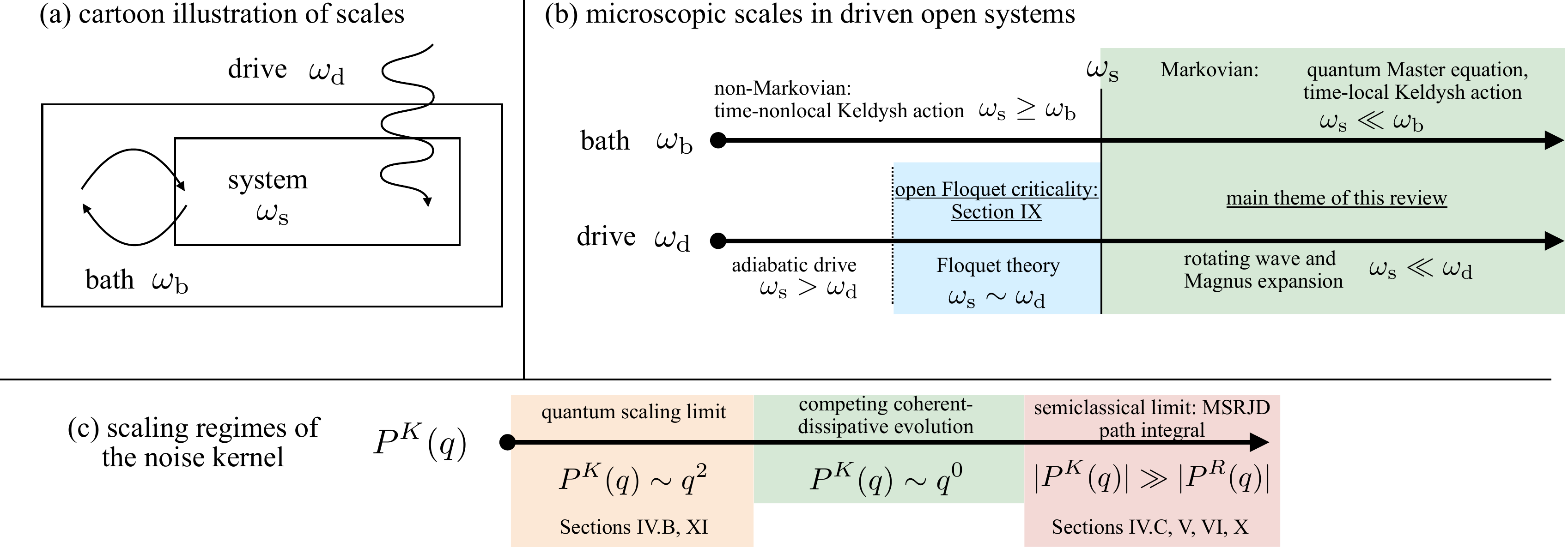}
  \caption{Illustration of typical environmental and drive scales and the
    corresponding regimes of driven-dissipative many-body systems. (a)~This review focusses on universal phenomena emerging in
    rapidly driven systems with Markovian dissipation. An excursion on
    universality in slowly driven systems, utilizing the Kibble-Zurek argument
    and the Floquet formalism, is presented in
    Sec.~\ref{sec:slowly-rapidly-driven-open-systems}. Extensive reviews of
    (open) Floquet many-body systems are provided in
    Refs.~\cite{Bukov2015,Eckardt2017,MoriReview2023}. For Non-Markovian systems
    we refer to the reviews in
    Refs.~\cite{deVegaReview,BreuerNonMarkovian}. (b)~Illustration of the
    emergence of the typical system and bath scales. (c) Scaling
    regimes of the environment-induced noise kernel with the quantum scaling
    limit and semiclassical Martin-Siggia-Rose-Janssen-de Dominicis (MSRJD)
    limit illustrated as particular cases.}
  \label{fig:scales}
\end{figure*}

The manifestations of universality surveyed in this review occur in the generic
setup shown in Fig.~\ref{fig:scales}(a): It consists of a system, coupled to a
bath and driven by a coherent classical field. Typical time scales of the
dynamics of the system, the bath and the drive are denoted by
$\omega_{\mathrm{s}}, \omega_{\mathrm{b}}$ and $\omega_{\mathrm{d}}$,
respectively. In this review, we focus on Markovian quantum dynamics, which
occur in the parameter regime
$\omega_{\mathrm{b}} \approx \omega_{\mathrm{d}}\gg \omega_{\mathrm{s}}$, see
Fig.~\ref{fig:scales}(b), and are described by a quantum master equation in
Lindblad form \cite{Lindblad1976, Gorini1976, breuer2002oxford,
  gardiner2004springer}, or, equivalently, by a time-local Keldysh action
\cite{Schwinger1960, Schwinger1961, Mahanthappa1962, Bakshi1963, Bakshi1963a,
  Keldysh1965, Kamenev2023, Altland2010a, Sieberer2016a, Thompson2023}. We
emphasize that $\omega_{\mathrm{s}}$ can be a placeholder for an entire
hierarchy of scales of the driven open quantum many-body system; their
sequence determines the physics discussed throughout this review. 

A first important delineation occurs between Markovian and non-Markovian
regimes, and results from comparing $\omega_{\mathrm{s}}$ to
$\omega_{\mathrm{b}}$, see Fig.~\ref{fig:scales}(b). Paradigmatic examples of
non-Markovian quantum dynamics \cite{deVegaReview,BreuerNonMarkovian}, such as
the Caldeira-Leggett model \cite{CaldeiraLeggett1983a, CaldeiraLeggett1983b,
  WeissDissipative} and the spin-boson model \cite{Leggett1987}, are concerned
with systems that are coupled to baths in thermodynamic equilibrium. Integrating
out such a bath leads to a time-nonlocal Keldysh action. However, any finite
temperature $T$ gives rise to a decoherence time scale $\sim 1/T$; at even
larger time scales, the system appears Markovian again. This is the origin of
the ``quantum critical fan'' appearing at finite temperatures above a quantum
critical point: At any nonzero temperature, classical (finite temperature)
asymptotic scaling behavior emerges.

In contrast, a bath at zero temperature results in a scale-free Keldysh action,
describing non-Markovian dynamics on all scales. This leads to qualitative
modifications, e.g., of critical behavior at phase transitions (see
Sec.~\ref{sec:organ-princ-beyond}). A comprehensive analysis of modifications of
\emph{nonequilibrium} universality due to such non-Markovian quantum dynamics is
still outstanding. However, also Markovian quantum dynamics can host an analog
of zero temperature states, in the form of dark states (see
Secs.~\ref{sec:quantum-criticality}, \ref{sec:imp}, and \ref{sec:fermions})---in
both cases, the system is stabilized in a pure quantum state.

We further focus generally on rapidly driven systems $\omega_{\text{s}}\ll \omega_{\text{d}}$, where the drive frequency does not interfere with the typical system time scales, and where treating the driving in a rotating wave approximation or in first-order Magnus expansion is well justified. This regime is well motivated physically: universality manifests itself on the largest distances and on the largest time scales. Thus the relevant long-wavelength modes act on effective time-scales much slower than the drive. In Sec.~\ref{sec:slowly-rapidly-driven-open-systems}, we discuss how the dynamical regime with  $\omega_{\text{s}}\sim \omega_{\text{d}}$ is connected to the two limiting cases $\omega_{\text{s}}\ll \omega_{\text{d}}$ and $ \omega_{\text{d}}=0$ with respect to universality, i.e., via the paradigmatic Kibble-Zurek mechanism and the activation of subleading scaling corrections.

With its focus on universal phenomena in driven open quantum matter, this review
complements related works: Several surveys focus on physical realizations of
driven open quantum systems, which can be structured into platforms made of
photons \cite{Hartmann_2008,Noh_2016,Noh_2017,Ozawa_2019},
atoms \cite{muller2012aamop,Mivehvar_2021,Browaeys_2020}, and
solids \cite{Carusotto2013,Vool_2017,Blais_2021,Huebener2021}. On the
methodological side, recent progress regarding numerical techniques and
simulation methods for open quantum many-body systems has been reviewed in
Refs.~\cite{Daley2014, Weimer2021RMP}, while the field theory approach has been
surveyed in Refs.~\cite{Sieberer2016a, Thompson2023}.

There are also several instances of universal nonequilibrium behavior in quantum
many-body systems which are not touched upon here. A first important delineation
concerns open vs.~closed nonequilibrium systems. Clearly, nonthermal stationary
or quasi-stationary states can also exist in closed quantum systems. One
paradigmatic example is the absence of thermalization in disordered interacting
systems, known as many-body
localization \cite{Abanin2019,MBLReviewNandkishore,MBLReviewAlet,MBLReviewAbanin}. Further
examples in this class are provided by certain gauge theories \cite{Brenes2018}
as well as Hilbert space fragmentation \cite{Moudgalya2022,Sala2020} and scar
states \cite{Turner_2018,Schecter2019} which, however, can also occur in open
systems \cite{Buca_2019, Li2023, Wang2024}. Moreover, various instances of
closed systems governed by unitary dynamics without energy conservation have
been studied. This concerns, for example, Floquet systems with periodically time
dependent Hamiltonians, which can be trapped in long-lived states before
generically heating up at long times \cite{Abanin2016, Mori2016, Kuwahara2016,
  Abanin2017, Eckardt2017, Moessner2017, Harper2020,
  Seetharam2019,zhu2019dicke}. Such systems can show collective phenomena, such
as the formation of time crystals \cite{khemani2019brief,yaormp2023}. Also the
study of unitary dynamics in random quantum circuits has attracted
attention. Circuit dynamics serve as models exhibiting generic aspects of
thermalization, entanglement dynamics, as well as information spreading and
scrambling \cite{Nahum2017,Skinner2019,Li2019b,Fisher2023,Haferkamp2022,Keyserlingk2018,Khemani2018}.

We also do not touch upon the time evolution of many-body systems---our notion
of ``nonequilibrium'' refers to the nonthermal nature of the stationary states
of driven open quantum systems. Of course, the temporal evolution of closed or
open systems following, e.g., a quench of system parameters, realizes a
different kind of nonequilibrium condition. Quenches, or more generally
nonthermal initial states, have been studied intensely for closed systems~(see
\cite{Polkovnikov2011, Eisert2015, Calabrese2016}, and some of the instances in
the preceding paragraph). Beyond the thermalization expected at asymptotically
long times \cite{DAlessio2016}, interesting universal aspects of the transient
temporal regimes such as turbulence \cite{Frisch1995}, prethermalized
plateaus \cite{Berges_2004}, or nonthermal fixed
points \cite{Berges2008,Erne_2018,Pruefer_2018,marino2022dynamical} have been
revealed. The dynamics of driven open quantum systems have so far received far
less attention, and we will come back to this point in the Perspectives in
Sec.~\ref{sec:outlook}.

\section{Description of driven open systems: from Lindblad to Keldysh}
\label{sec:Lindblad-to-Keldysh}

We begin by introducing the theoretical framework to describe driven open
quantum matter. Our microscopic starting point is a quantum master equation in
Lindblad form. We discuss different methods to access universal properties,
including Keldysh field theory, which is the focus of this review. Finally, as
an important example and point of reference, we introduce a model of driven open
Bose-Einstein condensation.


\subsection{Microscopic scales and Lindblad description}

The state of a driven open quantum system is specified by its density matrix
$\hat{\rho}$, and the dynamics are described by a quantum master equation in
Lindblad form \cite{Lindblad1976, Gorini1976, breuer2002oxford,
  gardiner2004springer},
\begin{equation}
  \label{eq:meq}
  \partial_t \hat{\rho} = \mathcal{L} \hat{\rho} = - \imag \left[ \hat{H},
    \hat{\rho} \right] + \mathcal{D} \hat{\rho}.
\end{equation}
Intrinsic dynamics of the system are captured by the commutator of $\hat{\rho}$
with the system Hamiltonian $\hat{H}$, and the dissipator $\mathcal{D}$ encodes
the coupling of the system to external reservoirs. The dissipator takes the
characteristic Lindblad form,
$\mathcal{D} \hat{\rho} = \hat{L} \hat{\rho} \hat{L}^{\dagger} - \frac{1}{2}
\left\{ \hat{L}^{\dagger} \hat{L}, \hat{\rho} \right\}$,
where $\hat{L}$ is called a Lindblad or quantum jump operator. For a many-body
system, there are typically several decay channels---e.g., when the system is
coupled to several baths. Then, the dissipator contains a sum over different
types of jump operators. The Lindblad form of the dissipator ensures that the
time evolution generated by the Liouville superoperator $\mathcal{L}$, which is
also known as the Liouvillian or the Lindbladian, is Hermiticity- and
trace-preserving, and completely positive. Derivations of master equations in
the many-body context and a discussion of the relevant approximations can be
found in Refs.~\cite{Diehl2008, Pichler2010, hoening2012critical, Marcos2012,
  Tomadin2012, Caballar2014, Goldstein2018}.


\subsection{Models and methods}
\label{sec:models-methods}

As anticipated in Sec.~\ref{sec:introduction}, a systematic derivation of a
long-wavelength effective description of driven open quantum matter is enabled
by reformulating the quantum master equation~\eqref{eq:meq} as a
Lindblad-Keldysh field theory \cite{Sieberer2016a, Thompson2023}. In general,
field theories and associated methods such as the renormalization group (RG),
allow us to develop a deep understanding of the physics underlying emergent
universal behavior: for example, by clarifying the role of microscopic
symmetries, identifying emergent symmetries, introducing suitable collective
variables, establishing mappings between different models, predicting scaling
forms and calculating critical exponents approximately or, in specific cases,
exactly, based, e.g., on exact scaling relations. A complementary approach is to
directly measure universal properties such as critical exponents in numerical
simulations. Here, the key challenge is that quantitatively accurate results
require large system sizes \cite{Rota2017, Biella2018, WeimerTWA, Jin2021,
  Li2022, Verstraelen2023}.


The loss of nonuniversal information that is inherent in effective field theory
and RG methods implies that a precise determination of the steady-state phase
diagrams of microscopic model systems such as the the dissipative XYZ
\cite{Lee2013} and Ising models \cite{Lee2012}, or the driven-dissipative
Bose-Hubbard model \cite{LeBoite2013, Carusotto2013}, is often beyond the scope
of field theories. For this purpose, a variety of mostly numerical but also
analytical techniques have been applied and newly developed, including
self-consistent approximations, mean-field theories, numerical RG approaches, as
well as variational, phase-space, and tensor-network methods \cite{Kessler2012,
  Lee2013, Degenfeld-Schonburg2014, Finazzi2015, Jin2016, Rota2017, Biella2018,
  Jin2018, Casteels2018, WeimerVariational, WeimerVariationalII, WeimerTWA,
  Weimer2021RMP, Verstraelen2023}.

Due to our focus on universality, we will often formulate models in terms of
effective bosonic field theories rather than as microscopic, e.g., spin
models. For example, the real bosonic field theories of Secs.~\ref{Sec:DP},
\ref{sec:SOC}, \ref{sec:competing}, and~\ref{sec:firstorder} apply to spin
models with Ising symmetry. Similarly, the complex bosonic theory introduced
below and discussed further in Secs.~\ref{sec:driven-open-BEC-2D-1D}
and~\ref{sec:bosonic-driven-open-criticality} also arises as effective
long-wavelength description of limit-cycle phases that emerge from a Hopf
bifurcation \cite{Lee2011, Ludwig2013a, Jin2013, Chan2015, Schiro2016}. For all
of these models, a formulation in terms of their common long-wavelength
effective field theory is an important step toward a qualitative and
quantitative description of universality.

\subsection{Driven open Bose-Einstein condensation}
\label{sec:driven-open-condensate}

To illustrate key concepts in the theory of driven open quantum many-body
systems, we shall often refer to a model of short-range interacting bosons in a
$d$-dimensional spatial continuum, which are subjected to single-particle
pumping and loss as well as two-body loss. This model describes driven open
Bose-Einstein condensation \cite{Sieberer2013, Sieberer2014}, e.g., of
exciton-polaritons \cite{Carusotto2013}, which is discussed in detail in
Sec.~\ref{sec:driven-open-BEC-2D-1D}. Its pedagogical value stands in the direct
comparison with the equilibrium BEC transition; furthermore, upon including
minimal modifications (change of dimensionality, or of jump operators involved)
we can explore several of the distinct notions of driven-open criticality
discussed in this review.

The Hamiltonian of the model reads
\begin{equation}
  \label{eq:H-driven-open-condensate}
  \hat{H} = \int \diff \mathbf{x} \left[ \hat{\psi}^{\dagger}(\mathbf{x})
    \left( - \frac{\nabla^2}{2 m} \right) \hat{\psi}(\mathbf{x}) + u_c
    \hat{\psi}^{\dagger}(\mathbf{x})^2 \hat{\psi}(\mathbf{x})^2 \right],
\end{equation}
where $\hat{\psi}(\mathbf{x})$ and $\hat{\psi}^{\dagger}(\mathbf{x})$ are
annihilation and creation operators for bosons at position
$\mathbf{x} \in \R^d$. These bosons have mass $m$ and interact with strength
$u_c$. Single-particle incoherent pumping and loss and two-body loss are
described by the jump operators
$\hat{L}_p(\mathbf{x}) = \sqrt{\gamma_p} \hat{\psi}^{\dagger}(\mathbf{x})$,
$\hat{L}_l(\mathbf{x}) = \sqrt{\gamma_l} \hat{\psi}(\mathbf{x})$, and
$\hat{L}_t(\mathbf{x}) = \sqrt{2 u_d} \hat{\psi}(\mathbf{x})^2$,
respectively. Consequently, the dissipator takes the form
\begin{multline}
  \label{eq:D-driven-open-condensate}
  \mathcal{D} \hat{\rho} = \int \diff \mathbf{x} \left[ \gamma_p \left(
      \hat{\psi}^{\dagger}(\mathbf{x}) \hat{\rho} \hat{\psi}(\mathbf{x}) -
      \frac{1}{2} \left\{ \hat{\psi}(\mathbf{x})
        \hat{\psi}^{\dagger}(\mathbf{x}), \hat{\rho} \right\} \right) \right. \\
  + \gamma_l \left( \hat{\psi}(\mathbf{x}) \hat{\rho}
    \hat{\psi}^{\dagger}(\mathbf{x}) - \frac{1}{2} \left\{
      \hat{\psi}^{\dagger}(\mathbf{x}) \hat{\psi}(\mathbf{x}), \hat{\rho} \right\} \right) \\
  \left. + 2 u_d \left( \hat{\psi}(\mathbf{x})^2 \hat{\rho}
      \hat{\psi}^{\dagger}(\mathbf{x})^2 - \frac{1}{2} \left\{
        \hat{\psi}^{\dagger}(\mathbf{x})^2 \hat{\psi}(\mathbf{x})^2, \hat{\rho}
      \right\} \right) \right].
\end{multline}
Other concrete models of bosonic and fermionic driven open quantum many-body
systems are introduced in Secs.~\ref{Sec:DP}--\ref{sec:fermions}
below. Condensation of the field $\hat{\psi}(\mathbf{x})$ is indicated by a
finite value of the condensate amplitude
$\psi(t, \mathbf{x}) = \langle \hat{\psi}(t, \mathbf{x}) \rangle$ in the steady
state. Here, the time-dependent expectation value is defined as
$\langle \hat{\psi}(t, \mathbf{x}) \rangle = \tr \! \left(
  \hat{\psi}(\mathbf{x}) \hat{\rho}(t) \right)$,
where $\hat{\rho}(t) = \e^{\mathcal{L} \left( t - t_0 \right)} \hat{\rho}(t_0)$
is the formal solution of the master equation. To understand the mechanism that
underlies the driven open condensation transition, we consider a mean-field
description, which we derive in two steps: First, we obtain the equation of
motion of the condensate amplitude by taking the temporal derivative of the
relation
$\psi(t, \mathbf{x}) = \tr \! \left( \hat{\psi}(\mathbf{x}) \hat{\rho}(t)
\right)$;
and second, we implement a mean-field approximation by replacing field operators
by the classical condensate field $\psi(t, \mathbf{x})$ according to
$\hat{\psi}(t, \mathbf{x}) \to \psi(t, \mathbf{x})$ and
$\hat{\psi}^{\dagger}(t, \mathbf{x}) \to \psi^{*}(t, \mathbf{x})$. This leads to
\begin{equation}
  \label{eq:driven-open-GPE}
  \imag \partial_t \psi = \left[ - \frac{\nabla^2}{2 m} + 2 u_c \abs{\psi}^2 +
    \frac{\imag}{2} \left( \gamma_p - \gamma_l - 4 u_d \abs{\psi}^2 \right)
  \right] \psi,
\end{equation}
which, for a closed system with $\gamma_p = \gamma_l = u_d = 0$, reduces to the
Gross-Pitaevskii equation \cite{Pitaevskii2003}. Instead, for a driven open
system with $\gamma_p > \gamma_l$, the above equation is solved by
$\psi(t, \mathbf{x}) = \psi_0 \e^{- \imag \mu t}$ with
$\abs{\psi_0}^2 = (\gamma_p - \gamma_l)/(4 u_d)$ and
$\mu = 2 u_c \abs{\psi_0}^2$. Physically, this solution describes the formation
of an oscillating condensate when net single-particle gain for
$\gamma_p - \gamma_l > 0$ is balanced by two-body loss with density-dependent
loss rate $4 u_d \abs{\psi_0}^2$.


As anticipated in Sec.~\ref{sec:introduction}, a theoretical description of
universal emergent phenomena in driven open quantum many-body systems such as
critical behavior at the driven open condensation transition is facilitated by
reformulating the quantum master equation~\eqref{eq:meq} in terms of an
equivalent Keldysh field theory. We next describe how this is achieved.

\subsection{Keldysh field theory for driven open systems}
\label{sec:keldysh-field-theory}

Keldysh field theory provides a functional integral representation of the
quantum master equation~\eqref{eq:meq}. The starting point for the construction
of the functional integral representation is the formal solution of the master
equation,
$\hat{\rho}(t) = \e^{\mathcal{L} \left( t - t_0 \right)} \hat{\rho}(t_0)$. A
crucial difference between the Hamiltonian evolution of a pure state
$\ket{\Psi(t)}$ and the Lindbladian evolution of a mixed state $\hat{\rho}(t)$
lies in the fact that the superoperator $\mathcal{L}$ acts on the density matrix
simultaneously from the left- and right-hand sides. This necessitates a
modification of the usual construction of the functional integral for the time
evolution operator
$\hat{U}(t, t_0) = \e^{-\imag \hat{H} \left( t - t_0 \right)}$. For
concreteness, we consider a single bosonic or fermionic field mode, described by
annihilation and creation operators $\hat{\psi}$ and $\hat{\psi}^{\dagger}$,
respectively. These operators obey the canonical commutation relation
$\left[ \hat{\psi}, \hat{\psi}^{\dagger} \right]_{\zeta} = \hat{\psi}
\hat{\psi}^{\dagger} - \zeta \hat{\psi}^{\dagger} \hat{\psi} = 1$,
where $\zeta = + 1$ for a bosonic and $\zeta = - 1$ for a fermionic mode. Then,
the time evolution operator for a pure state can be represented as a functional
integral over a single complex or Grassmann field $\psi$. The required
modification to describe the dynamics of the density matrix is the doubling of
fields $\psi \to \psi_{\pm}$, which is best understood by considering first the
case without dissipation, i.e., when $\mathcal{D} = 0$ in Eq.~\eqref{eq:meq}:
Then
$\hat{\rho}(t) = \e^{\mathcal{L} \left( t - t_0 \right)} \hat{\rho}(t_0) =
\hat{U}(t, t_0) \hat{\rho}(t_0) \hat{U}(t, t_0)^{\dagger}$,
where the operator $\hat{U}(t, t_0)^{\dagger} = \hat{U}(t_0, t)$, if applied to
a state vector, would describe evolution that proceeds backward in time. For
this reason, the time evolution of $\hat{\rho}(t)$ is commonly visualized as two
branches emanating from $\hat{\rho}(t_0)$: A \textit{forward branch} that
corresponds to $\hat{U}(t, t_0)$ and is described, in the functional integral to
be specified below, by integration over the field $\psi_+$; and a
\textit{backward branch} that corresponds to $\hat{U}(t, t_0)^{\dagger}$ and is
described by integration over the field $\psi_-$. In the Keldysh partition
function, which is defined as the trace of the density matrix,
$Z = \tr(\hat{\rho}(t)) = \tr(\hat{\rho}(t_0)) = 1$, the two branches are
connected at time $t$ to form a closed contour.

\subsubsection{Keldysh partition function and Lindblad-Keldysh action}
\label{sec:keldysh-part-funct-lindblad-action}

The derivation of the functional integral representation of the Keldysh
partition function is carried out in detail in
Appendix~\ref{sec:derivation-Lindblad-Keldysh}. For a single bosonic or
fermionic mode, which at time $t_0$ is in the state $\hat{\rho}(t_0)$, and the
dynamics of which include dissipation with a single jump operator $\hat{L}$, we
obtain the following expression for the Keldysh partition function:
\begin{equation}
  \label{eq:Keldysh-partition-function}
  Z = \int \Diff [\psi_+^{*}, \psi_+, \psi_-^{*}, \psi_-] \, \e^{\imag S}
  \braket{\psi_+(t_0) | \hat{\rho}(t_0) | \zeta \psi_-(t_0)}.
\end{equation}
The integration is over complex and Grassmann fields for bosons and fermions,
respectively. Grassmann fields on the backward branch are defined with a sign
$\zeta = - 1$ which is required to represent the trace in the Keldysh partition
function in terms of coherent states as is discussed in
Appendix~\ref{sec:derivation-Lindblad-Keldysh}. The Lindblad-Keldysh action is
given by
\begin{equation}
  \label{eq:S-Keldysh}
  S = \int_{t_0}^t \diff t' \left( \psi_+^{*}(t') \imag \partial_t \psi_+(t') -
    \psi_-^{*}(t') \imag \partial_t \psi_-(t') - \imag \mathcal{L}_{+-}(t') \right).
\end{equation}
As detailed below, to describe steady states of driven open systems, we will
typically consider the limit $t_0 \to - \infty$ and $t \to \infty$. The
Lindbladian that generates the dynamics of the system is encoded in
\begin{multline}
  \label{eq:L-plus-minus-regularized}
  \mathcal{L}_{+-}(t) = - \imag \left( H_+(t) - H_-(t) \right) + L_+(t)
  L_-^{\dagger}(t) \\ - \frac{1}{2} \left( L^{\dagger}_+(t) L_+(t_-) +
    L^{\dagger}_-(t) L_-(t_+) \right),
\end{multline}
where the functions $H_{\pm} = H(\psi_{\pm}^{*}, \psi_{\pm})$,
$L_+ = L(\psi_+^{*}, \psi_+)$, $L_- = L(\zeta \psi_-^{*}, \zeta \psi_{-})$,
$L_+^{\dagger} = L^{\dagger}(\psi_+^{*}, \psi_{+})$, and
$L_-^{\dagger} = L^{\dagger}(\zeta \psi_-^{*}, \zeta \psi_-)$ can be obtained
from the respective operators $\hat{H}$, $\hat{L}$, and $\hat{L}^{\dagger}$,
which we assume to be normal ordered, simply by replacing the creation operators
by $\psi_{\pm}^{*}$, and the annihilation operators by $\psi_{\pm}$, with an
additional sign $\zeta = - 1$ for Grassmann fields on the backward
branch. Infinitesimal time shifts are introduced in $t_{\pm} \to t \pm 0^+$ to
keep track of the original order of operators as detailed in
Appendix~\ref{sec:derivation-Lindblad-Keldysh}. These shifts are required only
to regularize certain classes of integrals that occur in diagrammatic
expansions, and can safely be neglected otherwise.

From Eq.~\eqref{eq:L-plus-minus-regularized}, we can read off the following
simple rule that allows us to straightforwardly translate from the Lindblad
master equation formalism to Keldysh field theory: In the master
equation~\eqref{eq:meq}, operators that act on the density matrix from the left-
and right-hand side lead to contributions on the forward and backward branches,
respectively, with an additional sign $\zeta = - 1$ for Grassmann fields on the
backward branch. Indeed, this rule does not only apply to the action, but also
to expectation values as follows directly from the construction of the Keldysh
field integral described in Appendix~\ref{sec:derivation-Lindblad-Keldysh}. When
there is no dissipation, the Keldysh action is the sum of two independent
contributions that correspond to the forward and backward branches, and the
branches are coupled only at $t_0$ through the matrix element of the initial state
in Eq.~\eqref{eq:Keldysh-partition-function}, and at $t$ through taking the
trace. In the presence of dissipation, the forward and backward branches are
coupled at all times through the term $L_+ L_-^{\dagger}$.

\subsubsection{Keldysh rotation: classical and quantum fields}
\label{sec:keldysh-rotat}

For finite initial and final times $t_0$ and $t$, respectively, the Keldysh
functional integral Eq.~\eqref{eq:Keldysh-partition-function} describes the full
dynamics of the system and is thus fully equivalent to the Lindblad
equation~\eqref{eq:meq}. Expectation values can be evaluated as functional
integrals. For example, the expectation value
$\langle \hat{\psi}(t) \rangle = \tr(\hat{\psi} \hat{\rho}(t))$ of a bosonic
field $\hat{\psi}$ at time $t$ is given by
\begin{equation}
  \langle \psi_+(t) \rangle = \int \Diff[\psi_+^{*}, \psi_+, \psi_-^{*}, \psi_-] \,
  \psi_+(t) \e^{\imag S} \braket{\psi_+(t_0) | \hat{\rho}(t_0) | \psi_-(t_0)},
\end{equation}
where, according to the rule that is formulated at the end of the previous
section, the operator $\hat{\psi}$ that is acting on $\hat{\rho}(t)$ from the
left translates to the field $\psi_+(t)$ on the forward branch. Applying this
rule to both sides of the equation
$\tr(\hat{\psi} \hat{\rho}(t)) = \tr(\hat{\rho}(t) \hat{\psi})$, which follows
from the cyclic property of the trace, we find
$\langle \psi_+(t) \rangle = \langle \psi_-(t) \rangle$. That is, expectation
values of bosonic fields on the forward and backward branches are
identical. This redundancy of the Keldysh formalism is eliminated by working
with symmetric and antisymmetric superpositions of the fields $\psi_{\pm}$,
which are referred to as classical and quantum fields, respectively, and defined
through the Keldysh rotation:
\begin{equation}
  \label{eq:bosonic-Keldysh-rotation}
  \psi_c = \frac{1}{\sqrt{2}} \left( \psi_+ + \psi_- \right), \qquad \psi_q =
  \frac{1}{\sqrt{2}} \left( \psi_+ - \psi_- \right).
\end{equation}
A finite expectation value
$\langle \hat{\psi}(t) \rangle = \langle \psi_c(t) \rangle/\sqrt{2}$ of a
bosonic field, which can signify, e.g., Bose-Einstein condensation in the model
described by Eqs.~\eqref{eq:H-driven-open-condensate}
and~\eqref{eq:D-driven-open-condensate}, is captured by the classical field;
this justifies a posteriori the nomenclature: condensation is a mean-field
behaviour with subleading quantum fluctuations. In contrast, the quantum field
describes fluctuations, and its expectation value vanishes by definition,
$\langle \psi_q(t) \rangle = 0$. Also fermionic fields cannot acquire a finite
expectation value---in other words, Grassmann variables are never
classical. Nevertheless, also for fermions it is meaningful to perform a Keldysh
rotation to classical and quantum fields. This is because the abovementioned
redundancy of the Keldysh formalism manifests also for even-order correlation
functions that are in general nonzero for both bosons and fermions. In
particular, such correlation functions vanish when the largest time argument
belongs to a quantum field as we discuss in more detail in
Appendix~\ref{sec:greens-funct-keldysh} for the example of the Green's functions
introduced below in Sec.~\ref{sec:retarded-advanced-Kledysh-GFs}.  However,
while for bosons the fields $\psi_{\pm}$ and $\psi_{\pm}^{*}$ are related by
complex conjugation, the corresponding Grassmann fields for fermions are
independent. Therefore, they can be transformed in an arbitrary manner. This
freedom is often exploited to introduce different forms of the Keldysh rotation
for bosons and fermions \cite{Kamenev2023, Altland2010a}. Here, instead, we opt
to keep the presentation symmetric by defining
$\psi_c^{*} = (\psi_+^{*} + \psi_-^{*})/\sqrt{2}$ and
$\psi_q^{*} = (\psi_+^{*} - \psi_-^{*})/\sqrt{2}$ also for fermions.

We have argued that the cyclic property of the trace motivates introducing
classical and quantum fields. More fundamentally, the cyclic property of the
trace is related to the general property that the Keldysh action vanishes when
$\psi_+ = \psi_-$ and thus $\psi_q = 0$. This property is part of what is known
as the causality structure of the Keldysh action \cite{Kamenev2023}. To see how
it comes about, we start from the conservation of probability that is expressed
through $\partial_t \tr( \hat{\rho}(t) ) = 0$, and which follows indeed from
using $\partial_t \hat{\rho} = \mathcal{L} \hat{\rho}$ and the cyclic property
of the trace. But $\tr( \hat{\rho}(t) ) = Z$ is just the Keldysh partition
function. Therefore, taking the temporal derivative of
Eq.~\eqref{eq:Keldysh-partition-function} and expressing the Keldysh action
Eq.~\eqref{eq:S-Keldysh} as
$S = \int_{t_0}^t \diff t' \, s(\psi_+(t'), \psi_-(t'))$, conservation of
probability implies that
$0 = \partial_t Z = \imag \langle s(\psi_+(t), \psi_-(t)) \rangle = \imag
\langle s(\psi_+(t), \psi_+(t)) \rangle$,
where in the last equality we have used the redundancy of the Keldysh formalism
explained above. As can be seen in the explicit form of the Keldysh action in
Eq.~\eqref{eq:S-Keldysh}, conservation of probability as expressed through
$\partial_t Z = 0$ is ensured by the property of the Keldysh action
$s(\psi_+, \psi_+) = 0$. In other words, the Keldysh action vanishes when
$\psi_q = 0$.

\subsubsection{Retarded, advanced, and Keldysh Green's functions}
\label{sec:retarded-advanced-Kledysh-GFs}

The two-point functions of classical and quantum fields, which are the retarded,
advanced, and Keldysh Green's functions, are basic elements of Keldysh field
theory:
\begin{equation}
  \label{eq:Greens-functions}
  \begin{split}    
    G^R(t, t') & = - \imag \langle \psi_c(t) \psi_q^{*}(t') \rangle = - \imag
    \theta(t -
    t') \langle [\hat{\psi}(t), \hat{\psi}^{\dagger}(t')]_{\zeta} \rangle, \\
    G^A(t, t') & = - \imag \langle \psi_q(t) \psi_c^{*}(t') \rangle = \imag
    \theta(t' - t) \langle [\hat{\psi}(t), \hat{\psi}^{\dagger}(t')]_{\zeta} \rangle, \\
    G^K(t, t') & = - \imag \langle \psi_c(t) \psi_c^{*}(t') \rangle = - \imag
    \langle [
    \hat{\psi}(t), \hat{\psi}^{\dagger}(t') ]_{-\zeta} \rangle,
  \end{split}
\end{equation}
where the Heaviside step function is defined by $\theta(t) = 1$ for $t > 0$ and
$\theta(t) = 0$ for $t < 0$, and
$[\hat{A}, \hat{B}]_{\zeta} = \hat{A} \hat{B} - \zeta \hat{B} \hat{A}$. For
driven open systems, two-time expectation values in the operator formulation are
defined through the quantum regression theorem \cite{Gardiner2014}. The
equivalence between the expressions for Green's functions in the operator and
Keldysh formalisms is established in Appendix~\ref{sec:greens-funct-keldysh}.

In general, the Green's functions depend on both time arguments $t$ and
$t'$. However, in the applications of Keldysh field theory discussed in this review, our main interest is in systems
that reach a steady state due to coupling to external reservoirs, and in
properties of the steady state. A straightforward way to access these properties
is to shift the initial time $t_0$ to the distant past, i.e., take the limit
$t_0 \to - \infty$. In the steady state, there is no memory left of the initial
state, and, therefore, the matrix element
$\braket{\psi_+(t_0) | \hat{\rho}(t_0) | \zeta \psi_-(t_0)}$ in
Eq.~\eqref{eq:Keldysh-partition-function} can be omitted. Consequently, the
Green's functions depend only on the difference $t - t'$. To account for the
fact that this difference can become arbitrarily large, it is convenient to
consider the limit of an infinite final time $t \to \infty$ in
Eq.~\eqref{eq:Keldysh-partition-function}.

\subsubsection{From a single to many modes}

So far, we have considered only a single field mode. The generalization to many
modes in a spatially extended many-body system is straightforward. For many-body
system in a spatial continuum or on a lattice, the Keldysh action in
Eq.~\eqref{eq:S-Keldysh} includes an integration over spatial coordinates or a
summation over lattice sites, respectively. As a specific example, the Keldysh
action for the Hamiltonian in Eq.~\eqref{eq:H-driven-open-condensate} and the
dissipator in Eq.~\eqref{eq:D-driven-open-condensate} is the sum of two terms,
$S = S_H + S_D$, which encode coherent and dissipative contributions to the
dynamics, respectively:
\begin{equation}
  \label{eq:S-H-driven-open-condensate}
  S_H = \sum_{\sigma = \pm} \sigma \int_{t, \mathbf{x}} \left[
    \psi_{\sigma}^{*} \left( \imag \partial_t + \frac{\nabla^2}{2 m} \right)
    \psi_{\sigma} - u_c \left( \psi_{\sigma}^{*} \psi_{\sigma}
    \right)^2 \right],
\end{equation}
where
$\int_{t, \mathbf{x}} = \int_{-\infty}^{\infty} \diff t \int \diff \mathbf{x}$,
and
\begin{multline}
  \label{eq:S-D-driven-open-condensate}
  S_D = - \imag \int_{t, \mathbf{x}} \left( \gamma_p \left[ \psi_+^{*} \psi_- -
      \frac{1}{2} \left( \psi_+^{*} \psi_+ + \psi_-^{*} \psi_- \right) \right]
  \right. \\ + \gamma_l \left[ \psi_+ \psi_-^{*} - \frac{1}{2} \left( \psi_+^{*}
      \psi_+ + \psi_-^{*} \psi_- \right) \right] \\ \left. + 2 u_d \left\{
      \psi_+^2 \psi_-^{* 2} - \frac{1}{2} \left[ \left( \psi_+^{*} \psi_+
        \right)^2 + \left( \psi_-^{*} \psi_- \right)^2 \right] \right\} \right).
\end{multline}

\section{Equilibrium vs.~nonequilibrium stationary states}
\label{sec:equilibrium-vs-nonequi-steady-states}

A defining signature of thermodynamic equilibrium is that the state of a system
in equilibrium does not change when the system is isolated from its
environment \cite{Ebeling2005}. This condition is clearly violated for a quantum
many-body system that is subjected to a time-periodic classical driving field
and coupled to a dissipative environment: Absorption of energy from the drive
causes the system to heat up indefinitely, and dissipation to the environment is
necessary to stabilize a nontrivial state. Microscopically, the systems that are
the focus of this review are exactly such periodically driven system coupled to
a bath, but in the limit of high-frequency driving. This limit is captured by
the rotating wave approximation, which amounts to taking the average over the
fast driving frequency. Consequently, the explicit time dependence disappears,
and time-translation symmetry is restored on the temporally coarse grained scale
described by the Lindblad evolution in Eq.~\eqref{eq:meq}. For such systems, it
is, therefore, not clear \textit{a priori} whether they will reach thermodynamic
equilibrium, or a nonequilibrium stationary state. As we discuss in the
following, a clear-cut identification of the violation of equilibrium conditions
is possible through a symmetry criterion for the Keldysh
action \cite{Sieberer2015, Aron2018, altland2021prx, Haehl2016, Haehl2017a,
  Haehl2017, Crossley2017, Glorioso2017, Haehl2018}.

\subsection{Thermal equilibrium and the fluctuation-dissipation relations}

How can one decide whether a quantum system is in thermal equilibrium or in a
nonequilibrium steady state? Making this distinction is not possible by focusing
solely on static properties of the steady state, such as expectation values of
observables or equal-time correlation functions. Instead, it is necessary to
consider also dynamics in the steady state in the form of unequal-time
correlation functions. Thermal equilibrium is achieved when the dynamics are
unitary and governed by the same time-independent Hamiltonian $\hat{H}$ that
defines also the thermal Gibbs state
$\hat{\rho}_{\beta} = \left. \e^{- \beta \hat{H}} \middle/ \tr \! \left( \e^{-
      \beta \hat{H}} \right) \right.$
at inverse temperature $\beta=1/T$. More generally, thermal equilibrium requires
that the \emph{generator of the dynamics} also determines the \emph{statistical
  weights} of the state. Otherwise nonequilibrium conditions are expected.

To formalize the distinction between thermal equilibrium and nonequilibrium
stationary states, one may first note that any positive semidefinite Hermitian
operator $\hat{\rho}$ can be written in the form of $\hat{\rho}_{\beta}$ for
some Hamiltonian $\hat{H}$ that is defined as the logarithm of
$\hat{\rho}$. This applies, in particular, to the steady states of the dynamics
described by a master equation in Lindblad form Eq.~\eqref{eq:meq}. However,
thermal equilibrium requires the dynamics to be generated by the same
Hamiltonian $\hat{H}$ that also determines the state. To test whether the
dynamics obey thermal equilibrium conditions, it is sufficient to consider the
linear response of the system to weak time-dependent perturbations. This
response is encoded in unequal-time correlation functions in the unperturbed
state. For the example of a single bosonic or fermionic mode $\hat{\psi}$, we
consider two-point functions such as
$\langle \hat{\psi}^{\dagger}(t) \hat{\psi}(0) \rangle$, where, for a system in
thermal equilibrium,
$\hat{\psi}(t) = \e^{\imag \hat{H} t} \hat{\psi} \e^{- \imag \hat{H} t}$ and
$\langle \dotsc \rangle = \tr( \dotsc \hat{\rho}_{\beta})$. Thermal equilibrium
implies the validity of the fluctuation-dissipation relation (FDR), which is thus
a necessary requirement for equilibrium conditions:
\begin{equation}
  \label{eq:FDR}
  G^K(\omega) = \left( 1 + 2 \zeta n_{\zeta}(\omega) \right) \left( G^R(\omega) -
    G^A(\omega) \right),
\end{equation}
with the Bose or Fermi distribution function
$n_{\zeta}(\omega) = \left. 1 \middle/ \left( \e^{\beta \omega} - \zeta \right)
\right.$,
where as above $\zeta = 1$ for bosons and $\zeta = -1$ for fermions. In a
stationary state, the Green's functions in Eq.~\eqref{eq:Greens-functions}
depend only on the difference of time arguments. Then, the Fourier transform,
e.g., of the Keldysh Green's function, is defined as
$G^K(\omega) = \int_{-\infty}^{\infty} \diff t \, \e^{\imag \omega t} G^K(t)$.

Studying the response to perturbations that couple to products of bosonic or
fermionic operators leads to generalizations of the FDR for higher-order
correlation functions, and the validity of the full hierarchy of generalized
FDRs can be taken as the defining property of thermal equilibrium. Clearly, this
definition is not very practical and one may rely instead on an equivalent
symmetry criterion.

\subsection{Thermal equilibrium as a symmetry of the Keldysh action}
\label{sec:thermal-symmetry}

In field theories, exact relationships between correlation functions can often
be attributed to the symmetries of the action. This is also the case for the
FDR. In fact, the FDR can be seen as a consequence of the Ward-Takahashi
identity which is associated with the symmetry of the Keldysh action under a
specific field transformation on the forward and backward branches of the
Keldysh contour:
\begin{equation}
  \label{eq:thermal-symmetry}
  \mathcal{T}_{\beta} \psi_{l, \pm}(t) = \left( \pm \imag \right)^{(1 -
    \zeta)/2} \sum_{l'} \psi_{l', \pm}^{*}(- t \pm
  \imag \beta/2) U_{l', l}.
\end{equation}
We consider here a multicomponent bosonic or fermionic field $\hat{\psi}_l$,
where $l$ is the site index in a lattice system or may be a combined index
including, e.g., spin or sublattice degrees of freedom. For any two-time
function, invariance under $\mathcal{T}_{\beta}$ is equivalent to the
combination of the Kubo-Martin-Schwinger relation \cite{Kubo1957, Martin1959}
with time reversal $\hatcal{T}$ \cite{Jakobs2010, Sieberer2015}, and, therefore,
implies the validity of the FDR. The time reversal transformation $\hatcal{T}$
is antiunitary, $\hatcal{T} \imag \hatcal{T}^{-1} = - \imag$, and defined by its
action on the fields,
$\hatcal{T} \hat{\psi}_l \hatcal{T}^{-1} = \sum_{l'} U_{l, l'}^{\dagger}
\hat{\psi}_{l'}$,
where for generality we include a unitary matrix $U$ that obeys
$U U^{*} = U^{*} U = \pm \id$ \cite{Sakurai2020, Ludwig2016}. If we further
assume that the Hamiltonian is invariant under time reversal,
$\hatcal{T} \hat{H} \hatcal{T}^{-1} = \hat{H}$, then invariance of the Keldysh
action under $\mathcal{T}_{\beta}$ implies a comprehensive hierarchy of
generalized FDRs for higher-order functions. This illustrates that the
\emph{thermal symmetry} of the Keldysh action, characterized by its invariance
under the transformation $\mathcal{T}_{\beta}$, can be considered as the
definition of thermal equilibrium conditions.

For systems that do not have time reversal symmetry, higher-order FDRs relate
correlation functions of the original system to those of its time-reversed
partner, in the sense that all external parameters are conjugated under time
reversal. In particular, magnetic fields are reversed. This is analogous to
Onsager relations which, when time reversal symmetry is broken, are replaced by
Onsager-Casimir relations \cite{Onsager1931, Casimir1945}. Therefore, the
validity of FDRs and the obedience of equilibrium conditions do not require
invariance of the Hamiltonian under time reversal.

As a concrete example, we consider a single bosonic or fermionic mode with
Hamiltonian $\hat{H} = \omega_s \hat{\psi}^{\dagger} \hat{\psi}$, which has time
reversal symmetry with $U = 1$. A thermal state of this mode is described by the
Keldysh action
\begin{multline}
  \label{eq:single-mode-equilibrium}
  S = \int_{\omega} \left( \psi_c^{*}(\omega), \psi_q^{*}(\omega) \right) \\
  \begin{pmatrix}
    0 & \omega - \omega_s - \imag \delta \\
    \omega - \omega_s + \imag \delta & \imag 2 \delta \left( 1 + 2 \zeta
      n_{\zeta}(\omega) \right)
  \end{pmatrix}
  \begin{pmatrix}
    \psi_c(\omega) \\ \psi_q(\omega)
  \end{pmatrix},
\end{multline}
where the shorthand notation
$\int_{\omega} = \int_{-\infty}^{\infty} \frac{\diff \omega}{2 \pi}$ is
used. This action is invariant under the transformation $\mathcal{T}_{\beta}$
for arbitrary values of the parameter $\delta$. The Green's functions that
correspond to a thermal state are obtained by taking the limit $\delta \to 0^+$
after calculating the functional integrals in
Eq.~\eqref{eq:Greens-functions} \cite{Kamenev2023, Altland2010a}.

The arguments above rely only on symmetry properties of the Keldysh action, and,
therefore, they apply equally to noninteracting and interacting theories. An
alternative formulation of the symmetry which is based on a different
implementation of time reversal in the Keldysh formalism is discussed for bosons
in Ref.~\cite{Sieberer2015} and for multicomponent fermions in
Ref.~\cite{altland2021prx}. The generalization from thermal equilibrium to
chemical equilibrium is readily implemented \cite{Sieberer2015, Aron2018}.

\subsection{Nonequilibrium and emergent equilibrium}\label{sec:emergenteq}

Having identified thermal symmetry of the Keldysh action as the defining
property of thermal equilibrium conditions, we define a nonequilibrium
stationary state as being described by an action that is not invariant under
$\mathcal{T}_{\beta}$. Different physical mechanisms can stabilize a
nonequilibrium stationary state, including the combination of a time-dependent
classical drive and the coupling to a heat bath, or the coupling to several
baths at different temperatures or chemical potentials. As a concrete example,
consider the driven open condensate described by
Eqs.~\eqref{eq:S-H-driven-open-condensate}
and~\eqref{eq:S-D-driven-open-condensate}. For direct comparison with
Eq.~\eqref{eq:single-mode-equilibrium} for a system in thermal equilibrium, we
focus on the Gaussian or quadratic part of the Keldysh action (see
Appendix~\ref{sec:Gaussian} for a general discussion of Gaussian Keldysh
actions). After the Keldysh rotation Eq.~\eqref{eq:bosonic-Keldysh-rotation},
the Gaussian action reads
\begin{equation}
  \label{eq:S-Gaussian}
  S_{\mathrm{Gaussian}} = \int_{\omega, \mathbf{q}} \left( \psi_c^{*}, \psi_q^{*} \right)
  \begin{pmatrix}
    0 & P^A(\omega, \mathbf{q}) \\ P^R(\omega, \mathbf{q}) & P^K
  \end{pmatrix}
  \begin{pmatrix}
    \psi_c \\ \psi_q
  \end{pmatrix},
\end{equation}
where for the integration over frequencies and $d$-dimensional momentum space,
we abbreviate
$\int_{\omega, \mathbf{q}} = \int_{-\infty}^{\infty} \frac{\diff \omega}{2 \pi}
\int \frac{\diff \mathbf{q}}{\left( 2 \pi \right)^d}$.
Furthermore,
$P^R(\omega, \mathbf{q}) = P^A(\omega, \mathbf{q})^{*} = \omega - q^2/(2 m) +
\imag r_d$
with $q = \abs{\mathbf{q}}$ and $r_d = (\gamma_l - \gamma_p)/2$. The key
difference to Eq.~\eqref{eq:single-mode-equilibrium} is that the Keldysh
component $P^K = \imag 2 \gamma$ with $\gamma = (\gamma_l + \gamma_p)/2$ is
frequency-independent. However, the nontrivial frequency dependence of the
distribution function in the Keldysh component of
Eq.~\eqref{eq:single-mode-equilibrium} is required for invariance under
$\mathcal{T}_{\beta}$. Therefore, the thermal symmetry criterion establishes
formally that the driven open condensate is in a nonequilibrium stationary
state. It is important to note that this analysis applies to the
\emph{microscopic} action and does not rule out that thermal equilibrium
conditions emerge upon coarse graining to obtain an effective long-wavelength
and low-frequency description. This is indeed the case for the driven open
condensate in sufficiently high spatial dimension, as discussed in
Sec.~\ref{sec:bosonic-driven-open-criticality}.

\section{Overarching principles in- and out of equilibrium}
\label{sec:organ-princ-beyond}

Field theories are ideally suited to describe emergent phenomena on long
distance and time scales, and, in particular, critical behavior at continuous
phase transitions. Key organizing principles that are fundamental to our
understanding of universal emergent behavior in thermal equilibrium, including
symmetries and the distinction between classical and quantum criticality, retain
their validity in systems that are driven out of equilibrium. Consequently,
their formulation can be adapted to the Keldysh contour formalism, where they
will likewise act as powerful guiding principles for theoretical analysis.

\subsection{Classical and quantum symmetries of the Keldysh action: symmetry
  breaking vs.~conservation laws}
\label{sec:class-quant-symm}

A global unitary continuous symmetry of a closed system entails the existence of
a conserved charge via the Noether theorem \cite{Zinn-Justin}. When the closed
system consists of two subsystems, i.e., the actual system of interest and a
bath, there is a fundamental distinction as to whether Noether charge can or
cannot be transferred between the system of interest and the bath. In the
Keldysh formalism, after integrating out the bath, the former case is referred
to as a \textit{classical symmetry,} while the latter case realizes a
\textit{quantum symmetry.} The distinction between these two types of symmetry
becomes important whenever one deals with an open system, irrespective of
whether it is driven or not. Here, we introduce these two types of symmetry
first for the example of an underlying continuous $\mathrm{U}(1)$ symmetry,
which allows us to obtain an intuitive physical picture in terms of conservation
laws. However, as we discuss further below, these concepts immediately carry
over to other types of continuous and discrete symmetries.

Classical and quantum symmetries can be illustrated using the example of a
single bosonic or fermionic mode coupled to a bath, as described by the
Hamiltonian $\hat{H} = \hat{H}_s + \hat{H}_{sb} + \hat{H}_b$, where the system
Hamiltonian reads $\hat{H}_s = \omega_s \hat{\psi}^{\dagger} \hat{\psi}$, and
the system-bath coupling and the bath Hamiltonian are given by
\begin{equation}
  \label{eq:H-sb-H-b}
  \hat{H}_{sb} = \sum_{\mu} g_{\mu} \left( \hat{L}^{\dagger}
    \hat{\phi}_{\mu} + \hat{\phi}_{\mu}^{\dagger} \hat{L} \right), \quad
  \hat{H}_b = \sum_{\mu} \omega_{\mu} \hat{\phi}_{\mu}^{\dagger}
  \hat{\phi}_{\mu},
\end{equation}
with $\hat{L} = \hat{\psi}$. The Hamiltonian $\hat{H}$ commutes with the total
number of excitations of the system and the bath,
$\hat{N} = \hat{N}_s + \hat{N}_b$, where
$\hat{N}_s = \hat{\psi}^{\dagger} \hat{\psi}$ and
$\hat{N}_b = \sum_{\mu} \hat{\phi}_{\mu}^{\dagger} \hat{\phi}_{\mu}$. Therefore,
$\hat{N}$ is conserved. Formally, we can regard $\hat{N}$ as the Noether charge
that is associated with the symmetry of the Hamiltonian under the unitary
transformation $\hat{G}_{\alpha} = \e^{\imag \alpha \hat{N}}$. For a finite
system-bath coupling $g_{\mu} \neq 0$, excitations, and, therefore, Noether
charge, can be transferred between the system and the bath. In contrast, when
$g_{\mu} = 0$, the numbers of excitations in the system and in the bath,
$\hat{N}_s$ and $\hat{N}_b$, respectively, are conserved individually. This is
reflected in the invariance of the Hamiltonian under both
$\hat{G}_{s, \alpha} = \e^{\imag \alpha \hat{N}_s}$ and
$\hat{G}_{b, \alpha} = \e^{\imag \alpha \hat{N}_b}$. The associated Noether
charges $\hat{N}_s$ and $\hat{N}_b$ pertain to the system and the bath,
respectively, and cannot be transferred between them.

These different types of symmetries---under only $\hat{G}_{\alpha}$ or under
both $\hat{G}_{s, \alpha}$ and $\hat{G}_{b, \alpha}$, which implies also
symmetry under
$\hat{G}_{\alpha} = \hat{G}_{s, \alpha} \hat{G}_{b, \alpha}$---are reflected in
the Keldysh action even after integration over the bath. Only the latter case of
a global continuous \textit{quantum symmetry} is associated with a conserved
Noether charge that pertains to the system alone. But also the former case of a
\textit{classical symmetry} has important physical consequences. For example,
the breaking of global continuous classical symmetries leads to the appearance
of Goldstone modes. The implications of continuous quantum and classical
symmetries are further discussed in Secs.~\ref{sec:quantum-sym}
and~\ref{sec:breaking-weak-Goldstone}. Before that, we elucidate the origin of
both continuous and discrete classical and quantum symmetries in the structure
of the Keldysh formalism and their equivalence to weak and strong symmetries of
Lindbladians \cite{Buca2012}.

\subsubsection{Classical and quantum symmetries}

The doubling of degrees of freedom in Keldysh field theory entails a
corresponding doubling of the symmetry group. This can be illustrated with the
functional integral representation of the time evolution operator
$\hat{U}(t, t_0) = \e^{-\imag \hat{H} \left( t - t_0 \right)}$ for pure
states. The corresponding action is obtained by keeping only the part of the
Keldysh-Lindblad action Eq.~\eqref{eq:S-Keldysh} that describes Hamiltonian time
evolution on the forward branch,
\begin{equation}
  \label{eq:S-propagator}
  S = \int_{t_0}^t \diff t' \left( \psi^{*}(t') \imag \partial_t \psi(t') -
    H(\psi^{*}(t'), \psi(t')) \right).
\end{equation}
For concreteness, consider the Hamiltonian of the bosonic many-body system given
in Eq.~\eqref{eq:H-driven-open-condensate}. The invariance of this Hamiltonian
under the $\mathrm{U}(1)$ phase-rotation transformation of field operators
$\hat{\psi}(\mathbf{x}) \mapsto \hat{G}_{\alpha}^{\dagger}
\hat{\psi}(\mathbf{x}) \hat{G}_{\alpha} = \e^{\imag \alpha}
\hat{\psi}(\mathbf{x})$
with $\hat{G}_{\alpha} = \e^{\imag \alpha \hat{N}}$ where
$\hat{N} = \int_{\mathbf{x}} \hat{\psi}^{\dagger}(\mathbf{x})
\hat{\psi}(\mathbf{x})$
is associated with the conservation of the number of particles $\hat{N}$. It
leads to invariance of the action $S$ in Eq.~\eqref{eq:S-propagator} under the
transformation of fields
$\psi(t, \mathbf{x}) \mapsto \e^{\imag \alpha} \psi(t, \mathbf{x})$. How does
this symmetry transfer to the Keldysh action in Eq.~\eqref{eq:S-Keldysh}? For a
closed system, the Keldysh action reduces to the sum of two copies that
correspond to the forward and backward branches of the Keldysh contour, and one
can perform independent transformations of the fields on each branch,
$\psi_{\pm}(t, \mathbf{x}) \mapsto \e^{\imag \alpha_{\pm}} \psi_{\pm}(t,
\mathbf{x})$.
The symmetry group of the action is thus enlarged to
$\mathrm{U}_+(1) \times \mathrm{U}_-(1)$, where branch indices are added for
clarity. Analogous arguments apply to symmetry groups other than
$\mathrm{U}(1)$, including discrete symmetries.

For closed systems, for which the contributions to the Keldysh action from the
forward and backward branches are independent and identical, the enlarged
symmetry group $\mathrm{U}_+(1) \times \mathrm{U}_-(1)$ does not yield
information beyond the original $\mathrm{U}(1)$ symmetry of the Hamiltonian. For
example, the explicit breaking of $\mathrm{U}_+(1)$ by adding terms to the
Hamiltonian implies that also $\mathrm{U}_-(1)$ is broken, since modifications
of the Hamiltonian affect the forward and the backward branches in the same
way. However, the distinction between symmetry transformations on the forward
and backward branches becomes meaningful and important for open systems, when
dissipative processes couple the forward and backward evolution. Then, a
fruitful perspective is opened up by distinguishing between \textit{classical}
and \textit{quantum symmetries}. In analogy to the distinction between classical
and quantum fields in the Keldysh rotation in
Eq.~\eqref{eq:bosonic-Keldysh-rotation}, in Ref.~\cite{Sieberer2016a}, the
generators of continuous classical and quantum symmetries are defined,
respectively, as symmetric and antisymmetric superpositions of the generators of
symmetry transformations on the forward and backward branches. For the case of
$\mathrm{U}(1)$ phase rotations
$\psi_{\pm}(t, \mathbf{x}) \to \e^{\imag \alpha_{\pm}} \psi_{\pm}(t,
\mathbf{x})$,
this definition of classical and quantum symmetries corresponds to the choices
$\alpha_+ = \alpha_-$ and $\alpha_+ = - \alpha_-$. To include also the case of
discrete symmetry groups, we employ in the following a more general definition
of classical and quantum symmetries.

Let us consider a many-body system with multicomponent field operators
$\hat{\psi}$ and a unitary transformation $\hat{G}$. We leave both the indices
of the field operators pertaining to internal and external degrees of freedom
and a potential dependence of $\hat{G}$ on continuous parameters implicit. The
action of $\hat{G}$ on the field operators can be represented through a unitary
matrix $G$ according to $\hat{G}^{\dagger} \hat{\psi} \hat{G} = G \hat{\psi}$.
This transformation of operators induces a transformation of complex or
Grassmann fields, which is also described by $G$. In particular, for the
functional integral representation of the propagator
$\hat{U}(t, t_0) = \e^{-\imag \hat{H} \left( t - t_0 \right)}$ which is governed
by the action $S$ in Eq.~\eqref{eq:S-propagator}, the fields transform as
$\psi \mapsto G \psi$. While we focus here an bosonic and fermionic theories,
the following considerations also apply to effective field theories that
describe emergent universal behavior in spin models. For example, critical
behavior at a continuous phase transition that breaks the Ising symmetry of a
spin model is described by a field theory for a single real field $\phi$, and
the Ising symmetry of the underlying spin model is described by
$\phi \mapsto G \phi = - \phi$ \cite{Sachdev2011}.

Extending the transformation $G$ to the Keldysh contour, we define a Keldysh
action as having a classical symmetry if it is invariant under
$\psi_{\pm} \mapsto G \psi_{\pm}$. That is, the fields on the forward and
backward branches are transformed in the same way. In contrast, we define a
quantum symmetry as requiring invariance of the Keldysh action under independent
transformations of the fields on the forward and backward branches. Clearly,
this includes the classical symmetry transformation
$\psi_{\pm} \mapsto G \psi_{\pm}$, but also two further possibilities:
$\psi_+ \mapsto G \psi_+$ and $\psi_- \mapsto \psi_-$ as well as
$\psi_+ \mapsto \psi_+$ and $\psi_- \mapsto G \psi_-$. These are all
possibilities for a discrete symmetry. If $G_{\alpha}$ depends on a set of
continuous parameters $\alpha$, the most general quantum symmetry transformation
reads $\psi_{\pm} \mapsto G_{\alpha_{\pm}} \psi_{\pm}$.

Consider now a set of transformations $G$ forming a group $\mathrm{G}$. If these
transformations describe a classical symmetry, the corresponding symmetry group
is again given by $\mathrm{G}_c = \mathrm{G}$; if the transformations describe a
quantum symmetry, the corresponding symmetry group is enlarged to
$\mathrm{G}_q = \mathrm{G}_+ \times \mathrm{G}_-$, where again branch indices
are added for clarity.

In our discussion of classical and quantum symmetries below, we will repeatedly
refer to the model of driven open Bose-Einstein condensation defined in terms of
the Hamiltonian and dissipation in Eqs.~\eqref{eq:H-driven-open-condensate}
and~\eqref{eq:D-driven-open-condensate}, respectively, with Keldysh action given
in Eqs.~\eqref{eq:S-H-driven-open-condensate}
and~\eqref{eq:S-D-driven-open-condensate}. In this model, particle number
conservation is broken explicitly by incoherent pumping and losses. When any of
the parameters $\gamma_p$, $\gamma_l$, and $u_d$ is nonzero, the quantum
symmetry group $\mathrm{U}_q = \mathrm{U}_+(1) \times \mathrm{U}_-(1)$ of the closed
system is reduced to $\mathrm{U}_c(1)$, corresponding to classical phase
rotations
$\psi_{\pm}(t, \mathbf{x}) \mapsto \e^{\imag \alpha} \psi_{\pm}(t, \mathbf{x})$.
Symmetry under the full group of quantum transformations $\mathrm{U}_q$
corresponding to
$\psi_{\pm}(t, \mathbf{x}) \mapsto \e^{\imag \alpha_{\pm}} \psi_{\pm}(t,
\mathbf{x})$ with arbitrary values of $\alpha_{\pm}$ is broken explicitly.

\subsubsection{Equivalence to weak and strong symmetries}

Classical and quantum symmetries of the Keldysh action are equivalent to weak
and strong symmetries of the Lindbladian
\cite{Buca2012,albert2014symmetries,PhysRevLett.125.240405}. This equivalence
results from the general rule, stated in Sec.~\ref{sec:Lindblad-to-Keldysh},
that fields on the forward and backward branches correspond to operators in the
Lindbladian that act on the density matrix from the left- and right-hand side,
respectively. Accordingly, the Keldysh action with fields transformed under a
classical symmetry as $\psi_{\pm} \mapsto G \psi_{\pm}$ corresponds to the
following Lindbladian:
\begin{equation}
  \mathcal{L}_{\hat{G}, c}^{} \hat{\rho} = - \imag \left[ \hat{H}_{\hat{G}}^{},
    \hat{\rho} \right]
  + \hat{L}_{\hat{G}}^{} \hat{\rho} \hat{L}_{\hat{G}}^{\dagger} - \frac{1}{2}
  \left\{ \hat{L}_{\hat{G}}^{\dagger} \hat{L}_{\hat{G}}^{}, \hat{\rho} \right\},
\end{equation}
where for simplicity we consider a single jump operator $\hat{L}$, and we write
$\hat{H}_{\hat{G}} = \hat{G}^{\dagger} \hat{H} \hat{G}$, and
$\hat{L}_{\hat{G}} = \hat{G}^{\dagger} \hat{L} \hat{G}$. The classical symmetry
of the Keldysh action implies that also the transformed Lindbladian should be
identical to the original Lindbladian, $\mathcal{L}_{\hat{G}, c} = \mathcal{L}$,
or, equivalently,
\begin{equation}
  \label{eq:weak-symmetry}
  \mathcal{L}(\hat{G} \hat{\rho} \hat{G}^{\dagger}) = \hat{G}
  \mathcal{L}(\hat{\rho}) \hat{G}^{\dagger}.
\end{equation}
This is the defining relation for a weak symmetry.

Applying the same logic, one can see that invariance of the Keldysh action under
the quantum symmetry $\psi_+ \mapsto G \psi_+$ and $\psi_- \mapsto \psi_-$,
which is one particular choice of the various possibilities to transform the
fields $\psi_{\pm}$ independently, implies that the transformed Lindbladian,
\begin{equation}
\label{eq:lindtrans}
  \mathcal{L}_{\hat{G}, q}^{} \hat{\rho} = - \imag \left( \hat{H}_{\hat{G}}^{}
    \hat{\rho} - \hat{\rho} \hat{H} \right) + \hat{L}_{\hat{G}}^{} \hat{\rho}
  \hat{L}^{\dagger} - \frac{1}{2} \left( \hat{L}_{\hat{G}}^{\dagger}
    \hat{L}_{\hat{G}}^{} \hat{\rho} + \hat{\rho} \hat{L}^{\dagger} \hat{L} \right),
\end{equation}
has to be identical to the original Lindbladian,
$\mathcal{L}_{\hat{G}, q} = \mathcal{L}$. Notice that in
Eq.~\eqref{eq:lindtrans} only the operators that act on the density matrix from
the left are transformed under the symmetry $\hat{G}$. For generic choices of
$\hat{H}$ and $\hat{L}$, this implies that
\begin{equation}
  \label{eq:strong-symmetry}
  \left[ \hat{G}, \hat{H} \right] = 0, \qquad \left[ \hat{G}, \hat{L} \right] =
  0.
\end{equation}
These are the defining equations for a strong symmetry.

The above reasoning applies both to continuous and discrete
symmetries. Returning to the example of driven open Bose-Einstein condensation
and $\mathrm{U}_q$ symmetry, we note that the second equality in
Eq.~\eqref{eq:strong-symmetry} is violated for the jump operators in
Eq.~\eqref{eq:D-driven-open-condensate} that describe particle gain and loss.

We focus here on symmetries of the Lindbladian, i.e., of the generator of
dynamics. Alternatively, weak and strong symmetries can also be defined for
mixed states $\hat{\rho}$~\cite{DeGroot2021}. Then,
Eqs.~\eqref{eq:weak-symmetry} and~\eqref{eq:strong-symmetry} for weak and strong
symmetries are replaced by $\hat{G} \hat{\rho} \hat{G}^{\dagger} = \hat{\rho}$
and $\hat{G} \hat{\rho} = \hat{\rho} \hat{G} = \e^{\imag \theta} \hat{\rho}$,
respectively. As an example, we consider again particle number conservation with
$\hat{G}_{\alpha} = \e^{\imag \alpha \hat{N}}$. The strong symmetry condition is
met with $\theta = \alpha N$ if $\hat{\rho}$ has a definite number of particles
$N$, i.e., if $\hat{\rho}$ is a possibly incoherent superposition of pure states
with the same number of particles. For a weak symmetry, $\hat{\rho}$ is
block-diagonal, with each block corresponding to a different value of $N$.

\subsubsection{Quantum symmetries,  conservation laws, and slow hydrodynamic modes}\label{sec:quantum-sym}

Global continuous quantum symmetries of both open and closed systems are
associated, through the Noether theorem, with conservation
laws \cite{Sieberer2016a}. An important example is again given by $\mathrm{U}_q$
phase rotation symmetry, which implies conservation of the number of
particles. Clearly, incoherent pumping and losses as in
Eq.~\eqref{eq:D-driven-open-condensate} break particle number
conservation. Formally, the absence of particle number conservation becomes
manifest as the absence of $\mathrm{U}_q$ symmetry of the Keldysh action in
Eq.~\eqref{eq:S-D-driven-open-condensate}. 

Here we discuss two examples, where quantum symmetries are realized in driven
open quantum matter:

\paragraph*{Heating to infinite temperature.}

One important example of particle number conserving dissipation, which is
relevant, in particular, for experiments with cold atoms in optical lattices
\cite{Pichler2010}, is given by dephasing with
$\hat{L}(\mathbf{x}) = \sqrt{\gamma} \hat{\psi}^{\dagger}(\mathbf{x})
\hat{\psi}(\mathbf{x})$
\cite{marino2012relaxation, Cai2013, Poletti2012, Poletti2013, Fischer2016,
  Levi2016}. Notably, the jump operator that describes dephasing is Hermitian,
$\hat{L}(\mathbf{x}) = \hat{L}^{\dagger}(\mathbf{x})$, which generically leads
to heating to infinite temperature. For Hermitian jump operators, the dissipator
can be rewritten as a double commutator,
$\mathcal{D} \hat{\rho} = \hat{L} \hat{\rho} \hat{L}^{\dagger} - \frac{1}{2}
\left\{ \hat{L}^{\dagger} \hat{L}, \hat{\rho} \right\} = - \frac{1}{2} \left[
  \hat{L}, \left[ \hat{L}, \hat{\rho} \right] \right]$.
Consequently, the infinite temperature state
$\hat{\rho}_{\infty} = \hat{1}/\tr(\hat{1})$ is a steady state of the master
equation~\eqref{eq:meq}. Furthermore, when
$\left[ \hat{H}, \hat{L} \right] \neq 0$, which is the case for dephasing and
the generic bosonic Hamiltonian given in
Eq.~\eqref{eq:H-driven-open-condensate}, $\hat{\rho}_{\infty}$ is typically also
the unique steady state.

As key implications of the conservation law, several works have reported
algebraic behavior in the dynamics approaching the infinite temperature
state. For example, Refs.~\cite{Poletti2012,Poletti2013} find instances of
anomalous diffusion in the Fock space of bosonic modes with a fixed total
particle number. Furthermore, Ref.~\cite{Cai2013} reports algebraic behavior in
the decoherence of an XXZ magnet with a universal power law $\sim t^{-1.58}$,
which can be traced back to $\mathrm{U}_q$ symmetry under rotations about the
$z$-axis.

\paragraph*{Cooling to pure states.}

Heating to infinite temperature can be replaced by the opposite behavior of
cooling into a pure state by means of dissipation. This is accomplished through
bilocal, yet also particle number conserving Lindblad operators in the absence
of a Hamiltonian. For example, for a bosonic system on a one-dimensional (1D)
lattice with lattice sites $l \in \{ 1, \dotsc, L \}$, $\mathrm{U}_q$ symmetric
and thus number conserving cooling into a perfectly phase-coherent condensate is
obtained through purely dissipative evolution with Lindblad operators
$\hat{L}_l^{} = \sqrt{\gamma} \left( \hat{\psi}_l^{\dagger} + \hat{\psi}_{l +
    1}^{\dagger} \right) \left( \hat{\psi}_l^{} - \hat{\psi}_{l + 1}^{}
\right)$ \cite{Diehl2008}.
The pure state reached as the dynamical fixed point of this evolution for $N$
particles, $\ket{D} \sim \left( \sum_l \hat\psi^\dag_l \right)^N \ket{0}$ where
$\ket{0}$ is the vacuum state, is an example of a dark state (see
Appendix~\ref{sec:dark}). Further examples will be encountered in
Secs.~\ref{sec:ddbose}, and~\ref{sec:fermionpuremix}.

Also in this case, the conservation law has observable physical
consequences. For example, Ref.~\cite{Iemini_2016} has found diffusive behavior
on top of a dissipatively stabilized, number conserving superfluid of fermions. \\

Finally, we briefly touch upon the fate and impact of continuous global quantum
symmetries in long-wavelength effective theories. In spatially extended systems,
local conservation of the Noether charge associated with a quantum symmetry is
expressed through a continuity equation and entails the existence of a gapless
hydrodynamic mode. (Notable exceptions occur when the associated transport
coefficient vanishes, as is the case for (topological) insulators at zero
temperature, or in the presence of anomalies, which are associated to
conservation laws of subsystems on the bare level of description
\cite{peskin1995westwood,fujikawa2004oxford}). Crucially, charge conservation is
a consequence of quantum symmetry of the action on the microscopic level. A
long-wavelength effective theory is derived through integrating out
short-wavelength modes, which can act as a bath for the long-wavelength
modes. In particular, Noether charge can be transferred between the short- and
long-wavelength modes. Therefore, integrating out short-wavelength modes
generates dissipative contributions to the action that break the microscopic
quantum symmetry. Indeed, the paradigmatic classical models of critical dynamics
with conserved quantities do not exhibit such a symmetry
\cite{Hohenberg1977}. However, the continuity equation remains preserved as an
exact property of the theory, even if its transport coefficients receive RG
corrections, including a possible emergent irreversibility of the hydrodynamics
\cite{Haehl2016, Haehl2017a, Haehl2017, Crossley2017, Glorioso2017, Haehl2018}.

\subsubsection{Spontaneous breaking of continuous classical symmetries and
  dissipative Goldstone theorem}
\label{sec:breaking-weak-Goldstone}

While the $\mathrm{U}_q$ symmetry of the Keldysh action specified by
Eqs.~\eqref{eq:S-H-driven-open-condensate}
and~\eqref{eq:S-D-driven-open-condensate} is broken explicitly by particle gain
and loss, the remaining $\mathrm{U}_c(1)$ symmetry can be broken spontaneously
through the formation of a condensate in the steady state, corresponding to a
nonvanishing expectation value of the classical field,
$\langle \psi_c(t, \mathbf{x}) \rangle \neq 0$. In general, the spontaneous
breaking of a global continuous classical symmetry leads to the occurrence of a
massless mode, i.e., a mode whose dispersion relation $\omega(\mathbf{q})$ obeys
$\omega(\mathbf{q}) \to 0$ when the momentum goes to zero, $\mathbf{q} \to 0$.
The existence of such a Goldstone mode is an exact property of the Keldysh
theory and serves as an important benchmark for approximate calculations of the
excitation spectrum \cite{Sieberer2016a}. To illustrate these points, we now
analyse fluctuations around a mean-field condensate.

Evaluating the Keldysh functional integral for a driven open condensate in a
mean-field approximation corresponds to neglecting fluctuations around the
solutions to the classical field equations given by
$\delta S/\delta \psi_c^{*} = 0$ and $\delta S/\delta \psi_q^{*} = 0$. The
former is solved by $\psi_q(t, \mathbf{x}) = 0$, and the latter reproduces
Eq.~\eqref{eq:driven-open-GPE} for the operator expectation value
$\psi(t, \mathbf{x}) = \langle \hat{\psi}(t, \mathbf{x}) \rangle$, the only
difference being a factor $\sqrt{2}$ in the definition of the condensate
amplitude, which is introduced in the Keldysh rotation,
Eq.~\eqref{eq:bosonic-Keldysh-rotation}. We thus find
$\psi_c(t, \mathbf{x}) = \psi_0(t, \mathbf{x}) = \psi_0 \e^{-\imag \mu t}$,
where, for $\gamma_p > \gamma_l$, the mean-field condensate amplitude is given
by $\psi_0 = \sqrt{- r_d/u_d}$ with $r_d = (\gamma_l - \gamma_p)/2$ and
$\mu = u_c \psi_0^2$. Here, without loss of generality, we choose $\psi_0$ to be
real and positive. In contrast, for $\gamma_p < \gamma_l$, there is no
condensation in the steady state, $\psi_0 = 0$. To study fluctuations around the
mean-field condensate, we parameterize the classical and quantum fields as
$\psi_c(t, \mathbf{x}) = \left( \psi_0 + \delta \psi_{c, 1}(t, \mathbf{x}) +
  \imag \delta \psi_{c, 2}(t, \mathbf{x}) \right) \e^{-\imag \mu t}$
and
$\psi_q(t, \mathbf{x}) = \left( \delta \psi_{q, 1}(t, \mathbf{x}) + \imag \delta
  \psi_{q, 2}(t, \mathbf{x}) \right) \e^{- \imag \mu t}$,
and we expand the action to second order in $\delta \psi_{\nu, i}$ for
$\nu = c, q$ and $i = 1, 2$. We find that fluctuations of $\delta \psi_{c, 1}$
are gapped, which implies that terms that involve derivatives of this field can
be neglected as they do not affect the low-frequency and low-momentum form of
the theory qualitatively. Then, after integrating out $\delta \psi_{c, 1}$ and
$\delta \psi_{q, 1}$, the Keldysh partition function takes the form
\begin{equation}
  Z = \int \Diff[\delta \psi_{c, 2}, \delta \psi_{q, 2}] \, \e^{\imag S'},
\end{equation}
where the action is given by
\begin{multline}
  S' = 2 \int_{\omega, \mathbf{q}} \delta \psi_{q, 2}(- \omega, - \mathbf{q})
  \left[ \frac{u_d}{u_c} \left( \imag \omega - D q^2 \right) \delta \psi_{c,
      2}(\omega, \mathbf{q}) \right. \\ \left. + \imag \left( \gamma - 2 r_d
    \right) \left( 1 + \frac{u_d^2}{u_c^2} \right) \delta \psi_{q, 2}(\omega,
    \mathbf{q}) \right].
\end{multline}
We write $q = \abs{\mathbf{q}}$, and for the integration over frequencies and
$d$-dimensional momentum space, we abbreviate
$\int_{\omega, \mathbf{q}} = \int_{-\infty}^{\infty} \frac{\diff \omega}{2 \pi}
\int_{\R^d} \frac{\diff \mathbf{q}}{\left( 2 \pi \right)^d}$.
The parameters in the action are $D = u_c/(2 m u_d)$ and
$\gamma = (\gamma_l + \gamma_p)/2$. From the action $S'$, we can read off the
retarded Green's function:
\begin{equation}
  G^R(\omega, \mathbf{q})
  = \int_{t, \mathbf{x}} \, \e^{\imag \left( \omega t - \mathbf{q} \cdot
      \mathbf{x} \right)} G^R(t, \mathbf{x}) = \frac{u_c}{2 u_d \left( \imag
      \omega - D q^2 \right)}.
\end{equation}
The pole of the retarded Green's function at $\omega = - \imag D q^2$ encodes
the dispersion relation $\omega = \omega(\mathbf{q})$ of elementary
excitations. As anticipated, $\omega(\mathbf{q})$ describes a Goldstone mode
with $\omega(\mathbf{q}) \to 0$ for $\mathbf{q} \to 0$. The Goldstone mode
corresponds to fluctuations of $\delta \psi_{c, 2}$ and, thus, to fluctuations
of the phase of the condensate. However, since phase fluctuations are gapless,
they should not be assumed to be small. Therefore, the expansion in
$\delta \psi_{c, 2}$ is, in fact, not justified. Instead, to obtain a consistent
theory of fluctuations around the mean-field condensate, we must set
$\psi_c = \sqrt{\rho} \e^{\imag \left( \theta - \mu t \right)}$ where
$\theta \in [0, 2 \pi)$, and only fluctuations of $\rho$ around the mean field
$\rho_0 = \psi_0^2$ can safely be assumed to be small. A full treatment of
nonlinear fluctuations of the condensate phase is given in
Ref.~\cite{Sieberer2016a}, and leads to the Kardar-Parisi-Zhang (KPZ) action, or
equivalently, the famous KPZ equation \cite{Kamenev2023,
  Tauber2014a}. Originally, the KPZ equation was used to describe the dynamics
of growing interfaces, where the dynamical variable is the noncompact height of
the interface \cite{Kardar1986}. The consequences of an emergent KPZ action for
the compact phase variable $\theta$ are discussed in
Sec.~\ref{sec:driven-open-BEC-2D-1D}.

The Goldstone mode that is associated with the breaking of the $\mathrm{U}_c(1)$
symmetry of a driven and open condensate propagates diffusively with diffusion
constant $D$. In contrast, the condensation transition in a closed system which
has $\mathrm{U}_q$ symmetry leads to the appearance of coherently propagating
sound waves with dispersion relation $\omega(\mathbf{q}) = \pm c q$, where $c$
is the speed of sound. In the models for critical dynamics with conservation
laws in thermal equilibrium, the existence of such coherently propagating
hydrodynamic modes is taken into account explicitly by introducing the
corresponding conserved densities as dynamical
variables \cite{Hohenberg1977}. While the existence of such modes is a
consequence of an underlying $\mathrm{U}_q$ symmetry, this symmetry is no longer
visible in the mesoscopic models used to analyze critical behavior as discussed
at the end of the previous section.

\subsection{Mixed vs.~pure states, classical vs.~quantum scaling}
\label{sec:mixedvspure}

Closed quantum many-body systems in thermal equilibrium can exhibit both
classical and quantum critical behavior---the former occurring in mixed thermal
states at finite temperature $T > 0$, and the latter in pure ground states at
$T = 0$ \cite{Sachdev2011}. As we explain in the following, the concepts of
classical and quantum criticality retain their validity in driven and open
systems. We base our discussion on an analysis of canonical scaling, first of
the propagators of the Gaussian theory, and then, in Sec.~\ref{sec:Limits}
below, of interaction vertices. Thereby, we highlight the analysis of canonical
scaling as an important tool both in and out of equilibrium.

For concreteness, we consider the driven and open condensate, whose Gaussian
action is given in Eq.~\eqref{eq:S-Gaussian}. In the retarded component, the
difference between loss and pump rates, $r_d = (\gamma_l - \gamma_p)/2$,
measures the distance from the mean-field condensation transition that occurs at
$\gamma_l = \gamma_p$. In contrast, in the Keldysh component, the noise caused
by incoherent pumping and gain adds up, $\gamma = (\gamma_l + \gamma_p)/2$. More
generally, by considering the retarded and Keldysh or noise components of the
inverse propagator (i.e., the matrix in the quadratic action
Eq.~\eqref{eq:S-Gaussian}), we are lead to distinguish two types of gaps, the
spectral gap and the noise gap, whose significance we discuss in the following.

\subsubsection{Spectral gap and criticality}\label{sec:specgap}

On the level of the Gaussian theory for a single field component, the dispersion
relation of elementary excitations is obtained by solving
$P^R(\omega, \mathbf{q}) = 0$ for $\omega = \omega(\mathbf{q})$. Propagators are
matrix-valued for multicomponent fields, and the equation to be solved in this
case is $\det \bigl( P^R(\omega, \mathbf{q}) \bigr) = 0$. The spectral or
dissipative gap is defined by
\begin{equation}
  \Delta_p = \min_{\mathbf{q}} \left( - \Im(\omega(\mathbf{q})) \right).
\end{equation}
That is, the dissipative gap is the minimal distance of the dispersion relation
from the real axis in the complex $\omega$ plane. Instead, for a closed system
with a purely real dispersion relation, the spectral gap should be defined in
terms of the minimal distance from the imaginary axis \cite{altland2021prx}, and
it is just the usual mass term that should be tuned to zero to access
criticality \cite{cardy1996scaling,Zinn-Justin,Kamenev2023}. The necessity to
employ different definitions of the spectral gap in open and closed systems
results from the noncommutativity of limits: In an open system, in the
infinite-time limit, the dynamics generically reach a stationary state. The
dissipative gap characterizes the rate at which the system relaxes back to the
stationary state after a perturbation. In contrast, the dynamics are drastically
different in a closed system, when the coupling to external reservoirs is set to
zero before the long-time limit is considered. Then, the relevant time scale for
perturbations to decay is determined by the excitation gap above the ground
state.

For our example of a driven and open condensate, on the Gaussian level, the
spectral gap is thus given by $r_d$ in the symmetric phase, and the mean-field
critical point at $r_d = 0$ corresponds to a zero of the spectral gap. The
calculation of corrections to mean-field theory in a loop expansion involves
integrals over products of Green's functions \cite{Zinn-Justin}. At the
mean-field critical point $r_d = 0$, the retarded and advanced Green's functions
exhibit canonical critical scaling,
$G^R(\omega, \mathbf{q}) = G^A(\omega, \mathbf{q})^{*} = 1/P^R(\omega,
\mathbf{q}) = 1/[\omega - q^2/(2 m)] \sim 1/q^2$,
which can lead to infrared (IR) divergencies, i.e., divergencies of momentum
integrals for $q \to 0$. In the framework of the RG, such IR divergencies drive
the RG flow at criticality towards the universal scaling regime which encodes
corrections to mean-field critical exponents \cite{cardy1996scaling}. In
contrast, a finite spectral gap implies the absence of zero modes of $P^R$,
i.e., $\abs{\omega(\mathbf{q})} \neq 0$ for all momenta $\mathbf{q}$, and,
consequently, the absence of infrared divergencies.

\subsubsection{Noise gap: classical and quantum scaling}\label{sec:noisegap}

While a nonvanishing spectral gap results from the absence of zero modes of
$P^R(\omega, \mathbf{q})$, a nonvanishing noise gap signifies the absence of
zero modes of $P^K(\mathbf{q})$. For the sake of generality, we consider here a
momentum-dependent Keldysh component. However, for Markovian dissipation,
$P^K(\mathbf{q})$ does not depend on frequency. Denoting the eigenvalues of
$P^K(\mathbf{q})$ by $\imag 2 \gamma(\mathbf{q})$, we define the noise gap as
\begin{equation}
  \Delta_n = \min_{\mathbf{q}} \abs{\gamma(\mathbf{q})}.
\end{equation}
A finite or vanishing value of the noise gap decides whether the criticality
that is induced by a vanishing spectral gap is classical or quantum,
respectively. 

To see that this is the case, it is instructive to compare the noise component
$P^K = \imag 2 \gamma$ of the driven and open condensate to the one of a bosonic
system coupled to a bath in thermal equilibrium,
$P^K(\omega) = \imag 2 \gamma \omega \coth(\beta \omega/2)$ \cite{Kamenev2023}.
In contrast to a system that is subjected to Markovian dissipation, the Keldysh
component $P^K(\omega)$ of a system in thermal equilibrium depends on the
frequency $\omega$.  Depending on the temperature, $P^K(\omega)$ exhibits
different behaviors:

\paragraph*{Classical scaling.}

In the high-temperature limit, we obtain
$P^K(\omega) \sim \imag 4 \gamma T \sim q^0$. That is, as in the case of the
driven open condensate, the Keldysh component is frequency- and
momentum-independent and exhibits a finite noise gap. If, under these
conditions, a phase transition is induced by closing the spectral gap, the
system will exhibit classical criticality. Therefore, we refer to
$P^K(\omega) \sim q^0$ as \textit{classical scaling.} Indeed, as explained
above, critical behavior is driven by fluctuations on small frequency and
momentum scales, $\omega, q \to 0$. For the system in thermal equilibrium, the
excitation energies of such fluctuations are much smaller than $T$. Therefore,
these fluctuations are effectively exposed to a high temperature and behave
classically. As we show explicitly in Sec.~\ref{sec:semiclassical-limit} below,
the same applies to systems out of equilibrium, with temperature replaced by
Markovian noise level.

\paragraph*{Quantum scaling.}

In the limit of zero temperature,
$P^K(\omega) \sim \imag 2 \gamma \abs{\omega}$ is frequency-dependent and
vanishes for $\omega \to 0$. We refer to the scaling
$P^K(\omega) \sim \omega \sim q^z$ as \textit{quantum scaling}, where for a
system with parabolic dispersion relation the dynamical exponent $z$ takes the
value $z = 2$. At $T = 0$, the system in thermal equilibrium is in its pure
ground state, and tuning the spectral gap to zero induces a quantum phase
transition with associated quantum critical behavior. A concrete example of such
quantum criticality in a driven and open system is discussed in
Sec.~\ref{sec:quantum-criticality}.

The above argument establishes the connection between quantum scaling and zero
temperature. As we show in the following on the level of the bosonic Gaussian
theory in Eq.~\eqref{eq:S-Gaussian} for which the canonical scaling is
developed, quantum scaling is also directly related to the purity of the
state. To that end, we consider the Hermitian covariance matrix, given by the
equal-time Keldysh Green's function,
\begin{equation}
  \label{eq:Corr_Mat_main}
  \Gamma(\mathbf{x}, \mathbf{x}') = \langle \{ \hat{\psi}(\mathbf{x}),
  \hat{\psi}^{\dagger}(\mathbf{x}')\} \rangle = \imag G^K(t, \mathbf{x}, t,
  \mathbf{x}').
\end{equation}
For a mixed state, the eigenvalues $\xi$ of the Hermitian bosonic covariance
matrix $\Gamma$ satisfy $\abs{\xi} \geq 1$; in contrast, for a pure state, the
eigenvalues $\xi$ have unit modulus, $\abs{\xi} = 1$
\cite{Barthel2022}. Furthermore, the covariance matrix squares to the identity,
$\Gamma^2 = \id$, which is equivalent to
$\tr \! \left( \hat{\rho}^2 \right) = 1$. The covariance matrix of a system with
translational invariance obeys
$\Gamma(\mathbf{x}, \mathbf{x}') = \Gamma(\mathbf{x} - \mathbf{x}')$ and is,
therefore, diagonal in momentum space,
\begin{equation}
  \Gamma(\mathbf{q}) = \int_{\mathbf{x}} \e^{- \imag \mathbf{q} \cdot
    \mathbf{x}} \Gamma(\mathbf{x}) = \imag \int_{\omega} G^K(\omega,
  \mathbf{q}).
\end{equation}
According to the abovementioned properties of the covariance matrix, the
absolute value of $\Gamma(\mathbf{q})$ is constant for a pure state,
$\abs{\Gamma(\mathbf{q})} = 1$. This condition is compatible only with quantum
scaling of $P^K$. To see that this is the case, note that
$G^K(\omega, \mathbf{q}) = - G^R(\omega, \mathbf{q}) P^K(\omega) G^A(\omega,
\mathbf{q}),$
where
$G^R(\omega, \mathbf{q}) = 1/P^R(\omega, \mathbf{q}) = G^A(\omega,
\mathbf{q})^{*}$.
Therefore, canonical critical scaling $G^{R, A}(\omega, \mathbf{q}) \sim q^{-2}$
implies $G^K(\omega, \mathbf{q}) \sim q^{-4}$ and
$G^K(\omega, \mathbf{q}) \sim q^{-2}$ for classical and quantum scaling of
$P^K(\omega)$, respectively. With $\diff \omega \sim q^2$, we find that
classical scaling leads to $\Gamma(\mathbf{q}) \sim q^{-2}$, while quantum
scaling results in $\Gamma(\mathbf{q}) \sim q^0$. In the classical case,
momentum modes are strongly occupied at low momenta in a bosonic system. But
this is not the case for pure states: As claimed above, only quantum scaling is
compatible with a pure state with $\abs{\Gamma(\mathbf{q})} = 1$.

For simplicity, we have considered here a scalar bosonic field, and we have
assumed the system to be invariant under continuous spatial
translations. However, the argument generalizes to vector fields, and to fields
defined on a lattice. Gaussian states and the covariance matrix for these cases,
both for bosons and fermions, are discussed in
Appendix~\ref{sec:Gaussian}. Fermionic Gaussian states will play a key role in
our discussion of topological phase transitions out of equilibrium in
Sec.~\ref{sec:fermionpuremix}. There, in Eq.~\eqref{eq:purity-gap}, we will
introduce yet another type of gap, the purity gap, which measures the distance
of the eigenvalues of the covariance matrix from zero, and distinguishes pure
from mixed states. This definition of the purity gap applies only to fermions,
for which the eigenvalues $\xi$ of the covariance matrix are bounded as
$\abs{\xi} \leq 1$.

\subsection{Scaling arguments: deterministic and semiclassical limit}
\label{sec:Limits}

Scaling arguments are fundamental tools in both equilibrium and nonequilibrium
statistical physics. They help assess the relevance of operators at long
wavelengths and enable controlled approximations, such as the $1/N$ expansion in
$\mathrm{O}(N)$ models. In driven open quantum systems, the Keldysh field
integral treats quantum and statistical fluctuations equally, resulting in a
complex microscopic description. Here, scaling arguments are systematically
applied to reduce complexity and derive an effective long-wavelength theory that
governs the dynamics in relevant limiting cases.

There are two important limits in which subleading fluctuations can be
disregarded systematically: the deterministic limit and the semiclassical
limit. The \emph{deterministic limit} corresponds to an expansion around a
macroscopically occupied field configuration such as a condensate wave function,
using an $N^{-1/2}$ expansion, where $N$ is the total number of particles. This
limit intentionally omits both quantum and statistical fluctuations, resulting
in a non-Hermitian evolution equation equivalent to a mean-field approximation
of the quantum master equation. Instead, the \emph{semiclassical limit} employs
canonical power counting to neglect subleading quantum fluctuations in systems
in which statistical fluctuations are dominant. This occurs, e.g., near critical
points at finite noise level, and both in and out of equilibrium as discussed in
Sec.~\ref{sec:mixedvspure}.

\subsubsection{Deterministic limit and relation to non-Hermitian physics}
\label{sec:deterministic-limit}

The Keldysh partition function
$Z = \int \Diff [\psi_c^{*}, \psi_c^{}, \psi_q^{*}, \psi_q^{}] \, \e^{\imag S}$
is an integral over field configurations, whose respective importance is
weighted by the exponential $e^{\imag S}$. In contrast, classical field theories
are controlled by a single deterministic field configuration that extremizes the
action $S$. Such \emph{classical} configurations often yield a dominant
contribution to the Keldysh field integral. The deterministic limit focuses
exclusively on this classical configuration and discards fluctuations. It arises
as a \emph{scaling limit} and is particularly relevant in the presence of
condensation phenomena like Bose-Einstein condensation or spontaneous
magnetization, when a single quantum state is macroscopically occupied by
bosonic degrees of freedom in the presence of weak noise and weak interactions.

For concreteness, consider the driven open condensate from
Sec.~\ref{sec:driven-open-condensate}. Rewriting its action $S = S_H + S_D$ in
Eqs.~\eqref{eq:S-H-driven-open-condensate}
and~\eqref{eq:S-D-driven-open-condensate} in Keldysh coordinates yields
\begin{multline}
  S = \int_{\omega, \mathbf{q}} \left( \psi_c^{*}, \psi_q^{*} \right)
  \begin{pmatrix}
    0 & P^A(\omega, \mathbf{q}) \\
    P^R(\omega, \mathbf{q}) & P^K
  \end{pmatrix}
  \begin{pmatrix}
    \psi_c \\ \psi_q
  \end{pmatrix}
  \\
  + \int_{t, \mathbf{x}} \left\{ 4\imag u_d \abs{\psi_c \psi_q}^2 - \left[ \left(
        u_c -\ii u_d \right) \psi_q^\ast
      \psi_c^\ast\left(\psi_c^2+\psi_q^2\right) + \mathrm{c.c.} \right]
  \right\}.
\end{multline}
The quadratic part, including the propagators $P^{R, A, K}$, is specified in
Eq.~\eqref{eq:S-Gaussian}. Consider now the case of weak
nonlinearities and a macroscopically occupied mode, i.e., a finite 
expectation value of the classical field $ \langle\psi_c\rangle$. An important example is
Bose-Einstein condensation,
$\langle\psi_c\rangle\sim\sqrt{N/V}$, with $N$ the particle number and $V$ the system
volume. In momentum space
\begin{gather}
  \psi_c (t, {\bf q}=0) \sim  \braket{\psi_c (t,{\bf q}=0)}  = \sqrt{V}\langle\psi_c\rangle\sim \sqrt{N} , \\
  \psi_c (t,{\bf q}\neq 0) \sim N^0, \quad \psi_q (t,{\bf q}) \sim N^0.
\end{gather} 
The quantum field cannot condense, hence its scaling. The condensed mode is
singled out according to
$\psi_c (t,{\bf q})\equiv \sqrt{V}\langle\psi_c\rangle\delta ({\bf q}) + \delta\psi_c(t,{\bf q})$. In real space the
Fourier transform
\begin{eqnarray}
  \psi_c (t,{\bf x}) = \frac{1}{\sqrt{V}} \int_\mathbf{\mathbf{q}}e^{i
  \mathbf{q} \cdot \mathbf{x}} \psi_c (t, {\bf q}) = \langle\psi_c\rangle  + \delta\psi_c(t,{\bf x})
\end{eqnarray}
reveals the scaling of the condensate $\langle\psi_c\rangle\sim N^{0}$, while
the fluctuations and the quantum fields scale as
$\delta\psi_c(t,{\bf x}), \psi_q(t,{\bf x})\sim N^{-1/2}$. This consideration is
readily generalized to inhomogeneous condensate fields
$\langle\psi_c\rangle\to\langle\psi_c(t,{\bf x})\rangle$ that vary smoothly in
space-time on scales much larger than the microscopic length and time scales.

Drawing the thermodynamic limit $V\to \infty$ at constant density $N/V$, both
$\psi_q$ and $\delta \psi_c$ yield subleading contributions. Expanding to
leading order in $N^{-1/2}$ yields an action
\begin{align}\label{eq:scalingact2}
  S=\int_{t, \mathbf{x}}\Big\{\psi_q^\ast\big[P^R-(u_c -\ii u_d)|\langle\psi_c\rangle|^2\big]\langle\psi_c\rangle + \text{c.c.}\Big\},
\end{align}
\textit{linear} in the quantum field.  Integration over $\psi_q, \psi_q^\ast$ is
then performed exactly: The resulting $\delta$-functional constrains the
dynamics of the condensate to the saddle point equation
$0=\delta S/\delta\psi_q^\ast(x)$, reminiscent of a classical, i.e.,
fluctuationless field theory. For the driven open condensate, this reproduces
the dissipative Gross-Pitaevskii equation~\eqref{eq:driven-open-GPE}, i.e., the
mean-field approximation. Quantum \emph{and} statistical fluctuations are
discarded, yielding a non-Hermitian equation of motion.

The field integral expresses the approximation as a controlled expansion around
a macroscopic condensate. It provides justification for scenarios where a
macroscopically large order parameter meets weak nonlinearities, such as, e.g.,
realized in atomic condensates at low temperatures \cite{Carusotto2013}, or in
photonic non-Hermitian systems \cite{Ozawa_2019}. It also provides criteria for
when the approximation breaks down: An important case is a second order phase
transition where the macroscopic order parameter vanishes,
$\langle \psi_c \rangle \to0$. Then, the expansion is no longer justified and
quantum and statistical fluctuations need to be incorporated.

Another common but different approximation that neglects statistical
fluctuations starts from the Lindblad equation~\eqref{eq:meq} and discards the
jump terms by setting $\hat L \rho \hat L^\dagger\to 0$. This yields the
evolution equation
\begin{align}
  \partial_t\hat\rho=-i \left(\hat
  H_{\text{eff}}^{\phantom{\dagger}}\hat\rho-\hat\rho\hat H_{\text{eff}}^\dagger
  \right),
\end{align} 
with an \emph{effective} non-Hermitian Hamiltonian
$\hat H_{\text{eff}}=\hat H-\frac{\imag}{2} \hat{L}^\dagger \hat{L}$. In this
case, the evolution neglects any feedback from the environment on the state, and
the system approaches a pure state $\hat\rho\to|\psi_0\rangle\langle\psi_0|$,
corresponding to the eigenstate
$\hat H_{\text{eff}}|\psi_0\rangle= \varepsilon |\psi_0\rangle$ with the largest
imaginary part of $\varepsilon$. The corresponding equation of motion, however,
is not probability conserving,
$\partial_t \tr(\hat{\rho}) \neq 0$. In Keldysh field theory, this
manifests in a violation of the causality condition
$\left. S_{\text{eff}}[\psi_+, \psi_-]\right|_{\psi_+\to\psi_-}\neq0$ for the action
associated with the non-Hermitian Hamiltonian $\hat H_{\text{eff}}$.

The non-Hermitian evolution generated by $\hat{H}_{\mathrm{eff}}$ thus does not
reflect a single particular field configuration $\psi_c, \psi_q$ and it does not
extremize the action. Instead of describing the probabilistically most likely
evolution, i.e., the largest contribution to the partition function $Z$, the
non-Hermitian Hamiltonian evolution rather selects a trajectory which results
from a \emph{rare} sequence (zero jumps) of system-bath interactions. This
interpretation is particularly transparent when modeling the environment as
performing measurements (or general positive operator-valued measures
\cite{NielsenChuang}) on the system. Then the effective Hamiltonian yields a
measurement trajectory in which measurement outcomes of a particular type have
been discarded, e.g., via postselection
\cite{TurkeshiZeroClick,GullansPostSel,PhysRevA.35.198, Gardiner2015}. Even
though such rare trajectories do in general not reflect the behavior of the
ensemble average, analyzing $\hat H_{\text{eff}}$ may still provide valuable
information regarding the response of the system or on potential dynamical
instabilities. This has been successfully exploited to explore the non-Hermitian
band structure in topological systems of photons \cite{Bergholtz2021,
  Ashida2020, Topo_nonHerm_rev,Ozawa_2019}. Care, however, needs to be exerted
regarding the interpretation of the effective Hamiltonian. Consider, for
instance, bosons or fermions with single-particle loss and pump from
Sec.~\ref{sec:driven-open-condensate}, i.e.,
$\hat L_p=\sqrt{\gamma_p}\hat\psi^\dagger$, $\hat L_l=\sqrt{\gamma_l}\hat \psi$
and $\hat H_0=\omega_0\hat\psi^\dagger\hat\psi$. Then disregarding the jump
terms in the evolution yields
\begin{equation}
  \begin{split}
    \hat{H}_{\mathrm{eff}} &=\omega_0\hat\psi^\dagger\hat\psi- \imag
    \frac{\gamma_l}{2} \hat \psi^{\dagger} \hat\psi - \imag \frac{\gamma_p}{2}
    \hat\psi \hat \psi^{\dagger} \\
    &= \left[ \omega_0-\frac{\imag}{2} \left( \gamma_l +\zeta \gamma_p \right)
    \right]\hat \psi^{\dagger} \hat \psi - \frac{\imag \gamma_p}{2}.
  \end{split}
\end{equation}
Using the effective Hamiltonian to define a retarded Green's function in an
analogous way to Hamiltonian systems in equilibrium, i.e.,
$G^R_{\text{eff}}\equiv\left(\omega-\omega_{\text{eff}}+i0^+\right)^{-1}$, where
$\omega_{\text{eff}}$ are the eigenvalues of $\hat H_{\text{eff}}$, does not
provide the correct retarded Green's function. The latter is instead
$G^R=\left[ \omega-\omega_0+i \left( \gamma_l-\zeta \gamma_p \right)
\right]^{-1}$,
see Appendix~\ref{sec:Gaussian}, which differs from $G^R_{\text{eff}}$ by the
sign in front of the pumping term $\sim \gamma_p$. This difference in
single-particle Green's functions in a noninteracting system reflects that there
is no general relation between the eigenvalues and eigenvectors of the effective
Hamiltonian $\hat{H}_{\mathrm{eff}}$ and of those of the full Lindbladian
superoperator $\mathcal{L}$. Another notable consequence of the absence of such
a direct relation concerns exceptional points, which are parameter values for
which at least two eigenvalues become degenerate and the corresponding
eigenvectors coalesce, and which have recently attracted much interest in
optics, optoelectronics, plasmonics, and condensed matter physics
\cite{Ozdemir2019, Miri2019, Ashida2020, Bergholtz2021}. In general, exceptional
points of the full Lindbladian $\mathcal{L}$ including quantum jump terms are
not captured by a description in terms of $\hat{H}_{\mathrm{eff}}$
\cite{Minganti2019}.

\subsubsection{Canonical power counting and semiclassical limit}
\label{sec:semiclassical-limit}

In Sec.~\ref{sec:mixedvspure}, we have concluded that the driven open
condensation transition exhibits  classical critical behavior. This conclusion
was based on a comparison of the scaling behavior of the noise component of the
inverse propagator  for a driven open condensate and for a system in  equilibrium  at high temperature. As we
discuss next, a more formal and unifying argument that corroborates this conclusion can be
given in terms of canonical power counting. In general, canonical scaling
dimensions of the couplings that appear in an action determine the RG relevance
of the respective couplings for the low-frequency and long-wavelength dynamics
of the system.

\paragraph*{Semiclassical limit of the driven open condensation transition.}

The canonical momentum scaling dimensions of the
fields $\psi_c$ and $\psi_q$ and, therefore, of all couplings in the action, can
be inferred from the scaling of the inverse propagator and the condition that
the action is dimensionless, i.e., it does not scale with momentum. In
particular, at the mean-field critical point of the driven open
condensate, which is determined by the vanishing of the spectral gap $r_d$, the
retarded component scales as
$P^R(\omega, \mathbf{q}) = \omega - q^2/(2 m) \sim q^2$. In contrast, the noise
component is frequency- and momentum-independent, and, therefore, does not
scale, $P^K = \imag 2 \gamma \sim q^0$. This leads to
$\psi_c \sim q^{(d - 2)/2}$ and $\psi_q \sim q^{(d + 2)/2}$. While the canonical
scaling dimensions of the fields are determined by the Gaussian part of the
action in Eq.~\eqref{eq:S-Gaussian}, they in turn determine the relevance of
interaction vertices that are not part of the Gaussian theory. Therefore, power
counting has strong implications beyond the analysis of the Gaussian theory in
the previous sections. In particular, the canonical scaling dimensions of the
fields imply that local vertices with more than two quantum fields are
irrelevant in spatial dimensions $d > 2$ \cite{Sieberer2016a}. Dropping these
terms is equivalent to taking the semiclassical limit \cite{Kamenev2023,
  Altland2010a} and results in
\begin{multline}
  \label{eq:S-driven-open-condensate-semiclassical}
  S = \int_{t, \mathbf{x}} \left\{ \psi_q^{*} \left[ \imag \partial_t + \left(
        K_c - \imag K_d \right) \nabla^2 - r_c + \imag r_d \right] \psi_c +
    \mathrm{c.c.} \right. \\ \left. - \left[ \left( u_c - \imag u_d \right)
      \psi_q^{*} \psi_c^{*} \psi_c^2 + \mathrm{c.c.} \right] + \imag 2 \gamma
    \psi_q^{*} \psi_q \right\},
\end{multline}
where $K_c = 1/(2 m)$ and the term that is proportional to $r_c$ corresponds to
the transformation to a rotating frame,
$\psi_{\nu}(t, \mathbf{x}) \mapsto \e^{\imag r_c t} \psi_{\nu}(t, \mathbf{x})$
for $\nu = c, q$, and is included here for the sake of generality. Furthermore,
the diffusion term with coefficient $K_d$ is absent in the microscopic model,
but is added here as it will inevitably be generated upon integrating out
short-scale fluctuations \cite{Sieberer2013, Sieberer2014, Sieberer2016a}. The
action in the semiclassical limit is equivalent to a stochastic equation of
motion for the condensate field \cite{Sieberer2016a},
\begin{equation}
  \label{eq:driven-open-condensate-Langevin}
  \imag \partial_t \psi = \left[ - \left( K_c - \imag K_d \right) \nabla^2 + r_c
  - \imag r_d + \left( u_c - \imag u_d \right) \abs{\psi}^2 \right] \psi + \xi,
\end{equation}
with a Gaussian noise source $\xi$ that vanishes on average,
$\langle \xi(t, \mathbf{x}) \rangle = 0$, and whose fluctuations are determined
by the noise component of the inverse propagator,
$\langle \xi(t, \mathbf{x}) \xi^{*}(t', \mathbf{x}') \rangle = 2 \gamma \delta(t
- t') \delta(\mathbf{x} - \mathbf{x}')$.
Here and in the following, we omit the subscript ``$c$'' of the field that
appears in the Langevin equation, with the understanding that such equations
always apply to classical fields. Compared to the deterministic saddle point
equation discussed in the last subsection, the Langevin equation differs by the
addition of noise. This random force is the element that allows the system to
explore configurations beyond the deterministic path. The description in terms
of a Langevin equation establishes the formal connection to the paradigmatic
models of classical dynamical critical behavior, which are determined by systems
of coupled Langevin equations for the order parameter field and slow
hydrodynamic modes that are associated with conservation laws
\cite{Hohenberg1977}. However, a fundamental difference between the models of
Ref.~\cite{Hohenberg1977} and Eq.~\eqref{eq:driven-open-condensate-Langevin}
lies in the fact that the former correspond to systems in thermal equilibrium
whereas the latter describes a condensation transition in a nonequilibrium
steady state. Formally, this can be seen by noting that the Keldysh action
Eq.~\eqref{eq:S-driven-open-condensate-semiclassical} is, in general, not
invariant under the semiclassical limit of the transformation
$\mathcal{T}_{\beta}$ Eq.~\eqref{eq:thermal-symmetry}. The latter is a symmetry
of the Keldysh action only when $K_c/K_d = u_c/u_d$. We note that the analogous
condition $K_c/K_d = r_c/r_d$ for the two mass scales $r_c$ and $r_d$ can always
be satisfied in a suitably chosen rotating frame, i.e., by modifying $r_c$
through a transformation
$\psi_c(t, \mathbf{x}) \mapsto \psi_c(t, \mathbf{x}) \e^{- \imag \omega t}$. A
violation of the condition $K_c/K_d = u_c/u_d$ leads to a modification of
critical behavior at the condensation transition \cite{Sieberer2013,
  Tauber2013a, Sieberer2014, Sieberer2016a}. This is discussed in more detail in
Sec.~\ref{sec:bosonic-driven-open-criticality}.

\paragraph*{Absence of a semiclassical limit for models with quantum scaling.}

As stated above, quantum criticality can be realized for a system coupled to a
thermal bath at zero temperature, such that $P^K \sim \omega \sim q^2$, leading
to $\psi_c \sim \psi_q \sim q^{d/2}$.  Consequently, interaction vertices with
different numbers of classical and quantum fields but an equal total number of
fields are equally relevant, and a semiclassical description is not
applicable. The simultaneous vanishing of the spectral and noise gaps
corresponds to a double fine-tuning that is required to realize nonequilibrium
quantum criticality, see Sec.~\ref{sec:quantum-criticality}. This is analogous
to quantum phase transitions in thermal equilibrium, where one needs to
fine-tune not only a system parameter to its critical value, but also the noise
gap to zero, via $T \to 0$.

\section{Absorbing state phase transitions and directed
  percolation}\label{Sec:DP} 
 
We now start our overview of instances of universality in driven open systems,
starting with realizations of paradigmatic classes of nonequilibrium
universality using quantum simulators (see Tab.~\ref{fig:overview}).
An important class of genuine nonequilibrium phase transitions are absorbing
state phase transitions. They appear in a variety of classical dynamical
systems, e.g., in population models \cite{PopulationRev}, in
epidemics \cite{PastorSatorras2015} or in chemical
reactions \cite{OdorRev}. Here, the \emph{absorbing state} corresponds to a
macroscopic configuration which may be reached during the time evolution but
which can never be left. A drastic example is the extinction of an entire
species in population models.

An absorbing state phase transitions is a transition in the structure of the dynamics: an absorbing or inactive phase is separated from an active phase. In the former, the absorbing state is dynamically stable and attractive. It will be reached from any initial state in finite time, leading the dynamics to always terminate in this particular state. In contrast, in the active phase, the absorbing state remains a 
stationary solution of the evolution, \emph{but} it becomes dynamically unstable. The system traverses a large part of the accessible configuration space before reaching the absorbing state. This leads to the persistence of nontrivial dynamics up to exponentially large time scales $t\sim \exp(N)$, where $N$ is the number of degrees of freedom, e.g., the sites of a lattice, and $\exp(N)$ is the size of the configuration space explored by these degrees of freedom. In the thermodynamic limit, $N\rightarrow\infty$, the absorbing state thus becomes unreachable and the dynamics proceed indefinitely. 

An absorbing state breaks detailed balance, which becomes crucial close to the
absorbing state phase transition. It invalidates the description of the critical
dynamics in terms of an equilibrium ensemble. This may be illustrated in a
configuration space picture: when approaching the absorbing state, say the
extinction of a species, the accessible configuration space (the number of
individuals), becomes smaller and smaller until it shrinks down to a single
point. In such a scenario, not only the order parameter (the average number of
individuals) approaches zero, but also the fluctuations of the order parameter, i.e., its noise, vanishes continuously. The noise thus scales proportional to the order parameter itself (or powers thereof). In many
cases, this causes the noise gap discussed in Sec.~\ref{sec:noisegap} to become proportional
to the order parameter, $P^K\sim \phi$, when approaching the transition, and
to vanish at the critical point.

This poses a challenge for the field theory approach to absorbing state transitions: an order parameter-dependent noise, which vanishes at the transition and obeys scaling in the vicinity of the critical point is fundamentally different from the noise at equilibrium phase transitions. For the latter, noise is either governed by thermal or quantum fluctuations, and thus does not change its structure when tuning across the critical point, see Sec.~\ref{sec:noisegap}. This leads to characteristic and significant modifications of the dynamical action and the scaling dimensions of the fields, which are discussed in the following.

Though the majority of known absorbing state phase transitions appear in classical systems, the concept is general and can be extended to driven open quantum many-body dynamics. Consider a system of $N$ degrees of freedom, e.g., spins, bosons or fermions. Its quantum state $\hat\rho$ shall follow a Lindblad master equation of the form
\begin{equation}\label{eq:DSMaster}
    \partial_t\hat\rho=\mathcal{L}_1(\hat\rho)+\gamma\mathcal{L}_2(\hat\rho).
\end{equation}
Here $\mathcal{L}_{1,2}$ are two many-body Lindbladians and $\gamma$ is a dimensionless tuning parameter. An absorbing state $\hat\rho_D$ has zero statistical fluctuations, i.e., is a represented by a pure state $\hat\rho_D=|D\rangle\langle D|$, with wave function $|D\rangle$. It is the dark state of both Lindbladians, $\mathcal{L}_{1,2}(\hat\rho_D)=0$, and thus the stationary solution of Eq.~\eqref{eq:DSMaster}, independently of the value of $\gamma$ (see Appendix~\ref{sec:dark}). In order to realize a dynamical transition, one may assume that $\hat\rho_D$ is a repulsive or unstable fixed point of $\mathcal{L}_1$, while being an attractive or stable fixed point of $\mathcal{L}_2$. Thus $\mathcal{L}_2$ pushes any initial state $\hat\rho$ toward $\hat\rho_D$, while $\mathcal{L}_1$ pushes any initial state $\hat\rho\neq\hat\rho_D$ away from the dark state---an intuitive example is $\mathcal{L}_1(\hat\rho)=-i[\hat H,\hat\rho]$ for some Hamiltonian $\hat H$, and $\hat\rho_D$ being one of the eigenstates of the Hamiltonian.

The absorbing state phase transition roots in the competition of $\mathcal{L}_1$ and $\mathcal{L}_2$, tuned by the parameter $\gamma$, and deciding whether the dark state $\hat\rho_D$ will be reached or not in the thermodynamic limit. A possible way to construct an order parameter is to take a set of operators $\{\hat O_l\}$ to which $|D\rangle$ is an eigenstate, $\hat O_l|D\rangle=o_l|D\rangle$. Then $\hat\varphi=\tfrac{1}{N}\sum_l (\hat O_l-o_l)$ is zero and fluctuationless in the stationary state of the absorbing phase, while it is nonzero and fluctuating in the active phase. 

An elementary, paradigmatic model for absorbing state phase transitions is \emph{directed percolation} (DP); a review of general absorbing state phase transitions is provided in Refs.~\cite{Hinrichsen2000,OdorRev}. DP describes general absorbing state phase transitions of a real scalar order parameter field without any particular symmetries, disorder or long range couplings. Despite its theoretical simplicity, it has been experimentally observed only in certain classical, turbulent systems \cite{Takeuchi2007,Lemoult2016} and in quantum simulators with ultracold Rydberg atoms \cite{Gutierrez2017absorbing, HelmrichSOC, KlockeHydro}, see below.

\subsection{Rydberg atom lattice in the facilitation
  regime}\label{subsec:rydberg}  

Directed percolation, and variants of it, can be realized in driven-dissipative
Rydberg atom ensembles. In order to illustrate the microscopic origin of the
phenomenon, one may consider a $d$-dimensional array of atoms, e.g., created
with optical tweezers or in an optical lattice, where neighboring atoms are kept
at a fixed relative distance $R$, and their motional degree is frozen. The atoms
are then optically driven into an excited state with a high principal quantum
number (see, e.g., Refs.~\cite{SaffmanRev,AdamsRev,PfauRev} for reviews on the
microscopic physics of Rydberg atoms). This generates a lattice of two level
systems (ground and excited states), where the excited atoms interact with a
dipole-induced van der Waals interaction. In a pseudo-spin representation of the
atomic ground $\ket{\downarrow}$ and excited $ \ket{\uparrow}$ states, where
$\hat{\sigma}^x_l=\ket\uparrow\bra\downarrow_l+\ket\downarrow\bra\uparrow_l$ is
the Pauli matrix and ${\hat{n}_l=\ket{\uparrow} \bra{\uparrow}_l}$ is the
projector onto the excited state acting on the $l$-th atom, this is described by
the quantum Ising Hamiltonian
\begin{eqnarray}
  \label{eq:RydbergIsing}
  \hat H=\sum_{l}\left(-\Delta \hat{n}_l+\sum_{m} V_{l,m} \hat{n}_l \hat{n}_m+\frac{\Omega}{2}\hat{\sigma}^x_l\right).
\end{eqnarray}
The parameters are the detuning $\Delta$ of the drive frequency from the atomic
resonance and the Rabi frequency $\Omega$, set by the driving field
intensity. Excited states repel each other with the van der Waals interaction
$V_{l,m}=\frac{C_6}{|\mathbf{x}_l-\mathbf{x}_m|^6}$, where $C_6$ is the van der
Waals coefficient of the atom and $\mathbf{x}_l$ is the position of atom $l$ in
the lattice \cite{Weber_2015}.

The collective behavior of the Rydberg atoms can be modified by adjusting the
detuning $\Delta$. When $|\Delta|,|\Omega|\ll|V_{l,m}|$, the Hamiltonian suppresses the
simultaneous excitation of neighboring lattice atoms and the array is in the
blockade regime \cite{Chotia_2008,Schauss2015Crystal}. By tuning $\Delta$
($\Omega$) the ground state undergoes a classical (quantum),
$\mathds{Z}_s$-symmetry breaking phase transition from a paramagnet to an
ordered state (the integer $s$ is set by the range of the
blockade) \cite{Bernien2017,Labuhn2016}.

An intriguing dynamical regime, the \emph{facilitation regime} is realized in
the opposite case when the detuning $\Delta$ is large compared to the Rabi
frequency $|\Delta|\gg\Omega$. Then single atom excitations are energetically
unfavorable, i.e., significantly suppressed by a factor $\Omega^2/\Delta$. Atoms
can only transfer between the excited and the ground state when the mutual
interaction $\Delta=V_{l,m}$ induced through neighboring excited states
compensates the detuning. This regime is known as anti-blockade or facilitation
regime since excited atoms which push their neighbors into resonance act as
seeds that \emph{facilitate} the spreading of
excitations \cite{Ates2007,Schempp2014,Malossi2014,Urvoy2015,Lee2011,Lesanovsky2013,Carr2013,Gutierrez2017absorbing,Helmrich2018,HelmrichSOC,Goldschmidt2016,Marcuzzi2017,Lee2012,PfauRev,Heidemann2007,Weimer2008,Amthor2010,Gart2013,Simonelli2016,Morsch2016a,Lesa2014,Morsch2016,Pupillo2010,Buchhold2017,BuchholdFirstOrder,valencia2023rydberg}. On
time scales $t\lesssim |\Delta/\Omega^2|$, off-resonant processes can be perturbatively
eliminated, yielding the effective Hamiltonian
\begin{eqnarray}\label{eq:FacHam}
  \hat H = \frac{\Omega}{2} \sum_{\langle l,m\rangle} \hat{\sigma}_l^x \hat{n}_m,
\end{eqnarray}
where the sum runs over nearest neighbors $l,m$. The Hamiltonian \eqref{eq:FacHam} contains some simplifications, e.g., it neglects the effect of multiply occupied neighbors, which can cause geometric frustration effects, see Refs.~\cite{BuchholdFirstOrder,PhysRevE.94.052108}. However, it captures the essence of the dynamics: (i)~it decouples the total spin down state $\ket{\downarrow\downarrow\downarrow \cdots \downarrow}$ from all excited states and (ii)~it leads to the spreading of excitations from individual seeds (spin-up particles).

In addition to the Hamiltonian dynamics, excited atoms spontaneously emit
optical photons when they decay back into the ground state. This happens at a rate $\gamma$ and is modeled by the master equation for the density matrix $\hat\rho$:
\begin{eqnarray}\label{eq:FacMasterEq}
  \partial_t\hat\rho=-i \left[\hat H,\hat \rho\right]+\gamma\sum_l \left(2\hat L_l\hat\rho\hat
  L^\dagger_l-\left\{\hat L^\dagger_l \hat L_l,\hat\rho \right\} \right),
\end{eqnarray}
with Lindblad jump operators
$\hat{L}_l=\hat{\sigma}^-_l=\ket\downarrow\bra\uparrow_l$. The master
equation~\eqref{eq:FacMasterEq} always has a stationary solution
$\hat\rho_D=|D\rangle\langle D|$ with zero excitations
$\ket{D} =\ket{\downarrow\downarrow\downarrow \cdots \downarrow}$, as follows
from $\hat H\hat\rho_D=\hat L_l\hat\rho_D=0$. This is the absorbing state. For
strong $\Omega>\gamma$, the rapid facilitation pushes the system away from reaching
this stationary state. Instead, the dynamics explore the full $2^N$-dimensional
Hilbert space of $N$ pseudo-spins. The absorbing state still terminates the evolution but it is now reached with the same probability as any other accessible state in Hilbert space. Analogous to classical dynamical systems, it is thus approached only on
exponentially long times $t\sim \exp(N)$, and the system is in the active phase \cite{Hinrichsen2000}.

\subsection{Field theory for directed percolation}\label{sec:FieldTheoryDP}

The order parameter for the transition between the active and the absorbing
state is the excitation density $\langle \hat{n}_l\rangle$. In order to study
its dynamics in a field theory framework, one may either derive the Keldysh
field integral for the master equation \eqref{eq:FacMasterEq}, see,
e.g., \cite{Buchhold2017}, or construct the Heisenberg-Langevin equation for the
operators $\hat{n}_l$ and then promote the Langevin equation to a Keldysh path
integral via the Martin-Siggia-Rose-Janssen-de Dominicis functional
approach \cite{BuchholdFirstOrder,KlockeControl}. Both formulations are equivalent in
the semiclassical limit, i.e., when higher powers in the quantum fields can be
neglected, see Sec.~\ref{sec:semiclassical-limit}. This limit bears strong
similarities to other semiclassical approaches, such as, e.g., the truncated
Wigner formalism \cite{WeimerTWA, Mink2022} (see Ref.~\cite{Polkovnikov2010} for
a discussion). The Langevin equation approach is optimally
suited to demonstrate the dark state property of the physical dynamics, as
outlined below.

The dynamics of any pseudo-spin operator
$\hat O_\alpha= \hat{\sigma}^x_l, \hat{\sigma}^y_l, \hat{n}_l$ are governed by
an operator-valued Heisenberg-Langevin equation
$\partial_t \hat O_{\alpha}=\mathcal{L}^\dag(\hat O_\alpha)+{\hat
  \xi}_{\alpha}$.
The operator $\mathcal{L}^\dag(\hat O_\alpha)$ is deterministic and
$\hat \xi_{\alpha}$ represents a \emph{quantum
  noise} \cite{Scully}. $\mathcal{L}^\dag(\hat O_\alpha)$ describes the action
of the adjoint master equation,
\begin{eqnarray}
  \mathcal{L}^\dag(\hat O_\alpha) = i \left[\hat H,\hat
  O_\alpha\right]+\gamma\sum_l\left(2\hat L^\dagger_l \hat O_\alpha \hat L_l- \left\{\hat L^
  \dagger_l\hat L_l,\hat O_\alpha\right\}
  \right).
\end{eqnarray}
Since the time evolution is not unitary, $\hat O_\alpha$ experiences statistical
fluctuations, i.e.,
$\mathcal{L}^\dag (\hat O_\alpha \hat O_\beta) \neq \hat
O_\alpha\mathcal{L}^\dag (\hat O_\beta) +\mathcal{L}^\dag(\hat O_\alpha) \hat
O_\beta$.
The fluctuations are encoded in the quantum noise $\hat \xi_{\alpha}$. The noise
has zero mean and an operator-valued variance
$\hat\chi_{\alpha\beta}= \hat \xi_{\alpha}\hat \xi_{\beta}$, determined by the
Einstein relation
\begin{equation}
  \hat\chi_{\alpha\beta} =\partial_t\left(\hat O_\alpha \hat O_\beta\right)-\hat
  O_\alpha\mathcal{L}^\dag (\hat O_\beta)-\mathcal{L}^\dag (\hat O_\alpha) \hat O_\beta.
\end{equation}

The pseudo-spin coherences $\hat{\sigma}_l^{x,y}$ are exponentially suppressed by the spontaneous emission. Eliminating them in second order perturbation theory (adiabatic elimination) yields the Heisenberg-Langevin equation
\begin{eqnarray}\label{eq:QuantumContact}
\partial_t \hat{n}_l= -\gamma \hat{n}_l +\frac{\Omega^2}{\gamma} \left( 1-2
  \hat{n}_l \right) \left(\sum_{\langle m, l \rangle} \hat{n}_m \right)^2+\hat\xi_l,
\end{eqnarray}
where the sum is restricted to nearest neighbors of $l$ and the noise has
variance $\hat\xi_l\hat\xi_m= \gamma \delta_{l,m} \hat{n}_l$. Importantly, the
noise depends on the operator $\hat n_l$, representing multiplicative quantum
noise. The square over the sum of the nearest neighbor excitations is of purely
quantum mechanical origin. It enhances the effective Rabi frequency $\Omega$ by
a factor $\sqrt{N}$, where $N$ is the number atoms in a coherent superposition
of excited states, known as the Rydberg superatom \cite{Weber_2015}. This
quantum mechanical, nonlinear enhancement of the facilitation probability can
result in a first-order dark state phase transition \cite{Marcuzzi2016}, which
is discussed in Sec.~\ref{sec:dark_state_bistable}.

The evolution equation for the excited state densities $\hat{n}_l$ describes the
competition between the local decay of excitations with rate $\gamma$ and the
spreading of excitations to nearby lattice sites with rate $\Omega^2/\gamma$. It
is an operator-valued version of the classical \emph{contact
  process} \cite{Luebeck2006}. However, all operators commute, and thus one
obtains effective classical dynamics from Eq.~\eqref{eq:QuantumContact}.
Indeed, the absorbing state phase transition is in the directed percolation
universality class. The key ingredient for an absorbing state is that the
variance $\hat\chi$ is multiplicative, proportional to the order parameter.

The effective long-wavelength theory for the contact process is derived by
coarse graining, i.e., by averaging the density $\hat{n}_l$ over a suitable
number of lattice sites. In the operator-valued version, one needs to also take
the quantum mechanical average, accompanied by a mean-field decoupling of
products $\langle \hat{n}_l \hat{n}_m\rangle\rightarrow \langle
\hat{n}_l\rangle\langle \hat{n}_m\rangle$, which is perturbatively controlled by
the low excitation density close to the dark
state \cite{Marcuzzi2016,Buchhold2017,KlockeControl}. Introducing the
coarse-grained order parameter field,
\begin{equation}
  \phi(t, {\bf x})=\frac{1}{V_{{\bf x}}}
  \sum_{l\in V_{\bf x}}\langle \hat{n}_l(t)  \rangle,
\end{equation}
as the average density of excitations in a small volume $V_{\bf x}$ centered at
the continuum coordinate ${\bf x}$, yields the Langevin equation (we set
$\phi\equiv\phi(t, {\bf x})$)
\begin{equation}\label{eq:LangeDP}
  \partial_t\phi = \left( D \nabla^2 - m \right) \phi-\kappa_2\phi^2-\kappa_3\phi^3+\eta.
\end{equation}
Here $\phi=\langle\phi_c\rangle$ corresponds to the average of the classical field in the Keldysh framework and the noise $\eta=\eta(t, \mathbf{x})$ is normally distributed with zero mean and
variance
$\langle \eta(t,{\bf x}) \eta(t',{\bf x}') \rangle=\delta(t-t')\delta({\bf
  x}-{\bf x}')\gamma\phi(t,{\bf x})$. Thus
 the noise gap is set by the order parameter. The precise values of
$\kappa_2$, $\kappa_3$ and $m$ depend on the microscopic details, such as, e.g.,
dephasing \cite{HelmrichSOC,Helmrich2018} and the dimension.

When the strength of the leading-order nonlinearity is positive $\kappa_2>0$, it
dominates the relaxation close to the transition.  The cubic term $\sim \phi^3$
is then subleading for small values of $\phi$ and can be neglected
$\kappa_3\rightarrow0$, i.e., it does not modify the universal scaling at the
transition. The corresponding Langevin equation describes the field theory for
\emph{directed percolation} (DP). On the mean-field level, it undergoes a
continuous phase transition from an absorbing phase ($\phi=0, m>0$) to an active
phase ($\phi>0, m<0$) at $m=0$ (approximately at $\Omega=\gamma$). In contrast,
if $\kappa_2<0$, the field $\phi$ experiences a bistability and the absorbing
state phase transition becomes first order on the mean-field
level \cite{BuchholdFirstOrder,Buchhold2017,Marcuzzi2015}. At $\kappa_2=0$,
mean-field theory predicts a first-order phase transition, which is separated
from the continuous directed percolation transition by a tricritial point. The
universality class of this point, and the field theory describing it, is known
as \emph{tricritical directed percolation}
(TDP) \cite{GrassbergerTDP,Ohtsuki2987,Luebeck2006}.

\subsection{Universality and symmetries}
As it is common for continuous phase transitions, the universality class of the transition is determined by symmetries (or the absence thereof), and a unique set of critical exponents. 
The peculiar universal behavior of DP and TDP, and the manifestation of an absorbing state, may be better understood in a nonequilibrium field integral framework. From the Langevin Eq.~\eqref{eq:LangeDP}, one can derive the equivalent Keldysh action from the Martin-Siggia-Rose-Janssen-de Dominicis construction \cite{Buchhold2017} (see also Sec.~\ref{sec:semiclassical-limit}):
\begin{eqnarray}
  S=\int d \mathbf{x} \, dt \, \phi_q \left(\partial_t-D\nabla^2+m+\kappa_2\phi_c+\kappa_3\phi^2_c-\gamma\phi_q\right)\phi_c. \nonumber
\end{eqnarray}
For $\kappa_2>0$, the cubic nonlinearity can be set to zero as argued above, i.e., $\kappa_3=0$. Upon rescaling, the action is invariant under the transformation 
\begin{eqnarray}
\left( \phi_c(t), \phi_q(t) \right) \mapsto - \left( \phi_q(-t), \phi_c(-t) \right).
\end{eqnarray}
This invariance is known as rapidity-inversion symmetry and is characteristic for DP \cite{JANSSEN2005}. It has important consequences for the critical behavior. For instance, it implies that the classical field $\phi_c$, i.e., the order parameter, and the quantum field $\phi_q$, possess the same scaling dimension $\sigma$ ($\sigma=-d/2$ at the Gaussian fixed point). This is unconventional at a classical phase transition, and reminiscent of quantum critical behavior (see Sec.~\ref{Sec:quantumScaling}). At the same time, the term quadratic in the quantum fields $\sim \gamma \phi_c \phi_q^2 $ implies a global effective temperature $T_{\text{eff}}\equiv\gamma\phi_c$. It thus depends on the order parameter itself and obeys scaling close to the critical point. In addition, the symmetry imposes the same scaling for $\kappa_2$ and $\gamma$, and sets the upper critical dimension to be $d_u=4$. The rapidity-inversion symmetry guarantees another important relation: the causality condition $S[\phi_c,\phi_q=0]=0$ implies under rapidity inversion that $S[\phi_c=0,\phi_q]=0$. The latter is in fact generic for absorbing state phase transitions. It reflects the absence of fluctuations in the dark state $\phi_c=0$.  

In addition to featuring this peculiar symmetry, the exact critical exponents at
the directed percolation transition, i.e., at the Wilson-Fisher fixed point, are
strongly modified compared to the Gaussian fixed point. While such a strong
modification can be inferred from perturbative renormalization group
approaches \cite{Janssen1,BronzanDP,Janssen1981,WhitelamDP}, and is confirmed by
nonperturbative
formulations \cite{Canet2004,Canet2005,Canet2006,Gredat2014,BuchholdBackgroundField,Buchhold2017},
a precise determination of the critical exponents from such methods remains
challenging, see Ref.~\cite{Dupuis_2021} for an overview. Numerical simulations
of the Langevin equation~\eqref{eq:LangeDP} are likewise challenging due to the
multiplicative noise, requiring a specific split-step integration
scheme \cite{Dornic2005,KlockeControl,Dickman1994}. The most accurate values for
the critical exponents were obtained from numerical simulations of the discrete
contact process, i.e., the classical version of Eq.~\eqref{eq:QuantumContact},
see, e.g.,
Refs.~\cite{Jensen_1999,Jensen1992,Voigt1997,Grassberger1996,Luebeck2006} or the
review article Ref.~\cite{Hinrichsen2000}.

At the tricritical point, $\kappa_3$ is relevant
and the rapidity inversion symmetry is lost. The
scaling dimensions of $\phi_c, \phi_q$  are no longer pinned to each other. Two-loop perturbation theory reveals an upper critical dimension $d_u=3$ \cite{BuchholdFirstOrder, Ohtsuki2987}. Numerical simulations \cite{Luebeck2006} and a functional renormalization group approach \cite{BuchholdFirstOrder} showed, however, a non-Gaussian tricritical fixed point only exists in $d\ge 2$ dimensions, while in $d=1$ the transition in the classical model flows towards the directed percolation fixed point.  

\subsection{The quest for quantum directed percolation}\label{sec:DPquant}

The important role of absorbing state phase transitions, and of directed
percolation, in classical nonequilibrium systems naturally fuels the question,
whether a genuinely quantum mechanical version of absorbing state phase
transitions may exist. The discovery of absorbing states in facilitated Rydberg
ensembles has sparked intense research in this
direction \cite{Marcuzzi2015,Marcuzzi2016,Morsch2016,Gutierrez2017absorbing,Carollo2019,Gillman_2019,Lesanovsky_2019,Gillman2020,Gillman2021,Minjae2019,Jo_2021}.

The idealized absorbing state
$\ket{\downarrow\downarrow\downarrow \cdots \downarrow}$ is a classical product
state, and the Heisenberg-Langevin equation \eqref{eq:QuantumContact} in the
computational basis remain effectively classical. Thus, for this idealized
scenario, quantum correlations may be present on intermediate length and time
scales and affect the short-distance scaling behavior. At the longest distances,
however, coherent effects will unavoidably vanish when approaching the classical
dark state. This is confirmed by the evolution of the quantum coherence being gapped,
including at the critical point \cite{Buchhold2017}. Thus the transition in this
idealized model will become classical \cite{BuchholdFirstOrder}, and exhibits
the universal behavior of directed percolation. Recent experiments in the
facilitation regime
\cite{Morsch2016,Simonelli2016,Gutierrez2017absorbing,HelmrichSOC,KlockeHydro}
and on comparable quantum simulation platforms \cite{Chertkov2022} report
behavior that is in agreement with directed percolation. However, direct
simulations of the master equation \eqref{eq:FacMasterEq} in an iTEBD framework
reported a set of non-DP critical exponents in one spatial dimension
\cite{Carollo2019,Gillman_2019}. Identifying the origin of this quantum
dynamical behavior poses a questing to theory and challenges the idealized,
semiclassical model of directed percolation. Such ideas have been put forward in
recent works, e.g. in Refs.~\cite{Brady2024,Brady2024a}.

Also alternative formulations of absorbing state phase transitions in quantum
systems have been developed: The authors of
Refs.~\cite{Lesanovsky_2019,Gillman2020,Gillman2021} have studied quantum
generalizations of discrete cellular automata, where short-ranged quantum
correlations were observed in the vicinity of an absorbing phase
transition. Absorbing state phase transitions in long-range interacting Rydberg
systems were investigated in Refs.~\cite{Minjae2019,Jo_2021}. There a continuous
transitions in the DP universality class was found in two spatial dimensions. In
one dimension, static exponents agreed with those of DP but a crossover in the
temporal scaling from a non-DP behavior to DP was reported for specific initial
states \cite{Jo_2021}. Finally, a phase transition between two different types
of absorbing states in a Rydberg gas was discussed in Ref.~\cite{Carollo2022}
but the nature of the phase transition remains elusive. Despite these efforts,
understanding the fate of nonequilibrium quantum correlations in the presence of
decoherence, and confirming the existence of a quantum absorbing state phase
transition with unique, universal behavior at the largest distances remains a
challenge.

\section{Self-organized criticality and Rydberg experiments}\label{sec:SOC}

Scale invariant behavior appears ubiquitously in complex dynamical systems. Its
observation ranges from large (even astronomical) scales, e.g., in solar flare
activity, earthquakes, avalanches and disease spreading, to small scale devices,
such as neural networks and electrical circuits. Many of its realizations share
in common that, even though influenced by a variety of different external
parameters, they seem to be robustly attracted toward a scale invariant, and in
this respect \emph{critical}, state without the need for external fine-tuning.

In a seminal paper in 1987, Bak, Tang and Wiesenfeld \cite{Bak1987,Bak1988}
introduced the Abelian sandpile model in order to explain the emergence of scale
invariance in complex systems. Here, the use of the word `Abelian' means that the geometric order in which
local updates, i.e. the deposition of sand and the propagation of avalanches, are performed is irrelevant. Despite its simplicity, the model
features self-organizing dynamics, where the interplay between drive and
dissipation evolves the system toward a critical, scale-invariant state without
the need for fine-tuned external parameters. They termed the phenomenon
\textit{self-organized criticality} (SOC). The hallmark of SOC, witnessing its
critical behavior, is the frequent but random creation of excitation avalanches,
whose lifetimes and sizes follow a scale invariant
distribution \cite{Altshuler2004,Field1995,Swingle,Aschwanden2016,Turcotte1999,
  Levina2007,Zierenberg}.

\begin{figure}
  \includegraphics[width=246pt]{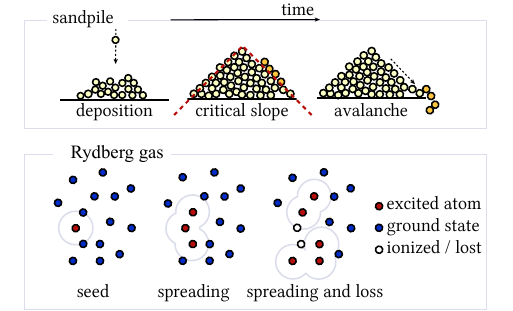}
  \caption{Top: Illustration of the sandpile model. Sand is piling up on a plate
    via slow deposition ($=$ drive). The sandpile reaches a critical slope at
    which friction and gravitation balance each other. Further deposition
    triggers scale invariant avalanches, which deplete at the boundaries ($=$
    dissipation). Bottom: Illustration of excitation spreading in a Rydberg gas
    in the facilitation regime. Excited atoms (red) act as seeds which
    facilitate the excitation of neighboring atoms that traverse the
    facilitation shell (gray circles). This leads to spreading of
    excitations. Ionization or loss of excited atoms yields a depletion of the
    atom density proportional to the number of excited atoms.}
\label{fig:Sandpile_Ryd}
\end{figure}

The appeal of the sandpile model roots in its simplicity. Its two-dimensional
version, for instance, can be illustrated by a simple toy model: take a plate of
circular shape, in the center of which some external drive deposits grains of
sand at a slow rate. A pile of sand fills up on the plate until a \emph{critical
  slope} is reached at which gravitation and friction balance each
other. Further deposition of sand generates a supercritical slope, which
eventually releases an avalanche of sand, rolling off the pile and dissipating
at the boundary of the plate, see Fig.~\ref{fig:Sandpile_Ryd}. Upon further
deposition the phenomenon repeats indefinitely, leading to a frequent
manifestation of excitation avalanches. This illustration reveals the mechanism
underlying SOC: the fine-tuning of external parameters required for conventional
critical points is replaced by a strict separation of time scales between
(i)~the systems internal dynamics (gravitation, friction) and (ii)~drive and
dissipation. For instance, a critical slope is only maintained when the
deposition of sand is much slower than its diffusion. Otherwise the pile would
enter a supercritical state, adding sand faster than it decays.

The work of Bak, Tang and Wiesenfeld fueled the research activity on
self-organized criticality. A number of similar models have been proposed in
order to describe SOC phenomena in a variety of systems. Notable examples
include the forest-fire model \cite{Drossel1992,Grassberger_forestfire}, the
Olami-Feder-Christensen model for
earthquakes \cite{Olami_Feder_Chris1,Grassberger_Olami}, the Bak-Sneppen model
for the coevolution of species \cite{Sneppen1,Sneppen2}, and modifications of
the original sandpile
model \cite{Turcotte1999,dickman2000,Aschwanden2016,pruessner2012self,Markovic2014}. Despite
ongoing research activity in SOC and the large number of potential
self-organized critical systems in Nature, there is no strict consensus on the
precise conditions that lead to truly self-organized critical behavior, i.e.,
scale invariance on all time and length scales, or whether one observes only
apparent critical behavior on intermediate scales (also termed self-organized
quasi-criticality) \cite{pruessner2012self,Bonachela2009,Bonachela2010}.

\subsection{Field theory for self-organized criticality}

Self-organized criticality is closely linked to absorbing state phase
transitions \cite{Bonachela2009,Bonachela2010,dickman2000,Munoz2018,Jensen2018,Watkins2016,pruessner2012self,Vespignani1998,Markovic2014}. Both
bear the same phenomenology at the critical point: although the stationary
density of excitations is zero or infinitesimal, a single, local seed can start
an extensive excitation avalanche growing over large parts of the system. Owing
to criticality, the size and lifetime of those avalanches follows a scale
invariant distribution function \cite{Munoz1999}.

From this viewpoint, SOC can be viewed as a phase transition between an
absorbing phase (corresponding to a subcritical slope and a vanishing density of
excitations) and an active phase (corresponding to a supercritical slope and a
nonzero density of excitations). By the interplay of drive and dissipation, the
system is constantly pushed toward the critical point. This can be achieved,
for instance, when the drive pushes the system from the absorbing to the active
phase, e.g., by depositing sand, and dissipation is \emph{conditioned} on the
active degrees of freedom, e.g., by removing excitation avalanches at the
boundary \cite{dickman2000}.  Thus, self-organized criticality is described by a
modified field theory for absorbing state phase transitions such as directed
percolation discussed in Sec.~\ref{Sec:DP}. Here the order parameter
$\phi(t, {\bf x})$ again represents the local density of excitations, e.g.,
forming an avalanche. To the conventional absorbing state field theory, SOC adds
the slope field $n(t,{\bf x})$, which acts as a local, \emph{dynamical mass}
$m\to m-n(t, {\bf x})$, i.e., the time-dependent slope around the critical
point. An important example is the extension of directed percolation to SOC, for
which the order parameter field evolves according to the Langevin equation
(cf.\ Eq.~\eqref{eq:LangeDP}) \cite{Bonachela2009,Bonachela2010,dickman2000,Munoz2018,Munoz2020}
(we set
$\phi\equiv \phi(t, {\bf x}), \eta\equiv\eta(t, {\bf x}), n(t, {\bf x})\equiv
n$)
\begin{equation}\label{eq:LangeSOC1}
\partial_t\phi = \left( D\nabla^2-m+n \right) \phi-\kappa\phi^2+\eta, 
\end{equation}
with a noise gap
$\langle \eta(t,{\bf x})\eta(t',{\bf x}') \rangle =\delta(t-t')\delta({\bf
  x}-{\bf x}')\gamma\phi(t,{\bf x})$.
The dynamical mass $n$ obeys the equation of motion
\begin{eqnarray}\label{eq:SlopeEq}
\partial_t n=\left(D \nabla^2-\mu\right)\phi+\delta.
\end{eqnarray}
One distinguishes several cases: (i)~For $\mu=\delta=0$, the total mass
$\int d \mathbf{x} \, n$ is conserved, which corresponds to the Manna model or
\textit{conserved directed percolation} \cite{Bonachela2009,Vespignani1998}. It
describes the dynamics of the Abelian sandpile model if the initial average
density is at the critical point. (ii)~The case $\mu>0, \delta=0$ corresponds to
a model with bulk dissipation, e.g., as in the forest fire model without
regrowth \cite{CLAR1999153}. The system may evolve toward a critical point and
displays transient critical behavior if $\mu$ is the smallest scale in the
theory. Asymptotically, it drops below the critical point and ends up in the
absorbing phase \cite{Jensen2018}, featuring a universal dynamics known as
dynamical percolation \cite{Janssen1981}. (iii)~For $\mu,\delta>0$ one finds a
reloading mechanism, similar to a forest fire model with regrowth. It pushes the
system into the active phase, with an average density $\phi=\delta/\mu$, which
acts as an effective mass scale. The cases (ii)~and (iii)~do not display true
critical behavior on all length scales due to the remaining scale $\mu$ or
$\delta/\mu$. These dynamics may appear self-organized critical on large time
and length scales \cite{KlockeControl} and thus has been termed self-organized
quasi-critical \cite{Bonachela2009}.

The universal scaling exponents of SOC are related to the size and lifetime of
excitation avalanches. Denoting by $s$ the maximum spatial extension of a given
avalanche and by $t_\ell$ its lifetime, and the respective distribution
functions by $P(s)$ and $P(t_{\ell})$, the scaling exponents $\alpha$, $\tau$,
and $\gamma$ are defined as follows:
\begin{equation}
  P(s)\sim s^{-\tau}, \quad P(t_\ell)\sim t_\ell^{-\alpha}, \quad \langle s\rangle_{t_\ell}\sim t_\ell^\gamma.
\end{equation}
The scaling exponents are unique for a unique absorbing state, e.g., for the
sandpile model and for the conserved directed percolation model, but have been
shown to vary for models with multiple absorbing
states \cite{Munoz1999,Bonachela2009}. This indicates that the cases (ii)~and
(iii)~defined above may not display unique scaling exponents, and instead
$\alpha$, $\tau$, and $\gamma$ may depend on $\delta,\mu$. However, no such
dependence was found in Ref.~\cite{KlockeControl}.

\subsection{Self-organized criticality in Rydberg atom ensembles}

In experiments with Rydberg gases in the facilitation regime, excitation avalanches have been observed in several different setups \cite{Goldschmidt2016,Boulier2017,Simonelli2016,RydAvalancheGross,HelmrichSOC,Ding2020}. Recently, signatures of self-organized criticality, i.e., the self-organization toward a unique density of atoms and a scale invariant distribution of excitation avalanches, have been reported \cite{HelmrichSOC, Ding2020, KlockeHydro}. At the heart of this connection is a conditioned loss  mechanism: excited atoms in the Rydberg state have a small probability to become ionized and to subsequently escape from the trap. The average number of potential facilitation partners thus decays, and the local decay rate is proportional to the number of excited, i.e., active, atoms. The average number of partners, i.e., the local Rydberg atom density $n(t, \mathbf{x})$, is the Rydberg-equivalent of the dynamical mass, and it experiences bulk dissipation as in Eq.~\eqref{eq:SlopeEq} through the loss of excited atoms \cite{Ding2020,HelmrichSOC,KlockeControl,KlockeHydro}. Thus we have the translation table
\begin{align}
   &\text{order parameter: } & \phi(t, \mathbf{x})=\frac{1}{V_{{\bf x}}}
\sum_{l\in V_{\bf x}}\langle \hat{n}_l(t) \rangle \text{ [excitation density]},\nonumber\\
&\text{dynamical mass: }&n(t,{\bf x})=\frac{1}{V_{{\bf x}}}
\sum_{l\in V_{\bf x}}\langle 1-\hat{n}_l(t) \rangle \text{ [atom density]}.\nonumber
\end{align}

In a typical experiment, a gas of Rydberg atoms is cooled down to low
temperatures ($\le 10 \, \mu$K) and confined in an optical trap, e.g., an
isotropic Gaussian potential $V\sim -\sum_l \exp(-\mathbf{x}_l^2/w^2)$ with
width $w$. The vector $\mathbf{x}_l$ is the position of atom $l$ with respect to
the origin of the trap. The atoms undergo thermal motion, and after a short
equilibration period the average atom density $n(t, {\bf x})$ follows
approximately the trapping potential. In this case, the Ising-type Hamiltonian
for the driven Rydberg ensemble in Eq.~\eqref{eq:RydbergIsing} does not depend
on discrete lattice positions but on the continuously varying spatial
coordinates of each atom $\mathbf{x}_l$. Facilitation dynamics are enabled by
applying an excitation laser with large detuning ($|\Delta|\gg\Omega$):
single-particle processes are strongly suppressed, while resonant excitations
take place in the facilitation shell, where the van der Waals interaction
$V_{lm}=C_6/|\mathbf{x}_l-\mathbf{x}_m|^6$ compensates the detuning $\Delta$,
see Fig.~\ref{fig:Sandpile_Ryd}.

In free space, the number of atoms contributing to facilitation during a time
interval $t$ is given by the number of atoms that are traversing the
facilitation shell of an excited atom during this interval. It thus increases
proportionally to the atom density, which after appropriate coarse
graining \cite{KlockeControl,KlockeHydro,HelmrichSOC} yields a modification to
the Langevin equation Eq.~\eqref{eq:LangeDP} (we set
$\phi\equiv\phi(t, \mathbf{x}), n\equiv n(t, \mathbf{x}), \eta\equiv \eta(t,
\mathbf{x})$)
\begin{equation}\label{eq:LangeSOC2}
  \partial_t \phi = \left( D\nabla^2-m+\kappa n/2 \right) \phi-\kappa\phi^2+\eta.
\end{equation}
Instead of a fixed mass $m$, which depends only on the coordination number, the spreading rate depends on the local atom density $n$. 
This is equivalent to the Langevin equation~\eqref{eq:LangeSOC1} for SOC with the dynamical mass $n\to\kappa n/2$ rescaled by a constant decay rate.

The evolution of the atom density is determined by two physical processes:
(i)~the motion of the atoms in the trap, which may be described by a continuity
equation \cite{KlockeHydro}, and (ii)~the total loss of atoms from
ionization. The latter is conditioned on the Rydberg state, i.e., it is
proportional to the excitation density $\phi$. It yields the evolution
\begin{eqnarray}\label{eq:SlopeEq2}
\partial_t n= \nabla \cdot\left( D\nabla+\mathbf{F} \right)n -\mu \phi.
\end{eqnarray}
This differs from Eq.~\eqref{eq:SlopeEq} in that $n$ generates
its own dynamics, composed of a homogeneous diffusion term $\sim  D$ and
an effective ``force'' $\mathbf{F}({\bf x})\sim \nabla V({\mathbf x})$, set by the trapping potential. 

For a homogeneous gas, $V(\mathbf{x})=0$,
Eqs.~\eqref{eq:LangeSOC1} and  \eqref{eq:SlopeEq2} yield a behavior similar to the
forest fire model without regrowth. The atom density $n$ at large times drops below its critical value, leaving a transient time window at which excitation avalanches can be observed. This sets a natural cutoff scale for the lifetime and the size of avalanches, which depends on the loss rate and the critical density of atoms \cite{KlockeControl, KlockeHydro}. 

\begin{figure}
  \includegraphics[width=246pt]{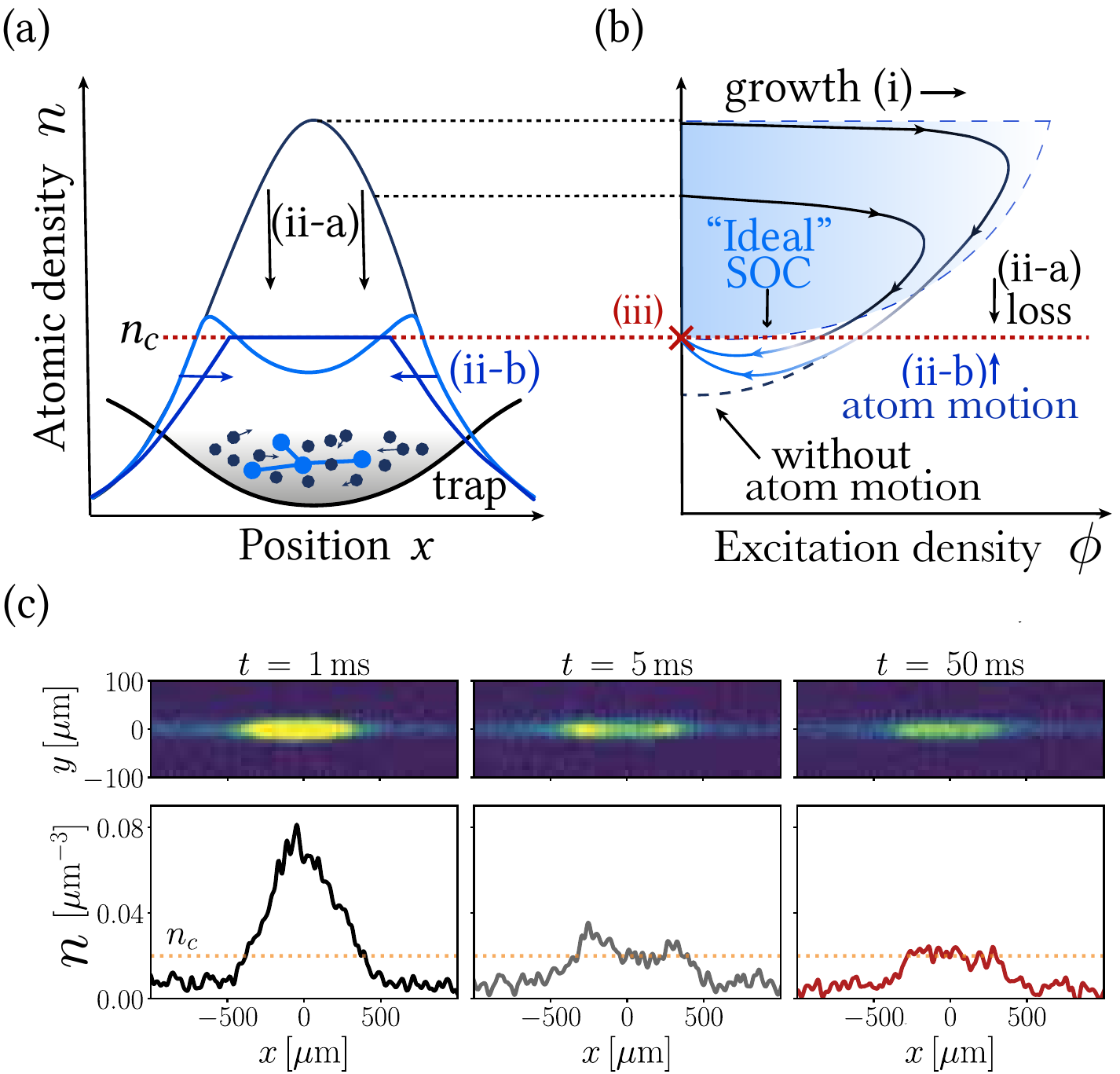}
  \caption{ Self-organized criticality in a gas of ultracold Rydberg atoms. (a)
    Driving an inhomogeneously trapped gas with an off-resonant laser induces
    the facilitated spreading of excitations (blue dots). (b) Evolution of the
    density of atoms $n$ ($\equiv n(t, \mathbf{x})$) and the excitation density
    $\phi$ ($\equiv \phi(t, \mathbf{x})$) from facilitation, loss and atomic
    motion. The initial supercritical state $n(t=0,\mathbf{x}) > n_c$ evolves through
    three stages: (i)~fast growth of excitations; (ii-a) self-organization into
    a critical density due to loss of particles; (ii-b) regrowth of the density
    in the center due to atomic motion from the flanks toward the center; (iii)
    the critical point is stabilized on transient time scales. (c) Upper panels:
    two-dimensional experimental absorption images. Lower panels: reconstructed
    atom densities along a 1D slice with $y = z = 0$. The flat-top coincides
    with the critical density. Figure adapted from Ref.~\cite{KlockeHydro}.}
\label{fig:RydbergSOCDens}
\end{figure}

In general, however, the trap is not homogeneous $V(\mathbf{x})\neq0$, and the
atoms tend to distribute themselves diffusively according to the trapping
profile. This may yield a competition between the atom motion, aiming to pile up
atoms in the center of the trap, and the loss from the Rydberg state, removing
atoms in the regimes of strong facilitation, i.e., of high density (large
dynamical mass) $n$. If the central region of the trapped gas is in the
self-organized critical regime, this leads to an observable, flat density
profile, with the average density being close to the critical value
$\langle n\rangle\approx \gamma/\kappa$, see
Fig.~\ref{fig:RydbergSOCDens}. Instead of dropping below the critical value
eventually, the density is replenished by atoms moving from the flanks of the
trapped gas to the center \cite{HelmrichSOC,KlockeHydro}. Over time, this leads
to a shrinking, or melting of the cloud, until no further facilitation is
possible. Both the formation of a "flat-top" density distribution and the
melting of the cloud have been identified in Rydberg experiments and in
theory \cite{KlockeHydro} and, in addition to the observation of avalanches,
provide further evidence of self-organized critical behavior in Rydberg
ensembles. A similar compensation mechanism has been identified in the atomic
vapour experiments \cite{Ding2020}.

Despite the range of theoretical results and the experimental evidence of a
scale invariant avalanche distribution (with measured exponent $\tau=1.37$), the
universality class of the Rydberg gas-variant of SOC, described by
Eqs.~\eqref{eq:LangeSOC2}, \eqref{eq:SlopeEq2}, has yet to be
determined. Although the order parameter dynamics are equivalent to both the
sandpile and the forest fire model, the evolution of the dynamical mass scale
differs from any previously proposed theory for SOC. One may see it as a
reversed scenario compared to the Abelian sandpile model: In the Rydberg setup,
bulk dissipation (the loss of excited atoms) is compensated by a boundary drive
(atomic currents from the flanks toward the center), whereas in the Abelian
sandpile model boundary dissipation (avalanches lost at the edges) is
compensated by bulk driving (the deposition of sand).

One may ask to what extent the dynamical mass can modify the universal behavior,
i.e., the critical exponents, of self-organizing systems. Deliberately
manipulating the dynamical mass may be a way to realize different classes of
self-organizing dynamics in open cold atom systems and to interpolate between,
e.g., the paradigmatic sandpile model and more complex scenarios such as the
self-organization in neural networks with tissue
growth \cite{Levina2007,Zierenberg}.

\subsection{Alternative sources of scale invariance without fine-tuning in cold
  atom dynamics}

Complex systems host further scenarios, besides SOC, which foster an emergent
scale invariant behavior without external fine-tuning. Potential sources are the
presence of conservation laws \cite{Grinstein1990,Lux2014}, hidden correlations
with the environment \cite{Schwab2014}, disorder \cite{Griffiths,Munoz2010}, or
dynamical bistabilities \cite{SOC_bistability}. Both disorder and dynamical
bistabilities can play a particularly important role for driven-dissipative
Rydberg gases. For instance, a \emph{dynamical Griffiths} phase \cite{Munoz2010}
has been experimentally observed in heterogeneous driven Rydberg
gases \cite{WintermantelNetwork} and theoretical works predict conditions for a
\emph{self-organized bistable}
regime \cite{SOC_bistability,BuchholdFirstOrder}. In both cases, scale invariant
behavior is enabled by microscopic quantum processes, while at large distances
it is governed by semiclassical dynamical equations.

\paragraph*{Self-organized bistability (SOB).}

SOB  refers to a scenario that displays a self-organization mechanism similar to SOC. However, SOB is based on an underlying discontinuous, first-order absorbing state phase transition. In this case, the self-organization mechanism does not push the system toward a critical point but rather to a bistability, which separates the absorbing from the active phase \cite{SOC_bistability}. Despite the different nature of the underlying phase transition, both SOC and SOB share similar characteristics, i.e., scale invariant avalanches, differing only in the precise values of the scaling exponents \cite{SOB2}. 

In Sec.~\ref{sec:FieldTheoryDP}, it was pointed out that the facilitation rate
of nearby atoms can be enhanced if multiple excited states are available
simultaneously by forming a coherent superposition (see the Rydberg superatom
\cite{Weber_2015}). This phenomenon was discussed for Rydberg atoms in a lattice
in Eqs.~\eqref{eq:QuantumContact} and \eqref{eq:LangeDP}. It yields an enhancement of the facilitation rate proportional to the cubic density $\phi^3$, see the discussion below Eq.~\eqref{eq:LangeDP}. In this case, the critical point is replaced by a bistable point, characteristic for a first-order phase transition. Although the probability of an atom traversing the facilitation shell of two or more Rydberg atoms simultaneously is strongly suppressed, it may become relevant in dense Rydberg plasmas, such as in Ref.~\cite{Ding2020}. Tuning the density of Rydberg atoms or their velocity distribution may thus enable to switch between the regimes of SOB and SOC. 

\paragraph*{Griffiths effects.}

Griffiths effects are responsible for the emergence
of scale invariant behavior in strongly heterogeneous systems, i.e., systems
that are subject to strong spatial fluctuations in their external
parameters. Typical examples are quantum and classical
magnets \cite{Voltja2005,Vojta2019,Griffiths} or networks with spatially
modulated connectivity \cite{Munoz2010}. Once the scale of the heterogeneity,
e.g., the disorder strength, becomes dominant over all other scales, the system
breaks up into disconnected or weakly connected clusters. Each cluster has its
own time- or energy scale, which largely depends on the (random) size of the
cluster. The dynamics of the system as a whole are then determined by the
ensemble of approximately independent and equally distributed spatial
clusters. In many cases, the ensemble displays scale invariant behavior through
the interplay between cluster size distribution and cluster lifetimes (see
below) even though each individual cluster is far from any critical
point \cite{Voltja2005,Griffiths}. As a consequence of the latter, the scaling
exponents are nonuniversal and vary with the external parameters.

For the dynamics on complex networks, the mechanism leading to Griffiths effects
is most transparent: Consider for simplicity a large network of nodes put on a
lattice. Any pair of neighboring nodes shall be connected by an edge with
probability $p$. When $p=1$, the network is homogeneous and has maximum
connectivity. Reducing $p<1$, the network becomes heterogeneous. Imagine a
dynamical process, e.g., the contact process in Eq.~\eqref{eq:QuantumContact},
on this network, such that the dynamics can only pass along connected
nodes. Weak heterogeneity then typically yields a perturbative renormalization
of the dynamical parameters. Once the heterogeneity of the network, however,
becomes significant, the connectivity of the nodes displays strong spatial
fluctuations. The dynamics are then governed by the network topology: if the
network is percolating, it undergoes a conventional absorbing state phase
transition, driven by the competition between spreading and decay of
excitations. For strong heterogeneity the network itself undergoes a phase
transition: it changes from a percolating network into a nonpercolating one,
which hosts only absorbing states.

The nonpercolating network breaks into a set of disconnected clusters without a
giant connected component. Each cluster has a finite size $N$ and, as a
consequence, can support an active phase only for a finite time scale
$\tau_N\sim\exp(\alpha N)$ for some $\alpha>0$. Assuming a distribution function
$P(N)\sim\exp(-\beta N), \,\beta>0$ of cluster sizes, which decays exponentially
for large $N$, the density of excitations on the network is
$\phi(t)\sim \int dN P(N)\exp(-t/\tau_N)\sim t^{-\beta/\alpha}$. It features a
scale invariant relaxation with a nonuniversal, continuously varying
exponent. Such dynamics are common for complex networks and appears, e.g., in
marketing \cite{Bampo2008effects}, finance \cite{Peckham2014contagion},
informatics \cite{Kephart1992directed}, and traffic
flow \cite{saberi2020simple}. It is a manifestation of a dynamical Griffiths
phase \cite{Voltja2005,Munoz2010,Arruda2020}.

Griffiths phases with a variable decay exponent and signatures of heterogeneity have been observed in gases of Rydberg atoms \cite{WintermantelNetwork,Brady2024a}.
In a gas, the atoms move freely and  facilitation becomes a dynamical process. It takes place whenever ground state atoms pass the facilitation shell of an excited atom. In order to understand the origin of heterogeneity, consider two atoms with relative velocity $\mathbf{v}$. The relative distance $\mathbf{x}$ between them is $\mathbf{x}=\mathbf{v}t+\mathbf{x}_0$. Let $\mathbf{x}_0$ be on the facilitation shell,  $\Delta=C_6/|\mathbf{x}_0|^6$. Close to $\mathbf{x}=\mathbf{x}_0$, this yields the Hamiltonian
\begin{eqnarray}
\hat H= \left( \frac{C_6}{|\mathbf{x}_0+\mathbf{v}t|^6}-\Delta \right)\sigma^z+\frac{\Omega}{2}\sigma^x\approx A t\sigma^z+\frac{\Omega}{2}\sigma^x,
\end{eqnarray}
with $A=\frac{6\Delta \mathbf{v}\mathbf{x}_0}{|\mathbf{x}_0|^2}$.  This is
reminiscent of the Landau-Zener problem: the particle undergoes adiabatic
passage if it remains in the instantaneous eigenstate. Here, the adiabatic
transition corresponds to a `spin-flip', i.e., to a transition from the atomic
ground state to the Rydberg state (or vice versa). The adiabatic transition
takes place when the relative velocity of the atoms is below a parameter
dependent Landau-Zener velocity $v_{\text{LZ}}$. This imposes an additional
velocity constraint $|\mathbf{v}|\le v_{\text{LZ}}$. The likeliness for the
constraint to be fulfilled depends on the velocity distribution of the atoms. If
the probability $p(|\mathbf{v}|\le v_{\text{LZ}})$ of finding a particle with
velocity $|\mathbf{v}|\le v_{\text{LZ}}$ is small, a large fraction of atoms is
excluded from facilitation, even though they may be spatially close to an
excited state. This is equivalent to removing bonds from a facilitation network
and to introducing heterogeneity. Depending on the thermal velocity
distribution, i.e., the average kinetic energy, the Rydberg atoms form a
heterogeneous, effective, facilitation network. The evolution of the Rydberg ensemble can then be described by a heterogeneous rate equation \cite{WintermantelNetwork} or, on a mean-field level, by a modified Langevin equation \cite{Brady2023a}, which take into account the spatially fluctuating excitation probabilities. The dynamics are then
reminiscent of epidemic spreading on complex
networks \cite{Viboud2016,PastorSatorras2015,Chowell2016mathematical,Chowell2016characterizing,Perez-Espigares2017,Watts1998collective,Buono2013}.
 
This discussion shows that the origin of scale invariant behavior on transient
and on asymptotic time scales can have various origins in driven quantum
systems. What all of the systems share in common is the presence of a
fluctuationless dark state. This manifestly modifies the structure of the
Keldysh action and results in nonequilibrium scaling behaviors. The construction
of scale invariance without fine-tuning, i.e., dynamics featuring a noiseless
field configuration and the corresponding vanishing of the action, may be seen
as a blueprint for the design of novel types of scale invariant quantum dynamics
under drive and dissipation. Conversely, it can be used as a guiding principle
to engineer scale invariant dynamics in experiments with driven-dissipative
quantum systems, in order to study paradigms of nonequilibrium
physics \cite{Griffithsbrain}.

\section{Driven open condensates in low spatial dimensions}
\label{sec:driven-open-BEC-2D-1D}

The Mermin-Wagner theorem states that two-dimensional (2D) equilibrium systems
with short-range interactions and at finite temperature cannot exhibit order
that spontaneously breaks a continuous global symmetry \cite{Mermin1966}. For
example, in condensates of interacting bosons in 2D, off-diagonal long-range
order is reduced to algebraically decaying correlations due to strong
fluctuations of the phase of the condensate. This is a manifestation of the
general phenomenon that long-wavelength fluctuations, which lead to the
emergence of universal behavior at large scales, are enhanced in low spatial
dimensions. A particular aspect of fluctuations of the phase $\theta$ of a
condensate lies in its compactness; that is, the property that the phase takes
values within a finite interval, $\theta \in [0, 2 \pi)$, whereby the endpoints
of that interval are identified with each other. Compactness implies that the
phase field can host topological defects: vortices in 2D and phase slips or
space-time vortices in 1D. The proliferation of such topological defects induces
the Kosterlitz-Thouless (KT) transition in 2D \cite{Kosterlitz1973} and the
resistive transition in narrow superconducting 1D channels \cite{Langer1967,
  McCumber1970a}. These results concern systems in thermal equilibrium. Here, we
address the question of which novel phase transitions and universal phenomena
can be induced out of equilibrium by gapless phase fluctuations and topological
defects. This question is of particular relevance for quantum fluids of
exciton-polaritons, which have emerged as highly versatile laboratories to study
nonequilibrium condensation phenomena in one and two spatial
dimensions \cite{Deng2010, Carusotto2013}.

\subsection{Long-wavelength theory of driven open condensates}
\label{subsec:longwave}

\begin{figure}
  \centering
  \includegraphics[width=246pt]{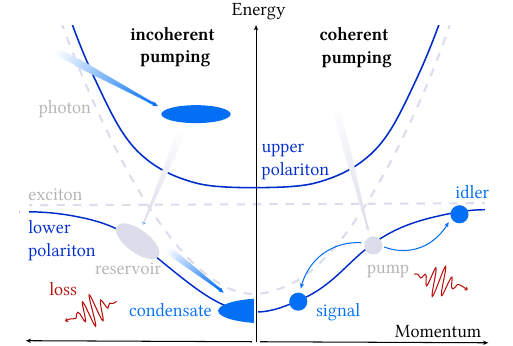}
  \caption{Exciton-polariton dispersion relation and pumping schemes. The
    coherent coupling of photons and excitons leads to the formation of two
    bands that are called lower and upper polaritons. Incoherent pumping:
    Excitations which are injected at high energies undergo relaxation through
    complex scattering processes and eventually condense at the bottom of the
    lower polariton branch. Coherent pumping: A laser is tuned close to the
    inflection point of the lower polariton dispersion relation. Pairs of
    coherently excited polaritons in the pump mode scatter parametrically into
    the signal and idler modes. In both pumping schemes, time-continuous pumping
    is required to compensate losses due to cavity leakage.}
  \label{fig:XPs}
\end{figure}

Exciton-polaritons are hybrid quasiparticles that emerge from the strong
coupling between light and matter excitations in semiconductor
microcavities \cite{Carusotto2013}. Specifically, the direct coherent coupling
between cavity photons and excitons or bound electron-hole pairs results in the
formation of two bands that are called the upper and lower polaritons, as
illustrated in Fig.~\ref{fig:XPs}. From their excitonic component, the
polaritons inherit screened Coulomb interactions; the leakage of photons through
the cavity mirrors leads to a finite lifetime of exciton-polaritons, which
necessitates a constant replenishment of their population through a laser drive.

Two possibilities to implement the laser drive are illustrated in
Fig.~\ref{fig:XPs}: In the \textit{incoherent pumping scheme,} the laser
frequency is chosen to resonantly populate highly excited states. The polaritons
generated in this way undergo relaxation through complex scattering processes
before they condense in the lower polariton band. Through these scattering
processes, all coherence of the exciting laser is lost, and the spontaneous
formation of a phase-coherent condensate breaks the $\mathrm{U}_c(1)$ symmetry
of incoherently pumped systems discussed in Sec.~\ref{sec:class-quant-symm}. In
contrast, in the \textit{coherent pumping scheme,} the coherence of the laser is
imprinted on the population of lower polaritons that is excited resonantly by
tuning the laser frequency close to the lower polariton dispersion. The coherent
laser drive of the lower polariton field explicitly breaks the $\mathrm{U}_c(1)$
symmetry, in particular, for the pump mode that is driven resonantly. However,
as illustrated in Fig.~\ref{fig:XPs}, parametric scattering of polaritons in the
pump mode can lead to a high occupation of two additional modes, the signal and
idler. While the phase of the pump mode is locked to the laser, these scattering
processes are invariant under rotations of the relative phase of the signal and
idler modes, and polariton condensation in the signal mode under coherent
pumping corresponds to the spontaneous breaking of this $\mathrm{U}(1)$
symmetry. Therefore, in both pumping schemes, the spontaneous breaking of a
$\mathrm{U}(1)$ phase rotation symmetry leads to the emergence of a gapless
Goldstone mode that is represented by a phase field $\theta$. Below, we review
the derivation of the effective long-wavelength description of driven open
condensates of exciton-polaritons in terms of this diffusive Goldstone mode. As
commonly done in the literature on exciton-polaritons \cite{Carusotto2013}, we
employ a description of the dynamics of driven open condensates in terms of
Langevin equations, which as explained in Sec.~\ref{sec:semiclassical-limit}
corresponds to the semiclassical limit of Keldysh field theory.

\subsubsection{Incoherent pumping scheme and mapping to the isotropic compact
  KPZ equation}
\label{sec:incoherent-pumping}

For incoherent pumping, a phenomenological description of the condensate
dynamics can be given in terms of two coupled evolution equations for the lower
polariton field $\psi(t, \mathbf{x})$ and the excitonic reservoir density 
$n_R(t, \mathbf{x})$. The first equation is a generalized stochastic
Gross-Pitaevskii equation that incorporates losses as well as stimulated
scattering from the reservoir into the condensate, and the second equation is a
rate equation for the reservoir density,
\begin{equation}
  \label{eq:condensate-reservoir}
  \begin{split}
    \imag \partial_t \psi & = \left( \omega_{\mathrm{LP}}(\unitvec{q}) -
      \frac{\imag}{2} \gamma_{\mathrm{LP}}(\unitvec{q}) + g_{\mathrm{LP}} \abs{\psi}^2 + 2 g_R n_R +
      \frac{\imag}{2} R n_R \right) \psi + \xi, \\
    \partial_t n_R & = P - \left( \gamma_R + R \abs{\psi}^2 \right) n_R.
  \end{split}
\end{equation}
Here, $\omega_{\mathrm{LP}}(\mathbf{q})$ is the dispersion of lower polaritons
illustrated in Fig.~\ref{fig:XPs}, and $\unitvec{q} = - \imag \nabla$ is the
momentum operator. The dispersion relation can be shaped through band
engineering \cite{Schneider2017}. To obtain a long-wavelength description of the
condensate dynamics, an expansion of $\omega_{\mathrm{LP}}(\mathbf{q})$ for
small momenta suffices:
$\omega_{\mathrm{LP}}(\mathbf{q}) = \omega_0 + q^2/(2 m_{\mathrm{LP}})$ with
$q = \abs{\mathbf{q}}$. This also applies to the momentum-dependent loss rate,
which can be expanded as
$\gamma_{\mathrm{LP}}(\mathbf{q}) = \gamma_{\mathrm{LP}, 0} +
\gamma_{\mathrm{LP}, 2} q^2$.
The momentum dependence of the loss rate is often omitted but can be crucial to
ensure the stability of the condensate \cite{Baboux2018}. Furthermore,
$g_{\mathrm{LP}}$ is the polariton-polariton interaction strength, and the term
$2 g_R n_R$ describes repulsive interactions between polaritons and reservoir
excitons. We assume that the rate of pumping $P$ into the excitonic reservoir is
spatially homogeneous on the scales of interest. Reservoir excitons either relax
into the polariton condensate by stimulated scattering with rate $R$, or they
decay through other channels at the rate $\gamma_R$.  Finally, as in
Eq.~\eqref{eq:driven-open-condensate-Langevin}, $\xi(t, \mathbf{x})$ describes
Gaussian noise induced by loss and laser drive. The noise field has a vanishing
mean value and its fluctuations are given by
$\langle \xi(t, \mathbf{x}) \xi^{*}(t', \mathbf{x}') \rangle = \left(
  \gamma_{\mathrm{LP}, 0} + R n_R \right) \delta(t - t') \delta(\mathbf{x} -
\mathbf{x}')$.

Through adiabatic elimination of the excitonic reservoir, which is valid in the
absence of dynamical instabilities \cite{Bobrovska2014, Bobrovska2015,
  Bobrovska2018, Baboux2018, Bobrovska2019, Vercesi2023}, the generalized
Gross-Pitaevskii equation~\eqref{eq:condensate-reservoir} can be reduced to a
Langevin equation for the condensate field of the form of
Eq.~\eqref{eq:driven-open-condensate-Langevin}. The spatially homogeneous and
noiseless mean-field solution of this Langevin equation takes the form
$\psi = \sqrt{\rho_0} e^{i \theta}$ where $\rho_0 = - r_d/u_d = - r_c/u_c$, with
the second equality being satisfied in a suitably chosen rotating
frame. Fluctuations of the condensate density are gapped and can be integrated
out. The dynamics of the condensate phase are then found to be described by the
compact KPZ (cKPZ) equation \cite{Grinstein1993, Grinstein1996, Altman2015,
  Gladilin2014, Ji2015, He2015},
\begin{equation}
  \label{eq:KPZ}
  \partial_t \theta = D \nabla^2 \theta + \frac{\lambda}{2} \left( \nabla
    \theta \right)^2 + \eta,
\end{equation}
where
\begin{equation}
  \label{eq:KPZ-nonlinearity}
  \lambda = - 2 K_c \left( 1 - \frac{K_d u_c}{K_c u_d} \right).
\end{equation}
Formally, Eq.~\eqref{eq:KPZ} is equivalent to the noncompact KPZ
equation. However, solutions to the cKPZ equation can also contain discontinuous
jumps by multiples of $2 \pi$ as are encountered, e.g., in 2D upon encircling a
vortex.

As explained in Sec.~\ref{sec:semiclassical-limit}, the Keldysh action that
corresponds to the Langevin equation~\eqref{eq:driven-open-condensate-Langevin}
obeys the thermal symmetry introduced in Sec.~\ref{sec:thermal-symmetry} only if
$K_c/K_d = u_c/u_d$, which implies $\lambda = 0$. For a pumped nonequilibrium
system, these relations correspond to unphysical fine-tuning. Generically,
$\lambda \neq 0$, and $\lambda$ serves as a single-parameter measure of the
strength of nonequilibrium conditions. The sign of $\lambda$ is irrelevant
because it can be absorbed in a redefinition of $\theta$ and the noise field
$\eta$. In contrast, the diffusion coefficient $D > 0$ has to be positive to
ensure dynamical stability. The diffusion coefficient $D$ and the strength
$\Delta$ of the noise field $\eta$, which has vanishing mean and fluctuations
$\langle \eta(t, \mathbf{x}) \eta(t', \mathbf{x}') \rangle = 2 \Delta \delta(t -
t') \delta(\mathbf{x} - \mathbf{x}')$, are given by
\begin{equation}
  D = K_c \left( \frac{K_d}{K_c} + \frac{u_c}{u_d} \right), \quad \Delta 
  = \frac{\gamma}{2 \rho_0} \left( 1 + \frac{u_c^2}{u_d^2} \right).
\end{equation}
We have outlined the derivation of the cKPZ equation~\eqref{eq:KPZ} from the
Langevin equation for the condensate, but the cKPZ equation can also be obtained
by integrating out density fluctuations within the Keldysh
formalism \cite{Sieberer2016a}. Furthermore, the coupled
equations~\eqref{eq:condensate-reservoir} can be mapped to the cKPZ equation
without adiabatic elimination of the excitonic
reservoir \cite{Fontaine2021}. This leads to a renormalization of the parameters
of the cKPZ equation, which does not affect the long-wavelength behavior but can
be significant for quantitative comparisons to numerics and
experiments. Conversely, if one is interested only in the correct form of the
long-wavelength description and not in quantitative accuracy of the
coefficients, it is possible to derive Eq.~\eqref{eq:KPZ} from only a few
fundamental principles: (i)~Spontaneous breaking of a $\mathrm{U}(1)$ symmetry
leads to the existence of a gapless Goldstone mode, and the evolution equation
of a gapless mode may contain only derivative terms. Furthermore, in an expansion in
derivatives as appropriate in the long-wavelength and low-frequency limit, we
can restrict ourselves to the lowest-order and thus most relevant terms that are
allowed by symmetry. (ii)~Rotational symmetry implies that there are no terms
that contain only a single spatial derivative such as
$\mathbf{B} \cdot \nabla \theta$. (iii)~In equilibrium, a term such as
$\left( \lambda/2 \right) \left( \nabla \theta \right)^2$, which cannot be
written in the form $\delta \mathcal{F}/\delta \theta$, i.e., as the variational
derivative of a free energy functional $\mathcal{F}[\theta]$, is
forbidden. However, out of equilibrium it is allowed. In particular, this term
contains the same number of spatial derivatives as the diffusion term
$D \nabla^2 \theta$ and is, therefore, equally relevant. (iv)~In the absence of
particle number conservation, the leading temporal derivative is of first order,
which leads to diffusive dynamics of the phase when $\lambda = 0$. Combining
(i)--(iv), we are led to Eq.~\eqref{eq:KPZ}. Instead, if the number of particles
is conserved, there is a coherently propagating sound mode, which is correctly
described by exchanging the first derivative with respect to time by a second
derivative \cite{Hohenberg1977}.

\subsubsection{Coherent pumping scheme and mapping to the anisotropic compact
  KPZ equation}

In the coherent pumping scheme, lower polaritons are excited directly through a
driving field
$F(t, \mathbf{x}) = F_0 \e^{\imag \left( \mathbf{q}_p \cdot \mathbf{x} -
    \omega_p t \right)}$
with amplitude $F_0$ and pump wave vector $\mathbf{q}_p$, whereby
$q_p = \lvert \mathbf{q}_p \rvert$ is typically chosen close to the inflection
point of the lower polariton dispersion relation shown in Fig.~\ref{fig:XPs},
and $\omega_p = \omega_{\mathrm{LP}}(\mathbf{q}_p)$. While the derivation of the
cKPZ equation for incoherent pumping presented in the previous section is
formally valid in any spatial dimension and relevant for exciton-polaritons in
2D and 1D, for the case of coherent excitation, we focus on 2D systems. Our
starting point is again a Langevin equation for the lower polariton field,
\begin{equation}
  \label{eq:coherently-driven-condensate}
  \imag \partial_t \psi = \left( \omega_{\mathrm{LP}}(\unitvec{q}) -
    \frac{\imag}{2} \gamma_{\mathrm{LP}, 0} + g_{\mathrm{LP}} \abs{\psi}^2 \right)
  \psi + F + \xi,
\end{equation}
where the noise strength is determined by the loss rate,
$\langle \xi(t, \mathbf{x}) \xi^{*}(t', \mathbf{x}') \rangle =
\gamma_{\mathrm{LP}, 0} \delta(t - t') \delta(\mathbf{x} - \mathbf{x}')$.
The driving field $F$ explicitly breaks the symmetry of
Eq.~\eqref{eq:coherently-driven-condensate} under phase rotations
$\psi \mapsto \psi \e^{\imag \theta}$.\footnote{The noise field $\xi$ does not
  break this symmetry. A global phase factor can be absorbed in a redefinition
  of $\xi$, which leaves the statistics of $\xi$ invariant.} However, as we
discuss next, there is an emergent $\mathrm{U}(1)$ symmetry in the dynamics of
the three dominant field modes, the pump, signal, and idler.

Depending on the strength of the external pump power, there are two different
regimes: In the pump-only regime, only a single mode $\psi_p$ with wave vector
$\mathbf{q}_p$ and frequency $\omega_p$ set by the external drive is occupied
substantially. By contrast, in the optical parametric oscillator (OPO) regime,
pairs of polaritons in the pump mode scatter parametrically to other modes,
which can thereby also become highly populated. In the simplest case
illustrated in Fig.~\ref{fig:XPs}, there are two additional highly populated
modes, the signal $\psi_s$ and the idler $\psi_i$. The respective momenta
$\mathbf{q}_s$ and $\mathbf{q}_i$ and frequencies $\omega_s$ and $\omega_i$ are
determined by the conditions $\mathbf{q}_s + \mathbf{q}_i = 2 \mathbf{q}_p$ and
$\omega_s + \omega_i = 2 \omega_p$. This situation is described by a three-mode
ansatz for the lower polariton field $\psi$,
\begin{equation}
  \label{eq:three-mode-ansatz}
  \psi(t, \mathbf{x}) = \sum_{m = s, p, i} \psi_m(t, \mathbf{x})
  \e^{\imag \left( \mathbf{q}_m \cdot \mathbf{x} - \omega_m t \right)},
\end{equation}
where the amplitudes $\psi_m(t, \mathbf{x})$ incorporate long-wavelength, slow
fluctuations around the carrier waves
$\e^{\imag \left( \mathbf{q}_m \cdot \mathbf{x} - \omega_m t \right)}$.
Therefore, inserting this ansatz in Eq.~\eqref{eq:coherently-driven-condensate},
filtering in momenta and frequencies around the carrier waves, and keeping only
parametric scattering processes, yields three coupled equations for the amplitudes
$\psi_m(t, \mathbf{x})$. A stability analysis shows that the three-mode ansatz
is dynamically stable within a restricted range of pump powers \cite{Zamora2017,
  Dunnett2018}. Outside of this range, parametric scattering leads to the
population of more than two additional modes. Here, we focus on the regime in
which the three-mode ansatz is stable.

Both in the pump-only and the OPO regime, the phase of the pump mode is locked
to the phase of the driving laser. Via parametric scattering, the laser drive
fixes also the sum of the phases of the signal and idler modes, but not their
difference. Consequently, the equations for the amplitudes
$\psi_m(t, \mathbf{x})$ are invariant under a simultaneous phase rotation of
signal and idler modes, $\psi_s \mapsto \psi_s \e^{\imag \theta}$ and
$\psi_i \mapsto \psi_i \e^{-\imag \theta}$. This U(1) symmetry is broken
spontaneously when the pumping strength is varied to cross the threshold to the
OPO regime \cite{Stevenson2000, Baumberg2000, Tartakovskii2002, Wouters2006},
and the associated Goldstone mode governs the long-wavelength dynamics of the
OPO condensate. To obtain the corresponding long-wavelength theory, we proceed
in two steps: First, we integrate out massive fluctuations, i.e., the density
fluctuations of the modes $\psi_m$ and the phase fluctuations apart from the
relative signal-idler phase $\theta = \theta_s - \theta_i$ \cite{Zamora2017}.
Second, we expand the lower polariton dispersion
$\omega_{\mathrm{LP}}(\mathbf{q})$ around each mode with momentum $\mathbf{q}_m$
up to second order in the gradient $\unitvec{q} = - \imag \nabla$,
\begin{equation}
  \label{eq:omega-LP-expansion}
  \omega_{\mathrm{LP}}(\mathbf{q}_m + \unitvec{q}) \approx
  \omega_{\mathrm{LP}}(\mathbf{q}_m) + \mathbf{B}_m \cdot \unitvec{q} +
  \frac{1}{2} \unitvec{q}^{\transpose} L_m \unitvec{q},
\end{equation}
where $\mathbf{B}_m = \nabla_{\mathbf{q}} \omega_{\mathrm{LP}}(\mathbf{q}_m)$
and
$L_m = \nabla_{\mathbf{q}} \nabla_{\mathbf{q}}^{\transpose}
\omega_{\mathrm{LP}}(\mathbf{q}_m)$.
The first-order contribution $\mathbf{B}_m \cdot \unitvec{q}$ can be removed by
transforming to a frame of reference that moves at a finite velocity
$\mathbf{v}$ according to
$\theta(t, \mathbf{x}) \to \theta(t, \mathbf{x} + \mathbf{v} t)$. At second
order, the spatial anisotropy imposed on the system by the pump wave vector
$\mathbf{q}_p$ manifests in two distinct eigenvalues of the matrix
$L_m$. Consequently, the long-wavelength description of the Goldstone mode
$\theta$ is given by an anisotropic cKPZ equation,
\begin{equation}
  \label{eq:anisotropic-KPZ}
  \partial_t \theta = \sum_{i = x, y} \left[ D_i \partial_i^2 \theta +
    \frac{\lambda_i}{2} \left( \partial_i \theta \right)^2 \right] + \eta,
\end{equation}
where $D_x \neq D_y$ and $\lambda_x \neq \lambda_y$. A convenient measure for
the degree of spatial anisotropy is provided by the anisotropy parameter
$\Gamma$ which is defined as
\begin{equation}
  \label{eq:anisotropy-parameter}
  \Gamma = \frac{D_x \lambda_y}{D_y \lambda_x}.
\end{equation}
In particular, when $\Gamma = 1$, the anisotropic cKPZ
equation~\eqref{eq:anisotropic-KPZ} can be reduced to the isotropic KPZ
equation~\eqref{eq:KPZ} through an anisotropic scale transformation. The sign of
$\Gamma$ distinguishes between the weakly and strongly anisotropic regimes for
$\Gamma > 0$ and $\Gamma < 0$, respectively. For OPO polaritons, the parameters
$D_{x, y}$ and $\lambda_{x, y}$ depend in a nontrivial way on the microscopic
system parameters and can be varied substantially by changing the pump strength,
the pump momentum $\mathbf{q}_p$, and detuning between photons and excitons at
zero momentum \cite{Zamora2017}. In particular, it is possible to reach both the
strongly anisotropic regime, and a regime in which the nonequilibrium strengths
$\lambda_{x, y}$ become dominant.

\subsection{KPZ universality in the absence of topological defects}
\label{sec:kpzuniv}

The isotropic and anisotropic KPZ equations have originally been introduced to
describe the kinetic roughening of growing interfaces \cite{Kardar1986,
  Wolf1991, Krug1997, Halpin-Healy1995}. However, physical manifestations of KPZ
universality occur in a much wider variety of systems \cite{Takeuchi2018}
ranging from liquid crystals \cite{Takeuchi2010} to burning
paper \cite{Maunuksela1997}, the growth of bacterial colonies \cite{Allen2019},
and urban skylines \cite{Najem2020}, but also in the growth of entanglement in
random quantum circuits \cite{Nahum2017} and in spin transport \cite{Wei2022,
  Keenan2022}.  All of these examples are described by the noncompact KPZ
equation. In contrast, the long-wavelength dynamics of driven open condensates
and also of a variety of other systems ranging from driven vortex lattices in
disordered superconductors \cite{Aranson1998}, to polar active
smectics \cite{Chen2013}, synchronization in oscillator arrays \cite{Lauter2017,
  Gutierrez2022}, and limit-cycle phases that emerge from a Hopf
bifurcation \cite{Lee2011, Ludwig2013a, Jin2013, Chan2015, Schiro2016}, are
described by the cKPZ equation. Before we discuss deviations from KPZ
universality due to compactness, we review predictions for the expected scaling
in the condensate coherence based on an RG analysis of the noncompact KPZ
equation \cite{Kardar1986, Wolf1991, Chen2013}.

\subsubsection{RG flow of the KPZ equation in 1D, 2D, and 3D}
\label{sec:RG-flow-KPZ}

In the absence of topological defects, the compactness of the phase field
$\theta$ can be neglected. Then, the KPZ equation~\eqref{eq:KPZ} can be brought
to a dimensionless form through the following rescaling, in $d$ spatial
dimensions, where $\Lambda$ is the ultraviolet (UV) momentum cutoff:
\begin{equation}
  \label{eq:KPZ-rescaling}
  \begin{aligned}
    \mathbf{x} & \to \frac{\mathbf{x}}{\Lambda}, & t & \to
    \frac{t}{\Lambda^2 D}, \\ \theta & \to \theta
    \sqrt{\frac{\Delta}{\Lambda^{2 - d} D}}, & \eta & \to \eta
    \sqrt{\Lambda^{2 + d} D \Delta}.
  \end{aligned}
\end{equation}
In the rescaled form of the KPZ equation, the only remaining independent
parameter is the dimensionless nonequilibrium strength defined as
$g = \Lambda^{d - 2} \lambda^2 \Delta/D^3$ \cite{Tauber2014a}. Its canonical
scaling dimension $2 - d$ determines the large-scale behavior of the theory:
\begin{itemize}
\item In 3D, the dimensionless nonequilibrium strength $g$ is irrelevant, and,
  for sufficiently small microscopic values of $g$, effective equilibrium is
  emergent at large scales (see the discussion in Sec.~\ref{sec:emergenteq}).
\item In 2D, the canonical scaling dimension of $g$ vanishes, and the loop
  correction renders the nonequilibrium strength marginally relevant.
\item In 1D, $g$ is relevant and grows under renormalization until the canonical
  flow is balanced by the loop correction at a strong-coupling fixed
  point.
\end{itemize}
1D and 2D condensates are thus unstable against small nonequilibrium
perturbations that occur on a microscopic scale---these perturbations will grow
under coarse graining. Consequently, the long-distance scaling properties of
these systems are strongly modified under nonequilibrium conditions. In
particular, KPZ scaling becomes manifest in the spatial coherence as measured
through the correlation function or Keldysh Green's function
$C(t, \mathbf{x}) = \langle \psi(t, \mathbf{x}) \psi^{*}(0, 0) \rangle \approx \rho_0 \e^{- \langle \left( \theta(t, \mathbf{x}) -
    \theta(0, 0 )\right)^2 \rangle/2}$,
where in the approximation we have neglected fluctuations of the density and
treated fluctuations of the phase to leading order in a cumulant expansion.  The
RG analysis of the KPZ equation leads to \cite{Tauber2014a}
\begin{equation}
  \label{eq:KPZ-scaling}
  \langle \left( \theta(t, \mathbf{x}) - \theta(0, 0) \right)^2 \rangle \sim
  \begin{cases}
    r^{2 \chi} & \text{for } r \to \infty, \\
    t^{2 \beta} & \text{for } t \to \infty,
  \end{cases}
\end{equation}
where $r = \abs{\mathbf{x}}$. That is, spatial and temporal fluctuations of the
condensate phase $\theta$ are determined by the roughness exponent $\chi$ and
the growth exponent $\beta$, respectively. In 2D, numerical studies indicate
that $\chi \approx 0.39$ and $\beta \approx 0.24$ (see, e.g.,
Ref.~\cite{Gomes-Filho2021} and references therein); in 1D, the exponents
take the exact values $\chi = 1/2$ and $\beta = 1/3$. The roughness and growth
exponents are interdependent through the scaling relation $\chi + z = 2$ which
holds at any nontrivial and finite fixed point of $g$, and where
$z = \chi/\beta$ is the dynamical exponent \cite{Tauber2014a}. For the
condensate correlation function, Eq.~\eqref{eq:KPZ-scaling} implies
stretched-exponential decay both with spatial and temporal distances. This
behavior is in stark contrast to the algebraic decay of correlations in 2D
condensates in thermal equilibrium.

The RG analysis provides predictions not only for the universal scaling behavior
of the correlation function but also for the length and time scales beyond
which KPZ scaling is established. In 2D, the marginality of the dimensionless
nonequilibrium strength $g$ implies that the RG flow starting from a small
microscopic value $g_0 \ll 1$ reaches the strong-coupling regime in which KPZ
scaling emerges only on exponentially large scales that are greater than
\begin{equation}
  \label{eq:L-KPZ}
  L_{\mathrm{KPZ}} = a \, \e^{8 \pi/g_0},
\end{equation}
where $a = 1/\Lambda$ is the microscopic length scale on which the RG flow is
initialized. In contrast, in 1D, the nonequilibrium strength is relevant, and,
therefore, the KPZ scaling regime is reached at a scale
$L_{\mathrm{KPZ}} \sim 2 \pi a/g_0$ that is only algebraically large in
$g_0 \ll 1$ \cite{Nattermann1992}.

Recently, some aspects of KPZ physics have been observed in synthetic magnetic
systems \cite{Wei2022}. Importantly, these systems are operated in one dimension
and in thermodynamic equilibrium; in 1D, the KPZ nonlinear term, driving the
characteristic universality, is accidentally compatible with equilibrium
conditions \cite{Kamenev2023, VanBeijeren2012, Kulkarni2013}, but this is not
the case in higher spatial dimensions. Therefore, 2D KPZ universality is a
unique promise of nonequilibrium platforms such as those discussed above. We
return to this point in Sec.~\ref{sec:obs2d}.

\subsubsection{RG flow of the anisotropic KPZ equation in 2D}

A new feature of the anisotropic KPZ equation~\eqref{eq:anisotropic-KPZ} in 2D
is that the nonequilibrium strength $g$ becomes irrelevant for sufficiently
strong anisotropy. That is, by making the system strongly anisotropic, effective
thermal equilibrium and algebraic quasi-long-range order can be restored on
large scales. This behavior is encoded in the RG flow of the nonequilibrium
strength $g = \lambda_x^2 \Delta/(D_x^2 \sqrt{D_x D_y})$ and the anisotropy
parameter $\Gamma$ defined in
Eq.~\eqref{eq:anisotropy-parameter} \cite{Wolf1991, Chen2013},
\begin{equation}  
  \begin{split}    
    \frac{d g}{d \ell} & = - \frac{g^2}{32 \pi} \left( \Gamma^2 + 4 \Gamma - 1
    \right), \\ \frac{d \Gamma}{d \ell} & = -\frac{\Gamma g}{32 \pi} \left( 1 -
      \Gamma^2 \right),
  \end{split}
\end{equation}
where $\ell$ is the logarithm of the running momentum cutoff. The RG flow is
markedly different in the regimes of weak and strong anisotropy which correspond
to $\Gamma > 0$ and $\Gamma < 0$, respectively: For weakly anisotropic systems,
spatial isotropy and KPZ scaling as described by Eq.~\eqref{eq:KPZ-scaling} are
restored on large scales; in contrast, for strong anisotropy, the RG flow is
attracted to a fixed point at $g = 0$ and $\Gamma = - 1$. Since the
nonequilibrium strength vanishes at this fixed point, effective thermal
equilibrium is established. This leads to logarithmic spatial growth of phase
fluctuations \cite{Chen2013},
\begin{equation}
  \label{eq:logarithmic-scaling}
  \langle \left( \theta(0, \mathbf{x}) - \theta(0, 0) \right)^2 \rangle \sim
  \ln \! \left( \lambda_y x^2 + \lambda_x y^2 \right),
\end{equation}
resulting in algebraic quasi-long-range order. As noted below
Eq.~\eqref{eq:omega-LP-expansion}, for coherently driven exciton-polaritons,
these results apply in a moving frame of reference. Spatial correlations are not
affected by the transformation back to the original frame. However, for temporal
correlations this transformation leads to a crossover between different
stretched exponential and algebraic scaling regimes for weak and strong
anisotropy, respectively \cite{Zamora2017}.

We emphasize that the mechanism of asymptotic thermalization described above
hinges on strong spatial anisotropy and is, therefore, profoundly different from
asymptotic thermalization in the isotropic KPZ equation in 3D, which results
from the irrelevance of $g$ as discussed in Sec.~\ref{sec:RG-flow-KPZ}, and also
from the other examples for the emergence of equilibrium at large scales
mentioned in Sec.~\ref{sec:bosonic-driven-open-criticality}.

\subsection{Vortex unbinding in two-dimensional driven open condensates}
\label{sec:vortex-unbinding}

The RG analysis that leads to the predictions for the spatiotemporal coherence
of driven open condensates given in Eqs.~\eqref{eq:KPZ-scaling}
and~\eqref{eq:logarithmic-scaling} ignores the possible occurrence of
topological defects. In the theoretical description of 2D condensates in
equilibrium, topological defects can be incorporated by employing an
electrodynamic duality which maps vortices to electric charges that form a 2D
Coulomb gas and rephrases the KT transition as a screening
transition \cite{Ambegaokar1978, Ambegaokar1980, Cote1986, Minnhagen1987}: At
low temperatures, pairs of vortices and antivortices remain tightly bound due to
their attractive interaction; but above the KT critical temperature, strong
fluctuations screen the vortex-antivortex interaction at large distances. Then,
unbound vortices can move freely through the system and destroy quasi-long-range
order. This electrodynamic duality has been extended to isotropic and
anisotropic driven open condensates in Refs.~\cite{Wachtel2016, Sieberer2016b}
and Ref.~\cite{Sieberer2018b}, respectively.

In the absence of vortices, universal KPZ scaling is established beyond the
exponentially large scale $L_{\mathrm{KPZ}}$ Eq.~\eqref{eq:L-KPZ}. A key
prediction of the electrodynamic duality is the existence of another
exponentially large scale $L_v$ associated with the proliferation of vortices in
isotropic systems. Therefore, the relative size of $L_{\mathrm{KPZ}}$ and $L_v$
determines the observability of KPZ universality in isotropic driven open
condensates. We will return to this point in Sec.~\ref{sec:obs2d} below.

\subsubsection{Electrodynamic duality}

In the dual description, topological defects and smooth fluctuations of the
condensate phase $\theta$ are represented through the coupled dynamics of a gas
of point charges and electromagnetic fields. The electric field $\mathbf{E}$ is
defined as
\begin{equation}
  \mathbf{E} = - \unitvec{z} \times \nabla \theta,
\end{equation}
where $\unitvec{z}$ is a unit vector normal to the 2D plane that supports the
condensate. An overdamped version of Faraday's law relates the electric field to
the magnetic field $\mathbf{B}$,
\begin{equation}
  \nabla \times \mathbf{E} + \frac{1}{D} \mathbf{B} = 0.
\end{equation}
Here, we set $D = D_x = D_y$, which can be achieved through an anisotropic
rescaling of the units of length in Eq.~\eqref{eq:anisotropic-KPZ}. From the
definition of the electric field in terms of the condensate phase and Faraday's
law it follows that $\mathbf{B} = B \unitvec{z}$, which in turn implies
$\nabla \cdot \mathbf{B} = 0$. Vortices act as sources of the electric field
according to Gauss' law,
\begin{equation}
  \nabla \cdot \mathbf{E} = \frac{2 \pi}{\varepsilon} n,
\end{equation}
where
$n(t, \mathbf{x}) = \sum_{\alpha} n_{\alpha} \delta(\mathbf{x} -
\mathbf{x}_{\alpha}(t))$
is the density of vortices with charges $n_{\alpha}$ at positions
$\mathbf{x}_{\alpha}(t)$, and $\varepsilon$ is the dielectric constant that
describes screening of the electromagnetic fields due to bound vortex-antivortex
pairs. Furthermore, Amp\`{e}re's law incorporates the KPZ nonlinearities
$\lambda_{x, y}$, the noise $\eta$, and the vortex current
$\mathbf{j}(t, \mathbf{x}) = \sum_{\alpha} n_{\alpha} \frac{\diff
  \mathbf{x}_{\alpha}(t)}{\diff t} \delta(\mathbf{x} -
\mathbf{x}_{\alpha}(t))$,
\begin{equation}
  \label{eq:Ampere}
  \nabla \times \mathbf{B} - \varepsilon \partial_t \mathbf{E} = 2 \pi
  \mathbf{j} + \unitvec{z} \times \nabla \left( \sum_{i = x, y} \frac{\lambda_i}{2} E_i^2 + \eta
  \right).
\end{equation}
For $\mathbf{j} = 0$, Amp\`{e}re's law reduces to the anisotropic cKPZ
equation~\eqref{eq:anisotropic-KPZ}. The above nonlinear and overdamped Maxwell
equations can also be derived systematically by discretizing the cKPZ equation
on a lattice and performing a modified Villain
transformation \cite{Sieberer2016b}.

To obtain a complete long-wavelength description of the condensate dynamics, the
Maxwell equations have to be complemented by an equation of motion for the
vortices. Under the assumption of overdamped dynamics, this equation is
\begin{equation}
  \frac{\diff \mathbf{x}_{\alpha}}{\diff t} = \mu n_{\alpha}
  \mathbf{E}(\mathbf{x}_{\alpha}) +\boldsymbol{\xi}_{\alpha},
\end{equation}
where $\mu$ is the phenomenologically introduced vortex mobility.  The
stochastic force $\boldsymbol{\xi}_{\alpha}$ has vanishing mean, and its
correlations are given by
$\langle \xi_{\alpha, i}(t) \xi_{\alpha', j}(t') \rangle = 2 \mu T
\delta_{\alpha, \alpha'} \delta_{i, j} \delta(t - t')$,
where $T$ can be interpreted as an effective vortex temperature which, for
systems in thermal equilibrium, is identical to the strength of the noise $\eta$
that enters Amp\`{e}re's law. 

\subsubsection{Noise-activated vortex unbinding in isotropic systems}

To determine the stability of the bound state of a vortex-antivortex pair at
distance $\mathbf{x}$, we have to study the screening of the vortex-antivortex
interaction due to the polarization of bound pairs of smaller size. Under the
assumption that the vortices have low mobility $\mu/D \ll 1$, which should not
affect our results qualitatively, the vortex-antivortex interaction can be found
in the dual description by solving an electrostatic problem with fixed vortex
positions. In doing so, we treat the KPZ nonlinearity perturbatively.

We first consider isotropic systems with $\lambda_x = \lambda_y$ in
Eq.~\eqref{eq:Ampere}. Then, the vortex-antivortex interaction is a conservative
force that can be derived from the potential \cite{Wachtel2016}
\begin{equation}
  \label{eq:vortex-interaction}
  V(\mathbf{x}) = \frac{1}{\varepsilon} \ln(r/a) - \frac{\lambda^2}{6
    \varepsilon^3 D^2} \left( \ln(r/a)^3 + \frac{3}{4} \ln(r/a)^2 \right),
\end{equation}
where $a$ is a microscopic cutoff. That is, at second order in $\lambda$, the
logarithmic 2D Coloumb potential acquires a \emph{repulsive} correction that
becomes dominant on distances greater than
\begin{equation}
  \label{eq:L-v}
  L_v = a \, \e^{2 D/\lvert \lambda \rvert},
\end{equation}
where we have set the dielectric constant to its microscopic value
$\varepsilon = 1$. The perturbative expansion, Eq.~\eqref{eq:vortex-interaction},
in powers of $\lambda \ln(r/a)$ breaks down at distances $r \gg L_v$. However,
studies of the complex Ginzburg-Landau equation, corresponding to
Eq.~\eqref{eq:driven-open-condensate-Langevin} without noise, indicate that the
repulsive vortex-antivortex interaction on distances $r \gtrsim L_v$ is
exponentially suppressed for $r \gg L_v$ \cite{Aranson1998}. Consequently, even
without taking the screening of the vortex-antivortex interaction due to bound
pairs into account, we are led to conclude that the bound state of a
vortex-antivortex pair is only metastable, and any finite vortex temperature
leads to vortex unbinding. In experiments, the vortex temperature, which we have
introduced phenomenologically, is determined by the intrinsic parameters of the
system and the external drive and is generically nonzero. Based on the present
analysis we should then expect a finite density $\sim 1/L_v^2$ of unbound
vortices. This expectation is corroborated by an RG approach that accounts for
the renormalization of the dielectric constant $\varepsilon$ due to the
polarization of bound vortex-antivortex pairs \cite{Wachtel2016}. Upon
integrating out bound pairs on gradually increasing length scales, the
dielectric constant is found to flow to ever larger values, signaling the
screening of the vortex-antivortex interaction and the presence of unbound
vortices.

\subsubsection{Stabilization of the ordered phase through strong spatial
  anisotropy}

\begin{figure}
  \centering
  \includegraphics[width=246pt]{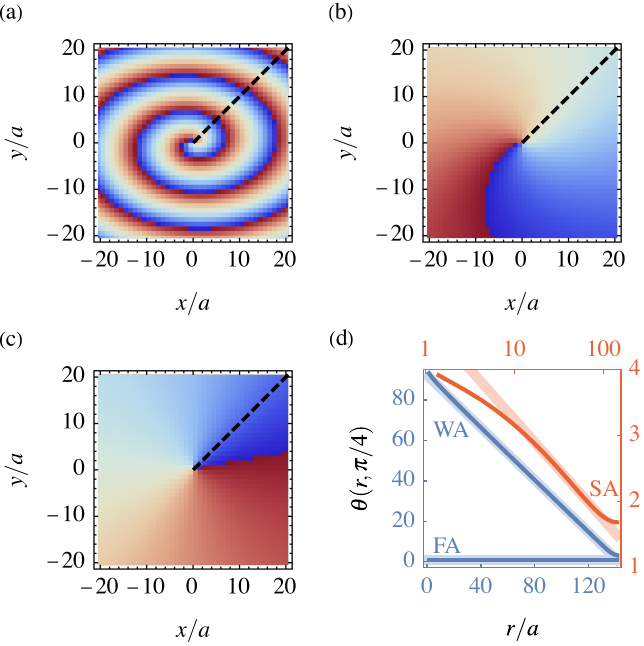}
  \caption{A single vortex in the anisotropic cKPZ equation. Colors from blue to
    red encode $\theta \in [0, 2 \pi)$. (a)~Weakly anisotropic (WA) regime with
    $\lambda_x/(2D) \approx 0.9$ and $\lambda_y/(2D) \approx 0.4$, leading to a
    pronounced spiral structure. (b)~Strongly anisotropic (SA) regime with
    $\lambda_x/(2D) \approx 0.9$ and $\lambda_y/(2D) \approx -0.4$, where the
    radial dependence of the vortex far field is much weaker. (c)~Fully
    anisotropic (FA) configuration with
    $\lambda_x/(2D) = - \lambda_y/(2D) \approx 0.7$. The vortex field does not
    depend on the radial coordinate. (d)~The radial dependence of the vortex
    field along the dashed black lines in (a--c) is $\sim r$ for WA,
    $\sim \ln(r)$ for SA, and $\sim \mathrm{const.}$ for FA. Straight lines are
    fits to the numerical data. Figure adapted from Ref.~\cite{Sieberer2018b}.}
  \label{fig:anisotropic_vortices}
\end{figure}

The interplay between nonequilibrium conditions and strong spatial anisotropy
leads to a stark modification of the vortex-antivortex
interaction \cite{Sieberer2018b}. This can be understood already by examining
the structure of a single vortex shown in Fig.~\ref{fig:anisotropic_vortices}:
In the weakly anisotropic regime characterized by
$\Gamma = \lambda_y/\lambda_x > 0$, vortices emit waves in the radial direction
away from the vortex core, leading to a spiral structure \cite{Aranson2002}; the
radial wave vector vanishes in the strongly anisotropic regime with
$\Gamma < 0$, which results in a much less pronounced spiral structure with a
logarithmic instead of linear dependence of the vortex field on the distance
from the vortex core; and in the fully anisotropic case corresponding to
$\Gamma = - 1$, the radial component of the vortex field vanishes altogether.
The exponential screening of the vortex-antivortex interaction on distances
$r \gg L_v$ in the isotropic case can be understood as a consequence of the
radially emitted wave \cite{Aranson1998}. Therefore, the absence of a radially
emitted wave for fully anisotropic vortices implies that the interaction is
\emph{not} screened in this case, and that a phase with bound vortex-antivortex
pairs can be stable. Further evidence in support of this expectation is provided
by a perturbative calculation of the vortex-antivortex interaction, which is
valid on scales $r \lesssim L_v$, and shows that strong spatial anisotropy can
render the correction to the Coulomb potential in
Eq.~\eqref{eq:vortex-interaction} \emph{attractive} \cite{Sieberer2018b}.

\subsubsection{Observability of KPZ universality in two-dimensional driven open
  condensates}
\label{sec:obs2d}

Experimentally observing KPZ universality in 2D represents a major challenge
due to the scarcity of experimentally accessible 2D
interfaces \cite{Takeuchi2018}. The finding that driven open condensates belong
to the KPZ universality class opens up the intriguing perspective of viewing the
condensate phase as such an interface. However, the observability of KPZ
universality in driven open condensates critically depends on the absence of
unbound vortices. Therefore, we have to compare the characteristic length scales
$L_{\mathrm{KPZ}}$ and $L_v$ given in Eqs.~\eqref{eq:L-KPZ} and~\eqref{eq:L-v},
which determine the onset of KPZ scaling and the mean separation between unbound
vortices, respectively. Both are exponentially large in systems that are driven
only weakly out of equilibrium with $\abs{\lambda} \ll D$; but crucially,
$L_{\mathrm{KPZ}}$ and $L_v$ are exponentially large in
$D^2/\lvert \lambda \rvert^2$ and $D/\lvert \lambda \rvert$, respectively, such
that generically $L_v \ll L_{\mathrm{KPZ}}$, indicating that the proliferation
of free vortices always preempts the emergence of KPZ scaling. Surprisingly,
recent numerical studies show that this does not have to be the case, making the
observation of 2D KPZ universality with driven open condensates a tantalizing
possibility \cite{Mei2021, Deligiannis2022, Ferrier2022}.

Early experimental and numerical studies of the coherence properties of 2D
driven open condensates have focused on parameter regimes for which the scales
$L_{\mathrm{KPZ}}$ and $L_v$ are much larger than the considered system sizes,
and, therefore, have observed equilibrium-like behavior characterized by
algebraic quasi-long-range order and a vortex-unbinding transition to a
disordered phase \cite{Roumpos2012, Nitsche2014, Dagvadorj2015,
  Kulczykowski2017, Caputo2018, Comaron2018, Gladilin2019, Zamora2020a,
  Comaron2021, Dagvadorj2021, Dagvadorj2022}. More recently, numerical studies
have started exploring the far-from-equilibrium regime in which KPZ universality
is expected to emerge. Indeed, in Ref.~\cite{Mei2021}, spatial and temporal
correlations have been found to be governed on large scales by KPZ scaling
exponents in incoherently pumped exciton-polaritons in a regime of weak
interactions and weak noise. On short scales, algebraic order is observable even
in the absence of interactions---in stark contrast to Bose gases in thermal
equilibrium. These coherence properties have been observed in simulations
without vortices, while the presence of vortices in the initial state has been
found to lead to the proliferation of vortices in the steady state. In
Ref.~\cite{Deligiannis2022}, the formation of free vortices has been shown to be
strongly suppressed in spatially discrete condensates in a lattice of
micro-pillars. This setup has also enabled the experimental observation of KPZ
universality in 1D as detailed below \cite{Fontaine2021}. In 2D, vortex creation
is suppressed for incoherent pump strengths far above the condensation
threshold, and even for experimentally realistic noise strengths. Under these
conditions, KPZ universality has been observed both in spatial and temporal
correlations and in the distribution of phase fluctuations. Furthermore, KPZ
universality of correlations and phase fluctuations has also been demonstrated
numerically in coherently pumped exciton-polaritons in an experimentally
realistic parameter regime identified in Ref.~\cite{Zamora2017}, where
$g = \lambda^2 \Delta/D^3 \gtrsim 1$ such that $L_{\mathrm{KPZ}} \approx a$, and
with weak spatial anisotropy \cite{Ferrier2022}. Remarkably, the strongly
nonequilibrium regime considered in these simulations has been found to be
stable against the proliferation of vortices, even though Eq.~\eqref{eq:L-v}
suggests that $L_v \approx a$ for the chosen parameters. The mechanism that
underlies this increased stability against vortex proliferation is not fully
understood but appears to be rooted in the interplay between fluctuations of
the amplitude and the phase of the condensate that is not captured by the
adiabatic elimination of density fluctuations leading to the cKPZ equation. This
assumption is supported by a numerical study of the cKPZ equation on a lattice,
which has confirmed key predictions of the electrodynamic duality, including the
proliferation of vortices in the steady state in weakly anisotropic systems
beyond a length scale that depends exponentially on the inverse of
$\abs{\lambda}$, and the stability of the ordered phase in strongly anisotropic
systems \cite{Zamora2020}. Repulsive vortex-antivortex interactions in isotropic
systems have also been observed in simulations of the full complex condensate
field \cite{Gladilin2017}, and have been found to strongly decrease the rate of
vortex-antivortex annihilation \cite{Fontaine2021, Gladilin2019a}.

The numerical studies summarized above show that using the condensate phase as
an interface which grows according to the KPZ equation is a promising route
toward the experimental observation of KPZ universality in 2D, and provide
clear guidance concerning the optimal choice of parameters. Another
intriguing aspect of driven open condensates, which deserves further numerical
and experimental investigations, is the stabilization of algebraic
quasi-long-range order through strong spatial anisotropy.

\subsection{One-dimensional driven open condensates}
\label{sec:1D-driven-open-condensates}

\begin{figure}
\includegraphics[width=246pt]{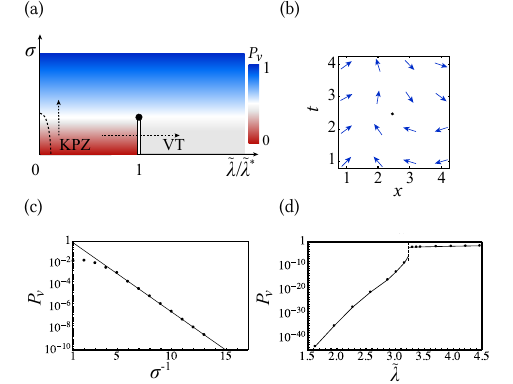}
\caption{(a) Schematic phase diagram of a 1D driven open condensate in terms of
  the rescaled nonequilibrium strength $\tilde{\lambda}/\tilde{\lambda}^{*}$ and
  the noise strength $\sigma$ (denoted by $\gamma$ in
  Eq.~\eqref{eq:driven-open-condensate-Langevin}). The color scale encodes the
   density of vortices in space-time, $P_v$. At low nonequilibrium and noise strengths,
  condensate autocorrelations exhibit KPZ universality. As $\tilde{\lambda}$ is
  increased beyond the critical value $\tilde{\lambda}^{*}$, a first-order
  transition to a vortex turbulent phase (VT) is crossed. The first-order
  transition line terminates in a second order critical point. (b) Phase
  configuration for a phase slip at $t \approx 2.5$. The equivalent
  interpretation as a space-time vortex is apparent. (c) Noise-activated
  behavior $P_v \sim \e^{- E_v/\sigma}$ along the vertical dashed line in
  (a). $E_v$ is the effective energy of a space-time vortex. (d) The space-time
  vortex density exhibits a discontinuous jump when $\tilde{\lambda}$ crosses
  $\tilde{\lambda}^{*}$ along the horizontal dashed line in (a). Figure adapted
  from Ref.~\cite{He2017}.}
\label{fig:space-time-vortex-crossover-turbulence}
\end{figure}

A crucial difference between the noncompact and the compact KPZ equations lies
in the fact that the former depends on a single tuning parameter, which is given
by the dimensionless nonequilibrium strength $g$, while the latter is determined
by two independent parameters. Indeed, the compactness of $\theta$ renders its
rescaling in Eq.~\eqref{eq:KPZ-rescaling} meaningless. Accordingly, after
rescaling $\mathbf{x}$ and $t$ as in Eq.~\eqref{eq:KPZ-rescaling}, and the noise
field as $\eta \to \eta \Lambda^2 D$, we are left with two independent
parameters, which can be chosen as the dimensionless nonequilibrium strength and
the dimensionless noise strength defined as $\tilde{\lambda} = \lambda/D$ and
$\tilde{\sigma} = \Lambda^{d - 2} \Delta/D$, respectively. As shown in
Ref.~\cite{He2017} for a 1D driven open condensate, both parameters are
associated with new phenomena beyond the physics of the noncompact KPZ equation:
The noise strength gives rise to a time scale that terminates the time window
during which KPZ scaling can be observed in autocorrelation functions; and
increasing the nonequilibrium strength $\tilde{\lambda}$ induces a first-order
transition to a new phase of space-time vortex turbulence. The phase diagram of
1D driven open condensates, obtained from numerical simulations of the Langevin
equation~\eqref{eq:driven-open-condensate-Langevin} for the condensate field, is
illustrated schematically in
Fig.~\ref{fig:space-time-vortex-crossover-turbulence}(a). Another mechanism by
which KPZ scaling can disappear is revealed through simulations of the coupled
dynamics of a condensate of lower polaritons and an excitonic reservoir as
described by Eq.~\eqref{eq:condensate-reservoir} \cite{Vercesi2023}: At high
pump powers, the adiabatic elimination of the reservoir, which underlies the
description of the dynamics in terms of a single evolution equation for the
condensate field, breaks down. Then, strong fluctuations of the condensate and
reservoir densities go along with the gradual closing of the temporal window of
KPZ scaling.

\subsubsection{KPZ universality in one-dimensional driven open condensates}

At weak noise and nonequilibrium strength, the fluctuations of the phase field
$\theta$ in the vicinity of any given space-time point are small, and,
therefore, the compactness of $\theta$ can be ignored. Then, the correlation
function or Keldysh Green's function
$C(t - t', x - x') = \langle \psi(t, x) \psi^{*}(t', x') \rangle$ is expected to
exhibit KPZ scaling as described by Eq.~\eqref{eq:KPZ-scaling}. In 1D, the
roughness exponent takes the value $\chi = 1/2$ for any value of the KPZ
nonlinearity $\lambda$ in Eq.~\eqref{eq:KPZ}, including, in particular, the case
$\lambda = 0$.  However, an unambiguous signature of a significant KPZ
nonlinearity, which in turn roots in the breaking of the thermal symmetry in a
driven open condensate as discussed in Sec.~\ref{sec:incoherent-pumping}, is
provided by the stretched-exponential decay of the autocorrelation function
$C(t, 0) \sim \exp(- c t^{2 \beta})$ with $\beta = 1/3$ for the KPZ universality
class. In contrast, $\beta = 1/4$ for the Edwards-Wilkinson universality class
corresponding to diffusion described by Eq.~\eqref{eq:KPZ} with $\lambda = 0$.
KPZ scaling has been observed in numerical studies of the Langevin
equation~\eqref{eq:driven-open-condensate-Langevin} for incoherently pumped
driven open condensates in the regime of weak noise, defined by the absence of
space-time vortices on the spatiotemporal extent of the
simulation \cite{Gladilin2014, Ji2015, He2015}. Further evidence comes from
simulations of the coupled dynamics of the condensate and reservoir as described
by Eq.~\eqref{eq:condensate-reservoir} \cite{Fontaine2021, Vercesi2023}, and
from numerics for 1D photonic cavity arrays \cite{Amelio2020}.

The spatiotemporal coherence of exciton-polariton condensates is determined by
the second moment of the distribution of phase fluctuations, which exhibit
universal KPZ scaling as described by Eq.~\eqref{eq:KPZ-scaling}. Signatures of
KPZ dynamics beyond scaling are contained in the full distribution of the phase,
which follows the Tracy-Widom form during the growth regime of KPZ
dynamics \cite{Fontaine2021, Squizzato2018, Deligiannis2020} before it crosses
over to Baik-Rains statistics in the stationary KPZ
regime \cite{Squizzato2018}. Moreover, subclasses of KPZ universality, which
share the same scaling exponents but are characterized by different variants of
the phase distribution and the spatial correlations of phase fluctuations,
are accessible in systems with external confinement \cite{Deligiannis2020}.

Recently, KPZ universality has been demonstrated in an experimental study of a
1D condensate of exciton-polaritons \cite{Fontaine2021}. Observing KPZ
universality requires sufficiently spatially extended condensates to avoid
finite-size effects, and this requirement poses a challenge to experiments since
large condensates are unstable toward fragmentation \cite{Smirnov2014,
  Bobrovska2014, Liew2015, Daskalakis2015, Bobrovska2018, Estrecho2018}. A way
to overcome this challenge is provided by band engineering in a lattice system
to endow the polaritons with a negative effective
mass \cite{Baboux2018}. Performing experiments with exciton-polaritons in a Lieb
lattice of coupled micropillars \cite{Schneider2017} allowed
Ref.~\cite{Fontaine2021} to reach larger scales than earlier measurements of
coherence \cite{Roumpos2012, Fischer2014, Nitsche2014}.

Interferometrically probing the light field that leaks out of the cavity gives
direct experimental access to the condensate coherence as measured by the
absolute value of the first-order correlation function,
\begin{equation}
  g^{(1)}(\Delta t, \Delta x) = \frac{\langle \psi^{*} (t + \Delta
    t, x + \Delta x) \psi(t, x)\rangle}{\sqrt{\langle  \abs{\psi(t + \Delta
        t, x + \Delta x)}^2 \rangle \langle \abs{\psi(t, x)}^2\rangle}}.
\end{equation}
According to Eq.~\eqref{eq:KPZ-scaling},
$\lvert g^{(1)}(\Delta t, \Delta x) \rvert$ is expected to exhibit
stretched-exponential decay with $\chi = 1/2$ and $\beta = 1/3$. These values
are well reproduced by fits to the experimental data, leading to
$\chi_{\mathrm{exp}} = 0.51 \pm 0.08$ and
$\beta_{\mathrm{exp}} = 0.36 \pm 0.11$. In particular, only the KPZ value
$\beta = 1/3$ and not the Edwards-Wilkinson value $\beta = 1/4$ lies within the
$95\%$ confidence interval around $\beta_{\mathrm{exp}}$. Further experimental
evidence for KPZ universality is provided by a collapse of the data for
$-2 \ln \! \left( \kappa |g^{(1)}(\Delta t, \Delta x)| \right)/\Delta t^{2/3}$,
plotted against the rescaled coordinate $y = \Delta x/\Delta t^{2/3}$, onto the
KPZ scaling function $F = C_0 F_{\mathrm{KPZ}}(y/y_0)$. The values of
$F_{\mathrm{KPZ}}$ are tabulated \cite{Prahofer2004}, and $C_0$ and $y_0$ are
nonuniversal constants. These results demonstrate that 1D condensates of
exciton-polaritons indeed belong to the KPZ universality class, and provide
motivation for pursuing experiments to demonstrate KPZ universality in
2D.

\subsubsection{Destruction of KPZ scaling through noise activated space-time
  vortices}

When local fluctuations of the phase become large due to either strong noise or,
as detailed below, strong nonequilibrium conditions, compactness of the phase
becomes important also in 1D. In particular, large phase fluctuations can induce
phase slips. As illustrated in
Fig.~\ref{fig:space-time-vortex-crossover-turbulence}(b), phase slips are best
visualized in analogy to topological defects in 2D as space-time vortices.

Experimental and numerical observations of KPZ scaling of phase fluctuations
indicate that phase slips are rare at sufficiently weak noise. The space-time
density $P_v$ of phase slips can be computed by formalizing the equivalence
between phase slips and space-time vortices through the mapping of the
$1 + 1 \mathrm{D}$ KPZ equation to the static equilibrium description of a
smectic $A$ liquid crystal \cite{Golubovic1992, Golubovic1994a,
  *Golubovic1994}. This calculation shows that for weak noise, space-time
vortices exhibit activated behavior with $P_v \sim \e^{- E_v/T}$, where $E_v$
and $\sigma$ are the effective energy and temperature of a space-time
vortex \cite{He2017}. The associated mean temporal separation of space-time
vortices gives rise to an exponentially large crossover time scale $t_c$: KPZ
universality with stretched exponential decay of condensate autocorrelations
prevails below $t_c$, and is superseded by simple exponential decay beyond
$t_c$. This crossover has been observed in simulations of the Langevin
equation~\eqref{eq:driven-open-condensate-Langevin}, and the activated behavior
of $P_v$ is illustrated in
Fig.~\ref{fig:space-time-vortex-crossover-turbulence}(c).

\subsubsection{Transition to space-time vortex
  turbulence}\label{sec:1st_order_vortex}

Increasing the nonequilibrium strength $\tilde{\lambda}$ beyond a critical value
$\tilde{\lambda}^{*}$ triggers a dynamical instability and induces a first-order
transition to a phase of vortex turbulence, at any finite noise
level \cite{He2017}. A related dynamical instability has also been observed in
simulations of both 1D and 2D arrays of coupled phase
oscillators \cite{Lauter2017}, and in the coupled dynamics of the condensate and
excitonic reservoir as described by
Eq.~\eqref{eq:condensate-reservoir} \cite{Vercesi2023}. The phase transition is
signaled by a discontinuous jump of the vortex density $P_v$ as shown in
Fig.~\ref{fig:space-time-vortex-crossover-turbulence}(d). Note that $P_v$ is
finite on both sides of the transition. Therefore, the long-time decay of the
autocorrelation function is always exponential. The two phases are distinguished
by the density of vortices, in analogy to the liquid-gas transition in thermal
equilibrium. As illustrated schematically in
Fig.~\ref{fig:space-time-vortex-crossover-turbulence}(a), for increasing noise
strength, the first-order line terminates in a second order critical point,
where the height $\Delta P_v$ of the jump in the vortex density goes to zero,
and the derivative of $\Delta P_v$ with respect to the noise strength $\sigma$
diverges as $\sim \abs{\sigma - \sigma^*}^{-\kappa}$ with $\kappa \approx 0.63$.

An experimentally accessible signature of the transition is provided by the
momentum distribution function
$n(q) = \int \diff x \, \e^{- \imag q x} C(0, x)$ \cite{Carusotto2013}. For
large momenta $q$, the momentum distribution function behaves generically as
$n(q) \sim q^{- \gamma}$. However, crossing the transition, the value of the
exponent $\gamma$ jumps from $\gamma \approx 2$, which is characteristic for
noise activated vortices \cite{Gasenzer2012}, to $\gamma \approx 5$, reminiscent
of turbulence.

\section{Driven open criticality}
\label{sec:driv-open-crit}

In the previous section, we have discussed how strong gapless fluctuations can
prohibit spontaneous symmetry breaking in low spatial dimensions. However, in
higher dimensions, stable ordered phases with broken symmetries are possible. As
anticipated in the Introduction, the strongest form of universality will then
manifest in the critical behavior at the symmetry-breaking phase
transition. Here, we focus on novel critical behavior that results from
the breaking of thermodynamic equilibrium conditions on microscopic scales
(see Tab.~\ref{fig:overview}).

As a first example, in Sec.~\ref{sec:bosonic-driven-open-criticality}, we will
discuss dynamical criticality at the driven open condensation transition in 3D
\cite{Sieberer2013, Sieberer2014, Tauber2013a, Sieberer2016a}. The starting
point of our analysis is the Keldysh action in the semiclassical
limit~\eqref{eq:S-driven-open-condensate-semiclassical} or the equivalent
Langevin equation~\eqref{eq:driven-open-condensate-Langevin}. As explained in
Sec.~\ref{sec:incoherent-pumping}, on large length and time scales the Langevin
equation~\eqref{eq:driven-open-condensate-Langevin} can be mapped to the
isotropic compact KPZ equation~\eqref{eq:KPZ}. The canonical scaling analysis of
Sec.~\ref{sec:RG-flow-KPZ} has shown that the KPZ nonlinearity
Eq.~\eqref{eq:KPZ-nonlinearity}, which serves as a measure for the breaking of
equilibrium conditions as explained in Sec.~\ref{sec:semiclassical-limit}, is
irrelevant in 3D, indicating that in 3D driven open condensates, thermal
equilibrium emerges at large scales. Yet, the breaking of equilibrium conditions
on microscopic scales becomes manifest in a new critical exponent that describes
the approach to equilibrium. Such a modification of critical behavior is
possible for a complex order parameter field, but cannot occur for a real order
parameter field that describes Ising order \cite{Mitra2006, Maghrebi2016,
  WeimerTWA}. However, as we discuss in Sec.~\ref{sec:competing}, novel forms of
nonequilibrium criticality can occur in the coupled dynamics of two competing
Ising order parameters \cite{young2020nonequilibrium}.

\subsection{Bosonic driven open criticality}
\label{sec:bosonic-driven-open-criticality}

Dynamical critical behavior at the driven open condensation transition in 3D has
been studied using the functional RG \cite{Berges2012} and a field-theoretical
RG approach \cite{Tauber2014a} in Refs.~\cite{Sieberer2013, Sieberer2014} and
Ref.~\cite{Tauber2013a}, respectively. A review of these works is presented in
Ref.~\cite{Sieberer2016a}. Here, we will content ourselves with a short summary.

The RG analysis of the driven open condensation transition yields the following
results: 

\textit{(i)~Asymptotic thermalization:} Thermal equilibrium conditions as
expressed through the relation $K_c/K_d = u_c/u_d$ for the coefficients in
Eq.~\eqref{eq:S-driven-open-condensate-semiclassical} are violated explicitly at
the \emph{microscopic} level of the theory. However, symmetry of the Keldysh
action under $\mathcal{T}_{\beta}$ Eq.~\eqref{eq:thermal-symmetry} with an
effective inverse temperature $\beta$ \emph{emerges} at the RG fixed point that
describes the condensation transition. Consequently, the static and dynamical
critical exponents that govern the asymptotic decay of spatial and temporal
correlation functions at large spatial and temporal distances take the same
values as in the corresponding equilibrium problem. The emergence of thermal
equilibrium at large spatial and temporal scales is a rather general phenomenon
\cite{Mitra2006, Diehl2008, Diehl2010a, dalla2010quantum, DallaTorre2013,
  Oztop2012, Wouters2006, mitra2011mode, Mitra2012, Maghrebi2016}. However, for
the driven open condensation transition, this behavior is specific to 3D. In
lower spatial dimensions, coarse graining instead \emph{enhances} the violation
of equilibrium conditions as discussed in Sec.~\ref{sec:driven-open-BEC-2D-1D}
and Sec.~\ref{sec:ddbose}.

\textit{(ii)~Universal decoherence:} The fixed point that describes the driven
open condensation transition is purely dissipative, i.e., the real parts of the
couplings in the Keldysh action
Eq.~\eqref{eq:S-driven-open-condensate-semiclassical} vanish,
$K_c/K_d, r_c/r_d, u_c/u_d \to 0$. However, the RG flow toward this fixed point,
which describes decoherence at large scales, is governed by a new universal
critical exponent that is independent of the static and dynamical exponents, and
the value of which distinguishes equilibrium from nonequilibrium conditions. The
independence of the new exponent requires the existence of an additional
microscopic scale in the Gaussian action \cite{Goldenfeld1992}. Since the
spectral mass scale $- r_c + \imag r_d$ in
Eq.~\eqref{eq:S-driven-open-condensate-semiclassical} is complex, there are in
total four independent real microscopic parameters: $r_c$, $r_d$, $\gamma$, and
a possible coupling $f$ to an external field, corresponding to a term
$\int_{t, \mathbf{x}} f \left( j_c^{*} \psi_q + j_q^{*} \psi_c + \mathrm{c.c.}
\right)$
with source fields $j_c$ and $j_q$. Consequently, there are also four
independent critical exponents: the correlation length exponent $\nu$, the
anomalous dimension $\eta$, the dynamical critical exponent $z$, and the new
decoherence exponent. In contrast, the breaking of nonequilibrium conditions
cannot lead to the occurrence of a new independent exponent at the transition to
a phase with Ising order. Since the Langevin equation for a real field
describing Ising order can only contain real coefficients, its emergent
semiclassical description and the resulting critical behavior are identical to
those of a system in thermal equilibrium \cite{Mitra2006,
  Maghrebi2016}. Nevertheless, self-similar scaling with novel critical
exponents can still occur in the dynamics in an underdamped regime that precedes
relaxation to the thermal fixed point
\cite{paz2022driven,marino2022universality}, and novel nonequilibrium
universality is possible in the coupled dynamics of two competing Ising order
parameters as detailed below (Sec.~\ref{sec:competing}). In addition, very
recent work has shown that for bosonic models with $N$-component complex fields,
governed by the (classical) symmetry group $\mathrm{O}(N)\times \mathrm{U}(1)$,
the equilibrium fixed point is unstable for any $N \geq 2$. Instead, there exist
stable nonequilibrium fixed points hosting novel universal
behavior~\cite{Daviet_2024}. Such models arise naturally near the onset of time
crystalline order. Their phenomenology is similar to the critical dynamics of
coupled competing Ising order parameters, which are addressed below
(Sec.~\ref{sec:competing}).

\begin{figure}
  \includegraphics[width=246pt]{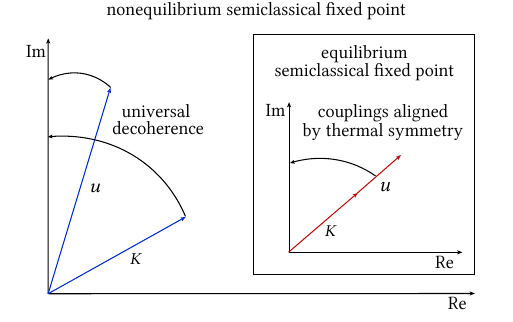}
  \caption{RG flow of the couplings $K = K_c + \imag K_d$ and
    $u = u_c + \imag u_d$ in the complex plane. Both in and out of equilibrium,
    the fixed point is purely dissipative with $K_c/K_d, \, u_c/u_d \to 0$, and
    asymptotic decoherence is governed by an independent critical
    exponent. However, in thermal equilibrium, the couplings are aligned in the
    complex plane due to the thermal symmetry of the Keldysh action. In
    contrast, out of equilibrium, the couplings are not aligned, and the
    exponent that describes universal decoherence takes a different value.}
\label{fig:semiclassical-RG-flow}
\end{figure}

The fact that the value of the new exponent at the driven open condensation
transition in 3D distinguishes equilibrium from nonequilibrium conditions can be
traced back to a qualitative difference between the RG flows of the ratios
$K_c/K_d$ and $u_c/u_d$ in and out of equilibrium, respectively: In equilibrium,
these ratios are locked onto each other and there is a single rate at which they
vanish asymptotically; in contrast, out of equilibrium, the flow of the ratios
is described by three distinct rates, and the new exponent corresponds to the
slowest one. This distinction is illustrated in
Fig.~\ref{fig:semiclassical-RG-flow} in terms of the RG flow of the complex
couplings $K = K_c + \imag K_d$ and $u = u_c + \imag u_d$. The microscopic
nonequilibrium conditions thus manifest even at the largest asymptotic scales in
a universal way.

\subsection{Driven-dissipative criticality with competing order parameters}
\label{sec:competing}

The competition between different order parameters allows us to escape emergent
thermalization at the critical point of driven-open systems.
This topic has first been explored through variational and mean-field methods in
Ising models on the lattice \cite{PhysRevA.95.042133} and in cavity
QED \cite{PhysRevLett.120.183603}.  

A paradigmatic example based on a field theory description has been proposed in Ref.~\cite{young2020nonequilibrium},
where the Langevin dynamics of two coupled   Ising order parameters
$\phi_{1, 2}$ is considered:
\begin{equation}\label{eq:multi}\begin{split}
    \gamma_1 \partial_t\phi_1=-\frac{\delta H_1}{\delta \phi_1} - g_{12}\phi_1\phi^2_2+\xi_1(t),\\
     \gamma_2 \partial_t\phi_2=-\frac{\delta H_2}{\delta \phi_2} - g_{21}\phi_2\phi^2_1+\xi_2(t).
    \end{split}
\end{equation}
In the equations above, $H_{1,2}$ are the quartic Hamiltonians  
\begin{equation}
  H_{\alpha}=\int_ \mathbf{x} \left( \frac{|\nabla
    \phi_{\alpha}|^2}{2}+\frac{r_\alpha}{2}\phi_{\alpha}^2+\frac{g_\alpha}{4}\phi_{\alpha}^4
  \right)
\end{equation}
with $\alpha=1,2$. The parameters $g_{12}$ and $g_{21}$ in Eq.~\eqref{eq:multi}
denote the mutual coupling of the two order parameters; finally, the noise terms
$\xi_{1,2}(t)$ are taken Gaussian and Markovian, with variance proportional to
the friction coefficients ${\gamma_{1,2} }$ in accordance with
fluctuation-dissipation relations.
As usual (see Sec.~\ref{sec:mixedvspure}) criticality is achieved by tuning the
spectral gaps of the two theories to zero: when $g_{12}, g_{21}\neq0$, the
$\Z_2 \times \Z_2$ symmetry of the total Hamiltonian can be spontaneously broken
with both order parameters undergoing critical behavior, in a similar fashion to
multi-critical points in conventional statistical
mechanics \cite{cardy1996scaling}.

\begin{figure}
\includegraphics[width=7cm]{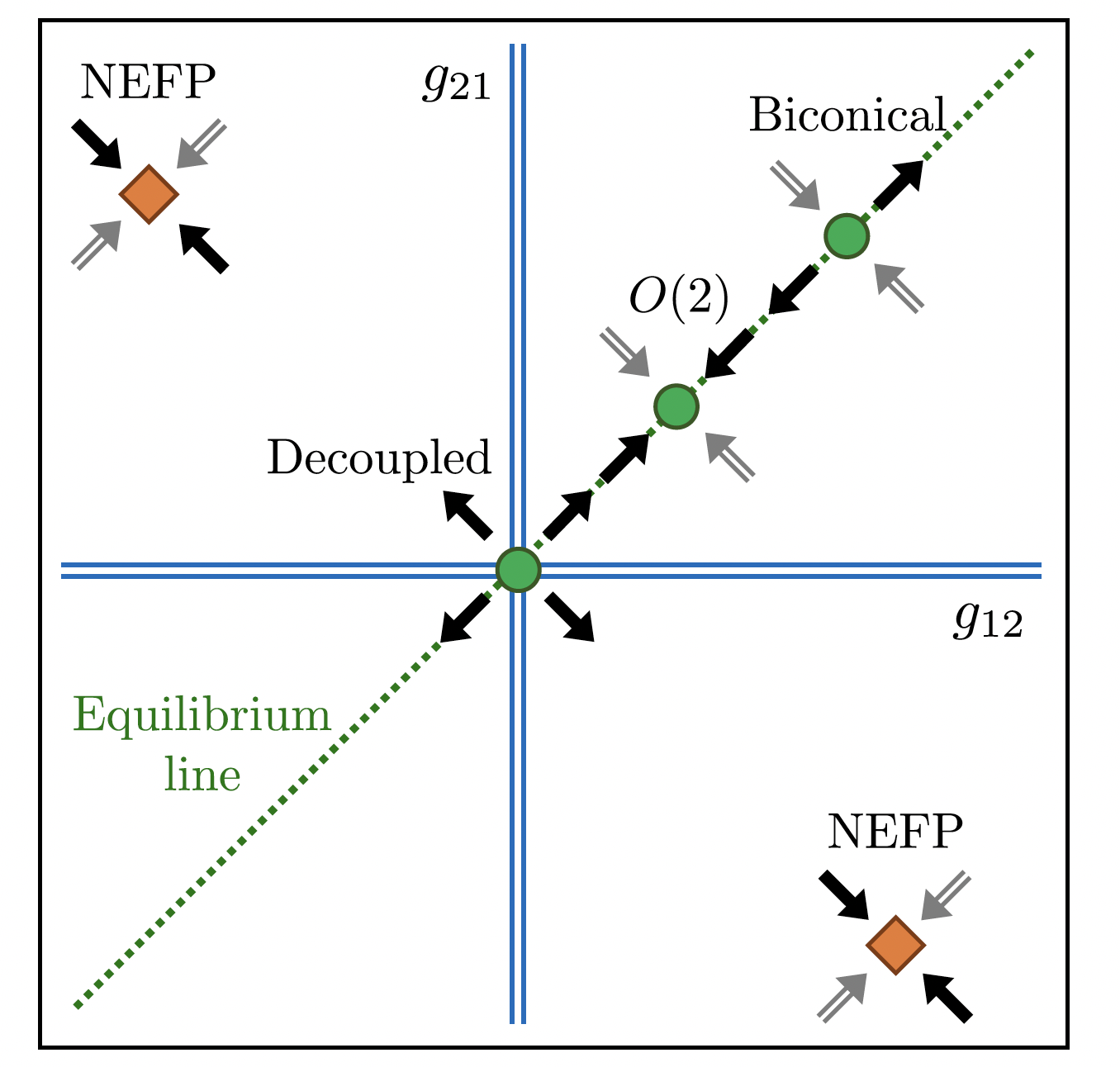}
\caption{ Portrait of the fixed points of two coupled critical Langevin
  models. When the microscopic couplings between the two order parameters are
  both positive, the RG flow is towards an attractive equilibrium
  $\mathrm{O}(2)$ theory. In contrast, when the flow is initiated with couplings
  of different signs, the flow is directed to one of two nonequilibrium fixed
  points (NEFP), which are multi-critical. Figure adapted from
  Ref.~\cite{young2020nonequilibrium}.  }
\label{fig:maghrebi}
\end{figure}

The key property of the dynamics of competing Ising order parameters with
$g_{12}, g_{21}\neq0$ is that the condition for an effective equilibrium
description is much more restrictive as compared to the case of a single order
parameter. This opens up a route to escape effective thermalization. Indeed,
obedience to the thermal symmetry of Sec.~\ref{sec:thermal-symmetry} would
require the dynamics to be governed by a single Hamiltonian, which can only
occur when $g_{12}=g_{21}=g$. In this case, the equations of
motion~\eqref{eq:multi} can be derived from
$H=H_1+H_2+\frac{g}{2}\int_\mathbf{x}\phi^2_1\phi^2_2$, and the steady-state
distribution of the Langevin dynamics is a Gibbs state, $\rho \sim e^{-H/T}$.
From an RG perspective, microscopic couplings can get renormalized in the flow
to their fixed points.  If the two coupling constants are both initialized with
positive values \cite{young2020nonequilibrium}, the RG flow will reach an
equilibrium fixed point in the $\mathrm{O}(2)$ universality class.
Figure~\ref{fig:maghrebi} shows in addition two unstable fixed points
corresponding to the decoupled limit ($g_{12}=g_{21}=0$) and to a biconical
fixed point \cite{PhysRevB.13.412}.

The situation becomes richer when the two microscopic couplings, $g_{12}$ and
$ g_{21}$, are initialized with opposite signs, and thus realize a strongly nonreciprocal
coupling \cite{young2020nonequilibrium}. As shown in Fig.~\ref{fig:maghrebi},
the RG flow is   attracted towards a pair of nonequilibrium fixed
points. These two fixed points have the same number of stable directions as the
equilibrium fixed points, i.e., the same number of RG relevant parameters tuned
to zero in order to approach criticality. Therefore, the RG flow can
discriminate between the equilibrium and nonequilibrium fixed points using   the
relative sign of the couplings as  control knob.

In the following we summarize the key properties of the infrared RG flow.

\textit{(i)~Discrete scale invariance:} 
An RG analysis to order $O(\epsilon^2)$ with $\epsilon = 4 - d$ the distance from the upper critical dimension $d_u=4$, reveals that the nonequilibrium fixed points are characterized by a discrete scale
invariance of response and symmetric correlation functions 
reminiscent of fractals. This should be contrasted with the ubiquity of
continuous scale invariance at equilibrium criticality and in most of the
nonequilibrium critical points discussed so far in this review.

\textit{(ii)~Breakdown of thermalization:} The nonequilibrium fixed points display an emergent effective
temperature (see Sec.~\ref{sec:bosonic-driven-open-criticality}), which scales
in the infrared ($q\to0$) as $T_{\mathrm{eff}}\propto q^{\eta-\eta'}$. Here,
$\eta$ and $\eta'$ are the anomalous dimensions of correlation and response functions,
respectively. Their mismatch $\eta \neq \eta'$ is a genuine signature of the
system being out of equilibrium since this rules out the possibility of having a
scale-independent temperature (see Sec.~\ref{sec:mixedvspure}). 

These two phenomena are also realized in the nonequilibrium criticality of $N$-component complex boson models~\cite{Daviet_2024,young2024}. A similar phenomenology is also found at the nonthermal quantum fixed point of
Sec.~\ref{sec:ddbose}, where it has been first identified; however, in that case the system retains quantum
coherence, while in the case discussed in this section the system is fully
classical. These exotic properties suggest that the criticality of driven
open systems can host new universal scenarios which have not been explored in non-equilibrium statistical mechanics so far, and can become a frontier of
exciting research in the years to come.

\section{Slowly and rapidly driven open systems}
\label{sec:slowly-rapidly-driven-open-systems}

So far, we have considered critical phenomena in nonequilibrium stationary
states, where no explicit time dependence was involved. In this section, we
address the question which novel universal phenomena emerge once such a time
dependence is introduced. To this end, we will consider systems that exhibit a
continuous phase transition and focus on the regimes of slow and fast drive. The
limit of an undriven system with periodic driving frequency $\Omega = 0$
corresponds to thermodynamic equilibrium, and the limit of an infinitely rapidly
driven and open system with $\Omega^{-1} = 0$ is captured by a time-independent
description in terms of a Lindblad equation. In the latter case, the combination
of drive and dissipation creates a nonequilibrium situation as discussed in
Sec.~\ref{sec:equilibrium-vs-nonequi-steady-states}, and critical behavior under
these conditions has been studied in
Sec.~\ref{sec:bosonic-driven-open-criticality}. An overview of the different
limits taken to derive the microscopic starting points for the extraction of the
universal macroscopic behavior is provided in Sec.~\ref{sec:introduction}.

Let us briefly anticipate the main results of the following discussion. The
vicinity of both limiting points $\Omega = 0$ and $\Omega^{-1} = 0$ exhibit a
similar phenomenology, characterized by a cutoff of the divergence of the
correlation length upon approaching the transition point. However, this common
trait is rooted in drastically different physical mechanisms, following from the
fact that slow and fast drives exert low- and high-energy modifications,
respectively. As a consequence, the critical exponents that determine the
scaling of the correlation length at distances below the cutoff are unmodified
in the slowly driven problem compared to the equilibrium case; yet the drive
allows one to \emph{activate} subleading scaling exponents \cite{Mathey2020a,
  Ladewig2020}. In contrast, the rapidly driven case exhibits an additional
independent critical exponent \cite{Mathey2019, Mathey2020}.

\subsection{Slowly driven systems: activating critical exponents}
\label{sec:slowdrive}

\subsubsection{Kibble-Zurek mechanism}

The basic physics of a slowly driven system close to a second order
finite-temperature or quantum phase transition is described by the Kibble-Zurek
mechanism \cite{kibble1976, Zurek1985, Polkovnikov2006,
  delCampo2014,dziarmaga2005dynamics,zurek2005dynamics}. Here the periodic
driving frequency $\Omega$ is taken so much smaller than the relaxation time
scale of the system, that its periodicity can be neclected, and the drive is
described by a (e.g. linear) ramp. It has been explored and validated in a broad
spectrum of experiments, e.g., recently with a Rydberg quantum
simulator \cite{Keesling2019}. More concretely, consider thus a slow ramp
through the critical point as described by a time-dependent tuning parameter
$g(t) = g_0 - V t$, where for the initial value $g(0) = g_0$ the system is deep
in the disordered phase and the critical point is reached for $g(t) = 0$.
Starting from $t = 0$, the time evolution is adiabatic as long as the relaxation
time $\tau$ of the system is small as compared to the rate of change of $g$
characterized by the time scale $t_g(t) = g(t)/\dot{g}(t) \approx g_0/V$. Close
to the critical point, however, the relaxation time starts to diverge as
$\tau \sim \xi^z$ where $\xi \sim g(t)^{- \nu}$ is the correlation
length. Adiabaticity is broken when $\tau \sim g(t)^{- \nu z} \approx t_g(t)$,
and the state of the system is frozen with correlations extending over a finite
length scale $\xi$, which obeys a universal scaling form as a function of the
drive velocity $v$ \cite{Polkovnikov2006, DeGrandi2010b, Barankov2008,
  Sen2008,zurek2005dynamics,dziarmaga2005dynamics}:
\begin{equation}
  \label{Eq:IntroKZMScaling}
  \xi \sim V^{-\frac{1}{z+1/\nu}}.
\end{equation}
Notably, the Kibble-Zurek mechanism gives access to the dynamical exponent $z$,
which is not available from the static correlation functions. However, as we
discuss below, this is just the tip of the iceberg of a more general
observation: By slowly driving the system in the vicinity of a phase transition,
it is possible to access the full spectrum of critical exponents.

In fact, there are infinitely many universal numbers defining the exponent
spectrum \cite{cardy1996scaling,Pelissetto2002,Tauber2014a}. The customary
exponents such as $\nu$ are associated with relevant couplings, and determine the
leading scaling behavior. Exponents which are associated with irrelevant couplings
provide only subleading corrections that are hard to resolve both in
experiments \cite{Pelissetto2002} and in theory \cite{Ferrenberg2018}. An exception
are conformal field theories, where a relation between the scaling dimensions of
operators and the energy spectrum has been established, making the subleading
exponents amenable to exact numerics \cite{Laeuchli2013}. Nevertheless, as we demonstrate
below, in slowly driven systems, these exponents become observable in the
 asymptotic scaling behavior of generic critical systems.

\subsubsection{Exact RG approach to slowly driven systems}

As discussed above, in the adiabatic regime, the equilibration time scale is
much shorter than the time scale of the drive. Time then only enters via the
parameters characterizing the system. Therefore, the full set of adiabatic
flow equations, which encodes the exact scaling dimensions of all couplings, is
obtained from its undriven counterpart by promoting the static parameters to
time-dependent ones, $k\partial_k \bvec{g}(t) =\boldsymbol{\beta}(\bvec{g}(t))$,
where $k$ is the running momentum cutoff, and we bundle the couplings and the
respective beta functions in vectors
$\bvec{g} = \bvec{{g}}({t}\,) = ({g}^1({t}\,),{g}^2({t}\,), \dotsc)$ and
$\boldsymbol{\beta}$. We consider a linear drive,
$\bvec{g}(t) = \bvec{g}_0 + \bvec{g}_1 t$. A standard
procedure \cite{Zinn-Justin, Tauber2014a} yields the flow equations for
dimensionless couplings $\mathbf{\hat{g}}_0$ and
$\mathbf{\hat{g}}_1$. Crucially, time must be rescaled like any other parameter,
leading to an additional contribution to the flow equations \cite{Mathey2020a}.

The phase transition is described by the Wilson-Fisher fixed point of the RG
flow of dimensionless couplings. To find the critical exponents, the set of
flow equations has to be linearized around this fixed point. We are interested in
the vicinity of the usual equilibrium fixed point where
$\bvec{\hat{g}}_{1,*} = 0$. In the adiabatic limit, the linearized flow
equations for $\mathbf{\hat{g}}_0$ and $\mathbf{\hat{g}}_1$ decouple. The
linearized flow of $\mathbf{\hat{g}}_0$ is described by a matrix $M$, whose
eigenvalues determine the full spectrum of critical exponents. In contrast, the
linearized flow of $\mathbf{\hat{g}}_1$ is determined by the matrix $M - z
\id$.
This shows that only the eigenvalues of $M$ and $z$ itself determine the
critical behavior of the adiabatic flow equations. That is, no new universal
information is contained in the slowly driven system---it just probes the
underlying equilibrium critical point, in analogy to the Kibble-Zurek result
Eq.~\eqref{Eq:IntroKZMScaling}. However, the shift by $z$ can lead to additional
relevant directions at the equilibrium Wilson-Fisher fixed point, whose physical
consequences will be discussed below.

Let us consider a concrete example of a two-parameter RG flow obtained by truncating
the infinite-dimensional matrix $M$ for an Ising magnet in zero external
field. The eigenvalues $\theta_i$ of $M$ are then sorted by magnitude and
labelled conventionally as $\theta_1 \equiv - 1/\nu$ and $\theta_2 = \omega$,
where $\nu > 0$ is the relevant exponent describing the scaling of the
correlation length according to $\xi \sim g^{\nu}$, and $\omega > 0$ is the
first irrelevant exponent \cite{cardy1996scaling}. Accordingly, the eigenvalues of $M - z \id$ are shifted
down to $-1/\nu - z$ and $\omega - z$. Typically, $z \approx 1$ for a linear
dispersion or $z \approx 2$ for a quadratic dispersion or diffusion. In both
cases, $z$ is larger in size than the typical value of $\omega$. For the example
of the Ising model in three dimension, $1/\nu \approx 1.54$ and
$\omega \approx 0.66$ \cite{juettner2017}, and the dynamical exponent
$z\approx 2.013$ for model A dynamics \cite{Zinn-Justin,Tauber2014a}. Therefore,
due to the slow drive, there are two additional relevant directions. As we
detail below, the first one leads to the Kibble-Zurek scaling in
Eq.~\eqref{Eq:IntroKZMScaling}, and the second one to new scaling behavior.

\subsubsection{Observable consequences}

\begin{figure}
  \includegraphics[width=246pt]{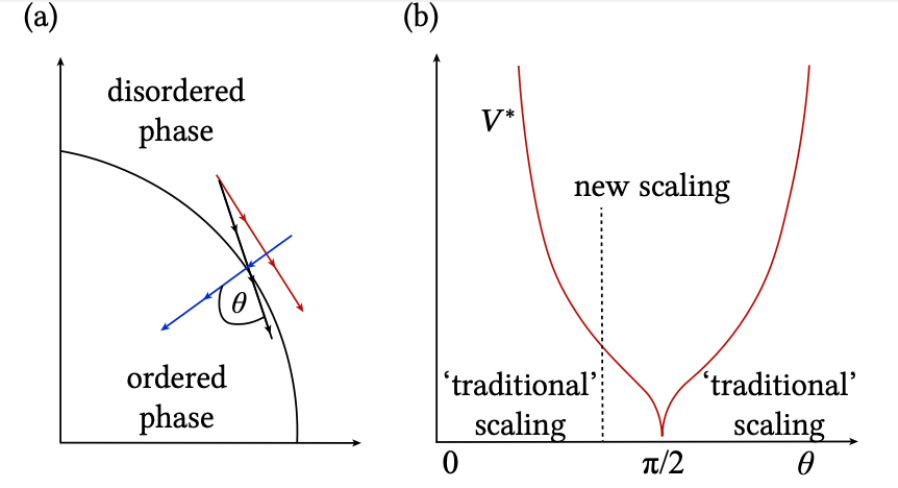}
  \caption{(a) Driving protocols: perpendicular (blue), parallel (red), and
    general (black) in a phase diagram with two experimental knobs (e.g.
    temperature and pressure). (b) Crossover velocity $V^*$ between different
    scaling regimes. For parallel drive $\theta =\pi/2$, the new scaling is
    observed even for small drive velocities.}
  \label{fig:rgexp}
\end{figure}

When a system is tuned close to a continuous phase transition, the RG flow of
dimensionless couplings will first approach the vicinity of the fixed point,
before being pushed out of this critical domain along one of the relevant
directions. In the above example of the two-parameter RG, there are three
relevant directions at the Wilson-Fisher fixed point. For simplicity, let us
assume that the matrix $M$ is diagonal, such that the couplings correspond to
the eigendirections emanating from the fixed point. Then, $-1/\nu - z$ and
$\omega - z$ are the exponents associated with the couplings $g_1^1$ and $g_1^2$,
respectively. The exiting of the critical domain along either direction leads to
the breaking of adiabaticity \cite{Ladewig2020}. Therefore, the scale at which
the critical domain is left determines the observable correlation length $\xi$
and will feature information on the respective exponents. The different
exponents can be accessed, or \emph{activated,} by approaching the phase
boundary along different directions. This is illustrated in
Fig.~\ref{fig:rgexp}(a): Tuning the experimental knobs at velocity $V$
perpendicular to the phase border (blue arrow) corresponds to a ramping of the gap and is
thus associated exclusively with $g_1^1$, i.e., $V \sim g_1^1$. In this case,
adiabaticity is broken and the scaling of the observable correlation length is
cut off at a scale $\xi$ that obeys the `traditional' Kibble-Zurek scaling
Eq.~\eqref{Eq:IntroKZMScaling}. In contrast, moving in parallel to the phase
border is associated exclusively with time-dependent irrelevant couplings, which
generically will have a component from the least irrelevant coupling $g_1^2$,
i.e., $V \sim g_1^2$ now. Then, the correlation length scales as
\begin{equation}
 \xi \sim V^{- \frac{1}{z -\omega}}.
\end{equation}
That is, the irrelevant exponent $\omega$ is now activated in the leading
asymptotic scaling behavior, and can thus be determined experimentally once $z$
is known.

A generic driving protocol is characterized by a velocity
$\mathbf{V} = V \left( \cos(\theta), \sin(\theta) \right)$ in the
two-dimensional phase diagram. In terms of the RG, this means that the couplings
$g_1^1$ and $g_1^2$ get mixed---which one `wins' in cutting off scaling depends
on the drive angle $\theta$, and there is a crossover velocity $V_*$ separating
two regimes of different asymptotic cutoff scaling, see
Fig.~\ref{fig:rgexp}(b). In a higher-dimensional phase diagram, more new scaling
directions can be revealed by suitably chosen drives \cite{Mathey2020a}.
 
These general findings are validated in the analytically solvable and
experimentally relevant transverse XY model in Ref.~\cite{Ladewig2020}. In
particular, it is shown that the driving protocol can be designed such that the
excitation density of defects hosts the full sequence of exponents in its
asymptotic scaling behavior.

\subsection{Rapidly driven systems: open Floquet criticality}

We now turn to the opposite limit and focus on periodically driven or Floquet
systems with fast driving. For closed and interacting systems, periodic driving
generically leads to heating. Even though the heating time scale is
exponentially large in the drive frequency, so that a quasi-stationary
prethermal state can emerge on intermediate time scales, such systems will
eventually heat up to infinite temperature \cite{Abanin2016, Mori2016,
  Kuwahara2016, Abanin2017, Eckardt2017, Moessner2017, Harper2020,
  Seetharam2019}. In contrast, coupling a Floquet system to a heat bath can
stabilize the entropy density at finite values, and at asymptotically long
times, observables become stationary up to a periodic time dependence with
frequency $\Omega$ imposed by the drive---the system synchronizes to the latter. Here, we focus on such \textit{open
  Floquet steady states.} The nature of symmetry-breaking phase transitions in
rapidly driven, open systems has been investigated in
Ref.~\cite{Mathey2019}. Technically, this amounts to the question of how to
renormalize an open Floquet system \cite{Mathey2020}. Prior RG-based work has
addressed this question in the slowly-driven limit \cite{DeSarkar2014,
  Nikoghosyan2016, Baoquan2016}, and numerical studies have focused on
intermediate driving
frequencies \cite{Korniss2000,Fujisaka2001,Buendia2008,lorenzo2017quantum}. Here, we concentrate
on the effect of fast, but not infinitely fast driving.

\subsubsection{Basic physical picture}

A key result of this analysis is that rapid weak drive prevents
the correlation length from diverging, in some phenomenological similarity to
the Kibble-Zurek scenario. Pristine criticality is thus absent and restored
only in the limit of infinitely rapid drive discussed in
Sec.~\ref{sec:bosonic-driven-open-criticality}. At first sight, such a drastic
modification of the nature of the phase transition might be unexpected---the
drive provides a fast scale $\Omega$, which should not affect the
long-wavelength critical properties. However, in the driven system
energy is conserved only $\text{mod}(\Omega)$, and the notion of `high' and
`low' energy is thus not well defined. Based on this fact
the system evades the common reasoning, as will be shown below. A picture that captures the essence of
the mechanism is instead obtained based on the synchronization of the relevant
spectral gap $r_d$ (see Sec.~\ref{sec:specgap}; we denote it as in Eq.~\eqref{eq:S-driven-open-condensate-semiclassical}, since we will specialize to this model below) to the drive, 
\begin{equation}
  r_d(t) = r_{d, 0} + \sum_{n\neq0} r_{d, n} \e^{- i n \Omega t}.
  \label{eq_mut}
\end{equation}
Even when the static system is critical for $r_{d, 0} \rightarrow 0$, the
synchronized mass still oscillates and is  dragged periodically across the phase
transition. This leads to a blurring of the phase boundary. In
particular, the divergence of the correlation length is cut off for any
$\Omega < \infty$. Only in the limit of infinitely fast driving, the rotating
wave approximation, in which all contributions with $n \neq 0$ are set to zero,
becomes exact. Then, criticality is restored.

\subsubsection{Keldysh-Floquet model}

Our microscopic starting point is a generalization of the model for driven open
Bose-Einstein condensation from Sec.~\ref{sec:Lindblad-to-Keldysh}, where now we
assume the Hamiltonian to be periodically time dependent,
$\hat{H}(t + 2\pi/\Omega) = \hat{H}(t)$. As discussed in
Sec.~\ref{sec:semiclassical-limit}, close to the condensation transition, a
semiclassical description applies. The corresponding Keldysh action is given in
Eq.~\eqref{eq:S-driven-open-condensate-semiclassical} but with time-dependent
coupling constants. Anticipating full decoherence in the critical semiclassical
model, all couplings can be approximated as being purely
imaginary \cite{Mathey2020}.

\begin{figure}
  \includegraphics[width=246pt]{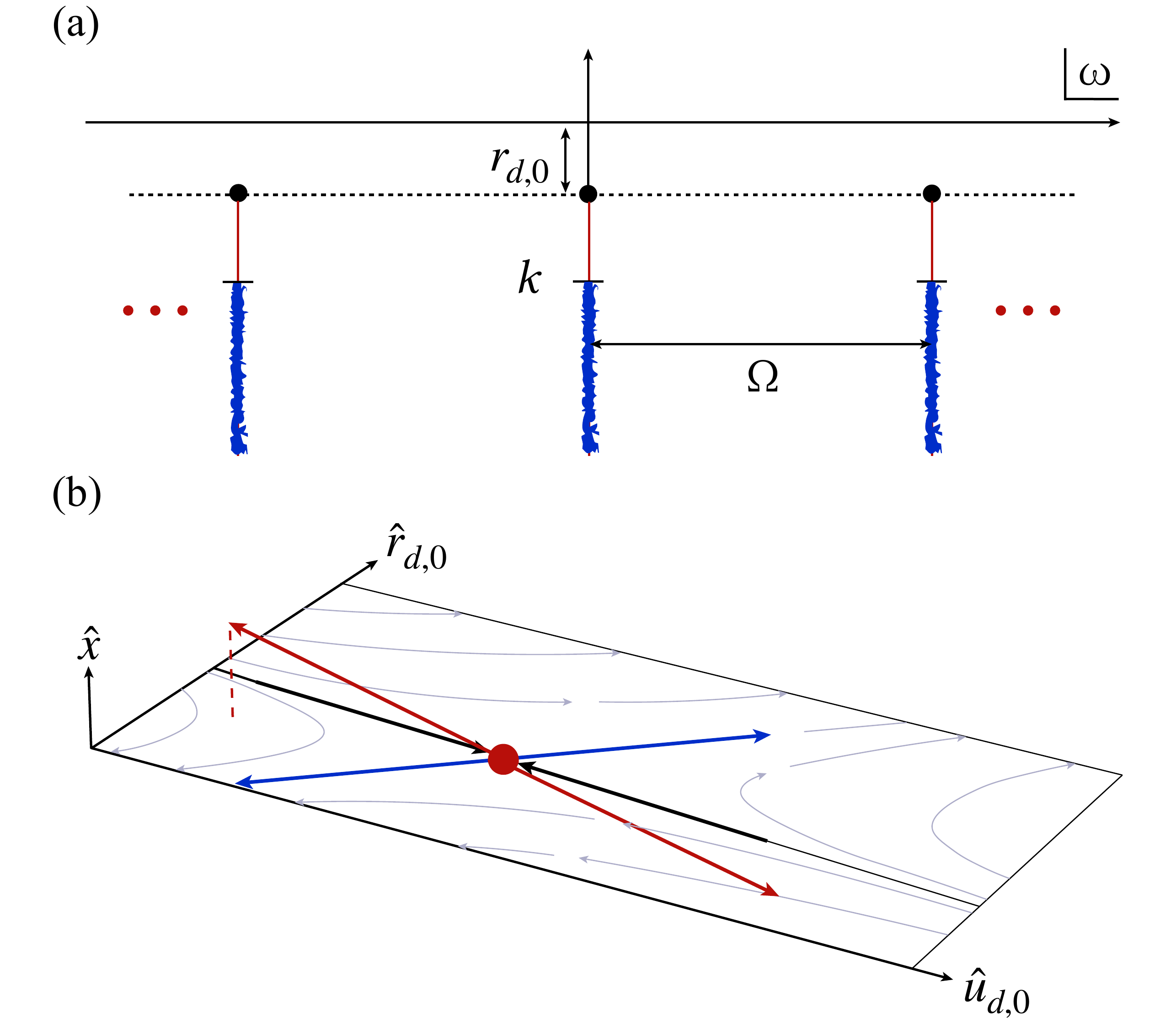}
  \caption{(a) The poles of the retarded Wigner Green's function are organized
    in lines (shown in red; parallel to the imaginary axis for the purely
    dissipative open Floquet problem considered here) and separated along the
    real axis by the drive frequency $\Omega$. The origin of the real axis is
    arbitrary, whereas the distance from the origin along the imaginary axis
    $r_{d,0}$ denotes the distance from the phase transition
    point. Renormalization up to a scale $k$ corresponds to integrating out
    modes along the lines of poles, indicated by blue shading. (b) RG flow of
    dimensionless couplings. The horizontal plane, spanned by the mass
    $\hat r_{d,0}$ and the interaction strength $\hat u_{d,0}$, hosts the usual
    Wilson-Fisher fixed point (red dot) with one relevant direction. The fast
    drive provides an additional new relevant direction
    $\hat x \sim \Omega^{-1}$.}\label{fig_poles}
\end{figure}

Key structures governing the behavior of the open Floquet system close to
criticality are contained in the single-particle Green's function $G(t, t')$
which, for explicitly broken continuous time-translation symmetry, depends on both $t$ and
$t'$. It is convenient to switch to the Wigner
representation \cite{Arrachea2005,wu2008,Stefanucci2008,Tsuji2008,Genske2015},
\begin{equation}  
  G \! \left( t + \tfrac{\tau}{2} ,t - \tfrac{\tau}{2} \right) = \sum_n \int
  \frac{d\omega}{2\pi} e^{-i (n \Omega t + \omega\tau) } G_n(\omega),
\end{equation}
where the discrete time translation invariance is encoded in a Fourier series
with fundamental frequency $\Omega$, while the dependence on the relative time
$\tau$ is represented through a Fourier transform as familiar from undriven
stationary states. Corrections to the rotating wave approximation can be
obtained through an expansion of $G_n(\omega)$ in powers of
$r_{d, n \neq 0}/\Omega$. As illustrated in Fig.~\ref{fig_poles}(a), the poles
of the retarded Wigner Green's function $G^R_n(\omega)$ form lines in the
complex plane, corresponding to different Floquet-Brillouin zones that are
separated along the real axis by integer multiples of $\Omega$. Under the
assumption of purely imaginary couplings, these lines are parallel to the
imaginary axis. Importantly, all lines of poles become critical
\textit{simultaneously} as $r_{d, 0} \to 0$.

\subsubsection{Renormalization of the open Floquet problem}

Renormalizing the open Floquet problem corresponds to integrating out modes
along the lines of poles in the complex plane, as indicated by the blue shading
in Fig.~\ref{fig_poles}. All lines of poles contribute with the same degree of
divergence, even though for weak driving with $r_{d, n \neq 0}/\Omega \ll 1$,
higher Floquet-Brillouin zones are suppressed. In an  $\epsilon = 4 - d$-expansion
close to the upper critical dimension $d_u = 4$, the flow equations for
dimensionless variables to order $O(\epsilon r_{d, n \neq 0}/\Omega)$ read
\begin{equation}
  \label{eq_dimensionless_rg_flow}
  \begin{split}
    k \partial_k \hat{u}_{d, 0} & = - \epsilon \hat{u}_{d, 0} + \frac{10
      S_d}{\left|1+\hat{r}_{d, 0}\right| \left( 1 + \hat{r}_{d, 0}\right)} \hat{u}_{d, 0} \left(\hat{u}_{d, 0} +\hat x\right) , \\
    k \partial_k \hat{r}_{d, 0} & = -2 \hat{r}_{d, 0} -
    \frac{4S_d}{\left|1+\hat{r}_{d, 0}\right|} \left( \hat{u}_{d, 0} + \hat x\right), \\
    k\partial_k \hat{x} & = -\epsilon \hat{x},
\end{split}
\end{equation}
with ${S_d = 2 \pi^{d/2}/[(d/2-1)! (2\pi)^{d}]}$ and dimensionless couplings
$\hat{r}_{d, 0} = k^{-2} \frac{r_{d, 0}}{K_{d, 0}}$ and
$\hat{u}_{d, 0} = k^{d-4} \frac{\gamma u_{d, 0}}{4 K_{d, 0}^2}$. The effect of
higher harmonics with $n \neq 0$ is subsumed in a new coupling $\hat{x}$. It can
be shown that this coupling receives no loop corrections at order
$O(\epsilon r_{d, n \neq 0}/\Omega)$, so it only shows dimensional running, and
the set of flow equations is closed \cite{Mathey2019}.

The flow equations~\eqref{eq_dimensionless_rg_flow} generalize those of time
translation invariant $\mathrm{U}(1)$ systems. Indeed, as shown in
Fig.~\ref{fig_poles}(b), the standard Wilson-Fisher fixed point emerges when
$\hat{x}=0$. The analysis shows that the periodic drive gives rise to a new
relevant coupling. In the absence of continuous time translation invariance, the
critical point is thus bicritical: Fine-tuning of two parameters is necessary to
reach it, and to reveal its critical scaling properties. In an experiment, one
will tune across the symmetry breaking phase transition at finite
$\hat{x} \sim \Omega^{-1}$. Then, the additional relevant direction provides a
finite correlation length. Consequently, this constitutes a symmetry-breaking
phase transition without asymptotic criticality.

This finding can be interpreted within the framework of fluctuation-induced
first-order transitions \cite{Coleman1973, Halperin1974, Fisher1974,
  Nelson1974}: The interplay of several modes that simultaneously become
critical changes the phase transition from a second to a first-order one. This occurs
without an explicit symmetry breaking as is usually present in first-order phase
transitions with a critical endpoint of higher symmetry. While in the
traditional instances of this scenario, the gapless modes are realized by a
critical mode coupled to Goldstone or gauge modes, here they are realized by the
poles of the different Floquet-Brillouin zones, all reaching criticality
jointly.

As usual, the critical exponents can be obtained from a linear stability
analysis of the RG flow equations close to the Wilson-Fisher fixed
point \cite{Mathey2019, Mathey2020},
\begin{eqnarray}
  -1/\nu = -2 + 2 \epsilon /5, \quad -\omega = \epsilon, \quad -1/\nu_d = -\epsilon.
\end{eqnarray}
The first two exponents are known and unmodified as compared to both the
undriven problem and the infinitely rapidly driven system, which shares the
static exponents with the equilibrium situation as discussed in
Sec.~\ref{sec:bosonic-driven-open-criticality}. In contrast, the exponent
$\nu_d$ is new, adding an independent relevant direction to the Wilson-Fisher
fixed point. The discussion can now be led in analogy to
Sec.~\ref{sec:slowdrive}: In the presence of two relevant couplings, the
observed scaling of the correlation length depends on the direction under which
the fixed point is approached. For example, when the system is infinitely
rapidly driven, $\hat x =0$, the correlation length diverges as
$\xi \sim r_{d, 0}^{-\nu}$. However, crossing the phase boundary with
$\hat{x} \neq 0$, the correlation length never diverges due to the repulsive
direction, and $\xi$ saturates to a finite value that scales as
$\xi \sim x^{-\nu_d}$ with the distance from the fixed point. In particular,
this implies that the new critical exponent $\nu_d$ can be observed by varying
$\Omega$. This is a key quantitative and universal prediction of this
analysis.

\subsubsection{Relation to the slowly driven limit}

The above discussion shows that the mechanism established here for a rapid drive
is distinct from the slowly driven Kibble-Zurek scenario investigated in
Sec.~\ref{sec:slowdrive}---despite the phenomenological similarity that close to
the limits of undriven and infinitely rapidly driven systems, scaling is cut
off. In RG language, this is due to the emergence of new relevant directions at
the respective fixed points. But the difference is also transparent in this
language: As we have discussed, in the Kibble-Zurek scenario, the drive provides
information on the underlying equilibrium critical point via the set of
equilibrium critical exponents, because the new relevant directions just host
information that is already encoded in the equilibrium fixed point. Instead, for
rapid drive we obtain a new independent exponent (actually, more than one, see
Ref.~\cite{Mathey2020}). This is rationalized by the fact that in the former
case, we deal with an \textit{infrared} modification of the critical physics: A
slow driving scale is introduced, so slow that the periodic functions in
Eq.~\eqref{eq_mut} can be expanded in powers of $\Omega$ and the periodicity is
never probed on the accessible time scales. Conversely, in the latter case, the
modification is in the \textit{ultraviolet:} A fast driving scale is introduced,
and the periodicity is crucial. Another way of seeing this is the presence of
the additional mass scales in Eq.~\eqref{eq_mut}, connecting back to the
intuitive picture of synchronized mass oscillations provided at the beginning of
this section. It is a basic but also fundamental insight of RG theory that it is
such ultraviolet scales that can modify and add critical exponents to
the observable phenomenology \cite{Goldenfeld1992}.

\section{Nonequilibrium first-order phase transitions}
\label{sec:firstorder}

First-order phase transitions appear in many important physical
scenarios \cite{Bloete1979}, including every day's boiling of water and hysteresis in ferromagnetic materials, and are characterized by a
discontinuous jump of static, thermodynamic observables. Despite the
discontinuous behavior, i.e., the absence of scaling in static
observables, first-order phase transitions
display universal behavior in a broader sense: systems undergoing a first-order
phase transition share a number of ubiquitously recurring
patterns. 

The hallmark of first-order phase transitions is the coexistence of two
macroscopically distinguishable, thermodynamic phases. In thermal equilibrium,
this coexistence corresponds to a bistability in the system's free energy, which
persists under coarse graining up to the largest distances. Hence, the distinguishability of the two phases implies that the
correlation length does not diverge at the  transition but remains finite.

A nondiverging correlation length gives rise to a peculiar phenomenology,
including structurally distinct dynamics that appear on different length
scales. This in turn leads to a variety of experimentally observable phenomena
such as scale invariant excitation avalanches \cite{Hysteresis1993}
(cf.\ crackling noise \cite{Crackling}) and intermittency between two
phases \cite{IntermittencyI,IntermittencyII}. The general phenomenology of first
order transitions is reviewed below. This is followed by a discussion of how the
transitions may be enriched if the dynamics are placed out of thermal
equilibrium \cite{Letscher2016, Melo2016,LesanovskyMetastability2015,
  Fruchart_2021,CheyneVitelli2022}.

\subsection{General phenomenology of first-order phase transitions}\label{sec:general_first_order}

We provide a brief summary of the phenomenology of first-order phase transitions
in thermal equilibrium from the perspective of effective field theory. For an
in-depth overview, we refer to the literature, e.g., the
reviews \cite{Binder_1987,RevModPhys.54.235,Berges_firstorder}. For an
elementary model, consider a real scalar order parameter field
$\phi(t, {\bf x})$ governed by a Langevin equation of the form (we set
$\phi\equiv \phi(t, \mathbf{x}), \eta\equiv \eta(t, \mathbf{x}))$
\begin{equation}\label{eq:bistabilityThermal}
  \partial_t\phi=D\nabla^2\phi-V'(\phi)+\eta,
\end{equation}
where $V'(\phi) = \diff V(\phi)/\diff \phi$, and $\eta$ is a Gaussian white
noise with average $\langle \eta(t, \mathbf{x}) \rangle =0$ and variance
$\langle \eta(t, {\bf x})\eta(t', {\bf x}') \rangle =
2 T \delta(t-t') \delta({\bf x}-{\bf x}')$.
This imprints thermal fluctuations with temperature $T$. In order to realize a
bistability, the potential $V(\phi)$ must take the form of a double- or
multi-well. For concreteness, consider a tilted $\phi^4$-potential
$V(\phi)=\frac{u_2}{2}\phi^2+\frac{u_4}{4}\phi^4-\frac{2\alpha \sqrt{u_2
    u_4}}{3}\phi^3$,
where the dimensionless parameter $\alpha$ tunes the strength of the
tilt.\footnote{This is equivalent to an Ising model in a magnetic field
 $h\sim \alpha(1-\alpha^2/\alpha_c^2)$.} For $u_2,u_4>0$ and
$\alpha>1$ the potential has two stable field configurations at $\phi=0$ and
$\phi=\varphi_f\equiv\sqrt{u_2/u_4}(\alpha+\sqrt{\alpha^2-1})$ and a separatrix
at $\phi=\sqrt{u_2/u_4}(\alpha-\sqrt{\alpha^2-1})$. The solution $\phi=0$ is
favored, i.e., is the global minimum, for $\alpha\le\alpha_c=3/\sqrt{8}$. We
introduce the order parameter $\varphi$ as the spatial average
$\varphi=\frac{1}{V}\int_V d \mathbf{x} \, \langle \phi \rangle $ over the
system volume $V$.

Mean-field theory predicts a first-order phase transition and coexistence of the
solutions $\varphi=0$ and $\varphi=\varphi_f$ at $\alpha=\alpha_c$, when the
global minimum of $V$ jumps from $\phi=0$ ($\alpha<\alpha_c$) to
$\phi= \varphi_f$ ($\alpha>\alpha_c)$. In the vicinity of this point, $\phi$ undergoes strong
fluctuations, performing local noise-induced
transitions between the two solutions. One distinguishes three different
dynamical regimes, according to the typical length scale on which they are
observed. The relevant, emergent length scale is the so-called droplet length $\xi_D$. It arises from the competition between noise and diffusion \cite{Langer1967,LangerNucleation} in an instanton-type calculus.

\emph{(i)~Short distance fluctuations}: On short distances
$|\mathbf{x}|\ll \xi_D$, the field $\phi$ performs fast (on time scales smaller than $1/T$ ) but small
fluctuations around one potential minimum. They are driven by the competition
between thermal noise $\sim T$ and diffusion $\sim D\nabla^2$. Diffusion
favors smooth configurations of $\phi$ in space and suppresses noise-induced
short-distance fluctuations. It prevents short-distance fluctuations from
reaching across the potential barrier, which are thus
`blind' to the double-well structure and only explore the
potential minima. This leads to a smooth renormalization of the potential
$V$ \cite{Berges_firstorder,BuchholdFirstOrder,Rudnick1975}.

\emph{(ii)~Droplet nucleation}: On sufficiently large distances
$|\mathbf{x}|\sim\xi_D$, thermal fluctuations may induce a spontaneous, local
but smoothly varying transition of the field $\phi$ from one minimum to the
other, $0\leftrightarrow\varphi_f$. This process is known as droplet
nucleation \cite{LangerNucleation, Elgart2004}. The spontaneously generated
droplet has a finite volume. Its size is determined by the competition between
diffusion, preferring extended, smooth configurations, and the fluctuation
strength, preferring local excitations. The droplet is bounded by a sharp domain
wall, which interpolates between the two different phases, e.g., between
$\phi=0$ and $\phi=\varphi_f$. The nucleation
corresponds to the formation of an instanton configuration of the field
$\phi$ \cite{LangerNucleation2,ColemanFalsevacuum}. It describes rare noise activation processes, which
lead to large fluctuations of the field $\phi$ at the droplet length scale
$\xi_D$. These fluctuations exceed the typical statistical fluctuations by
far (cf.\ the formation of steam bubbles in boiling water), which manifests in the
RG flow of the theory discussed below.

\emph{(iii)~Domain wall motion:} After nucleation, droplets either grow or
decay. On length scales larger than the droplet scale
$|\mathbf{x}|\gg\xi_D$, the growth is governed by the interface
between the two phases, i.e., by the motion of domain walls. On average, the
domain wall motion favors to decrease the potential energy. Hence, droplets grow
(shrink) if they correspond to the global (local) minimum of the potential. This
determines the thermodynamic phase of the system. At the transition, the energy
difference between both phases
vanishes, causing slow phase ordering kinetics \cite{Furukawa,Binder_1987}.

\subsection{First-order phase transitions out of
  equilibrium}\label{sec:1st_order_noneq}

How is the picture of first-order phase transitions modified out of equilibrium?
On a formal level, i.e., on the level of the Langevin equation, the modification
is similar to the case of second order phase transitions. In order to discuss
the possible scenarios, consider an $N$-component order parameter field
$\phi_a\equiv\phi_a(t, \mathbf{x})$, $a=1, \dotsc ,N$, which obeys the Langevin
equation
\begin{align}\label{eq:general_noneq_first}
 \partial_t\phi_a=D\nabla^2\phi_a+F_a(\boldsymbol{\phi})+\eta_a,
\end{align} where
$F_a(\boldsymbol{\phi})$ is the field-dependent force and
$\langle \eta_a(t, \mathbf{x}) \eta_b(t', \mathbf{x}') \rangle =
\chi_{ab}(\boldsymbol{\phi}(t, \mathbf{x}))\delta(t-t')\delta(\mathbf{x}-\mathbf{x}')$
is a Gaussian white noise, which imposes local field fluctuations with strength
$\chi_{ab}(\boldsymbol{\phi})$. In thermal equilibrium, two conditions are
simultaneously met \cite{Hohenberg1977,Sieberer2016a}: (i)~the deterministic force derives from a potential form
$V$, i.e., $F_a(\boldsymbol{\phi})\equiv-\partial_{\phi_a}V(\boldsymbol{\phi})$,
and (ii)~a flat noise spectrum
$\chi_{ab}(\boldsymbol{\phi})\equiv \delta_{a,b}T$ ensures a conventional
fluctuation-dissipation relation. Out of equilibrium, one generally
distinguishes the following two scenarios:

\emph{(i)~Absence of a potential form:} One way to break detailed balance, and
hence enforce nonequilibrium conditions, is the absence of a potential form for
the force
$F_a(\boldsymbol{\phi})\neq-\partial_{\phi_a}V(\boldsymbol{\phi})$. Bistability
at a first-order phase transition requires two (or more) field configurations
$\boldsymbol{\phi}_{1,2}$ to be solutions of
$\mathbf{F}(\boldsymbol{\phi}_{1,2})=0$. However, the transition cannot be
derived from an extremization principle, i.e., from the minimization of a global
free energy. Instead, the system will evolve into a flux equilibrium where the
sum of all possible paths connecting
$\boldsymbol{\phi}_1\leftrightarrow \boldsymbol{\phi}_2$ is
balanced \cite{Graham1970,Graham1980}.  One example discussed in
Sec.~\ref{sec:1st_order_vortex} is vortex turbulence, which arises in a compact,
single-component KPZ equation. A second important example are nonreciprocal
systems \cite{Fruchart_2021,CheyneVitelli2022,Gelhausen2018,Zelle2024}, where
first-order phase transitions correspond to an exceptional point in the system's
linear stability matrix
$A_{ab}\equiv \partial_{\phi_b}F_a(\boldsymbol{\phi})|_{\boldsymbol{\phi}=0}$. Such
situations describe, e.g., flocking and synchronization phenomena in active
matter or non-Hermitian
systems \cite{Fruchart_2021,CheyneVitelli2022,Hanai2020,Zelle2024}.

\emph{(ii)~Absorbing states:} Another way to impose nonequilibrium conditions is
via a state-dependent noise kernel $\chi(\boldsymbol{\phi})$. Under certain conditions,
such a noise kernel imposes the breaking of detailed balance akin to absorbing
state phase transitions and directed percolation in Sec.~\ref{Sec:DP}. For
simplicity, we discuss this case for a scalar field $\phi$. The state-dependence
of $\chi$ implies that the noise is biased. It induces strong fluctuations of
the field $\phi$ when $\chi(\phi)$ is large and weak fluctuations when
$\chi(\phi)$ is small. This may influence the stationary solution for the
optimal field configuration. Consider for instance two degenerate minima of the
potential $V(\phi_1)=V(\phi_2)$. Then of the two, the configuration with smaller
noise is the more stable one. For the case of state-dependent fluctuations, one
can define a modified free energy landscape $W(\phi)$ if and only if $W$ solves
the equation $\partial_{\phi} W(\phi)=-F(\phi)/\chi(\phi)$ for the entire
configuration space of $\phi$. If such a solution exists, then the system
fulfills an effective detailed balance with respect to $W(\phi)$. If no such
solution exists, then detailed balance is truly broken. This is the case when
$\chi(\phi)^{-1}$ is singular for some field configuration $\phi'$, e.g., when
$\chi(\phi')=0$. In particular, this is the case when a bistability occurs
between an active and an absorbing phase, i.e., a dark state for which
$\chi(\phi)=0$ at one of the bistable states \cite{Marcuzzi2016,
  BuchholdFirstOrder}. A setup where this scenario occurs has been introduced
for ultracold Rydberg gases in the facilitation regime discussed in
Sec.~\ref{sec:FieldTheoryDP}. Here, the quantum mechanical origin of the
facilitation mechanism introduces a bistability between a dark state and an
active state at high atom density \cite{Lee2011,PhysRevA.95.042133,Buchhold2017}.

\subsection{Dark state bistability}\label{sec:dark_state_bistable}

The emergence of dynamical bistable regimes has been explored and predicted for
a broad range of driven-dissipative systems, including gases or arrays of
Rydberg atoms \cite{PhysRevA.95.042133,
  WeimerBistabilityI,WeimerVariationalII,Marcuzzi2014,Lee2011,Lee2012,Kshetrimayum2017,WeimerVariational,Letscher2016,Carr2013,Malossi2014,Sibalic2016},
general driven two-level or spin systems
\cite{BistabilityTwoLevel,BistabilityTwoLevelII,Raghunandan2018,Maghrebi2016,Landa2020}
or cavity arrays and the driven-dissipative Bose-Hubbard model
\cite{LeBoite2013,Jin2013,Reeves2023,BenaryExperiment2022,FossFeig2017,BistableMagnon2018}. In
many cases, it has been pointed out, however, that on long wavelengths thermal
behavior emerges (see the discussion of emergent thermal symmetry,
Sec.~\ref{sec:emergenteq}). This leads to a constant noise kernel
$\chi(\phi)\sim \text{const.}$ in Eq.~\eqref{eq:general_noneq_first}. This
renders these transitions thermal or even transform the bistable region into a
second order transition, both of which show no sign of nonequilibrium
universality at large distances. Driven Rydberg atoms are ideal to prevent this
effective thermalization at large distances: the effective Hamiltonian in
Eq.~\eqref{eq:FacHam} together with dissipative decay of the excited states
implement the precise conditions for a robust dark state, and thus for a
nonequilibrium first-order phase transition
\cite{Lee2011,PhysRevA.95.042133,Buchhold2017,Kazemi2023,Letscher2017a}.

For a bistability separating an ordered phase from a dark state, the first-order
phase transition is significantly modified. As discussed in
Sec.~\ref{sec:FieldTheoryDP}, in the limit of vanishing off-resonant
excitations, i.e., when Eq.~\eqref{eq:FacHam} is the exact Hamiltonian, the dark
state $\hat\rho_D=|D\rangle\langle D|$ is a pure quantum state. In this state the
order parameter $\varphi$ and its fluctuations vanish (see
Appendix~\ref{sec:dark}). In contrast, the ordered (\emph{active}) phase yields
a nonzero value for the order parameter and its fluctuations, and resides in a
mixed state. For non-vanishing but strongly suppressed off-resonant excitations,
an almost pure dark state with product state structure emerges at long
wavelengths, where the mixedness of the state per volume vanishes
\cite{BuchholdFirstOrder}.

A first-order phase transition between an active and a dark state thus readily
dictates a set of modifications:

(1) \emph{Coexistence:} The dark state has a local product state structure,
e.g., ${\hat\rho_D=\prod_l |\downarrow\rangle\langle\downarrow|_l }$ for Rydberg
atoms (see Sec.~\ref{subsec:rydberg}). Coexistence between an active phase and a
dark state implies that the system hosts an extensive volume $V_D$ which is in
the dark state. Inside this volume, the order parameter and thus the
fluctuations vanish. Hence, the state of the total system is well approximated
by a product
$\hat\rho=(\hat\rho_{\text{active}})_{\bar V_D}\otimes(\hat\rho_D)_{V_D}$
between a pure state $\hat\rho_D$, constrained to the volume $V_D$, and a mixed
state $\hat\rho_{\text{active}}$ on its complement $\bar V_D$.

\begin{figure*}
  \includegraphics[width=\textwidth]{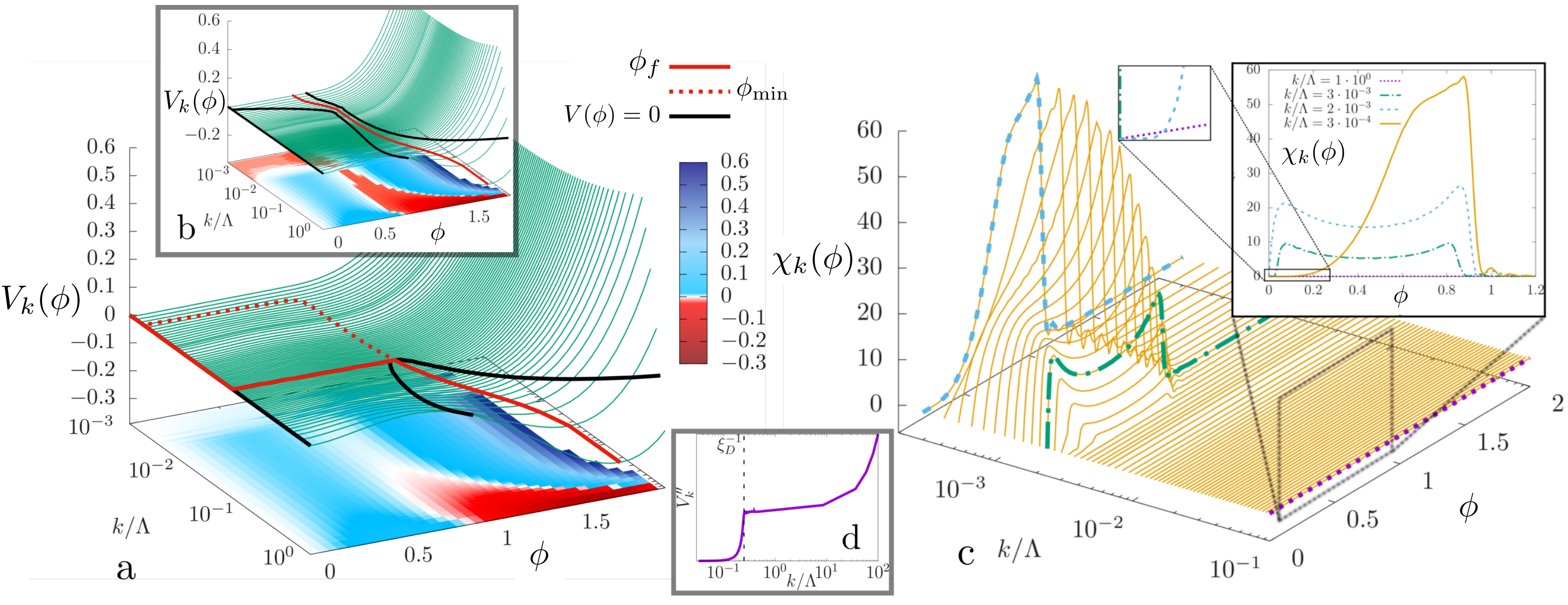}
  \caption{RG flow of the potential $V_k$ and noise kernel $\chi_k$ at a first
    order phase transition in $d=3$ dimensions. The microscopic model is given
    by the Langevin equation~\eqref{eq:bistabilityDark} with initial conditions
    $V_{\Lambda}=
    \tfrac{1}{2}\phi^2+\tfrac{1}{4}\phi^4-\tfrac{2\alpha}{3}\phi^3$
    and noise kernel $\chi_{\Lambda}=\tfrac{1}{10}\phi$. (a) $\alpha = 3.48$,
    (b) and (c) $ \alpha = 3.5$. The flow at large momenta $k\sim \Lambda$
    yields small corrections to the potential landscape: The minimum at $\phi>0$
    moves slightly and becomes nearly degenerate to the dark state at $\phi=0$
    in (a) and (b). (d)~The flow of $V''(0)$ provides an estimate for the change
    of the potential toward a convex function. At the droplet scale
    $k_D=\xi_D^{-1}$, it undergoes a sudden, sharp drop. (c) At this scale
    $\chi_k$ increases by three orders of magnitude. It assumes a bimodal form
    indicating large droplet fluctuations. At the largest distances fluctuations
    are suppressed again, confirmed by the inset. Figure adapted from
    Ref.~\cite{BuchholdFirstOrder}.}
\label{fig:FirstOrderFlow}
\end{figure*}

(2) \emph{Droplet nucleation:} Fluctuations are absent in the dark state volume
$V_D$. This prohibits noise activation from the dark state to the active
phase. Droplets appear exclusively in the active phase volume $\bar V_D$, This
leads to a unidirectional fluctuation pattern where local noise-induced
transitions can only follow the path
$\hat\rho_{\text{active}}\to\hat\rho_D$. Hence, one finds a maximally extended
hysteresis loop: once the global dark state has been reached, there is no return
to the active phase.

(3) \emph{Domain wall motion:} The domain walls experience a similar unidirectional fluctuation pattern. The presence (absence) of fluctuations in the active (inactive) phase makes the domain walls favor the dark state and extend the volume $\bar V_D$. This bias needs to be compensated by a nonzero difference in the potential energy. Hence, the noise $\chi$ and the potential $V$ can no longer be treated independently. 

Items (1)--(3) are relevant for the field theory approach to a first-order dark state phase transition. In particular, the interplay between the potential and the noise for both the domain wall motion and the droplet formation requires a treatment of the potential and the fluctuations on equal footing, i.e., an RG-approach in the Keldysh field integral framework. This approach provides an unambiguous identification of the three dynamical regimes (i)--(iii)~in Sec.~\ref{sec:general_first_order}, and their modifications (1)--(3) at a dark state transition \cite{BuchholdFirstOrder}.

\subsection{Field theory of first-order dark state phase transitions}

The dynamics at a dark state bistability with a scalar order parameter field $\phi\equiv\phi(t, \mathbf{x})$ and with a potential form can be examined in a Langevin framework as in Eq.~\eqref{eq:bistabilityThermal}. The dark state at $\phi=0$ is implemented by a modified noise kernel \cite{BuchholdFirstOrder,Canet2004} as in Eq.~\eqref{eq:general_noneq_first}: 
\begin{equation}\label{eq:bistabilityDark}
    \partial_t\phi=D\nabla^2\phi-V'(\phi)+\eta.
\end{equation}
Here $V$ is as defined above and the Gaussian noise $\eta$ has zero mean and
variance
$\langle \eta(t,{\bf x})\eta(t',{\bf x}') \rangle
=\chi(\phi(t, \mathbf{x}))\delta(t-t')\delta(\mathbf{x}-\mathbf{x}')$.
The noise kernel $\chi(\phi)$ is a positive, continuous function of $\phi$,
which vanishes in the dark state, $\chi(\phi=0)=0$. For concreteness, one may
assume $\chi(\phi)=\kappa\phi$ for some $\kappa,\phi>0$. This form of $\chi$ and
$V$ is inspired by the Rydberg setup discussed in Sec.~\ref{sec:FieldTheoryDP},
which yields Eq.~\eqref{eq:bistabilityDark} in the high density
regime \cite{Marcuzzi2015, Buchhold2017}.

The Langevin equation~\eqref{eq:bistabilityDark} describes the evolution of
the field $\phi$ on short distances. A conventional, perturbative RG approach
yielding the effective field theory on large distances is not applicable at a
first-order phase transition, due to the presence of a length scale and the
absence of universal scaling behavior. Nonperturbative approaches based on the
functional RG \cite{Wetterich:1992yh,Canet2004,Canet2005,Dupuis_2021} are more
promising, and have been successfully applied \cite{BuchholdFirstOrder,Berges_firstorder}. For a nonequilibrium
bistability as described by Eq.~\eqref{eq:bistabilityDark}, a convenient
implementation of the functional RG is performed in momentum space. The general
formulation for the RG transformations in the nonequilibrium Keldysh (or
Martin-Siggia-Rose-Janssen-de Dominicis) framework can be found, e.g., in
Refs.~\cite{Sieberer2016a,Canet2004,BuchholdFirstOrder}. At RG scale $k$, it
yields an effective theory for the field $\phi$ defined over momenta $q<k$ by integrating
out short distance fluctuations with momenta $q>k$.

The first-order transition can be approached in a Langevin truncation, which was
introduced in Ref.~\cite{BuchholdFirstOrder}. In this framework, each RG
transformation, i.e., each integration over short distance fluctuations,
modifies the potential $V$ and the noise kernel $\chi$. At RG scale $k$, the
dynamics are described by the Langevin equation \eqref{eq:bistabilityDark} with
$V \to V_k$ and $\chi \to \chi_k$. The potential $V_k$ and the noise kernel
$\chi_k$ transform under each RG step. A truncation of this form ensures that
the evolution of the distribution function of the field $\phi$ at each momentum
scale $k$ follows a regular Fokker-Planck equation, and thus remains a
well-defined, nonnegative probability distribution under coarse graining. In
general, one might also include a scale-dependent wave function renormalization,
$\sim Z_k\partial_t\phi$, and diffusion $\sim D_k\nabla^2\phi$, but their
evolution is expected to be negligible due to the finite correlation length
(verified in Ref.~\cite{BuchholdFirstOrder} for Eq.~\eqref{eq:bistabilityDark}).

The Keldysh action for the effective Langevin equation is
\begin{equation}
  S_k=\int_{t,{\bf x}} \phi_q \left[
    \left(\partial_t-D\nabla^2\right)\phi_c+
    V'_k(\phi_c)-\chi_k(\phi_c) \phi_q \right].
\end{equation}
This imposes a polynomial dependence on the quantum (response) field
$\phi_q$, but leaves the functional dependence on the classical field $\phi_c$
general. As mentioned above, such a truncation ensures that the theory at each
RG step can be mapped to an appropriate Langevin or Fokker-Planck
equation. Loosening this restriction, one may allow a more general dependence on
the $\phi_q$-fields as, e.g., discussed in
Refs.~\cite{Canet2004,Canet2005}. For the bistability in
Eq.~\eqref{eq:bistabilityDark}, such modifications remain, however, negligibly
small \cite{BuchholdFirstOrder}. One may expect this behavior to be general for
semiclassical theories whose distribution function is governed by a
Fokker-Planck equation at all relevant scales.

For the Langevin truncation, the RG equations for the (unrescaled) potential and noise vertex are \cite{BuchholdFirstOrder, Canet2005} (setting the diffusion constant $D=1$ and dropping the field dependence in the argument)
\begin{equation}
  \label{eq:FirstOrderFlow}
  \begin{split}
    \partial_k V^{(1)}_k&= U_{k,d}V_k^{(3)},\\
    \partial_k\chi_k&=U_{k,d}\left(\chi^{(2)}_k-4\frac{\chi^{(1)}_k
        V^{(3)}_k}{\left(k^2+V^{(2)}_k\right)}+\frac{\chi_k
        (V^{(3)}_k)^2}{2\left(k^2+V^{(2)}_k\right)^2}\right),
  \end{split}
\end{equation}
with
$U_{k,d}=-k^{d+1}
\chi_k\left[2^{d-1}\pi^{\frac{d}{2}}\Gamma\left(\frac{d}{2}\right)d\left(k^2+V^{(2)}_k\right)^2\right]^{-1}$.
At a first-order phase transition, the fields $\phi_c, \phi_q$ do not obey
canonical power counting. Thus a polynomial expansion of the potential and the
noise kernel cannot be justified. It is then necessary to keep the general
functional forms of $V_k$ and $\chi_k$, and to solve the flow equations
numerically.

\subsection{Renormalization group flow}

The phenomenology of the first-order phase transition, including the
modifications from the dark state, is reflected in the numerical solution of the
flow equations~\eqref{eq:FirstOrderFlow} for the bistability
equation~\eqref{eq:bistabilityDark} \cite{BuchholdFirstOrder}. Close to a first
order transition, the RG flow of the potential $V_k$ and noise $\chi_k$ reveal
regimes (i)--(iii)~from Sec.~\ref{sec:general_first_order}:

(i)~At short distances, i.e., at large momenta $k\gg \xi_D^{-1}$, where
$k=\Lambda\gg\xi_D^{-1}$ is the UV cutoff, the potential $V_k$ is weakly
renormalized (flattened) due to integrating over short wavelength fluctuations
(see initial stage of the RG flow for $k/\Lambda\ge0.1$ in
Fig.~\ref{fig:FirstOrderFlow}).

(ii)~At a sharp momentum scale $k_D\Lambda=\xi_D^{-1}\Lambda\approx0.1$, both
the potential $V_k$ and the noise kernel $\chi_k$ experience a sudden, strong
renormalization (witnessed by $V_k''$ in Fig.~\ref{fig:FirstOrderFlow}(d)). The
potential approaches a nearly degenerate bistability, $V_k(\phi=0)\sim
V_k(\phi=\varphi_f)$. The noise kernel $\chi_k$ assumes a bimodal structure with
peaks at $\phi=0,\varphi_f$, see Fig.~\ref{fig:FirstOrderFlow}(c). This
indicates strong fluctuations of the field between the two minima, i.e., the
onset of droplet nucleation at the scale $k_D=\xi_D^{-1}$.

(iii)~At momenta $k\ll k_D$, on distances much larger than the droplet scale
$\xi_D$, the noise kernel $\chi_k$ becomes flat $\chi_k(\phi)\approx0$ for
$\phi\in[0,\varphi_f)$. The corresponding Langevin equation
Eq.~\eqref{eq:bistabilityDark} is noiseless and describes deterministic domain
wall motion. Asymptotically ($k\to0$) the potential assumes a linear form
$V_k(\phi)\sim \phi$ and
vanishes at the bistability $V_k(\phi)=\chi_k(\phi)=0$ for $0<\phi<\varphi_f$.

This reconciles the field theory with the general phenomenology of first-order
phase transitions, and in particular with a dark state transition. Phase
coexistence between a dark state and an active state requires both the
deterministic dynamics generated by $V_k$ \emph{and} the noise $\sim\chi_k$ to
vanish. The renormalization group flow of the noise kernel $\chi_k$ and the
interplay between $V_k$ and $\chi_k$ is crucial to resolve this behavior.

Whether first-order dark state phase transitions can be observed in driven
quantum systems is a topic of ongoing research. Several possible scenarios have
been proposed \cite{Elgart2004}, ranging from Rydberg atoms in the facilitation
regime \cite{Marcuzzi2016,BuchholdFirstOrder,Buchhold2017,Sibalic2016,Letscher2016,Minjae2019},
to cellular automata \cite{Gillman2021}, or between two different dark
states \cite{Carollo2022}. However, first-order phase transitions are often
unstable against fluctuations in low dimensions \cite{BuchholdFirstOrder,
  Luebeck2006,Hinrichsen2000}. Thus, they are restricted to higher dimensions
$d\ge2$, where numerical simulations of quantum dynamics are challenging. This
is at the moment a major obstacle in   identifying   first-order dark state transitions.

\section{Quantum criticality in driven open systems}
\label{sec:quantum-criticality}

The previous sections have covered driven open quantum systems whose scaling
descriptions fall into extensions of thermal universality classes, or constitute
novel nonequilibrium fixed points \emph{per se}.
In each of the cases presented, the microscopic quantum nature of the problem is
levelled out on long wavelengths and becomes inconsequential for the effective
field theory description.
Here, instead, we will cover two instances of the quantum scaling limit
mentioned in Sec.~\ref{sec:semiclassical-limit}, marking the commencement of the third and last thematic part of this review (see Tab.~\ref{fig:overview}). We will focus on noise-driven
one-dimensional (1D) bosonic systems whose quantum features can persist over
long length and time scales.
Specifically, we will consider a Luttinger liquid driven by $1/f$ noise
(Sec.~\ref{sec:LLdriv}), and a driven open condensate subjected to diffusion
noise (Sec.~\ref{sec:ddbose}), i.e., Markovian noise whose variance scale
quadratically with momentum.
Both models illustrate how the formation of nonequilibrium quantum critical
states is determined by the interplay between dissipation and nonlinearities.
In the former case (\ref{sec:LLdriv}), scaling manifests in nonuniversal
exponents controlled by the noise strength, while in the latter case
(\ref{sec:ddbose}), scaling is governed by a quantum fixed point without
classical counterpart, leading to universal scaling exponents that define a
novel universality class.

\subsection{Quantum critical scaling of noise-driven Luttinger
  liquids}\label{sec:LLdriv}

We follow Refs.~\cite{dalla2010quantum,dalla2012dynamics,dalla2012noisy} and
consider a 1D bosonic wire, described as a Luttinger liquid with periodic
potential of strength $g$, and driven by a noise field $f(t)$ that couples to
fluctuations of the density. The Hamiltonian reads
\begin{multline}
  \label{eq:LL}
  \hat{H}(t)=\int \frac{dx}{u} \left[K \left( u\pi \partial_x
      \hat{\theta}(x) \right)^2+\frac{1}{K} \left( u\partial_x\hat{\phi}(x) \right)^2 \right. \\
  \left. \vphantom{\left( u\partial_x\hat{\phi}(x) \right)^2} -g \cos \! \left(
      2\hat{\phi}(x) \right)\right] -\frac{1}{\pi}\int \frac{dx}{u}
  f(t) \partial_x\hat{\phi}(x) + H_{\mathrm{bath}}(\hat{\phi}(x)),
\end{multline}
where $\hat{\theta}(x)$ and $\hat{\phi}(x)$ are Hermitian operators encoding,
respectively, the phase of the bosons and their long-wavelength density
fluctuations, satisfying canonical commutation
relations \cite{giamarchi2003quantum}, $K$ is the Luttinger parameter, and $u$
is the speed of sound. In Eq.~\eqref{eq:LL}, we have included the Hamiltonian of
a zero-temperature Ohmic bath, $H_{\mathrm{bath}}(\small {\hat{\phi}(x)})$,
linearly coupled to $\partial_x\hat{\phi}(x)$ and with friction $\eta$, which
serves to damp heating induced by the noise term $f(t)$.  The noise is assumed
to be Gaussian and with zero average. It has, however, a nontrivial dependence
on frequency, with fluctuations scaling as
$\langle f^*(\omega) f(\omega)\rangle=F/\omega$, on top of a subleading
contribution in the $\omega\to 0$ limit proportional to $ \eta i\omega$
(standard zero temperature noise of Caldeira-Leggett
bath \cite{Altland2010a}). This should be contrasted with the noise used for the
driven open Bose gas in Sec.~\ref{sec:ddbose} below, which is also Gaussian but
flat in frequency (i.e., Markovian), and with nontrivial dependence on momentum.

We first consider the model in the absence of nonlinearities ($g=0$). Then,
correlation ($G^K$) and response functions ($G^R$) of the Luttinger liquid obey
scaling behavior that explicitly violates the thermal symmetry of
Sec.~\ref{sec:thermal-symmetry}.
For instance, on long length and time scales, the correlation function of the
crystalline order parameter, $\hat O(x)=\cos(2\hat{\phi}(x))$, is described by
the scaling form
\begin{equation}
  \label{DalllaTscale}
G^K (t-t', x-x')\propto \left((x-x')^2-u^2(t-t')^2\right)^{- K \left( 1+\bar{F} \right)},
\end{equation}
with $\bar{F}=F/(u^2\pi^2\eta)$. 
The scaling of $G^K$ in Eq.~\eqref{DalllaTscale} is governed by a nonuniversal
exponent, and is a consequence of the number-conserving nature of the noise
drive ($f(t) \partial_x\hat{\phi}(x)$). This form of the drive preserves the
sound mode of the Luttinger liquid, which therefore remains gapless. In this
respect, the noise does not fundamentally alter the nature of the
zero-temperature Luttinger liquid fixed point, but only its scaling
exponents \cite{Diehl2010QuantumCN}. To derive Eq.~\eqref{DalllaTscale}, one has
to first take the limit $\eta\to0$ and, concomitantly, $F\to0$, with $\bar{F}$
fixed. In fact, for finite friction $\eta\neq0$, the quantum scaling of
Eq.~\eqref{DalllaTscale} would be valid up to time and length scales smaller
than $1/\eta$, where correlations start to decay exponentially.
 
In the case $g = 0$ we have considered so far, the model Eq.~\eqref{eq:LL} is
quadratic, and one can assign an independent temperature to each field
mode. Consequently, the state of the system is nonthermal, in a way reminiscent
of the generalized Gibbs ensemble in integrable
systems \cite{lange2017pumping,lange2018time}. In contrast, when $g\neq0$, the
nonlinearities encoded in the periodic potential enable the redistribution of
the energy injected by the noise among the different modes. As a result, the
modes reach a common effective temperature $T_{\mathrm{eff}} \propto g^2$. By the
fluctuation-dissipation relation, thermalization is accompanied by a damping term
which cuts the power-law scaling of $G^K$ in Eq.~\eqref{DalllaTscale}. This is
similar to the effect of temperature on equilibrium quantum critical points,
where a thermal de Broglie length delimits the scaling
regime \cite{Sachdev2011}.

\subsection{Quantum criticality of the 1D driven open Bose gas}
\label{sec:ddbose}

We now present the quantum scaling regime and universal properties of the driven
open Bose gas.
First, we discuss the implementation of the Markovian diffusion noise employed
to access the novel fixed point (Sec.~\ref{sec:implq}). We then focus on the
properties of the scaling limit from an RG perspective
(Sec.~\ref{Sec:quantumScaling}), and discuss its universal features
(Sec.~\ref{subsect:qFP}) in comparison to the 3D driven open Bose gas of
Sec.~\ref{sec:driven-open-condensate}. Finally, we compare the 1D driven open
Bose gas with the noise-driven Luttinger liquid discussed above
(Sec.~\ref{sec:comp-noisy-LL-1D-Bose-gas}).  

\subsubsection{Implementation of diffusion noise}
\label{sec:implq}

We consider a 1D variant of the driven open Bose gas introduced in
Eqs.~\eqref{eq:H-driven-open-condensate}
and~\eqref{eq:D-driven-open-condensate}, augmented by an additional jump
operator in the dissipator:
\begin{equation}\label{eq:diffL}
\hat{ L}(x)=\sqrt{\gamma_d}\partial_x\hat{\psi}(x).
\end{equation}
This term adds a diffusion contribution (quadratic in momentum) to the kinetic
coefficient and to the Markovian noise of the Bose gas in
Eq.~\eqref{eq:S-driven-open-condensate-semiclassical}.
Such quadratic-in-momentum scaling of the noise is reminiscent of model B in the
Hohenberg-Halperin classification \cite{Hohenberg1977}. However, the analogy
stops at this level, since the dynamics of model B is order-parameter
conserving, in contrast to the driven open Bose gas considered here.

A possible realization of the jump operator in Eq.~\eqref{eq:diffL} is offered
by an array of microwave cavities coupled to superconducting
qubits \cite{Blais_2021}.
This implementation relies on coupling each qubit to an anti-symmetric
combination of the photonic modes of its nearby cavities.
Integrating out the qubit dynamics leads to a loss term for the photons which is
proportional to such an anti-symmetric combination, and is converted to a spatial
gradient of a bosonic field in the continuum
limit \cite{marino2016quantum,Marcos2012}.

For $q\to0$, the infrared modes asymptotically decouple from this noise, since
it scales to zero with momentum, $\sim q^2$. In quantum optics language, this
circumstance is described by saying that the jump operator of
Eq.~\eqref{eq:diffL} supports dark states around $q = 0$ (see
App.~\ref{sec:dark} and Ref.~\cite{Marcos2012}).
The presence of such dark states is the physical reason for the appearance of
quantum critical scaling and of a universal regime without asymptotic
decoherence in an otherwise driven open system.  In the following, we formalize
this picture on the grounds of Keldysh field theory and an RG analysis.

\subsubsection{Nonequilibrium quantum scaling}\label{Sec:quantumScaling}
  
When translated into the Keldysh action, the jump operators in
Eq.~\eqref{eq:diffL} result in the following contribution to the Keldysh component
of the Gaussian action (see Eq.~\eqref{eq:S-Gaussian}):
\begin{equation}\label{eq:noisevari}
  P^K = i \left( 2 \gamma+\gamma_dq^2+ \dotsb \right).
\end{equation} 
As in Eq.~\eqref{eq:S-Gaussian}, $\gamma$ is given by the sum of one-body loss
($\gamma_l$) and pumping ($\gamma_p$) rates, $\gamma=
(\gamma_l+\gamma_p)/2$.
Similarly, the retarded component of the Gaussian action reads
\begin{equation}
  \label{eq:PR}
  P^R = Z\omega- \left( K_c-i\gamma_d \right) q^2+ir_d, 
\end{equation}
with $r_d= (\gamma_l-\gamma_p)/2$. The dynamical critical exponent here is
$z=2$. Indeed, as discussed in Sec.~\ref{sec:breaking-weak-Goldstone}, by tuning
$r_d\to0$ the condensation transition is accompanied by the onset of a diffusive
critical mode, with canonical dynamical exponent $z=2$. In Eq.~\eqref{eq:PR} we
have also introduced the wave function renormalization coefficient, $Z$, for
future purposes \cite{cardy1996scaling}. Notice that $Z=1$ in the microscopic
theory, see  Sec.~\ref{sec:driven-open-condensate}.

The form of $P^K$ in Eq.~\eqref{eq:noisevari} is the key to unveiling
nonequilibrium quantum critical behavior in the system.  As we tune
$\gamma_p\to\gamma_l$, the spectral gap of the driven open Bose gas closes
($r_d\to0$). If we tune simultaneously $\gamma_{p,l}\to0$, we enter a scaling
regime with both $P^R\sim q^2$ and $P^K\sim q^2$ scaling quadratically. This
realizes precisely the quantum scaling regime of
Sec.~\ref{sec:semiclassical-limit}.  Although formally analogous to the scaling
of a zero temperature quantum critical point, there are remarkable differences
as we will discuss in the following.
 
When the $R/A/K$ components of the Gaussian action scale with the same power of
momentum, $P^{R/A/K} \sim q^2$, both classical and quantum fields have the same
canonical scaling dimension, $\psi_c \sim \psi_q \sim q^{d/2}$.
By following the canonical power counting of Sec.~\ref{sec:Limits}, one readily
deduces that any quartic term, regardless of the number of quantum fields,
scales with momentum as $\sim q^{2-d}$. These terms are not restricted to obey
the thermal symmetry of Sec.~\ref{sec:thermal-symmetry}, and their precise
nature will be discussed in the next subsection.  This discussion sets $d=1$ as
the natural context to explore the Wilson-Fisher fixed point associated with
nonequilibrium quantum scaling. The canonical power counting would have been
completely different had the system supported a sound mode ($z=1$), as is the
case for the noise driven Luttinger liquid discussed above. Indeed, in contrast
to this latter case, sextic operators are marginal in $d=1$, with higher-order
nonlinearities  being RG irrelevant.

\begin{figure}
  \includegraphics[width=246pt]{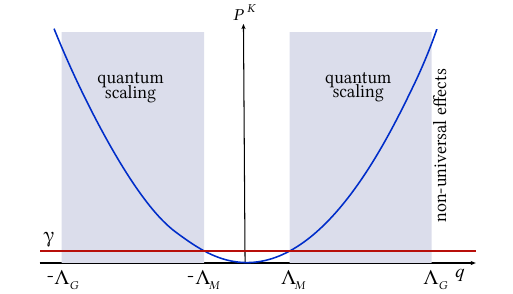}
  \caption{Diffusion noise, $\gamma_d q^2$ (blue line), and noise gap, $\gamma$
    (red line), as a function of momentum. For momentum scales
    $q\gtrsim \Lambda_G$, the renormalization group flow is dominated by
    nonuniversal effects, while below $\Lambda_G$ it is controlled by the
    nonequilibrium quantum fixed point. For $q\lesssim \Lambda_M$, the noise gap
    $\gamma$ dominates over the diffusion noise, and the quantum scaling regime
    is left. Figure adapted from
    Ref.~\cite{marino2016quantum}. }\label{figdiffusio}
\end{figure}

What is the extent of the quantum scaling regime in momentum space?
Quantum scaling can formally be accessed by setting the noise gap to
zero, $\gamma\to0$, or equivalently by working with momenta above the Markov scale
$q\gg\sqrt{\gamma/\gamma_d}\equiv\Lambda_M$, where diffusion noise dominates.
Accordingly, the quantum regime will be established between the two momentum
scales, $\Lambda_M\lesssim q \lesssim \Lambda_G$, where $\Lambda_G$ is the
Ginzburg scale \cite{Zinn-Justin}: for $q\lesssim \Lambda_G$ corrections to
Gaussian scaling become dominant, and the scaling behavior of the system is
governed by the Wilson-Fisher fixed point.
This situation is sketched in Fig.~\ref{figdiffusio}. It is analogous to the
conditions for observability of quantum scaling in equilibrium at finite
temperature. There, the relevant range of momenta is
$\Lambda_{dB}\lesssim q \lesssim \Lambda_G$, with $\Lambda_{dB}\sim T^{1/2}$ the
de Broglie momentum scale. Temperature, however, is a fixed energy scale,
protected against renormalization by the thermal symmetry. In contrast, in our
model, $\gamma$ can acquire sizable corrections under RG, which alters the bare
value of the Markov scale estimate, $\Lambda_M$.
The scales $\Lambda_{M,G}$ are nonuniversal and need to be determined for a
given microscopic model. Crucially, even after RG corrections are taken into
account, $\Lambda_M$ remains one order of magnitude smaller than $\Lambda_G$ in
the model specified above \cite{marino2016driven}.

As indicated in the previous paragraphs, the quantum scaling regime discussed
here pertains to a nonequilibrium Bose condensation transition in 1D.
However, the following argument shows that the effective dimension of the system
in the quantum scaling regime is, in fact, $D = 3$, enabling a Bose condensation
and symmetry-breaking phase transition at the point where the sign
of $(\gamma_l - \gamma_p)/(\gamma_l + \gamma_p)$ changes even though the spatial
dimension of the system is $d = 1$. For illustration, we consider the correction
to the dissipative spectral gap, $\Delta r_d$, but the result would remain
unaltered if we were to focus on other couplings. At one-loop order, the correction
is given by the tadpole diagram \cite{marino2016quantum}
\begin{equation}\label{tadopoleeq}
  \Delta r_d\propto  (u_c+iu_d) \int d\omega \, dq  ~\frac{P^K(\omega,q)}{P^R(\omega,q)P^A(\omega,q)},
\end{equation}
where $u_c$ and $u_d$ are the strength of interactions and two-body loss,
respectively (see Sec.~\ref{sec:driven-open-condensate}). In the quantum scaling
regime, $P^K$ scales quadratically, enlarging the momentum phase space by $q^2$
in Eq.~\eqref{tadopoleeq}, and mimicking an integration measure in three
dimensions which would also bring an extra quadratic factor ($d^3q\sim q^2 dq$).
As a result, the dimensionality of the problem is effectively enlarged by the
dynamical critical exponent, $D = d + z = 3$. The lack of particle number
conservation is crucial: A closed system would have $z=1$, which would still
prohibit a condensation transition in $d=1$.
 
When $\gamma\gg\gamma_d q^2$, this effective dimensional enlargement does not
hold anymore, since $P^K$ and $P^{R/A}$ will not scale alike ($P^K\sim q^0$,
$P^{R/A}\sim q^2$; classical scaling of Sec.~\ref{sec:mixedvspure}) and the
effective dimensionality of the system will shrink ($D\to d$).  In our setup,
this corresponds to $D\to1$, where no continuous symmetry breaking or quantum
criticality can occur. Therefore, for momenta $q\lesssim \Lambda_M$ (or length
scales $l>\Lambda^{-1}_M$), the quantum scaling regime is left, as marked in
Fig.~\ref{figdiffusio}.

\subsubsection{Properties of the nonequilibrium quantum fixed point}
\label{subsect:qFP}

At the microscopic scale, our model has the same nonlinear terms as the driven
open Bose gas of Eqs.~\eqref{eq:H-driven-open-condensate}
and~\eqref{eq:D-driven-open-condensate}: short-range contact interactions
($u_c$) and two-body losses ($u_d$).
As mentioned above, any quartic combination of fields which respects the
$\mathrm{U}_c(1)$ symmetry (see Sec.~\ref{sec:class-quant-symm}) will be
generated in the course of renormalization, irrespective of the number of
quantum fields involved. This is different from the semiclassical RG flow of the
3D driven open Bose gas, which retains only the vertex
$\left( u_c-iu_d \right) \psi^*_c\psi^*_q\psi^2_c$ (and its complex conjugate) as
RG-relevant nonlinearity.  The extra terms include the quantum counterpart of
this vertex ($\psi^*_c\psi^*_q\psi^2_q$), multiplicative Markovian noise
($\psi^*_q\psi_q\psi^*_c\psi_c$), and non-Gaussian Markovian noise fluctuations
$(\psi^*_q\psi_q)^2$. As mentioned above, since the system is open ($z=2$) and
the upper critical dimension is $d_u=2$, the hierarchy of RG-relevant
nonlinearities is truncated at fourth order. This is in contrast to isolated,
number conserving ($\mathrm{U}_q$ restored), low-dimensional quantum systems with $z=1$,
where all higher-order nonlinearities are equally
RG-relevant \cite{giamarchi2003quantum}.

The scaling solution of the RG flow in the quantum critical regime is governed
by a Wilson-Fisher fixed point with two repulsive directions, associated with
simultaneously tuning the spectral gap and the noise gap to
zero \cite{marino2016quantum}. This is analogous to thermal equilibrium, where
both spectral gap and temperature have to be tuned to zero in order to reach a
quantum critical point.
The nonequilibrium quantum fixed point is characterized by three key properties:

(1) \emph{Asymptotically broken thermal symmetry:} The fluctuation-dissipation
relation is not restored at infrared scales, since quartic nonlinearities
breaking the thermal symmetry of Sec.~\ref{sec:thermal-symmetry} have nonzero
fixed-point values.  This is in contrast to several instances of driven open
criticality, where an emergent effective temperature rules the occupation of
soft modes \cite{Mitra2006,DallaTorre2013,Maghrebi2016}, see also
Sec.~\ref{sec:mixedvspure} and~\ref{sec:driv-open-crit}. We have already
encountered in this review instances of criticality where the effective
temperature is momentum-dependent at infrared scales and therefore the system
spoils fluctuation-dissipation relations (Sec.~\ref{sec:competing}
and \cite{marino2022universality}). However, in these cases the fixed point is
classical (associated to a classical scaling solution).

\begin{figure}
  \includegraphics[width=246pt]{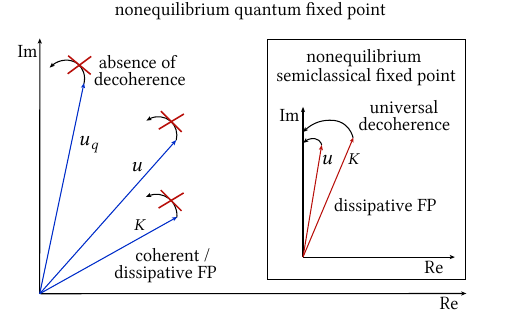}
  \caption{Comparison between the nonequilibrum quantum and classical fixed
    points (cf.\ Fig.~\ref{fig:semiclassical-RG-flow}). In the former case
    ($d=1$), the coherent and dissipative couplings flow under RG to a finite
    value, which indicates that the system has not fully decohered at the
    critical point (in figure, $u_q$ is the coupling of the quantum vertex). This is in contrast to the semiclassical fixed point in
    $d=3$, which displays decoherence at long-wavelengths, with only the
    dissipative couplings nonvanishing at the fixed point. Figure adapted from
    Ref.~\cite{marino2016quantum}.  }\label{figqcritic}
\end{figure}

(2) \emph{Absence of decoherence and RG limit cycle:} Coherent (spectral gap,
interactions, etc.) and dissipative couplings (noise gap, two-body losses, etc.)
scale with the same anomalous dimension in the infrared, with their ratios
approaching a constant. This signals the absence of decoherence at the
Wilson-Fisher fixed point in sharp contrast to the dissipative nature of the
semiclassical fixed point, where coherent couplings vanish in the infrared,
determining universal decoherence (see Sec.~\ref{sec:driv-open-crit} and
Fig.~\ref{figqcritic}). This is a signature of the dark states supported by the
jump operator, which are decoupled from noise in the infrared.  The simultaneous
presence of coherent and dissipative couplings at the fixed point is further
mirrored by slowly damped oscillations in the RG flow of the wave function
renormalization $Z$. Similar limit cycle phenomenology is found in coupled
Langevin-Ising models (Sec.~\ref{sec:competing}), and in cavity QED with
engineered dissipation \cite{seetharam2022dynamical,seetharam2022correlation}). 
  
(3) \emph{Absence of quantum-to-classical correspondence:} The quantum fixed
point is not only without equilibrium analog, but it also does not belong to the
same universality class as its 3D driven open counterpart, as follows from the
discussion above.
This violation of the quantum-to-classical correspondence between
low-dimensional ($d=1$) and high-dimensional ($d=3$) criticality is
rationalized by noting that the set of RG relevant operators at the
Wilson-Fisher fixed point is markedly different in the two cases.
Another way to appreciate such sharp discrepancy with a cornerstone of
equilibrium criticality \cite{Sachdev2011} is to notice that the equilibrium
symmetry is not restored in the quantum case, in contrast to asymptotic
thermalization of the semiclassical fixed point. This mismatch in the symmetry
properties leads to different universality classes.

\subsubsection{Comparison with noise-driven Luttinger liquids}
\label{sec:comp-noisy-LL-1D-Bose-gas}

The one dimensional driven-open Bose gas has similarities and important
differences to the noise-driven Luttinger liquid discussed in
Sec.~\ref{sec:LLdriv}.  The key difference is that by engineering Markovian
noise which scales quadratically with momentum, one can target the quantum
scaling limit. The $1/f$ noise used for the Luttinger liquid is not tailored to
 access a novel fixed point, and in fact the exponents are non universal in this
latter case.

In the model of the driven open Bose gas, particle number
conservation is broken by the presence of incoherent gain and loss of particles
  (the $\mathrm{U}_q$ symmetry of Sec.~\ref{sec:class-quant-symm} is
broken). These processes do not occur in the Luttinger liquid of
Eq.~\eqref{eq:LL}, which evolves under a time-dependent, number-conserving
Hamiltonian.
The different symmetry properties of the two models result in different
soft-mode dispersion relations at criticality (linear for the Luttinger liquid,
and quadratic for the open Bose gas), and assign the two systems to different
universality classes.

However, what the systems have in common is that the scaling regime of the
driven open Bose gas is also cut asymptotically by an emergent effective thermal
length ($\Lambda_M$), which is generated by the interplay of nonlinearities and dissipation.
 
\section{Universality in dissipative quantum impurities}
\label{sec:imp} 
 
This section is devoted to the interplay of local dissipative sources and
long-wavelength modes in low-dimensional quantum systems.
The problem of a quantum impurity, a situation where a few degrees of freedom
are locally coupled to an extended interacting quantum systems, is central to
the field of strongly correlated systems in
equilibrium \cite{giamarchi2003quantum,mahan2013many}.
It has been key to access prototypical instances of strong-coupling
phenomena, which would have been  hard to solve in generic
conditions, and has provided guidance for decades in solid state and atomic,
molecular, and optical
experiments \cite{rossini2021coherent}.

In a nutshell, distinguishing a few degrees of freedom (the impurity) from the
remainder of the system allows us to find nonperturbative phenomena where strong
correlations play a role, on the midway between a few-body problem and the full
many-particle one.
The many-body `environment' serves as a feed of correlations,
to which the impurity degrees of freedom couple, leading to a strong
intertwining of the
two.

In the realm of strongly correlated systems in equilibrium, examples range from
the disruptive effects of a local potential embedded in a Fermi sea (Anderson
catastrophe \cite{mahan2013many}), to the strong coupling of magnetic impurities
to fermionic or bosonic baths (Kondo effects \cite{pustilnik2004kondo}), and
encompass the dressing of static or moving particles in Fermi or Bose-Einstein
condensed environments (polarons \cite{grusdt2015new}).
This raises the question about the nonequilibrium counterparts of these iconic
cases of strongly correlated physics. Studying dissipative quantum impurities
elucidates the interplay of dissipation, interactions, and
correlations, which is at core of this review. Such impurities can be, for
example, local losses, incoherent pumping, or phase noise, embedded in an
extended interacting quantum many-body system.

We first consider solvable instances of dissipative impurities
(Sec.~\ref{sec:exact}). Then, we present a Keldysh approach to treat strongly
correlated regimes of interacting wires with local
losses~(Sec.~\ref{sec:MFKF}).

\subsection{Solvable models of dissipative impurities and the quantum Zeno effect}
\label{sec:exact}

A phenomenon that is central to the following sections is the quantum Zeno
effect \cite{koshino2005quantum,facchi2008quantum}. We will start by reviewing
its manifestation in solvable systems of dissipative impurities.

Let us consider a tight-binding model of fermions hopping along a one-dimensional chain with
one-body loss at the center of the chain, for reference at the lattice site
$l=0$ (for analogous setups in spin chains, see
Refs.~\cite{vznidarivc2010exact,PhysRevLett.123.230604,PhysRevLett.106.217206,PhysRevLett.107.137201,PhysRevLett.112.030603,PhysRevLett.121.030606,prosen2015matrix,tarantelli2022out}).
The corresponding Hamiltonian and jump operator for the dissipative impurity
site read
\begin{equation}\label{eq:fermidissip}
  \hat{H} = J \sum^L_{l=-L} \left( \hat{\psi}^\dagger_l\hat{\psi}_{l+1}^{} +
    \text{H.c.} \right), \qquad \hat{L}_0=\sqrt{\gamma}\hat{\psi}_0.
\end{equation}
In the dynamics starting from the ground state of the closed tight-binding
model, i.e., a filled Fermi sea, the effect of switching on the loss propagates
along the chain, reaching distances $d$ from the impurity on timescales
$d/J$ \cite{froml2020ultracold}. On smaller distances, a nonequilibrium
steady-state is formed. The rate at which fermions leave the system at the
impurity site is $r(\gamma)$. It scales as $r(\gamma)\propto \gamma$ for small
loss rate ($\gamma\ll J$), while,
for strong dissipation $\gamma\gg J$, the rate falls off to zero as
$r(\gamma)\sim
\gamma^{-1}$ \cite{krapivsky2019free,krapivsky2020free,tarantelli2022out}.
Due to particle-hole symmetry of the Hamiltonian, analogous results apply when local loss is
replaced by gain.

In order to rationalize the counterintuitive decay of $r(\gamma)$ for $\gamma\gg J$, we
employ a perturbative argument. The large energy scale set by the loss rate
splits the Hilbert space into fast and slow
sectors \cite{misra1977zeno,PhysRevA.41.2295}. All states with a particle on the
lossy site decay fast with a rate $\sim \gamma$. On the other hand, states with
zero occupancy on the dissipative site form the slow sector of the Hilbert
space. In second order perturbation theory in the hopping $J$ that couples the
slow and fast subspaces, one finds the effective decay rate of states in the
slow sector to be $r\sim J^2/\gamma$.
Another way to rephrase this result is to notice that for $\gamma \gtrsim J$,
the speed of particles replenishing the lossy site is smaller than the rate at
which they are lost into the environment, leading to a smaller net current
flowing toward the dissipative impurity. The dynamical decoupling induced by
dissipation finds numerous applications in the engineering of quantum simulators
based on ultracold gases and in the dissipative stabilization of quantum
information, and it is known as quantum Zeno
effect \cite{fischer2001observation,han2009stabilization,PhysRevLett.110.075301,garcia2009dissipation,zhu2014suppressing,Lihm2018}.
Figure~\ref{figfroml} shows the qualitative dependence of $r(\gamma)$. The RG
flow shown in the figure will be discussed in the next section.

To grasp the generality of the phenomenon, one could also consider a chain of
spins or fermions subjected to a local incoherent scatterer: for instance, the
Hamiltonian in Eq.~\eqref{eq:fermidissip} supplemented by a local time-dependent
potential $V=\xi(t) \hat{\psi}_0^{\dagger} \hat{\psi}_0^{}$ with $\xi(t)$
Gaussian white noise \cite{dolgirev2020non,PhysRevResearch.2.032003}. The
presence of a fast scale (i.e., dissipation rate), no matter if it is related to
loss/pumping or dephasing, will induce the same nonmonotonous behavior. In
Ref.~\cite{dolgirev2020non}, the system is initialized in the ground state of
Eq.~\eqref{eq:fermidissip} at fixed filling and then the noise is switched
on. After the local nonequilibrium state is formed around the impurity, one can
observe that the fraction of fermions left inside the Fermi sea grows and then
collapses to zero upon increasing the noise strength. This is analogous to the
dependence on $\gamma$ discussed above for the case of local losses. The quantum
Zeno effect can also manifest in the spreading of correlations or in
entanglement dynamics, as reported in studies involving integrable 1D systems
\cite{chaudhari2022zeno,
  alba2022noninteracting,alba2022unbounded,caceffo2022entanglement,d2022logarithmic}.

Experimentally, the quantum Zeno effect at dissipative impurities has been demonstrated in a wire of
ultracold bosons with an electron beam focused on a small region of space to
induce localized loss. In this setup, nonmonotonous behavior of the particle
current toward the region of localized dissipation has been
observed \cite{PhysRevLett.102.144101,PhysRevLett.110.035302,PhysRevLett.116.235302}. There
are a number of theory works modelling this scenario. They mostly resort to a
semiclassical description of dynamics or Bogoliubov theory, which are
appropriate in the presence of a
condensate \cite{will2022controlling,sels2020thermal,PhysRevResearch.3.013086}.
However, in a wire of interacting fermions, strong correlations could play a
decisive role by intertwining with the effect of local dissipation, leading to
nontrivial renormalization effects. We will discuss this case in detail in the
next section.

Fermionic dissipative impurities can be realized in ultracold atomic wires by
mimicking the two-leads scenario typical of mesoscopic physics
(experiments \cite{brantut2012conduction,PhysRevA.100.053605,Lebrat2019,huang2022superfluid};
theory \cite{visuri2022nonlinear,visuri2022symmetry,Visuri2023}). A tightly
focused beam at the center of the chain leads to spin-dependent particle losses,
with the reservoirs of different spin species used to probe transport across the
impurity by imposing a chemical potential difference.

\subsection{Local dissipation in interacting quantum wires: \\ the dissipative Kane-Fisher problem}
\label{sec:MFKF}

Our presentation of the interplay of local dissipation with the quantum
fluctuations of a 1D interacting fermionic system will be based on a close
analogy with the Kane-Fisher phenomenon. In the latter, a local coherent
scatterer acts as a relevant or irrelevant perturbation depending on the
repulsive or attractive character of the interactions in the wire prepared in
the ground state \cite{giamarchi2003quantum, PhysRevB.52.R8666,
  PhysRevLett.79.5086,PhysRevLett.76.3192, PhysRevLett.68.1220}.
Under renormalization, the strength of the impurity grows for repulsive
interactions, and the effective long-wavelength description corresponds to a
wire split into two decoupled segments. Instead, for attractive interactions,
the strength of the impurity diminishes, and the system `heals' itself from the
defect. Hence, the conductivity vanishes in the former case and becomes perfect
in the latter.
 
\subsubsection{Qualitative overview: Canonical power counting}\label{sec:RGimp0}

The starting, yet crucial, observation is that, after bosonization (see also
Sec.~\ref{sec:RGimp}), the scaling dimensions of a coherent and a dissipative
impurity in an interacting fermionic wire are the same. One would then expect,
at least on the level of canonical power counting, the same physical picture of
the Kane-Fisher problem. Indeed, the RG flow of the rescaled one-body loss rate,
$\bar{\gamma}=\gamma/\Lambda$, reads \cite{froml2019fluctuation}
\begin{equation}\label{eq:gamma}
    \frac{d \bar{\gamma}}{d \ell} = \left(1-g \right)\bar{\gamma},
\end{equation}
where $g$ is the Luttinger parameter of the wire. The canonical scaling dimension
of $\gamma$ can be   read off from Eq.~\eqref{eq:gammabos}, using that
the action is dimensionless. The explicit dependence on $g$ results, instead, from a
feature of the   momentum-shell RG integration, common also to the
unitary version of the Kane-Fisher problem (see
Refs.~\cite{froml2019fluctuation,PhysRevLett.79.5086,PhysRevLett.76.3192}). 

\begin{figure}
   \includegraphics[width=246pt]{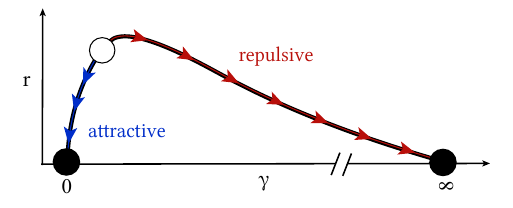}
   \caption{ Dependence of the escape rate $r$ on the microscopic loss rate at the impurity site
     $\gamma$. When $\gamma$ is large, $r$ drops to zero, which is a
     manifestation of the quantum Zeno effect. The red and blue arrows
     qualitatively portray the RG flow of an interacting fermionic wire with
     localized loss, leading to the two fixed points achieved for repulsive and
     attractive interactions, respectively. Figure adapted from
     Ref.~\cite{froml2019fluctuation}.}
  \label{figfroml}
\end{figure}

From this flow equation we
retrieve the physical picture discussed above. For attractive interactions in
the wire ($g>1$) dissipation is renormalized to zero and the impurity becomes
transparent; in contrast, for repulsive interactions ($g<1$), the dissipation
strength flows to infinity, invalidating perturbative RG. However, in this case,
we can approach the problem from a complementary angle, which leads to an RG
equation valid for the inverse of $\gamma$ \cite{froml2019fluctuation}. Its
rescaled value is defined as $\bar{D}=\Lambda/\gamma$ and obeys the RG flow
\begin{equation}\label{eq:D}
    \frac{d \bar{D}}{d \ell} = \left( 1-g^{-1} \right)\bar{D}.
\end{equation}
In this second case, one assumes a strong dissipative impurity, treating the
remainder of the chain as a perturbation. This is analogous to a similar
approach in the RG treatment of the original Kane-Fisher
problem \cite{Altland2010a,kane1997prb}, and to the perturbative argument outlined for the
Zeno effect in Sec.~\ref{sec:exact}.
The RG flow in Eq.~\eqref{eq:D} implies that for repulsive interactions ($g<1$),
the weak link is cut $\bar{D}\to 0$, while for attractive interactions,
$\bar{D}$ is renormalized to infinity, invalidating the RG treatment. This discussion is
summarized in Fig.~\ref{figfroml}. In the following, we will refer to the two
fixed points $\gamma \to \infty$ and $\gamma \to 0$ as fluctuation-induced
quantum Zeno (FIQZ) effect and fluctuation-induced transparency (FIT),
respectively. This nomenclature is due to the fact that no fine-tuning of
driving parameters is required to reach these two fixed point; it rather roots in
the presence of gapless modes in the wire.
These new universal regimes are purely fluctuation driven: it is the
nature of interactions in the chain ($g$ smaller or larger than one), which
determines the onset or lack of the Zeno effect.
   
\subsubsection{Keldysh field theory of the dissipative Kane-Fisher problem}
\label{sec:RGimp}

\begin{figure}
  \includegraphics[width=246pt]{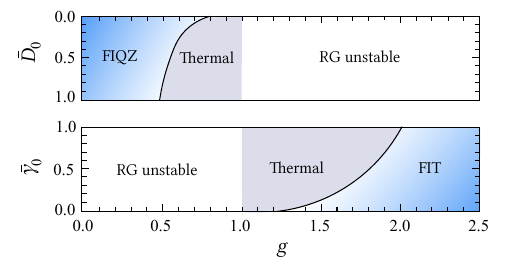}
  \caption{Phase diagram for a dissipative impurity in an interacting fermion
    wire from second order RG equations, for different values of the microscopic
    dissipation strength $\bar{\gamma}_0$ ($\bar{D}_0$ in the dual regime) and
    of the Luttinger parameter $g$. The blue-shaded area in the lower-right
    corner corresponds to values where the fluctuation-induced transparency
    (FIT) is visible, while the upper-left one corresponds to values where the
    fluctuation-induced quantum Zeno effect (FIQZ) is visible. The gray-shaded
    areas correspond to values where the effective temperature {suppresses}
    these effects.  The white regions correspond to values where the
    perturbative RG breaks down. Figure adapted from
    Ref.~\cite{froml2020ultracold}.}
  \label{fignoise}
\end{figure}

We now turn our attention to the interplay of fluctuations and localized
dissipation, going beyond canonical power counting.
The key new ingredient here is that, similar to the equilibrium impurity
problems, the interplay between
interactions and a continuum of gapless modes strongly renormalizes the couplings. The couplings still flow to
either of the two fixed points of Fig.~\ref{figfroml} (FIT or FIQZ). However, now the system may also enter a thermalizing phase where it heats up and no fluctuation-induced phenomena can establish. This is the physical picture
encapsulated in Fig.~\ref{fignoise}, which we substantiate in the following.

This case of study is, indeed, a clear example of emergent nonequilibrium physics in the
quantum regime resulting from the interplay of long-wavelength modes,
dissipation, and strong many-body interactions, which can be addressed with the
nonequilibrium RG approach widely employed in this review. We consider the
continuum Luttinger liquid description of a fermionic wire \cite{froml2020ultracold}
\begin{equation}
  \label{LL}
  \hat{H}=\frac{v}{2\pi}\int_x \left[ g \left( \partial_x\hat{\phi}
    \right)^2+g^{-1} \left( \partial_x\hat{\theta} \right)^2 \right],
\end{equation}
with $\hat{\theta}$ and $\hat{\phi}$ describing density and phase fluctuations,
and $v$ the velocity of sound.
At the center of the wire at $x=0$, there is a single-body fermion loss modeled
by the Lindblad jump operator, $\hat{L}(x=0)=\sqrt{\gamma}\hat{\psi}(x=0)$,
which in bosonization language is given by
\begin{equation}\label{bosoniz}
  \hat{\psi}(x=0) \approx e^{i(\hat{\phi}+\hat{\theta})|_{x=0}}+e^{i(\hat{\phi}-\hat{\theta})|_{x=0}}.
\end{equation}
Here we have retained only leading harmonics. Analogously to the treatment of
the usual Kane-Fisher problem \cite{PhysRevLett.68.1220}, one can integrate out
the chain at all points $x\neq0$ to obtain an effective action for the impurity,
reducing the extended system to a $0+1$-dimensional problem. This is feasible
analytically due to the quadratic nature of the Luttinger liquid
Hamiltonian~\eqref{LL}. Physically, the remainder of the chain acts as a bath
for the impurity degrees of freedom, which can then be treated in an RG
approach. We thus obtain a contribution to the Keldysh action at $x=0$ in the
form of an effective Caldeira-Leggett
bath \cite{CaldeiraLeggett1983a,CaldeiraLeggett1983b,WeissDissipative} which
describes the effect of the remainder of the chain on the impurity as an
environment:
\begin{equation}\label{eqgausslocaldissip}
  {S^B}|_{x=0}=\frac{i}{\pi}\int_{\omega} \left( \theta^*_c~\theta^*_q\right) \begin{pmatrix}
    0 & -\kappa\omega \\
    \kappa\omega & 2\kappa_0|\omega|+4\kappa T_{\text{eff}}
  \end{pmatrix}
  \begin{pmatrix}
    \theta_c \\ \theta_q
  \end{pmatrix}.
\end{equation}
In the expression above, the damping term in the retarded and advanced sectors
is proportional to $\kappa=1/g$, while $T_{\text{eff}}$ accounts for the
possibility of generating an effective temperature under RG transformations
(see Sec.~\ref{sec:LLdriv}), which is the noise gap in the language of
Sec.~\ref{sec:mixedvspure}.
While the damping coefficient $\kappa$ in the retarded and advanced sectors
can grow under renormalization, the corresponding term in the Keldysh sector
will remain at its microscopic value ($\kappa_0$) since perturbative corrections
cannot generate a nonanalytic function of the frequency ($\propto |\omega|$).

The master equation with jump operator Eq.~\eqref{bosoniz} adds to the free
Keldysh action in Eq.~\eqref{eqgausslocaldissip} the nonlinear term
\begin{multline}\label{eq:gammabos}
  S^I|_{x=0} = -2i\gamma\int_{t,x}\delta(x) \left[ \left( e^{i\sqrt{2}\phi_q} -
      \cos \! \left( \sqrt{2}\theta_q \right) \cos \! \left( \sqrt{2}\theta_c
      \right) \right)
  \right. \\
  \left. + \left( e^{i\sqrt{2}\phi_q} \cos \! \left( \sqrt{2} \theta_q \right) -
      1 \right) \right].
\end{multline}
Therefore, the total Keldysh action of the problem at the impurity site
$S|_{x=0}$ consists of the sum of the nonlinear term coming from local
dissipation, and of the environment action, $S|_{x=0}=S^I|_{x=0}+S^B|_{x=0}$.

Performing an RG analysis up to second order in $\gamma$ on the total action
$S|_{x=0}$, one finds a non-trivial RG flow of $\gamma$, $\kappa$ and
$T_{\mathrm{eff}}$ (the RG flow is initialized at zero effective temperature; an
analogous procedure can be carried out for the dual theory at large
$\gamma$). The first-order calculation reproduces the discussion above in
Sec.~\ref{sec:RGimp0}. At second order, fluctuations lead to a flow of $\kappa$
and, importantly, to the emergence of a finite effective temperature
$T_{\mathrm{eff}}$, corrupting the zero temperature scaling solution (recall
that the system is out of equilibrium, therefore the thermal symmetry does not
protect temperature from renormalization, see Secs.~\ref{sec:thermal-symmetry},
\ref{sec:driv-open-crit}, \ref{sec:quantum-criticality}). This allows one to
study quantitatively how the interplay of local heating and nonlinearity destroy
the two fluctuation-induced regimes, the FIT and FIQZ.
This analysis \cite{froml2020ultracold} is analogous to the comparison of the
Ginzburg and de Broglie scales in the context of quantum criticality, discussed
in detail in Sec.~\ref{sec:quantum-criticality}. Namely, for certain microscopic
values of $g$ and $\gamma$ (gray regions in Fig.~\ref{fignoise}), the length
scale at which thermal fluctuations are dominant is smaller than the typical
length where one can observe the renormalization of the dissipation strength;
this implies that the FIT and FIQZ are not observable in those parameter
windows. (In addition, there are the regions of instability of RG which match
the result at the level of canonical power counting, white in
Fig.~\ref{fignoise}.) A simple physical picture emerges: Interactions have to be
strong ($|g|$ large) to make these two effects visible in
experiment \cite{froml2020ultracold,muller2021shape}. In these regions, the
picture of Fig.~\ref{figfroml} remains qualitatively correct, yet quantitatively
altered by higher-order fluctuations.

A real-space RG calculation of transmission and reflection coefficients,
similarly to the traditional Kane-Fisher
problem \cite{matveev1993tunneling,yue1994conduction}, leads to conclusions
consistent with the momentum-space RG discussed above. The real-space RG,
working directly with fermionic degrees of freedom, has the advantage of being
more suitable for comparison with numerical approaches to the
problem \cite{froml2020ultracold, Wolff_2020}.

\section{Universality in fermion systems}
\label{sec:fermions}

In the previous section, we have made a first encounter with fermions in the
context of impurity problems, where the nonequilibrium perturbation was
restricted to a single point in space. Here, we continue our discussion of
fermions, but we return to setups of driven open quantum matter as discussed in
earlier sections, where out of equilibrium dynamics occur everywhere in
space. Quantum mechanics must be expected to play a crucial role, as by the
fermionic exclusion principle there is no simple classical (or deterministic)
limit; unlike bosons such as photons and bosonic atoms, macroscopic
single-particle macroscopic occupations are ruled out, giving rise to Maxwell-
or Gross-Pitaevskii type classical descriptions (see
Sec.~\ref{sec:deterministic-limit}).  Indeed, also the universal quantum
critical behavior of fermions---where fluctuations play a crucial role---differs
from its bosonic counterparts \cite{Boyack_2021}. We review elements of
universality in driven open fermion systems, and connect them to the basic
principles that underlie this review: pure vs.\ mixed states, the impact of
equilibrium vs.\ nonequilibrium generators of dynamics, and the presence of
strong symmetries.

A key additional ingredient, encountered only tangentially in the form of vortex
defects in Secs.~\ref{sec:vortex-unbinding} and~\ref{sec:1st_order_vortex} so
far, is topology. In particular, we will highlight that topology enables
universality, in that it leads to macroscopic behavior that is highly
insensitive to the microscopic generator of dynamics, and robust with respect to
the state of the system being pure or mixed in a sense made precise below.

To this end, we first discuss a scenario to create out-of-equilibrium analogs of
fermionic ground- and finite temperature states, which are equipped with a
nontrivial topology (Sec.~\ref{sec:fermionpuremix}), and thus give rise to
nonequilibrium analogs of topological insulators and
superconductors \cite{hasan2010rmp,QiReview2011,RyuReview2016}. This is achieved
by a suitable design of the jump operators. Pure stationary states can be
generated as dark states of Lindblad dynamics. When different sets of Lindblad
operators are put into competition, topological phase transitions ensue, which
can proceed in pure or mixed states. Although a comprehensive picture of such
phase transitions---and more generally, nonequilibrium phase transitions of
fermions---is still outstanding, one particular aspect becomes clear already:
Critical fermions can exist only in pure states, analogous to thermodynamic
equilibrium \cite{Sachdev2011}. We then continue to overview in
Sec.~\ref{subsec:univtopo} that topology (1)~`beats dynamics', i.e., it is
insensitive to whether generators of dynamics describe in- or out-of-equilibrium
situations, and that it also (2)~`beats mixedness', i.e., topological
quantization persists in suitable mixed state observables.  In both cases, a
universal topological gauge theory emerges, which describes physically
observable quantized (albeit generally nonlinear) responses, and entails a
bulk-boundary correspondence. We close with a discussion of fundamental discrete
symmetry operations such as time reversal and particle-hole transformations,
pointing out a dynamical fine structure in their implementation, depending
exclusively on whether the generator of dynamics obeys the thermal symmetry of
Sec.~\ref{sec:thermal-symmetry} or not (Sec.~\ref{subsec:symmclass}).

\subsection{Topological phase transitions of fermions out of equilibrium: pure
  vs. mixed states}
\label{sec:fermionpuremix}

The nonequilibrium analog of ground states of Hamiltonians are \textit{dark
  states} of Lindbladians: Both scenarios feature pure states, but are prepared
by dynamics respecting or violating equilibrium conditions, respectively, as per
the discussion in Sec.~\ref{sec:equilibrium-vs-nonequi-steady-states}. We have
encountered instances of dark states already in Secs.~\ref{sec:firstorder} and
\ref{sec:ddbose} (see also Appendix~\ref{sec:dark} for a more detailed
discussion). These states, represented by a pure density matrix
$\hat \rho_D = \ket{D}\bra{D}$, are annihilated by the Lindbladian in
Eq.~\eqref{eq:meq}, $\mathcal{L} \hat{\rho}_D = 0$, via satisfying the two
conditions
\begin{equation}
\label{eq:darkcondm}
\hat H \ket{D}= E\ket{D} \quad \text{and} \quad \hat L_l \ket{D} = 0 \quad \forall l.
\end{equation}
Here, $l$ is a multi-index, which will be made concrete below; we will consider
the case $\hat H =0$ for simplicity. Once such a state is reached, time
evolution stops and the system becomes stationary; when the dark state is unique
and no other stationary solutions exist, the dark state will be reached from an
arbitrary initial (and possibly mixed) state (see
Appendix~\ref{sec:dark}). This circumstance can be used for state preparation in
quantum optics \cite{Poyatos1996} and many-body
physics \cite{Diehl2008,Verstraete2009}, including for topologically ordered
states in complex spin systems \cite{Weimer_2010} and symmetry protected
topological order of fermions \cite{diehl2011np}.  Below we will focus on
universal aspects of a class of fermionic Lindbladians featuring a dark state
with such symmetry protected topological properties. Topological phase
transitions result from the competition between Lindbladians that stabilize
topologically distinct dark states. These transitions are nonequilibrium analogs
of quantum and classical topological phase transitions, proceeding in pure and
mixed states, respectively.

\subsubsection{Lindbladians with a topological dark state}

\begin{figure}
  \centering
  \includegraphics[width=246pt]{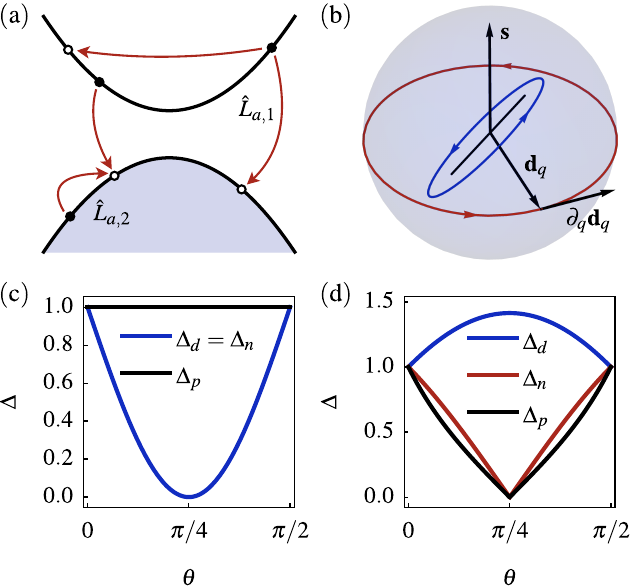}
  \caption{(a)~Action of Lindblad operators: $\hat L_{a,1}$ annihilate particles
    in the upper band to either recreate them in the lower band or redistribute
    them in the upper band. Similarly, $\hat L_{a,2}$ create particles in the
    lower band by transfer from the upper or redistribution in the lower band
    (adapted from Ref.~\cite{Tonielli2020}). (b)~Winding number and purity gap
    closing: Pure states are characterized by
    $|\mathbf{d}_q|=1 \Leftrightarrow \Gamma_q^2 = \id$ for all $q$. Then,
    $\mathbf{d}_q$ lies on a grand circle of the Bloch sphere in the presence of
    a chiral symmetry
    ($\mathbf{s}\perp \mathbf{d}_q\Leftrightarrow \{\Sigma,\Gamma_q\}=0$ for all
    $q$, red line). For mixed states, $|\mathbf{d}_q|<1$ (blue line). The
    closing of a purity gap then corresponds to a zero in $|\mathbf{d}_q|$ for
    at least one $q$ (black straight line). At such a point, the geometry of an
    ellipse, for which a winding number is well defined, collapses to a line. No
    winding can then be defined, which signals a topological phase
    transition. (c), (d)~Dissipative ($\Delta_d$), noise ($\Delta_n$), and
    purity gaps ($\Delta_p$) for topological quantum (c) and classical (d) phase
    transitions. $\Delta_d$ and $\Delta_n$ are measured in units of
    $\bar{\gamma}$, $\Delta_p$ is dimensionless. Discussion, see text.}
  \label{fig:topological_phase_transitions}
\end{figure}

For concreteness, we focus here on dark states representing two-banded
topological insulators in one dimension in a translationly invariant setting;
such models have been introduced in
Refs.~\cite{Goldstein2018,shavit2020prb}. This scenario can be generalized to
arbitrary dimension \cite{Huang_2022}. It also comprises topological
superfluids \cite{Diehl_2010,Yi_2012,diehl2011np,bardyn2012prl,bardyn2013njp,hoening2012critical},
see \cite{bardyn2013njp} for a review. Such topological dark states are
stabilized by a Lindbladian Eq.~\eqref{eq:meq} with Lindblad operators of the
form
\begin{equation}
  \label{eq:topdisslind}  
  \hat L_{a,1,i} = \hat\psi^\dag_{a,i} \hat l_{1,i}^{} ,\quad   \hat L_{a,2,i} =
  \hat \psi_{a,i}^{} \hat    l^\dag_{2,i}, \quad \hat l_{a,i} = \sum_{b,j}
  U_{ab,ij} \hat \psi_{b,j}.
\end{equation}
Here, $a, b \in \{ 1, 2 \}$ label the band index, and $i, j$ are lattice site
indices. The operators $\hat l_{a,i}$ are superpositions of the local
annihilation operators $\hat \psi_{a,i}$. They are related to the latter by the
unitary, canonical transformation $U$, which encodes the topological properties,
see below. When the transformation $U$ is not exactly local, the dark state
carries real-space entanglement, unlike the state considered in
Sec.~\ref{subsec:rydberg}. The operators in Eq.~\eqref{eq:topdisslind}
uniquely \cite{Kraus2008} stabilize the Gaussian, half filled dark state
\begin{align}\label{eq:topodark}
    \ket{D} = \prod_i \hat    l_{2,i}^\dag\ket{0}, \quad \hat L_{a,1,i} \ket{D} =\hat L_{a,2,i} \ket{D} = 0 \quad \forall \quad  a, i.
\end{align}
The annihilation of the dark state by the Lindblad operators happens due to
distinct mechanisms (see Fig.~\ref{fig:topological_phase_transitions}(a)):
$\hat L_{a,1}$ annihilates the dark state since there are no particles in
superposition $\hat l^\dag_{1,i}$, and $\hat L_{a,2,i}$ due to Pauli blocking
for particles in superposition $\hat l^\dag_{2,i}$. Once the state is
annihilated by the $\hat l_{1,i}$ and $\hat l^\dag_{2,i}$ pieces in the Lindblad
operators, the evolution stops. If the state is not annihilated, it is recycled
into the evolution until it is annihilated, by the uniqueness of the dark state
and the absence of other stationary solutions in the present problem.

The jump operators Eq.~\eqref{eq:topdisslind} conserve the total particle
number, $[\hat L_{a,1/2,i}, \hat N] =0$, as it should be for operators that
stabilize an insulating state (there is a strong $\mathrm{U}_q$ symmetry in
the associated field theory). Insisting on this property makes the Lindbladian
necessarily quartic in the field operators, and thus interacting. An alternative
version with Lindblad operators $\hat L_{a,1,i} = \hat l_{1,i}$ and
$\hat L_{a,2,i} =\hat l^\dag_{2,i}$ stabilizes the same dark state, but does not
conserve particle number, although it still exhibits a weak $\mathrm{U}_c(1)$
symmetry.

The number nonconserving operators can also be viewed as a mean-field
approximation to the full interacting Lindblad dynamics, valid at late times
when the dynamics can be linearized around the exactly known dark
state. Mean-field theory can be implemented within the operator formalism
\cite{bardyn2013njp} or in the functional integral representation
\cite{Tonielli2020}. The latter lends itself to extracting beyond mean-field
physics, see Sec.~\ref{subsec:univtopo} below. In a functional integral
representation, mean-field theory has the status of a one-loop self-consistent
Born approximation \cite{Altland2010a}. It results in a simple replacement of
the overall decay rate $\gamma \to \bar \gamma = 2 n \gamma$, where $n = 1/2$
for half filling as in Eq.~\eqref{eq:topodark}. The effective mean-field
Gaussian action with $l = (l_1,l_2)^{\transpose}$ and in the frequency-momentum
domain reads
\begin{equation}
  \label{eq:S-topo-mean-field}
  \begin{gathered}
    S =\int_{\omega,q}\left(l_c^{* \transpose}, 
      l_q^{* \transpose}\right)\left(
      \begin{array}{cc}
        0 & P^A(\omega,q) \\
        P^R(\omega,q) & P^K(q)
      \end{array}\right)\left(
      \begin{array}{c}
        l_c \\
        l_q 
      \end{array}\right),\\
    P^R(\omega,q) = \left( \omega+ \ii \bar \gamma_{q} \right) \id,\quad
    P^K({q}) =2\ii\bar \gamma_{q} \sigma_z,
  \end{gathered}
\end{equation}
with damping rate $\bar \gamma_q \geq 0$. Due to the weak $\mathrm{U}_c(1)$
symmetry of the number nonconserving mean-field theory, the action does not
contain pairing terms such as $l_{1, c} l_{1, q}$, which are included in the
general Nambu structure of Gaussian actions discussed in
Appendix~\ref{sec:Gaussian}. The action Eq.~\eqref{eq:S-topo-mean-field} will
allow us to discuss the basic properties of nonequilibrium topological phase
transitions. The noise matrix $P^K$ is not positive definite---unlike for
bosons, $P^K > 0$ is not required for convergence of the fermionic functional
integral. The positive eigenvalue of $P^K$ indicates that the modes $l_1$ are
empty; the negative eigenvalue indicates that the modes $l_2$ are occupied. This
results from the different ordering of $l_{1,q}^{*} l_{1,q}^{}$ vs.\
$l_{2,q}^{} l_{2,q}^{*} = -l_{2,q}^{*} l_{2,q}^{}$. Using
$l(\omega,q) = U_q \psi (\omega,q) $ with $\psi = (\psi_1,\psi_2)^{\transpose}$
in the Fourier representation (cf.\ Eq.~\ref{eq:topdisslind}), the action can be
expressed in terms of the original lattice fermion fields. In this basis,
$P^{R/A}(\omega,q)$ continue to be unit matrices, but
$P^K(q) \to 2\ii \bar \gamma U_{q}^\dag\sigma_zU_{q}$ becomes a general
anti-Hermitian matrix, with the structure of a free topological Hamiltonian up
to the imaginary prefactor. The topology is thus encoded in the noise matrix in
dissipatively stabilized topological states of fermions.

The above discussion applies in a translationally invariant setting and discards
edge mode effects, which lead to topological degeneracy and higher dimensional
dark spaces. We refer to
Refs.~\cite{bardyn2013njp,altland2021prx,Santos2020,Gau2020a,Gau2020b} for a
discussion of these aspects.

\subsubsection{Phase transitions}

Topological phase transitions are defined by an abrupt jump of an integer-valued
topological invariant as a function of parameter changes. We consider here a
simple model with a stationary state of Su-Schrieffer-Heeger type \cite{SSH1979}, equipped with
a chiral symmetry, and topologically characterized by a chiral winding number
invariant. The invariant can be constructed based on the covariance matrix
collecting the single-particle correlation functions
$\Gamma_{{q}} = \ii\int\frac{d\omega}{2\pi} G^K (\omega,{q})$, where
$G^K = - G^R P^K G^A$ can be read off Eq.~\eqref{eq:S-topo-mean-field}. The
winding number is then defined as
\begin{equation}
  \label{eq:winding}
  W = \int_{-\pi}^{\pi} \frac{d q}{2\pi} \tr \! \left( \Sigma \Gamma^{-1}_q \ii \partial_q \Gamma_q \right),
\end{equation} 
where the chiral symmetry is expressed by the constraint that there exist an
Hermitian matrix $\Sigma$ such that $\{\Sigma, \Gamma_q\}=0$ for all $q$. The
winding is illustrated in Fig.~\ref{fig:topological_phase_transitions}(b), where
we represent the matrices in terms of vectors according to
$\Gamma_q = \mathbf{d}_q\cdot\boldsymbol{\sigma}, \Sigma = \mathbf{s}\cdot
\boldsymbol{\sigma}$
($\boldsymbol{\sigma}$ collects the Pauli matrices). A winding number of $W
  = + 1$ corresponds to $\mathbf{d}_q$ encircling $\mathbf{s}$ once in
  counterclockwise direction upon $q = - \pi \to \pi$. For example, a model
defined with quasi-local operators
\begin{equation}
  \begin{pmatrix}
    \hat    l_{1,i}\\
    \hat l_{2,i}
  \end{pmatrix}
  = \frac{1}{\sqrt{2}}
  \begin{pmatrix}
    \hat \psi_{1,i+1} + \hat \psi_{2,i} \\
    - \hat \psi_{1,i+1} + \hat \psi_{2,i} \\
  \end{pmatrix},
\end{equation}
has the
covariance matrix $\Gamma_q = e^{-\ii q}\sigma^+ + e^{\ii q}\sigma^-$,
$\Sigma = \sigma^z$, and the winding $W=+1$. The opposite winding $W = -1$
results from operators $\hat r_{1,i}$ and $\hat r_{2,i}$, obtained upon
exchanging $\hat \psi_{1,i+1} \to \hat \psi_{1,i}$ and
$\hat \psi_{2,i}\to\hat \psi_{2,i+1}$ in the above equation.

Topological phase transitions with a jump $W = + 1 \to - 1$ can be generated by
putting the operators $\hat l$ and $\hat r$ into competition. To understand the
basic properties of these transitions, it is sufficient to study the linearized
model, described by the Gaussian action Eq.~\eqref{eq:S-topo-mean-field}. The
abrupt change of the topological invariant can occur in two ways: A first option
is a closing of the dissipative gap, which is analogous to the closing of the
energy gap in the ground state of Hamiltonian systems. A second one is the
closing of the noise and purity gaps. The latter case is illustrated in
Fig.~\ref{fig:topological_phase_transitions}(b), straight black line inside the
Bloch sphere. Indeed there are two qualitatively distinct transitions reflecting
these two options, which occur in pure or mixed states, respectively:

\textit{(i)~Topological quantum phase transition (in a pure dark state).} We set
the operators $\hat l$ and $\hat r$ into competition by means of a coherent superposition
of operators,
\begin{equation}
  \hat x_{a,i} = \cos(\theta) \hat l_{a,i} + \sin(\theta) \hat r_{a,i},
\end{equation}
with the limiting cases $\hat x= \hat l$ for $\theta =0$ and $\hat{x} = \hat{r}$
for $\theta = \pi/2$. For any value of $\theta \in [0, \pi/2]$, the Lindbladian
with jump operators $\hat{x}$ has a dark state given by
$\ket{D}=\prod_i \hat x^\dag_{2,i}\ket{0}$. Upon tuning $\theta = 0 \to \pi/2$,
a topological phase transition is crossed, where the winding number jumps from
$W=+1 \to -1$. The dissipative gap $\Delta_d$ and the noise gap
$\Delta_n$ (see Sec.~\ref{sec:mixedvspure}), which are determined by $P^R(\omega, q)$ and $P^K(q)$ given in
Eq.~\eqref{eq:S-topo-mean-field} directly in the diagonal basis, are
shown in Fig.~\ref{fig:topological_phase_transitions}(c). At the transition at
$\theta = \pi/4$, the gaps close, $\Delta_d = \Delta_n = 0$. The fact that the
steady state is a pure dark state for all values of $\theta$ is reflected in the
constant value $\Delta_p = 1$ of the purity gap, which is defined in terms of
the eigenvalues $\lambda_{a, q}$ of the covariance matrix,
\begin{equation}
  \label{eq:purity-gap}
  \Delta_p = \min_{a, q} \abs{\lambda_{a, q}}.
\end{equation}
This definition implies that a state with a vanishing purity gap features at
least one fermionic mode in a totally mixed (infinite temperature) state.

\textit{(ii)~Topological classical phase transition (in a mixed state).} Now we
consider a competition by means of an incoherent sum of operators, in terms of a
Lindblad generator
\begin{equation}
  \mathcal{L} = \cos(\theta) \mathcal{L}_l + \sin(\theta) \mathcal{L}_r,
\end{equation}
where $\mathcal{L}_{l}$ features Lindblad operators involving $\hat l$, and
analogously for $\mathcal{L}_{r}$; again we set $\hat H=0$. In this case, the
stationary state is known exactly only in the cases $\theta = 0,\pi/2$, where
the steady state is given by the respective dark states of opposite winding
number. In between, the stationary state is mixed. However, the chiral symmetry
is preserved \cite{altland2021prx}, and there must be a topological phase
transition. This behavior is captured by an effective linear model, with a
generator of dynamics which is a weighted direct sum of the $l$ and $r$ models
\cite{Huang_2022}. The dissipative, noise, and purity gaps are shown in
Fig.~\ref{fig:topological_phase_transitions}(d). The dissipative gap is finite
for $0\leq \theta\leq \pi/2$. However, both noise and purity gaps close at
$\theta =\pi/4$. At this point, the momentum modes $q=\pm\pi/2$ are in a fully
mixed state.  The winding number evaluates to $W=+1$ for $0\leq \theta <\pi/4$
and to $W = - 1$ for $\pi/4< \theta \leq\pi/2$.  In these regions, the purity
gap is open. The integrity of a winding number defined in Eq.~\eqref{eq:winding}
is verified straightforwardly under this condition. Indeed, intuitively, the
number of windings of the vector $\mathbf{d}_q$ is well-defined until the circle
is deformed to a line or point, which can only happen once the purity gap
closes.  The collapse of the purity gap is illustrated in
Fig.~\ref{fig:topological_phase_transitions}(b), as already mentioned above. In
summary, the transition happens when the noise and, consequently, the purity gap
close, while the dissipative gap remains open. This yields a topological phase
transition without thermodynamic signatures, such as divergent length or time
scales.

For fermions, the topological (i)~quantum and (ii)~classical phase transitions
are the only types of transitions that can occur: Since the spectrum of $\Gamma$
is bounded between $-1$ and $1$ by the Pauli principle, there cannot be a
transition where the dissipative gap closes, while the noise gap remains
open. Transitions with divergent length and time scales for the fermion degrees
of freedom are possible only in pure states. Topological phase transitions are
possible in mixed states, but they proceed without thermodynamic
signatures. This is a general consequence of Fermi statistics, applies in and
out of equilibrium, and includes interacting systems. The absence of critical
fermions in mixed states is seen perhaps more straightforwardly at thermodynamic
equilibrium: In a Euclidean functional integral formulation, the fermionic
Matsubara frequencies are characterized by odd integers,
$\omega_n = \left( 2n+1 \right)\pi/\beta$ with
$n\in \mathbb{Z}$ \cite{Zinn-Justin}, and do not allow for zero modes of the
inverse fermion Green's function at nonzero temperature.

The concept of purity gaps and their closing as enabling topological transitions
is not bound to nonequilibrium problems. However, at thermal equilibrium, the
circumstances under which a purity gap can close are more restrictive, since all
modes experience the same temperature by the principle of detailed balance. For Gaussian states, the discussion is straightforward: The
covariance matrix  is
$\Gamma = \tanh(\beta H)$, with inverse temperature $\beta$ and (first
quantized) Hamiltonian $H$. Zero modes of $\Gamma$, signalling the purity gap
closing, can only occur if there are zero modes of $H$ for finite temperature,
or for infinite temperature $\beta =0$ in the absence of zero modes of
$H$. Interacting systems can exhibit finite temperature topological phase transitions, driven the thermal activation of topological defects~\cite{HuangTT2024}.

We close with a comment on the nature of the pure state phase transitions in the
present models. Due to the specific choice of Gaussian dark states, all static
correlation functions in stationary state can be computed exactly, including at
the critical point. The corresponding static exponents are thus
Gaussian. However, since the generator of dynamics is interacting, this does not
imply that the dynamic exponent is Gaussian as well (one example where a
non-Gaussian dynamic and Gaussian static exponents are realized is provided by
the KPZ equation in one dimension, see~\cite{Tauber2014a,Kamenev2023}). A
general analysis of out-of-equilibrium fermion criticality is outstanding.

\subsection{Universality of topological response out of equilibrium}
\label{subsec:univtopo}

In equilibrium ground states, topology is encoded in the ground-state wave
function, i.e., in the state of the system. A fundamental question is then, how
topological structures generalize once the realm of ground states is left. This
may concern (i)~the breaking of equilibrium conditions while preserving the pure
state nature analogously to a ground state, and (ii)~giving up purity, in- and
out of equilibrium.

An overarching framework to address this question is topological field
theory \cite{PhysRevLett.62.82,PhysRevB.44.5246,PhysRevB.78.155134,PhysRevB.74.085308}. Based
on the interplay of topology and gauge structures, effective theories provide a
bridge between microphysics and observable
macrophysics \cite{PhysRevLett.62.82,PhysRevB.44.5246,PhysRevD.29.2366,shinsei2012prb1}.
Where gauge principles exist, they show a high level of robustness, including in
the presence of interactions \cite{shinsei2012prb1} or translational symmetry
breaking \cite{PhysRevB.93.075113}. Indeed, general adiabatic gauge principles
have been identified for nonequilibrium Lindblad dynamics in a series of seminal
works \cite{avron2012jsp, avron2011njp,avron2012cmp,alberta2016prx}. Here, we
specialize to the case of $\mathrm{U}(1)$ gauge theory, which also allows one to
directly study the physical consequences of the gauge structure in many-body systems. 

\subsubsection{Universality of topological response: equilibrium vs. nonequilibrium dynamics}

We consider the nonequilibrium analog of a Chern insulator in two dimensions,
described by a wave function $\ket{D}$. As above, this scenario is realized by
choosing Lindblad operators such that $\ket{D}$ is their dark state. Then the
stationary state $\ket{D}$ is equivalent to the equilibrium ground state of a
suitably chosen Hamiltonian, while the dynamics of the system differs from the
Hamiltonian evolution. More precisely, we require the following: $\ket{D}$
should

(i)~be unique, such that $\hat\rho_D =\ket{D}\bra{D}$ is a pure state; 

(ii)~be stable, so that local perturbations to the steady state relax at a finite minimal rate, i.e., there is
a dissipative (many-body) gap;  

(iii)~result as the stationary state of a particle number conserving dynamics,
implemented by the presence of a strong $\mathrm{U}_q$ symmetry.

This implements the conditions required by Laughlin’s gauge argument \cite{PhysRevB.23.5632}: the threading of a quantum
Hall annulus by a time varying magnetic flux can be adiabatic only if the bulk
state is nondegenerate and has a (many-body) spectral gap. Then, the insertion
of flux quanta will lead to the transfer of an integer number of charges from
one edge to the other, provided these charges cannot be lost (e.g., to a bath).

The above requirements are met by a choice of Lindblad operators of the same
structure as considered in the preceding section (working in the spatial
continuum here, and with overall damping rate $\gamma$):
$\hat L_{a,1}^{} = \hat \psi_{a}^{} \hat l_1^{}$ and
$\hat L_{a,2}^{} = \hat \psi_{a}^{} \hat l^\dag_2$, where in momentum space
($\hat l = (\hat l_1, \hat l_2)^{\transpose}$)
\begin{equation}\label{eq:regd}
  \hat l_\mathbf{q} = V_\mathbf{q}\hat\psi_\mathbf{q}, \quad V_\mathbf{q} = q_x \id + i q_y \sigma_z+ i m \sigma_y \equiv d^{1/2}_\mathbf{q} U_\mathbf{q}.
\end{equation}
The transformation $V$ is local in real space,
$V_\mathbf{x} = -i\partial_x \id - \partial_y \sigma_z + i m \sigma_y $, such
that the Lindblad operators $\hat L$ share this property, and
$d_\mathbf{q} = q^2 + m^2$ with $q = \abs{\mathbf{q}}$ is a normalization factor
introduced here to ensure that $U_\mathbf{q}$ is unitary. The dark state
targeted by these Lindblad operators coincides with the ground state of the
Hamiltonian
$\hat H = \int_{\mathbf{q}} \hat \psi^\dag_\mathbf{q} \mathbf{d}_\mathbf{q}
\cdot \boldsymbol{ \sigma} \hat \psi_\mathbf{q} = \int_{\mathbf{q}} \hat
l^\dag_\mathbf{q} \sigma_z \hat l_\mathbf{q}$,
with $\mathbf{d}_\mathbf{q} = (2m q_x,2mq_y,-m^2+ q^2)^{\transpose}$ such that
$d_{\mathbf{q}} = \abs{\mathbf{d}_{\mathbf{q}}}$. Similar to the discussion in
Sec.~\ref{sec:fermionpuremix}, the winding of the map
$\mathbf{q} \mapsto \mathbf{d}_\mathbf{q}$ defines a Chern number $\theta =-1$
for any $m\neq 0$.

A gauge field is introduced by minimal coupling in the field theory framework:  This is
enabled by the strong $\mathrm{U}_q$ symmetry associated with particle number
conservation, which translates to independent transformations
$\psi_\pm(t, \mathbf{x}) \mapsto e^{i\theta_\pm}\psi_\pm(t, \mathbf{x})$ on each
contour. Promoting global to local gauge invariance under transformations
$\theta_\pm(t, \mathbf{x})$ then requires us to introduce gauge fields appearing in
covariant derivatives $\partial_\mu \to \partial_\mu - i A_{\pm,\mu}$, where
  $\mu \in \{ t, x, y \}$. Up to terms that are topologically trivial, this
replaces the Lindblad operators by, e.g.,
\begin{equation}
  L_{1,1} \to L_{1,1} + \psi^\dag_1 \left( a_i l \right)_1 A_{\pm,i} + \dotsb, \quad a_i =
  \left(i \partial_{q_i} U_\mathbf{q} \right)U_\mathbf{q}^\dag,
\end{equation}
where $a_i$ ($i=x,y$) is a Berry connection represented in momentum space on the right,
encoding the topology of the problem. Indeed, in a mean-field theory for the
fermions as described above, followed by a gradient expansion of the fermionic
Gaussian functional integral to lowest order, a Chern-Simons theory on the
Keldysh contour emerges,
\begin{align}\label{eq:CS}
  S[A_\pm] &= \frac{\theta}{4\pi} \int_{t, \mathbf{x}} \left[
             \epsilon^{\mu\nu\rho} \left( A_{+, \mu +} \partial_\nu A_{+,\rho } -
             A_{-,\mu } \partial_\nu A_{-,\rho } \right) + \dotsb \right], 
             \nonumber\\
  \theta &= -\frac{\pi}{2} \int_\mathbf{q} \tr \! \left[ \sigma_z
           \left( \partial_{q_x} a_y -\partial_{q_y} a_x \right) \right] = -1,
\end{align}
where $\epsilon^{\mu \nu \rho}$ is the Levi-Civita symbol. Here,
$\partial_{q_x} a_y -\partial_{q_y} a_x$ is the Berry curvature associated with
$a_i$. Only the strong $\mathrm{U}_q$ symmetry allows for the existence of
two independent contour fields $A_\pm$, underlining the importance of particle
number conservation. The prefactor of the effective action is the Chern number,
describing a quantized topological response on top of a purely dissipative
bulk. With the topological gauge theory at hand, the bulk-boundary
correspondence follows from standard arguments \cite{stone2012prb,fujikawa2004oxford}. It entails the perhaps
counterintuitive results that reversible, underdamped chiral modes exist in a
dark subspace at an edge of a bulk undergoing purely irreversible, dissipative
dynamics.

The robustness of this result---i.e., the universality of the topological
response, irrespective to the nonequilibrium nature of the generator of
dynamics---relies on the dissipative gap. Here, the dissipative gap for single
fermion excitations is finite, and evaluates to $2\gamma n m$ within mean-field
theory. The problem also has a gap for local particle-hole excitations, since
the particle numbers of upper and lower band are not separately
conserved \cite{Tonielli2020,Nosov_2023,lyublinskaya2023diffusive}.

\subsubsection{Universality of topological response: pure vs. mixed states}

We now give up the constraint on the purity of the state, and study to what
extent topological structures persist in mixed stationary quantum states, which
occur as dynamical fixed points of equilibrium or nonequilibrium dynamics. Based
on the understanding of the dissipative gap as the counterpart of the energy gap
in closed systems, and the universality of the topological response for pure
states, we now choose a symmetry-based approach to the problem.  An adiabatic
long-wavelength topological $\mathrm{U}(1)$ gauge action can be constructed in
any dimension based on the following prerequisites \cite{Huang_2022},
generalizing those of the previous section:

(i)~The dynamics converge to a form dubbed Dirac stationary state:
$\hat\rho_s\sim e^{-\hat G}$, where $\hat G$ represents a (dimensionless) Dirac
operator in second quantized formulation. For example, in equilibrium,
$\hat G = \beta \hat H$, with
$\hat H = \int d \mathbf{x} \sum_{a, b} \hat \psi^\dag_a(\mathbf{x})H_{ab}\hat
\psi_b(\mathbf{x})$. The matrix $H$ is defined as
\begin{equation}
  H=\mathbf{d}\cdot\boldsymbol{\alpha}, \text{ with }\mathbf{d}=(-i\partial_{x_1},\dots,-i\partial_{x_2},\dots,-i\partial_{x_d}, m),\label{eq:DHamiltonian}
\end{equation} 
and the matrices $\alpha^i$ satisfy the Clifford algebra
$\{\alpha^i,\alpha^j\} = 2\delta^{ij}$. This form is generic near phase
transitions of weakly correlated systems including systems with dark
states \cite{Huang_2022}, and general enough to cover the universal topological
properties of symmetry protected quantum
matter \cite{ryu2010njp,RyuReview2016,altland2021prx}. We can thus work with the
parameterization $\hat G = \beta \hat H$ both in and out of equilibrium.

(ii)~There exist a fast microscopic time scale (finite spectral gap, realized as
an energy gap in equilibrium and as a dissipative gap in Lindblad dynamics) and
a finite purity gap.

(iii)~The charge conservation of the underlying dynamics ensures the existence of
a $\mathrm{U}(1)$ real-time response theory.

(iv)~The charge quantization ensures large gauge invariance \cite{dunne1999springer}.

As above, gauge fields are then introduced by minimal coupling. In
Ref.~\cite{Huang_2022}, an effective gauge action is derived in odd space-time
dimension from combining the above ingredients with the Atiyah-Singer index
theorem for Dirac operators. The result is a mixed-state generalization of
Chern-Simons theories. The even-dimensional cases are constructed via the
bulk-boundary correspondence resulting in mixed-state $\theta$-terms. We focus
here on the example of $2+1$ dimensions (see also \cite{deser1997prl,
  dunne1997prl} for a calculation in the Euclidean equilibrium functional
integral for this special case),
\begin{equation}
  \label{action_static_homogenous}
  \begin{split}
    S[A] & = \frac{\theta}{\pi}\, \int d \mathbf{x} \, \mathcal{I} \! \left(\int
      dt \, A_{q,t} \right) \epsilon^{ij}\partial_i A_{c,j}, \\ \mathcal{I}(a) &
    = 2\arctan \!  \left( \tanh \! \left( \beta \abs{m} \! /2 \right) \tan \!
      \left( a/2 \right) \right).
  \end{split}
\end{equation}
The classical and quantum components of the gauge field in this Keldysh gauge
action are defined as $A_{c} = (A_+ + A_-)/2$ and $A_{q} = A_+ - A_-$.
Furthermore, $\tanh(\beta \abs{m} \! /2)$ is the purity gap of the stationary
density matrix parameterized by the effective Dirac Hamiltonian
Eq.~\eqref{eq:DHamiltonian}, and $\theta$ is the Chern number.\footnote{For
  Dirac models, the actual calculation yields a half-integer Chern number, but
  this can be cured by a suitable UV renormalization. An example for such
  regularization is the term $\sim q^2$ in $\mathbf{d}_\mathbf{q}$ of
  Eq.~\eqref{eq:regd}. A further detail is that
  Eq.~\eqref{action_static_homogenous} is complete only if
  $A_{q,t}(t,\mathbf{x})=A_{q,t}(t)$, see Ref.~\cite{Huang_2022} for the full
  action.} In the pure state limit $\beta\to\infty$, the function $\mathcal{I}$
becomes the identity, $\mathcal I (a) \to a$. The action
Eq.~\eqref{action_static_homogenous} then coincides with the one in
Eq.~\eqref{eq:CS}, if a projection of $A_{q,t}(t,\mathbf{x})$ onto the
zero-frequency limit is performed in the latter
($A_{q,t}(t,\mathbf{x})\to \int dt \, A_{q,t}(t,\mathbf{x})/T$,
$T = \int dt$)---which is sufficient to reproduce the adiabatic physical
response. Conversely, for mixed states described by finite $\beta$, the action
still has a large gauge invariance: under a transformation
$A_{q,t}(t,\mathbf{x}) \to A_{q,t}(t,\mathbf{x}) + 2\pi n /T $ with integer $n$
in Eq.~\eqref{action_static_homogenous}, the action transforms as
$S\to S + 2\pi n \,\theta$. Thus, $e^{iS}$ remains invariant, taking the integer
quantization of the Chern number $\theta$ into account. This is how topology is
still encoded for mixed states. The large gauge invariance physically reflects
the quantization of particle number \cite{Tonielli2020,Huang_2022}. This
rationalizes the finding of a quantized nonlinear (in $A_{q,t}$) response,
described by Eq.~\eqref{action_static_homogenous}: particle number quantization
persists including for mixed states. We note that the linear response, obtained
by expanding $\mathcal{I}$ to first order in $A_{q,t}$, ceases to be quantized
for finite $\beta <\infty$, in accordance with expectation
\cite{dunne1999springer, wang2013prl}. Nevertheless, there are interferometric
physical observables witnessing topological quantization in mixed states, such
as the ensemble geometric phase \cite{bardyn2018prx}. In the framework of the
above topological gauge theory, they are represented as nonlinear responses.

Finally, we note that Eq.~\eqref{action_static_homogenous} features an
equilibrium symmetry, which is emergent (see Sec.~\ref{sec:emergenteq}). This is
rationalized by the simple dependence of the mixed state gauge action on the
purity gap, induced by a gap of $\hat G$, alone.

\subsection{Symmetry classification of open fermion matter}\label{subsec:symmclass} 

We finally leave the ground of topology, and turn to symmetry. Specifically, we
ask: How universal is the paradigmatic symmetry classification of fermion
matter \cite{altland1997prb}, when the realm of thermal equilibrium is left?
Here we point out a dynamical fine structure for the fundamental symmetry
operations, which universally distinguishes equilibrium and nonequilibrium
generators of dynamics.

For equilibrium ground states, Altland and Zirnbauer have provided a tenfold
symmetry classification of free Hamiltonians, in terms of the basic discrete
symmetry operations that are time reversal, particle-hole transformation, and
their concatenation, the chiral transformation \cite{altland1997prb}. This
symmetry classification has been harnessed for a topological
classification of free fermion
systems \cite{schnyder2008prb,kitaev2009aip,ryu2010njp}.

The classified object in this case is the first quantized, Hermitian Hamiltonian
matrix. This setting has been generalized to non-Hermitian matrices, leading to
a 38-fold symmetry classification based on the absence of
Hermiticity \cite{Bernard_2002,kawabata2019prx,Zhou_2019}. However, this 
does not yet incorporate fundamental constraints of physical evolution such as
probability conservation, complete positivity, particle statistics, and the
equilibrium vs.\ nonequilibrium nature of the generator of dynamics. A first
step in this direction was taken in Ref.~\cite{lieu2020prl} considering
single-particle fermion systems, pointing at a reduction from 38 classes down to
10. In Ref.~\cite{altland2021prx}, starting from the Fock space representation
of interacting open fermion dynamics and requiring the invariance of the
equation of motion for the density matrix, it was demonstrated that in fact
there is a fine-structure relating to whether the generator of dynamics obeys
detailed balance or not. In either case, there are 10 classes. The classes are
distinct from each other in the cases where time-reversal is involved, due to a
different implementation of this transformation in the time domain in either
case (the full quantum mechanical time reversal transformation consists of a
Fock space transformation, and an operation on the time parameter). This
concerns 7 out of 10 classes.  Refs. \cite{S__2023,Kawabata_2023} further refine the problem of symmetry classification with  focus on strong vs. weak symmetries, but discard the distinction between equilibrium and non-equilibrium generators. A full reconciliation of these different approaches is still outstanding.

These results are not at variance with the universal topological response in- and out of equilibrium discussed above: In that case, there is no time reversal symmetry involved. Whether and how more generally topology can level out the dynamical fine structure of discrete symmetries is an intriguing open question for future research.

\section{Perspectives}
\label{sec:outlook}

We have reviewed universal physics in driven open quantum matter, which
witnesses the breaking of detailed balance on the microscopic level in
macroscopic observables. Significant advances have been made in both experiment
and theory. Yet, the field presents a number of challenges and opportunities for
future research. In addition, there are intriguing links to other active areas
of physics. This concerns all of the major directions reported in this work.

\subsection{Realizations of paradigmatic nonequilibrium universality}  

Above, we have focused on distilling universal long-wavelength phenomena
starting from a given microscopic physics. This procedure has deepened our
understanding of the microscopic ingredients required to obtain a desired
macroscopic behavior. The time seems right to invert the logic, and move down in
wavelength again, to identify new experimental platforms for the exploration of
nonequilibrium universality. To give a concrete example: The challenge of
observing KPZ physics in two spatial dimensions is largely open. As we have
seen, the basic ingredients for macroscopic KPZ behavior in driven open quantum
matter are minimal and natural. Required are a broken phase rotation symmetry,
absence of particle number conservation, and the breaking of detailed
balance. Such a situation could be realized in various solid state or cold atom
setups. For example, in driven magnonic systems \cite{Manchon2019, Wang2020}, or
in cold atomic superfluid films sandwiched between particle reservoirs at
different chemical potentials \cite{brantut2012conduction,Krinner_2017}. More
generally, by a suitable engineering of ``nonequilibrium
simulators''---controlled setups for the quantitative study of paradigmatic
phenomena---novel platforms for nonequilibrium universality could be created.

A case in point are noisy intermediate-scale quantum platforms. These are
operated under conditions of drive and dissipation, and could be geared into
nonequilibrium simulators. For example, Kibble-Zurek scaling was observed on a
Rydberg quantum simulator \cite{Keesling2019}, directed percolation has been
implemented in a trapped ion quantum computer \cite{Chertkov2022}, and 1D KPZ
physics was reported on a noisy coupled transmon device \cite{Keenan2022}. An
arena for realizing the phenomena of epidemic spreading and self-organized
criticality has been established with ultracold Rydberg atoms
\cite{HelmrichSOC,WintermantelNetwork,Ding2020,Brady2024a}.

There is also great potential for condensed matter platforms. These realize open
systems due to the simultaneous presence of electronic, spin, and phononic
degrees of freedom, with vastly different characteristic time
scales. Generically, such systems relax quickly to thermal equilibrium due to
scattering processes, which overwrite an external weak drive
by lasers or terahertz radiation. However, there might be a generic way out of this scenario: For soft modes, a weak
breaking of equilibrium conditions can be sufficient to induce strong
nonequilibrium universal effects. One prominent example is again KPZ physics in low dimensional systems, which ensues at large scales for an infinitesimal violation of detailed balance. But there are more examples, like
the activation of Goldstone mode limit cycles by an arbitrarily weak driving of
an equilibrium ferrimagnet \cite{Zelle2024}. An intriguing general question
emerges from these examples: Under which conditions are equilibrium RG fixed
points destabilized and left in favor of nonequilibrium fixed points---of known,
but also novel kind? The answer would pave the way toward robust nonequilibrium
universality induced by a weak drive only, as required for solid state systems.

More broadly, searching for
such scenarios in a targeted way may create an intriguing link to the physics of active
matter \cite{RevModPhys.85.1143,Ramaswamy2010}: In fact, the overarching trait of
such systems is a breaking of detailed balance in an extensive way, i.e., at
every point in space (as opposed to more familiar solid state nonequilibrium
settings, such as voltage biases at the boundaries). This is precisely the
setting discussed throughout this review---many-body Lindbladians describe
active quantum matter.

\subsection{Novel nonequilibrium universality}\label{sec:novelneqpersp}

The quest for new universal structures keeps inspiring the field. Such
structures may be found in stationary state, but also in the dynamics of driven
open quantum systems.

\paragraph*{Non-Hermitian physics and universality near exceptional points.}

One promising direction for finding universality in stationary states is
provided by the field of non-Hermitian physics (see also
Sec.~\ref{sec:deterministic-limit}). A key phenomenon are exceptional points,
giving rise to degeneracies in the complex excitation spectra of generators of
dynamics, where reversible and irreversible processes appear on an equal
footing \cite{Ozdemir2019, Miri2019, Ashida2020, Bergholtz2021}. Exceptional
points have fueled a stream of research in condensed matter, atomic condensates,
and optics: On the one hand, they hold promises for applications, such as
sensing due to an enhanced response to external perturbations in their
vicinity \cite{Hodaei17,Chen17,Poli15, Ashida2020}. At the same time, on the
conceptual side, such points host novel topological phenomena, such as nodal
topological phases with open Fermi surfaces, or an anomalous bulk-boundary
correspondence \cite{Zeuner15,Zhou18,Weimann17,Xiao20,Helbig2020,Cerjan_2019,
  Bergholtz2021}.

Much of the focus has so far been devoted to effective single-particle physics
in the absence of noise. This can be a good approximation deep in stable phases
of matter. Yet recently, important connections to the
field of nonlinear dynamics \cite{Cross1993} and active
matter \cite{Fruchart_2021,CheyneVitelli2022} have been made. Transitions of
first and second order have been identified on the mean-field level. The time
seems mature to add the final layer of complexity, namely to include the
interplay of nonlinearities and of noise fluctuations. As has been emphasized
throughout this review, this becomes indispensable once gapless modes are
present, or when a system is brought close to a critical point with
long-wavelength fluctuations. Recently, first such studies have appeared,
addressing the coalescence of an exceptional- with a critical point
 \cite{Hanai2020,Zelle2024}. Such points host giant fluctuations, which
ultimately render the phase transition weakly first-order \cite{Zelle2024}. This
finding holds universally for critical exceptional points in the class of
nonequilibrium $\mathrm{O}(N)$ models, and is protected by the Goldstone theorem
in this case. Yet, other realizations of critical exceptionality are
conceivable, for example, in magnonic systems \cite{Wang_2020}. A systematic
analysis of their universal properties presents a challenge for future
research.

\paragraph*{Phases with limit cycles.}

The physics of limit cycles, or, more generally, spatiotemporal pattern
formation in nonlinear dynamics \cite{Cross1993}, stands in between stationary
states and nonequilibrium time evolution. Recently, the related effect of time
crystals in closed quantum systems has received a lot of attention
\cite{khemani2019brief,yaormp2023}. These phenomena are intimately connected to
the driven nature of the system. They can be rationalized as a transmutation of
scales: Time-periodic drive on the microscopic scale can surface as periodicity
in time on the macroscopic level. This remains true even if at an intermediate
scale of modelling (e.g., by a Lindblad equation), the system appears to be
invariant under continuous translations in time. A paradigmatic example is
provided by the Van der Pol oscillator, originally realized in electrical
circuits with active nonlinear elements \cite{doi:10.1080/14786440108564176}.
Limit cycles have recently been discussed for systems in the quantum realm
\cite{Walter_2014,Walter_2014b,Iemini_2018,Dutta_2019,Buca_2019,Buca_2019b,Tindall_2020,BucaGoold2020,Ben_Arosh_2021,Buca_2022,Zeng2023}. First
experiments in cavity quantum electrodynamics realize this phenomenology
\cite{Dogra_2019,Dreon_2022}, and might be scaled up to include an extensive
number of spatial degrees of freedom in the future
\cite{BucaDonner2022,Mivehvar_2021,kong2021melting}. A comprehensive
understanding of the phase transitions in such systems, which in part pass
through exceptional points \cite{Fruchart_2021,CheyneVitelli2022,Zelle2024},
remains yet to be established. The nature of the Goldstone modes associated to
the breaking of continuous time translations has been studied in
\cite{Hayata_2018}. It provides yet an additional route towards KPZ physics
\cite{Zelle2024}.


\paragraph*{Dynamics of driven open quantum systems.}

The identification of universal regimes in the time evolution of driven open
quantum matter has so far received relatively little attention, and
has mainly focused on one-dimensional atomic systems. An example concerns the
heating dynamics of atoms in optical lattices subject to phase noise from the
lattice laser \cite{Pichler2010, Poletti2012, Poletti2013, Cai2013,
  buchhold2015, Cai_2022, Buchhold2015b, Klocke2021}. Here particle number is conserved, rationalizing the existence of algebraically
slow and universal scaling regimes. Another example is provided by the critical
behavior of topological pumping phenomena at dynamical phase
transitions \cite{Sayyad2021, Starchl2022, Starchl2023}.
 
Beyond ultracold atomic systems, motivation for addressing dynamics comes from
pump-probe experiments in solid state physics
\cite{Basov2017a,Yusupov,delatorre2022,delatorrerev,liu2011,Beaud2014,Mehio2023,Sun2020}:
A system is photo-excited beyond the linear response regime, such that its
occupation number distribution becomes highly nonthermal, and then relaxes back
to its equilibrium state. As mentioned above, electronic, spin, and phonon
degrees of freedom relax on very different time scales, which enables a
description in terms of driven and open subsystems. These subsystem degrees of
freedom can experience various sources of universal behavior: Conservation laws
due to strong symmetries, gapless modes via the spontaneous breakdown of weak
symmetries, and (topological) defects \cite{Zong_2019}.

The impact of conservation laws on nonequilibrium time evolution has
been studied extensively in the context of closed systems. They give rise to
quasi-stationary turbulent behavior in terms of the energy or particle number
cascade \cite{Frisch1995}. There are even more exotic phenomena, such as
nonthermal fixed points:  Here, a dynamically evolving system gets trapped for
a long time before eventually reaching a thermal state, and time itself acts as
a scaling
variable \cite{Berges2008,Erne_2018,Pruefer_2018,rodriguez2022far}. While driven
open systems violate the conservation of energy, other quantities such as the
number of particles can be conserved, making room for possibly modified yet
universal dynamical scenarios.

Gapless modes due to the spontaneous breaking of weak symmetries are likewise expected to strongly influence dynamics.  Promising first steps have been taken in
Refs. \cite{Dolgirev_2020,Zong_2021,lang2023field}, studying the recovery dynamics in pump-probe settings, where an order parameter is coupled to a gapless mode continuum. When the occupation number distribution function has not yet reached its stationary value, there is a potential for
universal physics beyond linear response. Indeed, these works point out novel dynamical
scaling laws due to the bottleneck to equilibration provided by Goldstone modes.

Finally, defects---surfaces on which the order parameter vanishes---can also
give rise to slow dynamics. This has been explored extensively in the field of
phase ordering kinetics \cite{Bray_1994}. Defects can be present when the
relevant symmetry group is small enough: For example, for the group
$\mathrm{O}(N)$, the defect core defines a surface of dimension $d - N$, such as
domain walls in the Ising model ($N=1$) in arbitrary dimensions, or vortices and
vortex lines in the XY model ($N=2$) in $d=2$ and $d=3$. Such defects often have
a long-ranged shape. Therefore, they affect the long-distance equilibration
dynamics even when statistical fluctuations are neglected. So far, the focus has
been mainly on generators of dynamics which are compatible with thermodynamic
equilibrium. A novel ingredient specific to driven systems connects to the above
mentioned limit cycles: Out of equilibrium, defects with explicit time
dependence can be activated, such as Archimedean screws or other structures in
complex magnets \cite{del_Ser_2021,Lohani_2022}. Disentangling the interplay of
these distinct sources of dynamical universality theoretically, and exploring
them in experiment, is a challenging and rewarding direction of research.

\subsection{Nonequilibrium quantum phenomena}\label{sec:neqqu}

\paragraph*{Nonequilibrium fermion criticality. }

Fermionic systems do not possess a simple classical limit, and potentially host
novel instances of nonequilibrium universality in the quantum regime. While we
have discussed weak forms of fermionic nonequilibrium universality here---such
as the robustness of topological phenomena out of equilibrium---a scenario of
nonequilibrium fermionic criticality is still outstanding. Its key ingredients are visible though---as discussed in
Sec.~\ref{sec:fermionpuremix}, fermionic criticality is confined to pure states
both in and out of equilibrium. One route to preserve the pure state critical
scaling of a fermionic Green's function is then to address the short-time
dynamics after a quench, when the fermion distribution is in a prethermal
regime, giving rise to a new fixed point with critical
fermions \cite{Swingle}. In stationary states, a case in point are the
topological transitions proceeding in pure states far from
equilibrium \cite{bardyn2013njp,altland2021prx}. The infrared
enhancement at exceptional points in interacting
systems \cite{Schaefer_2022,Crippa_2022} tuned to criticality mentioned in Sec.~\ref{sec:novelneqpersp} above might give rise to
yet more exotic forms of fermion criticality without equilibrium counterpart. Another route towards universal phenomena in fermion systems can be the shaping a nonthermal distribution function by immersion into optical cavities \cite{Piazza_2014,Piazza_2014b} or Floquet drive \cite{shi2023floquet,Matsyshyn_2023}.

\paragraph*{Scaling near pure states.}

Other grounds to search for quantum effects out of equilibrium are provided by
scaling phenomena near pure states, beyond the fermionic systems described
above. One example is the quest for quantum generalizations
of the directed percolation universality class discussed in Sec.~\ref{sec:DPquant} (see also below for a different angle of approach), where
there is a pure absorbing state on one side of the phase transition; efforts
concentrate on designing models where coherence effects are nonnegligible near
the critical point
\cite{Carollo2019,Gillman2020,Carollo2022,Jo_2021,Buchhold2022}; a particularly promising symmetry-based approach has been put forward in~\cite{thompson2024}. Another
instance is the decay of a system toward the vacuum. When the decay is
nonlinear (e.g., two- or three-body losses), such decay dynamics proceed
algebraically slowly. This scenario was explored in the context of classical
reaction-diffusion systems, revealing non-Gaussian scaling behavior despite the
steady state being void of particles \cite{Cardy_1996,Cardy_1998}. Recently,
first studies of quantum mechanical systems described by Lindblad
equations with decay appeared, both in the reaction-limited (weak losses)
\cite{Horssen2015,Perfetto2022} and diffusion limited (strong losses, Zeno
regime) \cite{Rosso_2021,Rosso_2022,Rosso_2022b,Gerbino2023} regimes, in part
showing structurally different scaling behavior from their classical
counterparts.

More broadly, we point out that once the ground of thermodynamic equilibrium is
left, the borders between ``quantum'' and ``classical'' necessarily get
blurred. The reason is the absence of a universal distribution function imposed
by thermal symmetry, and thus of fixed scaling regimes associated with ground
($T=0$) and mixed ($T>0$) states for bosons or fermions, respectively. In fact,
the freedom of shaping nonequilibrium distribution functions could enable new
scaling regimes. At the same time, this discussion suggests a less strained
focus on such borders.  An intriguing, yet challenging, possibility would be to
extract entanglement measures for open quantum systems in order to characterize
the phase transitions discussed in this review analogously to previous work focusing on thermodynamic equilibrium \cite{osterloh2002scaling}. In that context, it has been
shown that entanglement can exhibit scaling behaviour when a quantum phase
transition is approached. Similarly to order parameter correlation
functions, such scaling behaviour can be smeared out by finite
temperature. Performing a similar analysis in the context of open quantum
systems, would require to distinguish classical and quantum inter-particle
correlations in a driven-open setting, and establishing whether one or both them
can exhibit universal features at criticality \cite{kazemi2023unpredictability}. A way to extract such behaviour
could be to compute the negativity
\cite{vidal2002computable,plenio2005logarithmic}, which is however a challenging
task even in simple spin systems. Building a connection between the scaling
behaviour of order parameter and negativity in the framework of Keldysh path
integrals could represent an important step forward in this direction.

\paragraph*{Monitored quantum systems.}

A recent frontier in the study of universal nonequilibrium phenomena involves monitored quantum systems \cite{Fisher2018,Li2019b,Skinner2019}. In this context, \emph{monitoring} refers to a scenario where quantum measurements are not merely employed to gather information from a system, but also to dynamically evolve the quantum mechanical wave function over time. The measurement projection introduces an irreversible, nonunitary evolution, potentially competing with the unitary elements generated by a Hamiltonian or unitary gates. In this framework, measurements can be harnessed both to build quantum correlations and entanglement among different subsystems, or to dismantle them by extracting local information. Regardless, the measurement-induced evolution yields individual, nonunitarily evolved, and pure quantum states, representing physical realizations of quantum trajectories.

A unique and universal phenomenon in monitored quantum systems are 
measurement-induced phase transitions (MIPTs). These occur in ensembles of pure
state wave functions exposed to combined unitary- and measurement dynamics in the
thermodynamic limit, and result from the competition resp.\ noncommutativity of
distinct quantum operations. MIPTs surface, for example, in a characteristic
change of the growth of the entanglement entropy in the system. In this sense,
they are analogous to quantum phase transitions. But there is also a major
difference to the latter: Due to the randomness of the recorded measurement
outcomes, trajectory averages which are linear in the quantum state
projector---and thus any standard quantum mechanical observable---behave as if
evaluated in an infinite temperature state. Such averages thus do not carry
physical information on the trajectory ensemble any more. A way out is presented
by state-dependent observables, which are nonlinear in the quantum state, such
as the entanglement entropy.

MIPTs appear to be generic in one-dimensional monitored quantum circuits
composed of unitary gates and Pauli
measurements \cite{Skinner2019,Li2019,Choi_2020,Gullans2020}; while some results
in two and more dimensions draw a similar
picture \cite{Lavasani2021,Sierant2022}. Very roughly, these are transitions
from an information scrambling phase in which trajectories feature
(sub-)extensive entanglement scaling, e.g., a volume law, into a quantum Zeno
phase, where trajectories obey an area law. The phases are connected by a second
order phase transition with an emergent conformal
invariance \cite{Zabalo2022}. In the limit of large onsite Hilbert space, where
each lattice site hosts a ``qudit'', the circuit dynamics has been mapped to
statistical mechanics models that undergo a percolation
transition \cite{Nahum2021,Bao2020,Jian2020,weinstein2022scrambling}. The
numerical results for circuits composed of local qubits are, however, partially
at variance with this prediction \cite{Zaballo2021} both for Haar random
circuits and Clifford circuits. The detailed understanding of the phase
transition therefore remains a topic of active research. Another class of
measurement-induced transitions occurs in monitored systems, for which the
unitary part of the dynamics is generated by a Hamiltonian. Particular attention
has been paid to free fermion systems. Depending on their symmetries, numerical
simulations predict critical weak-measurement regimes where the entanglement
entropy grows $S \sim L^{d-1} \ln(L)$ with system size $L$ in $d$ dimensions,
separated from strong-measurement regimes featuring area law scaling
$S\sim
L^{d-1}$ \cite{alberton2021enttrans,TurkeshiZeroClick,fava2023,bao2021symmetry,chahine2023,Poboiko2023,Poboiko2024,Cao2019,Starchl2024}.
Analytical mappings predict a variety of different universality classes, ranging
from localization \cite{Klocke2023,chahine2023,Poboiko2024} to weak
localization \cite{Poboiko2023}, weak anti-localization \cite{fava2023} and a KT
scenario \cite{buchhold2021effective}. The analytical approaches indicate an
intriguing connection between monitored quantum systems in $d$ dimensions and
Anderson localization in $d+1$-dimensional systems, i.e., a potential link
between nonequilibrium quantum systems in $d$ dimensions and the equilibrium
\emph{quantum} statistical mechanics in $d+1$ dimension---a new perspective that
yet needs to be explored.

This scenario opens up a number of challenges for the physics of driven open
quantum systems. First, the proper description of trajectory resolved monitored
systems requires the development of new theoretical tools: The measurement
process is nonlinear in the state, and in addition requires replica approaches
able to produce the relevant trajectory averages. In addition to the statistical
mechanics mappings mentioned above, first systematic field theory approaches
have been constructed to this
end \cite{buchhold2021effective,bao2021symmetry,Barratt_2022}, and exactly
solvable models are
emerging \cite{Nahum2020,Xuyang2022,Minoguchi2022,Klocke2023}. These represent
promising steps toward a better understanding of the universality classes of
monitored quantum systems. A second challenge is to make these transitions
visible in experiment. For Clifford circuits, exploiting their classical
simulability, both postselection \cite{Koh2023} and feedback \cite{Noel2022}
 reveal a transition in experiment. For monitored
fermions, a preselection strategy has been proposed
recently \cite{Buchhold2022}. It pulls the phase transition to the observable
level by performing active feedback chosen such that the universal properties of
the underlying measurement-induced transition are left unmodified. This has
revealed intriguing connections to quantum absorbing state
transitions \cite{Buchhold2022,Odea2022,sierant2023b,iadecola2023b,Piroli2023},
beyond the quantum directed percolation scenario touched upon above. This
sparks the hope that novel classes of such transitions between pure and mixed
states with distinct entanglement properties can be constructed. There are also
striking parallels of postselection strategies to reveal measurement-induced
transitions in experiments, and the physics of large deviations usually
considered in the classical context \cite{Garrahan_2009,Garrahan_2018}, which
remain to be explored. Another route for detecting measurement-induced
transitions in experiments is based on employing cross correlations, e.g., the
linear cross entropy, between the monitored system and a classically simulable
counterpart \cite{LiCrossEntropy,Garratt2023,GullansScalable}. Finally, the
insight that measurement-induced transitions can be associated with the maximum
threshold at which error correction schemes 
fail \cite{Choi_2020,LiFisher2021}, might prove useful in transferring concepts
from the (field) theory of driven open quantum systems to quantum information,
and vice versa.

More broadly, real world monitored quantum systems present a fresh perspective
on driven open quantum systems. They encompass both reversible dynamics, driven
by Hamiltonian evolution and unitary gates, and irreversible dynamics, involving
measurements, dissipative environments, as well as combinations of reversible
and irreversible operations through feedforward mechanisms. This inclusive
framework covers all fundamental aspects of quantum mechanical evolution. The
theoretical description of such systems faces new challenges, and Keldysh field
theory is one promising tool to address them. At the same time, this
new arena unlocks possibilities for realizing unconventional forms of universal
behavior in driven open quantum matter outside of previously anticipated
paradigms.

\begin{acknowledgements}
  We would like to acknowledge interactions and collaborations with numerous
  colleagues and students who have generously shared insights that appear in
  this review. We would particularly like to thank our collaborators on the
  lines of research reviewed here, O.\ Alberton, A.\ Altland, E.\ Altman, C.-E.\
  Bardyn, J. C.\ Budich, K.\ Chahine, A.\ Chiocchetta, E.\ Dalla Torre, R.\
  Daviet, O.\ Diessel, P.\ Dolgirev, R.\ Fazio, M.\ Fleischhauer, H.\ Fr\"oml,
  A.\ Gambassi, M.\ Gievers, L.\ He, M.\ Heyl, H.\ Hosseinabadi, D.\ Hsieh,
  Z.-M.\ Huang, K.\ Klocke, C.\ Kollath, B.\ Ladewig, J.\ Lang, I.\ Lesanovsky,
  M.\ Lukin, M.\ Marcuzzi, S.\ Mathey, Y.\ Minoguchi, Th.\ M\"uller, P.\ Rabl,
  A.\ Rosch, D.\ Roscher, S.\ Sachdev, M.\ Scherer, P.\ Strack, X.\ Sun, M. H.\
  Szyma\'nska, A.\ Silva, U. C.\ T\"auber, J.\ Toner, F.\ Tonielli, G.\ Wachtel,
  S.\ Whitlock, C.\ Zelle and P.\ Zoller.  We thank E.\ Santos for help with the
  figures. L.S.\ acknowledges support from the Austrian Science Fund (FWF)
  through the projects P 33741-N and 10.55776/COE1, and from the European Union
  - NextGenerationEU. M.B.\ and S.D.\ acknowledge support from the Deutsche
  Forschungsgemeinschaft (DFG, German Research Foundation) under Germany’s
  Excellence Strategy Cluster of Excellence Matter and Light for Quantum
  Computing (ML4Q) EXC 2004/1 390534769, and by the DFG Collaborative Research
  Center (CRC) 183 Project No. 277101999 - project B02, and S.D. additionally by
  the DFG Collaborative Research Center (CRC) 1238 Project No. 277146847 -
  projects C03, C04.
  J.M.\ acknowledges support by the Deutsche Forschungsgemeinschaft
  (DFG, German Research Foundation) through Project-ID 429529648, TRR 306
  QuCoLiMa (“Quantum Cooperativity of Light and Matter”), through the grant
  HADEQUAM-MA7003/3-1, and by the Dynamics and Topology
  Centre funded by the State of Rhineland-Palatinate.
\end{acknowledgements}

\appendix

\section{Derivation of the Lindblad-Keldysh action}
\label{sec:derivation-Lindblad-Keldysh}

Here we provide a detailed derivation of the Keldysh partition function for a
driven open system, focusing on the example of a single bosonic or fermionic
field mode $\hat{\psi}$. We pay special attention to the subtle issue of
operator ordering. As we argue in the following and have anticipated in
Eq.~\eqref{eq:L-plus-minus-regularized}, the proper way to keep track of
operator ordering is to introduce a temporal regularization of the Keldysh
action in the form of infinitesimal time shifts \cite{Sieberer2014}. The origin
of the time shifts is understood most transparently by performing the steps that
lead to the master equation~\eqref{eq:meq}, starting from a Hamiltonian
system-bath setting, in the Keldysh formalism \cite{Sieberer2014}. In the
following, we first briefly summarize this derivation and generalize it to
fermionic fields. Then, we present a derivation of the Keldysh partition
function starting directly from the master equation~\eqref{eq:meq}.

\subsection{Derivation from a system-bath setting}
\label{sec:Lindblad-Keldysh-from-sys-bath}

The Hamiltonian that describes the dynamics of a system coupled to a bath is of
the form $\hat{H} = \hat{H}_s + \hat{H}_{sb} + \hat{H}_b$. We assume that the
system Hamiltonian $\hat{H}_s$ is explicitly time-dependent due to a classical
driving field with frequency $\omega_0$, but becomes time-independent after a
transformation to a suitably chosen rotating frame, generated by an operator
$\hat{H}_0$. A paradigmatic example is given by a two-level system undergoing
Rabi oscillations. The system-bath coupling $\hat{H}_{sb}$ and the bath
Hamiltonian $\hat{H}_b$ are given in Eq.~\eqref{eq:H-sb-H-b}, where the bath
modes $\hat{\phi}_{\mu}$ with frequencies $\omega_{\mu}$ are bosonic when the
system operator $\hat{L}$ in $\hat{H}_{sb}$ is bosonic or even in fermionic
operators; in contrast, when $\hat{L}$ is odd in fermionic operators, the bath
modes are fermionic.

We perform a transformation to a rotating frame described by the time-dependent
unitary operator $\hat{U} = \e^{\imag \left( \hat{H}_0 + \hat{H}_b \right) t}$.
By assumption, the system Hamiltonian becomes time-independent in the rotating
frame; the bath Hamiltonian is removed by the transformation, and the
system-bath coupling becomes
\begin{equation}
  \hat{H}_{sb}' = \sum_{\mu} g_{\mu} \left(  \hat{L}^{\dagger}
    \hat{\phi}_{\mu} \e^{- \imag \left( \omega_{\mu} - \omega_0 \right) t} +
    \hat{\phi}_{\mu}^{\dagger} \hat{L} \e^{\imag \left( \omega_{\mu} - \omega_0
      \right) t} \right).
\end{equation}
We note that a more general form of the system-bath coupling contains also the
terms $\hat{L} \hat{\phi}_{\mu}$ and
$\hat{L}^{\dagger} \hat{\phi}_{\mu}^{\dagger}$. In the rotating frame, such
terms oscillate with frequency $\omega_{\mu} + \omega_0$, and are dropped in a
rotating wave approximation.

After bringing the system Hamiltonian as well as $\hat{L}$ and
$\hat{L}^{\dagger}$ to normal ordered form, we can construct the usual Keldysh
functional integral for the system-bath setup without
ambiguity \cite{Altland2010a, Kamenev2023}. We introduce an additional sign
$\zeta = - 1$ for fermionic fields on the backward branch. This sign results
from the representation of a trace in fermionic Fock space in the basis of
coherent states, and allows us to obtain a Keldysh action that satisfies the
usual causality structure described for the action in
Sec.~\ref{sec:keldysh-rotat} and for the Green's functions in
Appendix~\ref{sec:greens-funct-keldysh}. Integrating out the bath yields the
following contribution to the Keldysh action:
\begin{multline}
  \label{eq:Delta-S}
  \Delta S = - \sum_{\mu} g_{\mu}^2 \int_{t, t'} \left( L^{\dagger}_+(t),
    \zeta_b L^{\dagger}_-(t) \right) \e^{\imag \omega_0 \left( t - t' \right)} \\
  \times \hat{\sigma}_z
  \begin{pmatrix}
    G_{\mu}^{++}(t - t') & G_{\mu}^{+-}(t - t') \\
    G_{\mu}^{-+}(t - t') & G_{\mu}^{--}(t - t')
  \end{pmatrix}
  \hat{\sigma}_z
  \begin{pmatrix}
    L_+(t') \\ \zeta_b L_-(t')
  \end{pmatrix},
\end{multline}
where the Green's functions of the bath are given by
\begin{equation}
  \label{eq:bath-GFs}
  \begin{split}
    G_{\mu}^{+-}(t) & = - \imag \zeta_b n_{\zeta_b}(\omega_{\mu}) \e^{- \imag \omega_{\mu} t}, \\
    G_{\mu}^{-+}(t) & = - \imag \left( 1 + \zeta_b n_{\zeta_b}(\omega_{\mu}) \right)
    \e^{- \imag
      \omega_{\mu} t}, \\
    G_{\mu}^{++}(t) & = \theta(-t) G_{\mu}^{+-}(t) + \theta(t) G_{\mu}^{-+}(t), \\
    G_{\mu}^{--}(t) & = \theta(t) G_{\mu}^{+-}(t) + \theta(-t) G_{\mu}^{-+}(t).
  \end{split}
\end{equation}
In the above equations, the sign $\zeta_b = \pm 1$ refers to the bath:
$\zeta_b = + 1$ for a bosonic bath, and $\zeta_b = - 1$ for a fermionic
bath. Accordingly, the distribution function is
$n_{\zeta_b}(\omega) = \left. 1 \middle/ \left( \e^{\beta \omega} - \zeta_b
  \right) \right.$.
A possible additional sign is contained in the field variables $L_{\pm}$ and
$L_{\pm}^{\dagger}$. For example, for $\hat{L} = \hat{\psi}$, we obtain
$L_+ = \psi_+$, $L_- = \zeta \psi_-$, and $L^{\dagger}_+ = \psi_+^{*}$,
$L^{\dagger}_- = \zeta \psi_-^{*}$.

We assume that the bath remains in thermal equilibrium at all times---this is
essentially the Born approximation, and relies on weak system-bath
coupling. Furthermore, we wish to perform the Markov approximation, assuming
that the correlation time of the bath is short on typical time scales of
evolution of the system, such that we can replace $L_{\pm}(t') \to L_{\pm}(t)$
in Eq.~\eqref{eq:Delta-S}. However, we have to keep track of the temporal order
of $L_{\pm}^{\dagger}(t)$ and $L_{\pm}(t')$ that is encoded in the Heaviside
functions in Eq.~\eqref{eq:bath-GFs} as can be seen by introducing a new
integration variable $\tau = t - t'$ and rewriting Eq.~\eqref{eq:Delta-S} as
\begin{multline}
  \label{eq:Delta-S-tau}
  \Delta S = - \int_{t, \tau} \left[ L_+^{\dagger}(t) \left( \theta(-\tau)
      \Gamma^+(\tau) + \theta(\tau) \Gamma^-(\tau) \right) L_+(t - \tau)
  \right. \\ - L_-(t - \tau) \Gamma^+(\tau) L_+^{\dagger}(t) - L_+(t - \tau)
  \Gamma^-(\tau) L_-^{\dagger}(t) \\ \left. + L_-^{\dagger}(t) \left(
      \theta(\tau) \Gamma^+(\tau) + \theta(-\tau) \Gamma^-(\tau) \right) L_-(t -
    \tau) \right],
\end{multline}
where
\begin{equation}
  \begin{pmatrix}
    \Gamma^+(\tau) \\ \Gamma^-(\tau)
  \end{pmatrix}
  = \sum_{\mu} g_{\mu}^2 \e^{\imag \omega_0 \tau}
  \begin{pmatrix}
    G_{\mu}^{+-}(\tau) \\ G_{\mu}^{-+}(\tau)
  \end{pmatrix}.
\end{equation}
In the terms that are mixed in the branch index, we have exchanged the order of
$L_{\pm}^{\dagger}(t)$ and $L_{\pm}(t - \tau)$. If the jump operator is odd in
fermionic operators, this leads to an additional minus sign, which cancels
$\zeta_b$ in Eq.~\eqref{eq:Delta-S}.

Next, we consider the continuum limit of dense bath frequencies, centered around
$\omega_b \approx \omega_0$ with a large bandwidth $\vartheta$. We can then
replace the sum over bath modes $\mu$ by an integral,
\begin{equation}
  \begin{pmatrix}
    \Gamma^+(\tau) \\ \Gamma^-(\tau)
  \end{pmatrix}
  = - \imag \int_0^{\infty} \diff \omega \, \nu(\omega) g(\omega)^2 \e^{\imag
    \left( \omega_0 - \omega \right) \tau}
  \begin{pmatrix}
    \zeta_b n_{\zeta_b}(\omega) \\ 1 + \zeta_b n_{\zeta_b}(\omega)
  \end{pmatrix}.
\end{equation}
where $\nu(\omega)$ is the density of states of the bath. The key assumption
underlying the Markov approximation is that the product
$\nu(\omega) g(\omega)^2$ is a smoothly varying function of $\omega$ in the
vicinity of $\omega_0$. This implies that the bath correlation functions
$\Gamma^{\pm}(\tau)$ appearing in Eq.~\eqref{eq:Delta-S-tau} are sharply peaked
around $\tau = 0$. Therefore, keeping track of temporal order, we make the
replacement
$\theta(\pm \tau) L_{\pm}(t - \tau) \to \theta(\pm \tau) L_{\pm}(t_{\mp})$ where
$t_{\pm} \to t \pm 0^+$. We can now perform the integration over $\tau$ using
\begin{equation}
  \label{eq:FT-Heaviside}
  \begin{split}    
    \int_{- \infty}^{\infty} \diff \tau \, \e^{\imag \omega \tau} \theta(\pm
    \tau) = \pi \delta(\omega) \pm \imag \mathcal{P} \frac{1}{\omega},
  \end{split}
\end{equation}
where $\mathcal{P}$ denotes the principal value. This leads to
\begin{equation}
  \begin{split}
    \int_{-\infty}^{\infty} \diff \tau
    \begin{pmatrix}
      \Gamma^+(\tau) \\ \Gamma^-(\tau)
    \end{pmatrix}
    & = - \imag \gamma
    \begin{pmatrix}
      \zeta_b n_{\zeta_b}(\omega_0) \\ 1 + \zeta_b n_{\zeta_b}(\omega_0)
    \end{pmatrix}, \\
    \int_{-\infty}^{\infty} \diff \tau \, \theta(\pm \tau)
    \begin{pmatrix}
      \Gamma^+(\tau) \\ \Gamma^-(\tau)
    \end{pmatrix}
    & = - \imag \frac{\gamma}{2}
    \begin{pmatrix}
      \zeta_b n_{\zeta_b}(\omega_0) \\ 1 + \zeta_b n_{\zeta_b}(\omega_0)
    \end{pmatrix}
    \mp
    \begin{pmatrix}
      \Delta_+(\omega_0) \\ \Delta_-(\omega_0)
    \end{pmatrix},
  \end{split}
\end{equation}
where $\gamma = 2 \pi \nu(\omega_0) g(\omega_0)^2$, and
\begin{equation}
  \begin{pmatrix}
    \Delta_+(\omega_0) \\ \Delta_-(\omega_0)
  \end{pmatrix}
  = \mathcal{P} \int_0^{\infty} \diff \omega \, \frac{\nu(\omega)
    g(\omega)^2}{\omega_0 - \omega}
   \begin{pmatrix}
     \zeta_b n_{\zeta_b}(\omega) \\ 1 + \zeta_b n_{\zeta_b}(\omega)
   \end{pmatrix}.
\end{equation}
We thus find that $\Delta S = \Delta S_d + \Delta S_L$ is the sum of a
dissipative contribution and a Lamb shift, which are given by
\begin{multline}
  \label{eq:Delta-S-d}
  \Delta S_d = - \imag \gamma \int_t \left\{ \left( 1 + \zeta_b
      n_{\zeta_b}(\omega_0) \right) \vphantom{\frac{1}{2}} \right. \\ \times
  \left[ L_+(t) L_-^{\dagger}(t) - \frac{1}{2} \left( L_+^{\dagger}(t) L_+(t_-)
      + L_-^{\dagger}(t) L_-(t_+) \right) \right] \\ \left. +
    n_{\zeta_b}(\omega_0) \left[ L_+^{\dagger}(t) L_-(t) - \frac{1}{2} \left(
        L_+(t_+) L_+^{\dagger}(t) + L_-(t_-) L_-^{\dagger}(t) \right) \right]
  \right\},
\end{multline}
and
\begin{multline}
  \Delta S_L = - \int_t \left[ \Delta_+(\omega_0) \left( L_+^{\dagger}(t) L_+(t_+)
      - L_-^{\dagger}(t) L_-(t_-) \right) \right. \\ \left. - \Delta_-(\omega_0)
    \left( L_+^{\dagger}(t) L_+(t_-) - L_-^{\dagger}(t) L_-(t_+) \right)
  \right].
\end{multline}
For a bath at zero temperature, we can set $n_{\zeta_b}(\omega_0) = 0$. Then,
after absorbing $\sqrt{\gamma}$ in a redefinition of $L_{\pm}$,
Eq.~\eqref{eq:Delta-S-d} yields Eq.~\eqref{eq:L-plus-minus-regularized}. The
Lamb shift can be included in the Hamiltonian contribution to the action.

\subsection{Derivation from the quantum master equation}
\label{sec:Lindblad-Keldysh-from-MEQ}

The above derivation of the Keldysh action for a driven open system shows that
infinitesimal time shifts occur naturally in a Hamiltonian system-bath
setting. Moreover, above we have required normal order only of $\hat{H}$,
$\hat{L}$, and $\hat{L}^{\dagger}$, but not of the product
$\hat{L}^{\dagger} \hat{L}$ that also appears in the master
equation~\eqref{eq:meq}. These insights serve as a guideline for a direct
derivation of the Keldysh action from the master equation \cite{Sieberer2016a},
which we generalize here to fermionic systems.

A key ingredient in the construction of the functional integral representation
are coherent states. For bosons, a coherent state with amplitude $\psi \in \C$
is defined as $\ket{\psi} = \e^{\psi \hat{\psi}^{\dagger}} \ket{0}$, where the
vacuum state $\ket{0}$ obeys $\hat{\psi} \ket{0} = 0$. Defining fermionic
coherent states, on the other hand, requires us to employ Grassmann numbers that
form an anticommutative algebra. To unify the presentation for bosons and
fermions, we will denote also Grassmann numbers with $\psi$. Then, a fermionic
coherent state is given by
$\ket{\psi} = \e^{- \psi \hat{\psi}^{\dagger}} \ket{0}$. Both bosonic and
fermionic coherent states have the following properties: They are eigenstates of
the annihilation operator, $\hat{\psi} \ket{\psi} = \psi \ket{\psi}$; the
overlap of two coherent states is $\braket{\psi | \psi'} = \e^{\psi^{*} \psi'}$;
and coherent states span the bosonic or fermionic Fock space, as expressed by
the completeness relation
\begin{equation}
  \label{eq:identity-coherent-states}
  \hat{1} = \int \frac{\diff \psi^{*} \diff \psi}{\pi^{(1 + \zeta)/2}} \, \e^{-
  \psi^{*} \psi} \ket{\psi} \bra{\psi}.
\end{equation}
However, there is a fundamental difference between bosons and fermions: For
bosons, $\psi^{*}$ denotes the complex conjugate of $\psi \in \C$. In contrast,
the notation $\psi^{*}$ for a Grassmann number $\psi$ is merely
symbolic. Indeed, for fermions, $\psi$ and $\psi^{*}$ are strictly independent
variables \cite{Wegner2016}.

As anticipated in Sec.~\ref{sec:keldysh-field-theory}, the starting point for
the construction of the functional integral representation is the formal
solution of the master equation,
$\hat{\rho}(t) = \e^{\mathcal{L} \left( t - t_0 \right)} \hat{\rho}(t_0)$. We
split the evolution from $t_0$ to $t$ into $N$ discrete steps of duration
$\Delta t = \left( t - t_0 \right)/N$,
\begin{equation}
  \label{eq:meq-formal-solution}
  \hat{\rho}(t) = \e^{\mathcal{L} \left( t - t_0 \right)} \hat{\rho}(t_0) = \left(
    \e^{\mathcal{L} \Delta t} \right)^N \hat{\rho}(t_0) \approx \left( \mathcal{I} +
    \mathcal{L} \Delta t \right)^N \hat{\rho}(t_0),
\end{equation}
where $\mathcal{I}$ is the identity superoperator. We denote the density matrix
at time $t_n = t_0 + n \Delta t$ by $\hat{\rho}_n = \hat{\rho}(t_n)$. The
evolution during one time step is thus given by
\begin{equation}
  \label{eq:time-step}
  \hat{\rho}_{n + 1} \approx \left( \mathcal{I} + \mathcal{L} \Delta t \right) \hat{\rho}_n.
\end{equation}
To keep track of the order of operators, we further split each time step into
two, leading to a total of $2 N$ time steps. Details of this splitting are given
below. We construct a field integral representation of the time evolution by
inserting resolutions of the identity in terms of coherent states
Eq.~\eqref{eq:identity-coherent-states} in between consecutive time steps and
both to the left and to the right of the density matrix. In particular, for the
density matrix $\hat{\rho}_n$ at time $t_n$, which is reached after $2 n$ time
steps, we write
\begin{multline}
  \label{eq:density-matrix-coherent-state-representation}
  \hat{\rho}_n = \int \frac{d \psi_{+, 2 n}^{*} d\psi_{+, 2 n}}{\pi^{(1 + \zeta)/2}}
  \frac{d \psi_{-, 2 n}^{*} d \psi_{-, 2 n}}{\pi^{(1 + \zeta)/2}} e^{- \psi_{+,
      2 n}^{*} \psi_{+, 2 n} - \psi_{-, 2 n}^{*} \psi_{-, 2 n}} \\ \times
  \braket{\psi_{+, 2 n} | \hat{\rho}_n | \zeta \psi_{-, 2 n}} \ket{\psi_{+, 2 n}}
  \bra{\zeta \psi_{-, 2 n}}.
\end{multline}
As explained in Sec.~\ref{sec:Lindblad-Keldysh-from-sys-bath}, we have to
include a sign $\zeta = - 1$ for Grassmann variables on the backward
branch.

To obtain a representation of the full time evolution, we have to relate the
matrix element
$\braket{\psi_{+, 2 n + 2} | \hat{\rho}_{n + 1} | \zeta \psi_{-, 2 n + 2}}$,
appearing in the coherent state representation of $\hat{\rho}_{n + 1}$, to the
corresponding matrix element of $\hat{\rho}_n$. This can be done by inserting
Eq.~\eqref{eq:density-matrix-coherent-state-representation} in
Eq.~\eqref{eq:time-step}, leading us to consider the matrix element of the
superoperator $\mathcal{I} + \mathcal{L} \Delta t$ appearing in
Eq.~\eqref{eq:time-step},
$\braket{\psi_{+, 2 n + 2}| \left( \mathcal{I} + \mathcal{L} \Delta t \right) \!
  \left( \ket{\psi_{+, 2 n}} \bra{\zeta \psi_{-, 2 n}} \right) | \zeta \psi_{-,
    2 n + 2}}$,
which contains, in particular, the terms
$\braket{\psi_{+, 2 n + 2} | \hat{L}^{\dagger} \hat{L} | \psi_{+, 2 n}}$ and
$\braket{\zeta \psi_{-, 2 n}| \hat{L}^{\dagger} \hat{L} | \zeta \psi_{-, 2 n +
    2}}$.
When $\hat{L}$ and $\hat{L}^{\dagger}$ are normal ordered, their product
$\hat{L}^{\dagger} \hat{L}$ is in general not normal ordered. Therefore, we
insert additional resolutions of the identity in terms of
$\ket{\psi_{+, 2 n + 1}}$ and $\ket{\zeta \psi_{-, 2 n + 1}}$ between
$\hat{L}^{\dagger}$ and $\hat{L}$. In other matrix elements such as
$\braket{\psi_{+, 2 n + 2} | \hat{H} | \psi_{+, 2 n}}$ we also insert these
additional resolutions of the identity. The position, e.g., before or after
$\hat{H}$, is arbitrary, and we make the choice of applying the Hamiltonian
$\hat{H}$ and the jump operator $\hat{L}$ on the forward and $\hat{L}^{\dagger}$
on backward branch between steps $2 n$ and $2 n + 1$. We can then replace the
matrix elements of operators by functions of the coherent state amplitudes, both
for the Hamiltonian,
$H(\psi^{*}, \psi') = \braket{\psi | \hat{H} | \psi'}/\braket{\psi | \psi'}$,
and the jump operator,
$L(\psi^{*}, \psi') = \braket{\psi | \hat{L} | \psi'}/\braket{\psi | \psi'}$,
and its Hermitian conjugate,
$L^{\dagger}(\psi^{*}, \psi') = \braket{\psi | \hat{L}^{\dagger} |
  \psi'}/\braket{\psi | \psi'}$.
Specifically, following our convention for inserting resolutions of the
identity, we obtain the following matrix elements:
\begin{equation}
  \label{eq:regularized-matrix-elements}
  \begin{aligned}
    H_{+, n} & = H(\psi_{+, 2 n + 1}^{*}, \psi_{+, 2 n}), & H_{-, n} & = H(\zeta
    \psi_{-, 2 n}^{*}, \zeta \psi_{-, 2 n + 1}), \\ L_{+, n} & = L(\psi_{+, 2 n
      + 1}^{*}, \psi_{+, 2 n}), & L_{-, n} & = L(\zeta \psi_{-, 2 n + 1}^{*},
    \zeta \psi_{-, 2 n + 2}), \\ L^{\dagger}_{+, n} & = L^{\dagger}(\psi_{+, 2 n
      + 2}^{*}, \psi_{+, 2 n + 1}), & L^{\dagger}_{-, n} & = L^{\dagger}(\zeta
    \psi_{-, 2 n}^{*}, \zeta \psi_{-, 2 n + 1}),
  \end{aligned}
\end{equation}
in terms of which we can write the Lindbladian as
\begin{multline}
  \mathcal{L}_{+-, n} = - \imag \left( H_{+, n} - H_{-, n} \right) \\ + L_{+, n}
  L^{\dagger}_{-, n} - \frac{1}{2} \left( L^{\dagger}_{+, n} L_{+, n} +
    L^{\dagger}_{-, n} L_{-, n} \right).
\end{multline}
We thus find the desired relation between the matrix elements of the density
matrix at times $t_{n + 1}$ and $t_n$:
\begin{multline}
  \braket{\psi_{+, 2 n + 2} | \hat{\rho}_{n + 1} | \zeta \psi_{-, 2 n + 2}}
  \approx \int \frac{d \psi_{+, 2 n + 1}^{*} d \psi_{+, 2 n + 1}}{\pi^{(1 +
      \zeta)/2}} \\ \times \frac{d \psi_{-, 2 n + 1}^{*} d \psi_{-, 2 n +
      1}}{\pi^{(1 + \zeta)/2}} \frac{d \psi_{+, 2 n}^{*} d \psi_{+, 2
      n}}{\pi^{(1 + \zeta)/2}} \frac{d \psi_{-, 2 n}^{*} d \psi_{-, 2
      n}}{\pi^{(1 + \zeta)/2}} \\ \times \e^{\left( \psi_{+, 2 n + 2}^{*} -
      \psi_{+, 2 n + 1}^{*} \right) \psi_{+, 2 n + 1} + \left( \psi_{+, 2 n +
        1}^{*} - \psi_{+, 2 n}^{*} \right) \psi_{+, 2 n} + \psi_{-, 2 n}^{*}
    \left( \psi_{-, 2 n + 1} - \psi_{-, 2 n} \right)} \\ \times \e^{\psi_{-, 2 n
      + 1}^{*} \left( \psi_{-, 2 n + 2} - \psi_{-, 2 n + 1} \right) +
    \mathcal{L}_{+-, n} \Delta t} \braket{\psi_{+, 2 n} | \hat{\rho}_n | \zeta
    \psi_{-, 2 n}}.
\end{multline}
Iterating this relation yields a discrete-time field integral representation of
the Keldysh partition function $Z(t) = \tr(\hat{\rho}(t))$,
\begin{equation}
  Z(t) \approx \int \prod_{n = 0}^{2 N} \prod_{\sigma = \pm} \frac{\diff
    \psi_{\sigma, n}^{*} \diff \psi_{\sigma, n}}{\pi^{(1 + \zeta)/2}} \,
  \e^{\imag S} \braket{\psi_{+, 0}| \hat{\rho}(t_0) | \zeta \psi_{-, 0}},
\end{equation}
where the Lindblad-Keldysh action is given by
\begin{multline}
  \label{eq:S-Lindblad-Keldysh-discrete-time}
  S = \sum_{n = 0}^{2 N - 1} \frac{\Delta t}{2} \left( - \imag \frac{\psi_{+, n
        + 1}^{*} - \psi_{+, n}^{*}}{\Delta t/2} \psi_{+, n} - \imag \psi_{-,
      n}^{*} \frac{\psi_{-, n + 1} - \psi_{-, n}}{\Delta t/2} \right) \\ - i
  \left( \psi_{-, 2 N}^{*} \psi_{+, 2 N} - \psi_{+, 2 N}^{*} \psi_{+, 2 N} -
    \psi_{-, 2 N}^{*} \psi_{-, 2 N} \right) - \imag \sum_{n = 0}^{N - 1} \Delta
  t \mathcal{L}_{+-, n}.
\end{multline}
The first term in the second line follows from taking the trace to obtain the
Keldysh partition function $Z = \tr(\hat{\rho}(t)) = \tr(\hat{\rho}_N)$: Setting
$n = N$ in Eq.~\eqref{eq:density-matrix-coherent-state-representation}, and
denoting by $\{ \ket{\alpha} \}$ a basis of Fock space, we find that
$\tr(\hat{\rho}_N)$ contains the factor
\begin{multline}  
  \tr \! \left( \ket{\psi_{+, 2 N}} \bra{\zeta \psi_{-, 2 N}} \right) =
  \sum_{\alpha} \braket{\alpha | \psi_{+, 2 N}} \braket{\zeta \psi_{-, 2 N} |
    \alpha} \\ = \sum_{\alpha}\braket{\psi_{-, 2 N} | \alpha} \braket{\alpha
    | \psi_{+, 2 N}} = \braket{\psi_{-, 2 N} | \psi_{+, 2 N}} = \e^{\psi_{-, 2
      N}^{*} \psi_{+, 2 N}},
\end{multline}
where in the second equality we have exchanged the order of
$\braket{\alpha | \psi_{+, 2 N}}$ and $\braket{\zeta \psi_{-, 2 N} | \alpha}$,
leading to a sign change of $\psi_{-, 2 N}$ for the case of Grassmann fields,
and in the third equality we have used
$\hat{1} = \sum_{\alpha} \ket{\alpha} \bra{\alpha}$. To take the continuum limit
$N \to \infty$ and thus $\Delta t \to 0$, we write
$\psi_n = \psi(t_n) \to \psi(t)$. In each of the matrix elements $H_{\pm, n}$,
$L_{\pm, n}$, and $L^{\dagger}_{\pm, n}$ defined in
Eq.~\eqref{eq:regularized-matrix-elements}, for $N \to \infty$ we can treat the
fields as being evaluated at the same instant in time. Crucially, however,
$L_{+, n}$ is evaluated infinitesimally earlier than $L^{\dagger}_{+, n}$, which
can be accounted for by writing
$L^{\dagger}_{+, n} L_{+, n} \to L^{\dagger}(t) L(t_-)$ where
$t_{\pm} \to t \pm 0^+$. Similarly, $L_{-, n}$ is evaluated infinitesimally
later than $L^{\dagger}_{-, n}$, thus
$L^{\dagger}_{-, n} L_{-, n} \to L^{\dagger}_-(t) L_-(t_+)$. Therefore, in the
continuum limit, the action in Eq.~\eqref{eq:S-Lindblad-Keldysh-discrete-time}
reproduces Eq.~\eqref{eq:S-Keldysh}. The method of introducing additional
resolutions of the identity can be extended to subdivide products of operators
appearing in $\hat{H}$, $\hat{L}$, and $\hat{L}^{\dagger}$. This can be required
to accommodate antinormal order of jump operators \cite{Huang_2022}.

\subsection{Temporal order and causality}

We conclude this appendix by pointing out that the regularized representation of
the Lindbladian in the Keldysh action Eq.~\eqref{eq:L-plus-minus-regularized}
has the important property of obeying the causality structure of Keldysh
actions, while this is not the case if normal instead of temporal ordering of
the product $\hat{L}^{\dagger} \hat{L}$ is employed to construct the field
integral. As discussed in Sec.~\ref{sec:keldysh-rotat}, causality requires that
$\mathcal{L}_{+-} = 0$ when $\psi_+ = \psi_-$. Clearly, this is the case for
Eq.~\eqref{eq:L-plus-minus-regularized} when we ignore the infinitesimal time
shifts. Instead of introducing these time shifts through an additional time step
as described above, we could also have normal ordered the product
$\hat{L}^{\dagger} \hat{L}$. However, as we illustrate in the following for the
example of a dephasing process, normal ordering leads to a Keldysh action that violates
causality in the continuous time limit.

Dephasing of a single particle is described by the jump operator
$\hat{L} = \sqrt{\gamma} \hat{\psi}^{\dagger} \hat{\psi}$, where we consider
here bosonic creation and annihilation operators $\hat{\psi}^{\dagger}$ and
$\hat{\psi}$, respectively. The normal ordered form of the product
$\hat{L}^{\dagger} \hat{L}$ reads
$\hat{L}^{\dagger} \hat{L} = \gamma \left( \hat{\psi}^{\dagger}
  \hat{\psi}^{\dagger} \hat{\psi} \hat{\psi} + \hat{\psi}^{\dagger} \hat{\psi}
\right)$.
Therefore, normal order of $\hat{L}^{\dagger} \hat{L}$ leads to the following
modified form $\mathcal{L}_{+-}^{\mathrm{no}}$ of $\mathcal{L}_{+-}$ given in
Eq.~\eqref{eq:L-plus-minus-regularized}:
\begin{equation}
  \mathcal{L}_{+-}^{\mathrm{no}} = \mathcal{L}_{+-} + \frac{\sqrt{\gamma}}{2}
  \left( L_+ + L_- \right),
\end{equation}
where $L_{\pm} = \psi^{*}_{\pm} \psi_{\pm}$. Clearly, the normal ordered version
does not obey the causality structure discussed in
Sec.~\ref{sec:keldysh-rotat}, $\mathcal{L}_{+-}^{\mathrm{no}} \neq 0$ for
$\psi_+ = \psi_-$. We emphasize that for single-particle processes such as
single-particle loss with $\hat{L} = \sqrt{\gamma} \hat{\psi}$ or particle gain
with $\hat{L} = \sqrt{\gamma} \hat{\psi}^{\dagger}$, the time ordering and
normal ordering of $\hat{L}^{\dagger} \hat{L}$ yield the same result (apart from
an unimportant constant) in the functional integral. However, for higher order
terms, such as two-body gain
$\hat{L} = \sqrt{\gamma} \hat{\psi}^{\dagger} \hat{\psi}^{\dagger}$ or dephasing
$\hat{L} = \sqrt{\gamma} \hat{\psi}^{\dagger} \hat{\psi}$, only time ordering
yields a Keldysh action that obeys causality in the continuous time limit.

\section{Green's functions in the Keldysh formalism}
\label{sec:greens-funct-keldysh}

In this appendix, we summarize several important properties of the retarded,
advanced, and Keldysh Green's functions defined in
Eq.~\eqref{eq:Greens-functions}. We focus on open systems with time evolution
generated by a Lindbladian $\mathcal{L}$.

\subsection{From operators to fields}

In the operator formalism of second quantization, the second equalities in each
of the lines of Eq.~\eqref{eq:Greens-functions} define the retarded, advanced,
and Keldysh Green's functions. Here, we show how to obtain the corresponding
expressions in Keldysh field theory. We consider first the retarded Green's
function,
\begin{equation}
  \label{eq:G-R}
  G^R(t, t') = - \imag \theta(t - t') \left( \langle \hat{\psi}(t)
    \hat{\psi}^{\dagger}(t') \rangle - \zeta \langle \hat{\psi}^{\dagger}(t')
    \hat{\psi}(t) \rangle \right).
\end{equation}
Assuming that time evolution is generated by a Lindbladian $\mathcal{L}$, we
evaluate two-time averages using the quantum regression
theorem \cite{Gardiner2014}. For the time order $t > t' > t_0$ imposed by the
Heaviside step function, we obtain
\begin{equation}
  \label{eq:quantum-regression-1}
  \langle \hat{\psi}(t) \hat{\psi}^{\dagger}(t') \rangle = \tr \! \left(
    \hat{\psi} \e^{\mathcal{L} \left( t - t' \right)} \left(
      \hat{\psi}^{\dagger} \e^{\mathcal{L} \left( t' - t_0 \right)}
      \hat{\rho}(t_0) \right) \right) = \langle \psi_+(t) \psi_+^{*}(t')
  \rangle,
\end{equation}
where we have used the rule formulated at the end of
Sec.~\ref{sec:keldysh-part-funct-lindblad-action} to translate field operators
that multiply the density matrix from the left to fields on the forward
branch. Alternatively, using the cyclic property of the trace, we can write
\begin{equation}
  \label{eq:two-time-expectation-value-alternative}
  \begin{split}
    \langle \hat{\psi}(t) \hat{\psi}^{\dagger}(t') \rangle & = \tr \! \left(
      \e^{\mathcal{L} \left( t - t' \right)} \left( \hat{\psi}^{\dagger}
        \e^{\mathcal{L} \left( t' - t_0 \right)} \hat{\rho}(t_0) \right)
      \hat{\psi} \right) \\ & = \zeta \langle \psi_+^{*}(t') \psi_-(t) \rangle =
    \langle \psi_-(t) \psi_+^{*}(t') \rangle.
  \end{split}
\end{equation}
Taking the sum and difference of the two alternative representations of the
two-time average, we obtain the relations
\begin{equation}
  \label{eq:G-R-part-one}
  \langle \hat{\psi}(t) \hat{\psi}^{\dagger}(t') \rangle = \frac{1}{\sqrt{2}}
  \langle \psi_c(t) \psi_+^{*}(t') \rangle, \quad 0 = \langle \psi_q(t)
  \psi_+^{*}(t') \rangle.
\end{equation}
The latter equality is a manifestation of the causality structure of correlation
functions, discussed in full generality below, according to which any multi-time
expectation value vanishes when the largest time argument is associated with a
quantum field. Note that the sign $\zeta = - 1$ for Grassmann fields on the
backward branch in Eq.~\eqref{eq:two-time-expectation-value-alternative}, which
results from the representation of the trace in fermionic Fock space in terms of
coherent states, ensures causality for fermions.

For the second two-time average in Eq.~\eqref{eq:G-R}, again with
$t > t' > t_0$, the quantum regression theorem yields
\begin{equation}
  \label{eq:quantum-regression-2}
  \begin{split}
    \langle \hat{\psi}^{\dagger}(t') \hat{\psi}(t) \rangle & = \tr \! \left(
      \hat{\psi} \e^{\mathcal{L} \left( t - t' \right)} \left[ \left(
          \e^{\mathcal{L} \left( t' - t_0 \right)} \hat{\rho}(t_0) \right)
        \hat{\psi}^{\dagger} \right] \right) \\ & = \zeta \langle \psi_+(t)
    \psi_-^{*}(t') \rangle = \langle \psi_-^{*}(t') \psi_+(t) \rangle,
  \end{split}
\end{equation}
or, equivalently,
\begin{equation}  
  \langle \hat{\psi}^{\dagger}(t') \hat{\psi}(t) \rangle = \tr \! \left( \e^{\mathcal{L} \left(
        t - t' \right)} \left[ \left( \e^{\mathcal{L} \left( t' - t_0 \right)}
        \hat{\rho}(t_0) \right) \hat{\psi}^{\dagger} \right] \hat{\psi} \right) =
  \langle \psi_-^{*}(t') \psi_-(t) \rangle,
\end{equation}
which can be combined to
\begin{equation}
  \label{eq:G-R-part-two}
  \langle \hat{\psi}^{\dagger}(t') \hat{\psi}(t) \rangle = \frac{1}{\sqrt{2}}
  \langle \psi_-^{*}(t') \psi_c(t) \rangle, \quad
  0 = \langle \psi_-^{*}(t') \psi_q(t) \rangle.
\end{equation}
Therefore, using Eqs.~\eqref{eq:G-R-part-one} and~\eqref{eq:G-R-part-two}, the
retarded Green's function Eq.~\eqref{eq:G-R} can be written as
\begin{equation}
  \begin{split}
    G^R(t, t') & = - \frac{\imag}{\sqrt{2}} \theta(t - t') \left( \langle
      \psi_c(t) \psi_+^{*}(t') \rangle - \zeta \langle \psi_-^{*}(t') \psi_c(t)
      \rangle \right) \\ & = - \imag \theta(t - t') \langle \psi_c(t)
    \psi_q^{*}(t') \rangle = - \imag \langle \psi_c(t) \psi_q^{*}(t') \rangle.
  \end{split}
\end{equation}
In the last equality, we have used that according to
Eqs.~\eqref{eq:G-R-part-one} and~\eqref{eq:G-R-part-two}, the two-time
expectation value vanishes when the quantum field is inserted at a later time
than the classical field, and, therefore, we can omit the Heaviside step
function. This proves the first line in Eq.~\eqref{eq:Greens-functions}. The
derivations for the advanced and Keldysh Green's functions proceed analogously
after introducing a factor of $1 = \theta(t - t') + \theta(t' - t)$ to impose a
definite time order on each of the two-time averages appearing in the Keldysh
Green's function.

\subsection{Equal time arguments and Hermiticity}

Expressing the Green's functions as expectation values of operators is also
useful for deriving various exact properties. First, note that by setting
$t' = t \mp 0^+$ in, respectively, the retarded and advanced Green's functions
in Eq.~\eqref{eq:Greens-functions}, the Heaviside step functions evaluate to unity,
and the time arguments of the fields can be taken to be equal, which leads to
\begin{equation}
  \label{eq:G-R-G-A-sum}
  G^R(t, t - 0^+) = - G^A(t, t + 0^+) = - \imag.
\end{equation}
Furthermore, by using explicit representations of two-time averages as in
Eqs.~\eqref{eq:quantum-regression-1} and~\eqref{eq:quantum-regression-2}, and
preservation of Hermiticity as expressed through the relation
$\left( \mathcal{L} \hat{A} \right)^{\dagger} = \mathcal{L} \!  \left(
  \hat{A}^{\dagger} \right)$
for an arbitrary operator $\hat{A}$, it is straightforward to show that
\begin{equation}
  \label{eq:Greens-functions-Hermiticity}
  G^R = \left( G^A \right)^{\dagger}, \qquad G^K = - \left( G^K \right)^{\dagger}.
\end{equation}
Hermitian conjugation of the Green's functions amounts to complex conjugation,
interchanging the time arguments, and, for multicomponent fields, interchanging
the field indices.

\subsection{Causality structure}

As anticipated below Eq.~\eqref{eq:G-R-part-one}, a key property of Green's
functions that generalizes to multi-time correlation functions is what we refer
to as their causality structure,
\begin{equation}
  \label{eq:causality-structure}
  \langle \psi_q(t) \cdots \rangle = \langle \psi_q^{*}(t) \cdots \rangle = 0,
\end{equation}
where ``$\cdots$'' stands for an arbitrary product of fields with time arguments
smaller than $t$. To derive this property, we consider two ordered sequences of
times, $t_{+, N_+} > t_{+, N_+ - 1} > \dotsb > t_{+, 1}$ and
$t_{-, N_-} > t_{-, N_- - 1} > \dotsb > t_{-, 1}$, and initial and final times
$t_0 < t_{\pm, 1}$ and $t > t_{\pm, N_{\pm}}$, respectively. For operators
$\hat{a}$, $\hat{b}_1, \dotsc, \hat{b}_{N_+}$, and
$\hat{c}_1, \dotsc, \hat{c}_{N_-}$, that are each one of $\hat{\psi}$ or
$\hat{\psi}^{\dagger}$, we define a multi-time correlation function as
\begin{multline}
  \label{eq:multi-time-correlation-function}
  G(t, t_{+, 1}, \dotsc, t_{+, N_+}, t_{-, 1}, \dotsc, t_{-, N_-}) \\
  = \tr \! \left( \hat{a}(t) \hat{b}_{N_+}(t_{+, N_+}) \dotsb \hat{b}_1(t_{+,
      1}) \hat{\rho}(t_0) \hat{c}_1(t_{-, 1}) \dotsb \hat{c}_{N_-}(t_{-, N_-})
  \right).
\end{multline}
The explicit form of this correlation function depends on the relative order of
the times $t_{\pm, n}$. For example, for $N_+ = 1$, $N_- = 2$, and
$t_0 < t_{+, 1} < t_{-, 1} < t_{-, 2} < t$, we obtain
\begin{multline}
  G(t, t_{+, 1}, t_{-, 1}, t_{-, 2}) = \tr \! \left( \hat{a} \e^{\mathcal{L}
      \left( t - t_{-, 2} \right)} \left\{ \left[ \e^{\mathcal{L} \left( t_{-,
              2} - t_{-, 1} \right)} \right. \right. \right. \\
  \left. \left. \left. \times \left( \left\{ \e^{\mathcal{L} \left( t_{-, 1} -
                t_{+, 1} \right)} \left[ \hat{b}_1 \left( \e^{\mathcal{L} \left(
                    t_{+, 1} - t_0 \right)} \hat{\rho}(t_0) \right) \right]
          \right\} \hat{c}_1 \right) \right] \hat{c}_2 \right\} \right).
\end{multline}
The pattern of inserting evolution superoperators $\e^{\mathcal{L} t}$ and
operators $\hat{b}_n$ and $\hat{c}_n$ on the left- and right-hand-side of the
density matrix generalizes to arbitrary relative time orders and numbers of
operators. A field integral representation of the multi-time correlation
function can be obtained as outlined in
Appendix~\ref{sec:derivation-Lindblad-Keldysh}, whereby the operators
$\hat{b}_n$ and $\hat{c}_n$ are replaced by fields on the forward and backward
branch, $b_{n, +}$ and $c_{n, -}$, respectively. Due to the cyclic property of
the trace, the operator $\hat{a}$ can equivalently be placed on the forward and
backward branches,
$\langle a_+(t) \cdots \rangle = \zeta \langle \cdots a_-(t) \rangle = \langle
a_-(t) \cdots \rangle$,
where in the last equality we have used that for the correlation function to be
nonzero for fermions, ``$\cdots$'' must correspond to a product of an odd number
of operators. We thus obtain $\langle a_q(t) \cdots \rangle = 0$. Furthermore,
by moving, e.g., one of the operators $\hat{c}_n(t_{-, n})$ in
Eq.~\eqref{eq:multi-time-correlation-function} from the right- to the
appropriate position on the left-hand-side, keeping the time order intact, we
obtain the same relation $\langle a_q(t) \cdots \rangle = 0$ but with
$c_{n, -}(t_{-, n})$ replaced by $c_{n, +}(t_{-, n})$. Generalizing this
argument, we see that the relation $\langle a_q(t) \cdots \rangle = 0$ holds for
any choice of branch indices of the fields representing the operators
$\hat{b}_n$ and $\hat{c}_n$, which in turn implies
Eq.~\eqref{eq:causality-structure}.

\section{Gaussian states and Gaussian actions}
\label{sec:Gaussian}

A particularly important subset of nonequilibrium systems is formed by Gaussian
evolution protocols. In the operator formulation, this corresponds to
Lindbladian evolution where at most two creation or annihilation operators act
simultaneously on the state $\hat \rho$;
see \cite{eisert2010noise,prosen2008njp,prosen2010jsmte,Barthel2022} for
treatments in the operator formalism. In the framework of Keldysh field theory,
this is equivalent to an action which is at most quadratic in bosonic or
fermionic fields. As such, technically, Gaussian dynamics can be solved formally
exactly and thus represents an important example for the equivalence of the
operator-based Lindbladian evolution and the Keldysh field integral. In the
following, we demonstrate this equivalence and provide a simple general
framework for the dynamics of Gaussian systems. Physically, their significance
lies in the fact that many interacting models can be reduced approximately to
Gaussian models, by means of suitable mean-field approaches (see, e.g.,
Sec.~\ref{sec:fermionpuremix}).

Consider a bosonic (fermionic) system with $N$ discrete single-particle degrees
of freedom, e.g., lattice sites, lattice momenta or atomic orbitals, which are
labeled with an index $\alpha=1, \dotsc,N$. The states in the many-body Hilbert
space are generated by the bosonic (fermionic) creation and annihilation
operators $\hat\psi^\dagger_\alpha$ and $\hat\psi_\alpha$, respectively, with
$[\hat\psi_\alpha, \hat\psi^\dagger_{\alpha'}]_\zeta=\delta_{\alpha,\alpha'}$.
It is convenient to work in a Nambu basis and collect annihilation and creation
operators in one Nambu vector
$\hat A=(\hat\psi_1^{}, \dotsc, \hat\psi_N^{}, \hat\psi^\dagger_1, \dotsc,
\hat\psi^\dagger_N)^{\transpose}$ with $2N$ entries.

When the particle number is not conserved, it is further convenient to work in a
basis of real bosons or Majorana fermions by applying the unitary transformation
$\hat u_{\alpha}=\frac{1}{\sqrt{2}}(\hat\psi_\alpha+\hat\psi^\dagger_\alpha),
\hat
u_{\alpha+N}=\frac{i}{\sqrt{2}}(\hat\psi_\alpha-\hat\psi_{\alpha}^\dagger)$.
It is equivalent to multiplying the Nambu vector with a unitary matrix $\Sigma$,
\begin{align}
  \hat u=\Sigma \hat A, \quad \Sigma=\frac{1}{\sqrt{2}}\left(\begin{array}{cc}\mathds{1}_{N\times N}& \mathds{1}_{N\times N}\\ i\mathds{1}_{N\times N} & -i\mathds{1}_{N\times N}\end{array}\right).
\end{align}

Consider now a general Lindblad master equation
\begin{align}\label{eq:GaussLindblad}
    \partial_t\hat \rho=-i \left[ \hat{H},\hat{\rho} \right]+\sum_n \left(2\hat L_n\hat \rho \hat
  L^\dagger_n- \left\{ \hat L^\dagger_n \hat L_n,\hat \rho \right\}\right),
\end{align}
with linear jump operators $\hat L_n$ and Hamiltonian $\hat{H}$. In the case of a Gaussian evolution, we can parameterize the Hamiltonian
\begin{align}
    \hat H&=\frac{1}{2}\sum_{\alpha, \alpha'}\hat A^\dagger_\alpha H_{\alpha\alpha'}\hat A_{\alpha'}=\frac{1}{2}\hat A^{\dagger} H \hat A=\frac{1}{2}\hat u^{\transpose}\tilde H \hat u
\end{align}
in terms of the complex $2N\times 2N$ matrices $H$ or
$\tilde H=\Sigma H\Sigma^{\dagger}$. The boson and fermion exchange statistics
are encoded in $\tilde H^{\transpose}=\zeta \tilde H$. Furthermore, the Lindblad
operators can be written as
\begin{align}
    \hat L_n=\frac{1}{\sqrt{2}}\sum_{\alpha}l^{(n)}_{\alpha}\hat A_\alpha=\frac{1}{\sqrt{2}}(l^{(n)})^{\transpose}\hat A=\frac{1}{\sqrt{2}}(l^{(n)})^{\transpose}\Sigma^{\dagger} \hat u.
\end{align}
This yields a compact definition of the terms
\begin{align}
    \sum_n \hat L^\dagger_n \hat L_n&=\frac{1}{2}\sum_{\alpha,\beta}\hat A^\dagger_\alpha \hat A_\beta \left[\sum_{n}(l_{\alpha}^{(n)})^*l_{\beta}^{(n)}\right]= \frac{1}{2}\hat A^{\dagger} M\hat A=\frac{1}{2}\hat u^{\transpose}\tilde M \hat u,\nonumber\\
    \sum_n \hat L_n\hat \rho \hat L_n^\dagger&=\frac{1}{2}\sum_{\alpha\beta}\hat A_\beta \hat \rho \hat A^\dagger_\alpha M_{\alpha\beta}=\frac{1}{2}\sum_{\alpha\beta}\hat u_\beta\hat \rho \hat u_\alpha \tilde M_{\alpha\beta}.
\end{align}
Again, $\tilde M=\Sigma M\Sigma^{\dagger}$ is defined via the unitary
transformation $\Sigma$. The matrices $M$ and $\tilde M$ are
Hermitian. Therefore, we can express them in terms of their real and imaginary
parts
\begin{equation}
     M= D+iP, \quad 2D=M+M^*, \quad 2P=-i(M-M^*)
\end{equation}
and analogously for $\tilde M$, which implies $D^{\transpose}=D$ and $P^{\transpose}=-P$. 

When the state $\hat \rho(t)$ is Gaussian, all information is contained in the
single-particle covariance matrix that is equivalent to the equal-time Keldysh
Green's function defined in Eq.~\eqref{eq:Greens-functions},
\begin{equation}
  \Gamma_{\alpha, \beta}= \tr \! \left( [\hat A_\alpha,\hat A^\dagger_\beta]_{-\zeta} \hat
    \rho(t) \right) = \imag G^K_{\alpha, \beta}(t, t'),
\end{equation}
as also noted in Eq.~\eqref{eq:Corr_Mat_main} for a scalar bosonic field in a
spatial continuum. We denote the covariance matrix for real fields by
$ \gamma_{\alpha\beta}=\text{tr}([\hat u_\alpha, \hat u_\beta]_{-\zeta} \hat
\rho(t))$. Its equation of motion is
\begin{align}
    \partial_t\gamma=i \left( \tilde R\gamma-\gamma \tilde R^\dagger \right)-\tilde K,
\end{align}
with $R=H-iD, K=2P$ for fermions and $R=\sigma(H+P), K=2\sigma^{\dagger}
D\sigma, \sigma=i\sigma^y\otimes\mathds{1}_{N\times N}$ for bosons. It has the
solution
\begin{align}
\gamma(t)=e^{i\tilde Rt}\gamma(0)e^{-i\tilde R^{\dagger} t}-\int_{-t}^0 d\tau e^{-i\tilde R\tau}\tilde Ke^{i\tilde R^{\dagger} \tau}.
\end{align}
In the long time limit, this yields the matrix equation
\begin{align}
  \label{eq:GaussStatsol}
  \gamma(t\to\infty)=-\int_{-\infty}^{\infty} \frac{d\omega}{2\pi}\frac{1}{\omega-\tilde R}\tilde K\frac{1}{\omega-\tilde R^\dagger}.
\end{align}

With this preparation, we can readily write down the Keldysh field integral for
fermions and bosons. Following the conventional procedure, we define boson and
fermion fields on the $\pm$ branches. For convenience, we will collect them in a
$4N$-component vector $A=(A_+,A_-)^{\transpose}$ with
$A_{\pm}=(\psi_{1,\pm}, \dotsc, \psi_{N,\pm}, \psi_{1,\pm}^*, \dotsc,
\psi_{N,\pm}^*)$
and $ \psi_{\alpha,\pm}^*$ denoting the conjugate field (complex conjugate for
bosons, independent Grassmann field for fermions). This leads to an action
\begin{align}
S=\int_{-\infty}^\infty dt\,  A^{* \transpose} \mathcal{G}_{\pm}A  ,  
\end{align}
with the $4N\times 4N$ matrix in the $\pm$-basis ($\Sigma_z=\sigma_z\otimes\mathds{1}_{N\times N}$)
\begin{align}
    \mathcal{G}_{\pm}=\left(\begin{array}{cc}i\Sigma_z\partial_t-H+iM& -2i M^{\transpose}\\ 0_{2N\times2N}&  -i\Sigma_z\partial_t+H+iM\end{array}\right). 
\end{align}
For bosons, one can readily perform the transformation to real fields, yielding
\begin{align}
    \mathcal{G}_{\pm}^{(b)}=\left(\begin{array}{cc}i\sigma\omega-\tilde H+i\tilde D& -i \tilde D-\tilde P\\ -i\tilde D+\tilde P&  -i\sigma\omega+\tilde H+i\tilde D\end{array}\right), 
\end{align}
where we have exploited the symmetrization rules of $\tilde M$ and the boson
fields. Performing the conventional boson Keldysh rotation, we obtain the
inverse Green's function
\begin{equation}
  \begin{split}    
    G^{-1}&=\left(\begin{array}{cc}0& i\sigma\omega-\tilde H-\tilde P \\ i\sigma\omega-\tilde H+\tilde P& -2i\tilde D\end{array}\right)\\&=i\left(\begin{array}{cc}0& \sigma(\omega-\tilde R) \\ (\omega-\tilde R^\dagger)\sigma& -\sigma \tilde K\sigma \end{array}\right), 
  \end{split}
\end{equation}
which when inverted yields the result in Eq.~\eqref{eq:GaussStatsol} for bosons. The elements of the inverted matrix are the Green's functions 
\begin{align}
    G=\left(\begin{array}{cc}G^K& G^R\\ G^A& 0\end{array}\right)
\end{align}
as given in the main text in Eq.~\eqref{eq:Greens-functions}. 

For fermions, exploiting the anticommutation rules of the Majorana fields yields a Keldysh Green's function
\begin{align}
    G^{-1}&=\left(\begin{array}{cc}0& \omega-\tilde H-i\tilde D\\     \omega-\tilde H+i\tilde D & 2\tilde P\end{array}\right),
\end{align}
which again yields the result in Eq.~\eqref{eq:GaussStatsol} but for fermions.

\section{Dark states of Lindblad dynamics}
\label{sec:dark}

Dark states are pure-state dynamical fixed points of Lindbladians. We consider a
driven open quantum system evolving according to the quantum master equation
(see also Eq.~\eqref{eq:meq})
 \begin{equation}\label{eq:lindbladx}
    \partial_{t}\hat{\rho}\,= -i \left[\hat{H},\hat{\rho} \right] +
    \sum^M_{l=1}\gamma_{l} \left(
      2\hat{L}_{l}^{\,}\hat{\rho}\hat{L}_{l}^{\dagger} -
      \left\{\hat{L}_{l}^{\dagger}\hat{L}_{l}^{\,},\hat{\rho} \right\} \right).
\end{equation}
A \textit{dark state} $\ket{D}$, with associated pure state density matrix
$\hat \rho_D = \ket{D}\bra{D}$, is defined by the two conditions:
\begin{equation}
\label{eq:darkcond}
\hat H \ket{D}= E\ket{D} \quad \text{and} \quad \hat L_l \ket{D} = 0 \quad \forall l.
\end{equation}
Dark states are thus zero modes of the Lindbladian, or in more physical terms,
fixed points of Lindblad dynamics, $\mathcal{L} \hat\rho_D=0$, with the
additional property of being pure. They are called `dark' since they are not
affected by dissipation: for instance, in a quantum optics setup, a system in a
dark state is decoupled from the radiation field and will not emit any photons.

For the case $\hat{H} = 0$, the following two requirements have to be fulfilled
to ensure uniqueness of the dark state as an attractor of dynamics: (i)~The dark
subspace is one-dimensional, i.e., there is exactly one normalized dark state
which fulfills Eq.~\eqref{eq:darkcond}. (ii)~No stationary solutions other than
the dark state exist. A sufficient criterion for this situation to occur is
provided in Ref.~\cite{Kraus2008}.
  
Uniqueness is a key property, since, under this circumstance, the system will be
attracted toward the dark state for an arbitrary initial density matrix. That
is, starting from any $\hat{\rho}(t_0)$, the dynamics will purify the density
matrix:
\begin{equation}
  \hat \rho(t_0)= \sum_n p_n \ket{\psi_n}\bra{\psi_n} \overset{t\to \infty}{\longrightarrow} \ket{D}\bra{D} .
\end{equation}
In the case of a dark subspace (e.g., in the context of topological systems,
see Sec.~\ref{subsec:univtopo}) that is spanned by several linearly independent
states, the dynamics will still be attracted to this subspace as long as
condition (ii)~above is satisfied.

Comprehensive overviews of the concept of dark states in many-body systems and
their applications to modern quantum simulators can be found in
Ref.~\cite{muller2012aamop,Harrington_2022}.

\bibliography{bibliography}

\end{document}